\documentclass[manuscript,screen]{acmart}

\usepackage{graphicx}
\usepackage{dirtytalk}
\usepackage{colortbl}
\usepackage{wasysym}
\usepackage{twemojis}
\usepackage{utfsym}
\usepackage{fontawesome}
\usepackage{hyperref}
\usepackage{multirow}
\usepackage{tabularx}
\usepackage{adjustbox}
\usepackage{balance}
\usepackage{longtable}

\newcommand{\nocolonsubsection}[1]{\vspace{0.25cm}\noindent\hspace*{0.5em}\textit{#1}\vspace{0.1cm}}
\newcommand{\framedtext}[1]{%
\par%
\vspace{3pt}
\noindent\fbox{%
    \parbox{\dimexpr\linewidth-2\fboxsep-2\fboxrule}{#1}%
}%
}
\newcommand{\edits}[1]{\textcolor{black}{#1}}
\newcommand{\editss}[1]{\textcolor{black}{#1}}
\newcommand{\fix}[1]{\textcolor{black}{#1}}

\AtBeginDocument{%
  }

\setcopyright{acmlicensed}
\copyrightyear{2025}
\acmYear{2025}
\acmDOI{XXXXXXX.XXXXXXX}

\begin{document}

\title{It Is Giving Major Satisfaction: Why Fairness Matters for Software Practitioners}

\author{Emeralda Sesari}
\email{e.g.sesari@rug.nl}
\orcid{0000-0001-6586-3614}
\affiliation{%
  \institution{University of Groningen}
  \country{Netherlands}
}

\author{Federica Sarro}
\email{f.sarro@ucl.ac.uk}
\orcid{0000-0002-9146-442X}
\affiliation{%
  \institution{University College London}
  \country{United Kingdom}}

\author{Ayushi Rastogi}
\email{a.rastogi@rug.nl}
\orcid{0000-0002-0939-6887}
\affiliation{%
  \institution{University of Groningen}
  \country{Netherlands}
}

\begin{abstract}
Software practitioners often encounter workplace unfairness, such as unequal recognition and gender bias. While the link between fairness and job satisfaction has been established in other fields, its relevance to software professionals remains underexplored. This study examines how fairness perceptions relate to job satisfaction among software practitioners, focusing on both general trends and demographic-specific differences. We conducted an online survey of 108 software practitioners, followed by ordinal logistic regression to analyze the relationship between fairness perceptions and job satisfaction in software engineering contexts, with moderation analysis examining how this relationship varies across demographic groups. Our findings indicate that all four fairness dimensions (namely distributive, procedural, interpersonal, and informational fairness) significantly affect overall job satisfaction and satisfaction with job security. Among these, interpersonal fairness has the biggest impact. The relationship between fairness and job satisfaction is stronger for female, ethnically underrepresented, less experienced practitioners, and those with work limitations. Fairness in authorship emerged as an important factor for job satisfaction collectively, while fairness in policy implementation, high-demand situations, and working hours impacted specific demographic groups. This study highlights the role of fairness among software practitioners, offering strategies for organizations to promote fair practices and targeted approaches for certain demographic groups.
\end{abstract}

\ccsdesc[500]{Software and its engineering~Collaboration in software development}
\ccsdesc[500]{Social and professional topics~Socio-technical systems}

\keywords{Fairness, job satisfaction, human and social aspects, software engineering}


\maketitle

\section{Introduction}
In the field of Software Engineering (SE), practitioners frequently encounter fairness problems in various aspects of their work, such as problems in recognition of contributions \cite{casari2021}, 
gender bias \cite{Canedo2019}, and ethnicity bias \cite{Nadri2021}.
A study by Sesari et al.\cite{sesari2024understanding} explored these fairness problems by identifying different contexts in which they arise, based on the well-established Colquitt's  organizational fairness framework \cite{Colquitt2001}. By leveraging this framework, the authors were able to pinpoint specific areas within SE where fairness issues are most prevalent. The top five areas identified were income, treatment, allocation of work, recruitment, and working hours \cite{sesari2024understanding}.

Leveraging the same framework, German et al. \cite{German2018} developed a framework explaining fairness in the modern code review process to identify strengths and potential areas for improvement.  
For example, distributive fairness, which includes principles like equity \cite{Colquitt2001}, suggests that decisions about how work is evaluated and rewarded should consider both the person who made the contribution and the quality of that contribution. German et al. \cite{German2018} further studied fairness problems among developers involved in the OpenStack project by surveying them about the code review process. The survey results showed that many developers were concerned about fairness, especially regarding issues like equity, equal treatment, consistency, and preventing bias \cite{German2018}.
However, while these problems had been identified, the impact of facing such fairness issues on software developers remains unclear.

To better understand the broader implications of fairness in the workplace, Colquitt conducted a meta-analytic review of organizational justice literature, which included 120 separate meta-analyses of 183 empirical studies \cite{colquitt2001justice}. This comprehensive review examined the effects of organizational fairness—encompassing distributive, procedural, interpersonal, and informational fairness—on several outcomes, including job satisfaction. Colquitt’s findings demonstrate a significant link between perceptions of fairness and job satisfaction across diverse contexts.
For example, in the healthcare sector, fairness in managerial practices has been shown to significantly influence nurse retention and job satisfaction \cite{judge2004organizational}. In the manufacturing industry, procedural justice has been found to enhance employee commitment and reduce turnover intentions \cite{konovsky1991perceived}. Similarly, in educational settings, distributive fairness related to workload and rewards is a key predictor of teacher satisfaction \cite{Moorman1991}.
Given these findings across various fields, an important question arises: does the observed link between fairness and job satisfaction apply in the context of SE?

In the SE literature, enhanced job satisfaction among developers is known to improve employee attraction, retention, and well-being \cite{Fagerholm2014}. Furthermore, many researchers and companies assume that more satisfied developers tend to be more productive \cite{Storey2021}. Storey et al. \cite{Storey2021} developed a theory that links higher job satisfaction among software developers to increased perceived productivity. In their study, they also identified various social and technical factors that significantly impact job satisfaction, such as appreciation and rewards, work culture, and work life balance.


\edits{Building on this work, our study aims to explore the perception of fairness as a critical factor influencing developer job satisfaction.
 Based on prior organizational behavior research, we theorize that fairness perceptions influence job satisfaction. Colquitt’s meta-analytic review \cite{colquitt2001justice} found that perceptions of distributive, procedural, interpersonal, and informational fairness each significantly predict job satisfaction across industries. This relationship has been supported in diverse settings \cite{judge2004organizational,konovsky1991perceived,Moorman1991}. Drawing on these findings, and consistent with emerging SE research highlighting the importance of fairness issues \cite{German2018, sesari2024understanding}, we posit that developers who perceive higher fairness across these dimensions will report higher job satisfaction. Thus, fairness perceptions are conceptualized as key antecedents of job satisfaction in the SE context. }

\fix{While previous studies in SE have separately explored fairness-related challenges \cite{sesari2024understanding, German2018} and job satisfaction factors \cite{Fagerholm2014, Storey2021}, no study has directly linked developers' fairness perceptions to their job satisfaction.
 We address this critical gap by empirically investigating how fairness perceptions shape developers' satisfaction with their work.
 Through a survey among software practitioners, we measure fairness perceptions based on Colquitt’s organizational justice dimensions and analyze their relationship with job satisfaction outcomes. Our approach is consistent with recent SE research that emphasizes grounding empirical investigations in clear theoretical constructs and modeling perception-to-outcome pathways (e.g., \cite{verwijs2024agile, lambiase2024investigating}). Our findings aim to inform future empirical studies or organizational assessments focused on fairness-related workplace outcomes in SE and provide practical implications for fostering fair, satisfying work environments that support talent retention and developer well-being.}

 \editss{In this study, our primary objective is to examine how software practitioners' perceptions of fairness influence their job satisfaction. The main contributions of this paper are as follows: (1) we empirically demonstrate that fairness perceptions are significantly associated with job satisfaction among software practitioners; (2) we show that this relationship differs across demographic groups; and (3) we identify which fairness dimensions matter the most in specific SE contexts and for which demographic groups.}

\fix{The rest of the paper is structured as follows: Section~\ref{sec:background} provides background on fairness and job satisfaction in SE. Section~\ref{sec:framework} presents the conceptual foundations by defining the key constructs and explaining the hypothesized relationships behind our research questions. Section~\ref{Methods} outlines the methodology, including data collection, data processing and analysis strategies. Section~\ref{sec:findings} presents the empirical findings. Section~\ref{sec:discussion} discusses the implications of these findings. Section~\ref{sec:validity} outlines threats to validity, and Section~\ref{sec:conclusion} concludes the paper with key takeaways and future research directions.}

\section{Background}
\label{sec:background}

\subsection{Developer Job Satisfaction}
In our study, we draw from Storey et al.’s understanding of job satisfaction \cite{Storey2021}, which is rooted in Wright and Cropanzano’s definition \cite{wright2000psychological} who described job satisfaction as an internal state where an individual evaluates their job experiences positively or negatively.

Beyond job satisfaction, developer motivation is another important factor frequently discussed in the literature. Beecham et al. reviewed SE studies and identified several factors influencing motivation, along with signs of both motivation and demotivation \cite{Beecham2008}. Following this, Sharp et al. proposed a framework that includes motivators, outcomes, and the context of software engineers' work \cite{sharp2009models}. More recently, França et al. pointed out that factors such as career growth and autonomy are key to motivation, while also clarifying that motivation and job satisfaction are related \cite{França2012, Franca2013,França2014, francca2018motivation}.

Drawing inspiration from this research, Storey et al. incorporated relevant motivational factors into their study of developer satisfaction \cite{Storey2021}. Storey et al. identified numerous social and technical factors, such as appreciation, rewards, work culture, and work-life balance, that influence job satisfaction in software development \cite{Storey2021}. They also developed a theory linking higher job satisfaction to increased perceived productivity among software developers. 

In designing our survey, we took inspiration from their work, particularly the 44 factors they identified as influencing job satisfaction among software practitioners \cite{Storey2021}. While their research focused broadly on these factors, our study seeks to explore fairness as a critical aspect, aiming to understand how fairness impacts job satisfaction in SE.

\subsection{Fairness in Software Engineering}
In this research, we define fairness in alignment with the broader justice literature. Fairness, as explained by Colquitt et al. \cite{Colquitt2015} and Cuguero et al. \cite{Cuguero2013}, is a subjective assessment of whether organizational activities are appropriate or morally sound. Colquitt’s fairness theory divided fairness into several dimensions relevant to workplaces \cite{Colquitt2001}. \textbf{Distributive fairness} addresses the fairness of outcomes, like rewards and recognition, and how well these align with principles such as equity or equality \cite{ADAMS1965267, Deutsch1975, LEVENTHAL197691}. \textbf{Procedural fairness} concerns the fairness of the decision making processes, focusing on consistency, bias suppression, and opportunities for participation in decision-making \cite{Leventhal1980,thibaut_walker_1975}. Later work introduced \textbf{interpersonal fairness}, which refers to respectful and dignified treatment during interactions with leads or managers, and \textbf{informational fairness}, which focuses on transparency and the clear communication of decisions with leads or managers \cite{Greenberg1993}.

In the context of SE, fairness is increasingly seen as important in software development. German et al. \cite{German2018} used Colquitt’s fairness theory \cite{Colquitt2001} to develop a framework for making code reviews fairer. They highlighted the need for distributive fairness, which means reviews should consider the value of patches and decide whether to prioritize based on equity (i.e., importance of the contribution) or equality (i.e., treating all patches the same) \cite{German2018}. This includes setting basic standards for patch reviews and making sure newcomers are treated fairly.

For procedural fairness, the authors stressed the importance of having clear and consistent guidelines so that reviews are done without bias \cite{German2018}. 
Procedural fairness also involves giving the author a chance to respond to feedback and creating an appeals process for rejected patches that involves new, unbiased reviewers.
Interpersonal fairness is supported by codes of conduct, which aim to ensure respectful communication during reviews \cite{German2018}. These codes should also protect authors’ privacy by preventing personal information from being used improperly.
Informational fairness is maintained through the openness of code reviews in open-source projects \cite{German2018}. Reviewers should give clear reasons when they reject patches and inform authors if a review is delayed.


Sesari et al. further explored fairness concerns in SE through a review of developer posts on Stack Exchange sites, identifying key contexts where fairness issues arise \cite{sesari2024understanding}. These contexts include income,
treatment, allocation of work, recruitment, working hours, demand, evaluation, authorship, policy, and customer/client relations. Their findings showed that most concerns focused on procedural fairness, such as work allocation process that is applied consistently, interview procedure and performance review that is based on accurate information, and software testing that is free from confirmation bias \cite{sesari2024understanding}.

Interestingly, while Sesari et al. found that many of the fairness issues are not exclusively tied to protected attributes such as gender, gender-related concerns remain a frequently discussed topic in fairness-related posts \cite{sesari2024understanding}. These studies suggest that fairness issues in SE extend beyond technical processes, affect the broader social dynamics within development teams both as a collective and from different backgrounds.

\subsection{How Fairness and Job Satisfaction are Related}
The relationship between fairness and job satisfaction has been extensively studied in organizational justice literature. Research has consistently shown strong links between procedural fairness and overall job satisfaction \cite{Mossholder1998,Wesolowski1997}. Additionally, Masterson et al. found that procedural fairness was a stronger predictor of job satisfaction than interpersonal and informational fairness, though all were significant factors \cite{Masterson2000}.
To explore the broader impact of fairness in organizational settings, Colquitt’s meta-analysis of organizational justice research examined the outcomes of 183 empirical studies across multiple fields, confirming the strong influence of perceived fairness on factors like organizational commitment, citizenship behaviors, and job satisfaction \cite{colquitt2001justice}.

Fairness has been shown to have a major impact on job satisfaction across various industries. For instance, fairness in management practices significantly influences nurse retention and job satisfaction in healthcare \cite{judge2004organizational}.  In the manufacturing sector, procedural fairness enhances employee commitment and decreases turnover intentions \cite{konovsky1991perceived}. Similarly, in education, fairness in workload distribution and rewards strongly predicts teacher satisfaction \cite{Moorman1991}.

In the realm of SE, job satisfaction plays an essential role in improving employee retention, well-being, and productivity \cite{Fagerholm2014}.  Many researchers and industry leaders assume that higher job satisfaction is closely tied to improved productivity among developers \cite{Storey2021}. Storey et al. developed a theory linking increased job satisfaction to greater perceived productivity, identifying key factors like rewards, work culture, and work-life balance as crucial elements influencing satisfaction \cite{Storey2021}. 
Developers in agile teams report higher job satisfaction than those in non-agile environments due to increased autonomy, decision-making power, and involvement in projects \cite{melnik2006comparative}. These factors relate to procedural fairness, as agile environments provide transparency and involvement in decision-making, making developers feel their contributions are fairly recognized.

Personality traits like agreeableness and conscientiousness also influence job satisfaction \cite{Acuña2009}. Teams that allow developers to organize their work and minimize conflicts promote interpersonal fairness through respect and collaboration \cite{Acuña2009}. 
Psychological safety and clear team norms are also found to be key predictors of job satisfaction, which allow developers to voice concerns without fear of judgment \cite{Lenberg2018Psychological}. This fosters interpersonal fairness by creating a respectful environment, and informational fairness through well-communicated expectations, both of which contribute to job satisfaction. 
Building on this, our research explores whether perceptions of fairness are a key factor in developer job satisfaction, alongside other well-established factors.

\section{Conceptual Framework}\label{sec:framework}

\edits{In this section, we present the conceptual foundations of our study by defining the constructs and explaining the hypothesized relationships between them. We ground our framework in established organizational fairness and job satisfaction theories, while extending their application to SE.}

\subsection{Theoretical Constructs}

\edits{Our study centers on three major constructs: \textit{fairness perceptions}, \textit{job satisfaction}, and \textit{fairness experiences in SE contexts}.}




\editss{\textbf{Fairness Perceptions.} We draw on Colquitt’s framework of organizational justice~\cite{Colquitt2001}, which defines fairness along four dimensions: distributive fairness (perceived fairness of outcomes, such as rewards or recognition), procedural fairness (fairness of the processes used to determine those outcomes), interpersonal fairness (respectful and dignified treatment by others), and informational fairness (honesty and clarity in communications). These dimensions have been shown to significantly predict job satisfaction across various organizational settings~\cite{judge2004organizational, konovsky1991perceived, Moorman1991}. In our study, we conceptualize fairness perception as developers’ subjective evaluation of how fairly they are treated across these four dimensions. These perceptions are treated as predictors of job satisfaction.}

\editss{\textbf{Job Satisfaction.} Job satisfaction reflects the extent to which individuals feel fulfilled, valued, and content with their work. In SE, this is influenced by a combination of social and technical factors. Storey et al.~\cite{Storey2021} identified 44 job satisfaction factors specific to software practitioners, including appreciation, autonomy, work-life balance, and rewards. In this study, we treat job satisfaction as the main outcome variable. By doing so, we aim to examine how perceived fairness contributes to job satisfaction.}

\editss{\textbf{Fairness Experiences in SE Contexts.} While fairness perceptions are often assessed at a general level, we also examine how they are experienced in specific SE scenarios. Building on Sesari et al.~\cite{sesari2024understanding}, who identified key fairness contexts in SE, such as work allocation, recruitment, income, and evaluations, we include items that capture developers' fairness experiences across these contexts. This allows us to explore not just whether fairness matters, but also when it matters the most.}

\subsection{Framing the Research Questions}

\editss{\begin{description}
  \item[RQ1:] \textbf{Does fairness perception shape job satisfaction among software practitioners? If so, how?}
\end{description}}

\editss{This question is grounded in prior organizational behavior research, which suggests that fairness perceptions influence job satisfaction. Colquitt’s meta-analytic review~\cite{colquitt2001justice} found that perceptions of distributive, procedural, interpersonal, and informational fairness each significantly predict job satisfaction across industries~\cite{judge2004organizational, konovsky1991perceived, Moorman1991}. Drawing on these findings and emerging SE research on fairness~\cite{German2018, sesari2024understanding}, we posit that developers who perceive higher fairness across these dimensions will report higher job satisfaction.}

\editss{\begin{description}
  \item[RQ2:] \textbf{Does the relationship between fairness perceptions and job satisfaction change based on demographic groups? If so, how?}
\end{description}}

\editss{While RQ1 explores the general relationship, existing literature highlights demographic disparities in SE, a field predominantly young, White, and male. Studies have shown that minority groups often face unique challenges. For example, Trinkenreich et al.~\cite{Trinkenreich2022AnEI} identified that women in the software industry encounter issues such as work-life balance, the glass ceiling, and sexism. Canedo et al.~\cite{Canedo2019} found that women are underrepresented in software development and open source projects, often due to male-dominated workplace conditions. Nadri et al.~\cite{Nadri2021} reported that less than 10\% of contributions to open source projects come from visibly non-White developers. Similarly, Baltes et al.~\cite{Baltes2020} discussed age-related biases, noting that older developers are often perceived as less adaptable, which can hinder career progression. Given these disparities, we theorize that the majority’s perspective may not fully capture the experiences of minority groups. By examining perceptions across diverse demographics, we can ensure that our findings reflect a broader range of experiences.}

\editss{\begin{description}
  \item[RQ3:] \textbf{How do fairness perceptions influence job satisfaction in specific SE contexts?}
\end{description}}

\editss{While RQ1 and RQ2 reveal important patterns, they do not explain how fairness perceptions operate in specific SE contexts. Building on the work of Sesari et al.~\cite{sesari2024understanding}, who identified key scenarios in which fairness problems persist in SE, we further investigate how such issues manifest and impact job satisfaction in situational settings.}

\editss{\begin{description}
  \item[RQ3.1:] \textbf{How does fairness perception influencing job satisfaction differ across demographic groups in SE contexts?}
\end{description}}

\editss{Building on our findings from RQ2, where we observe whether the relationship between fairness perceptions and job satisfaction varies across demographic groups, it becomes essential to explore this dynamic in specific SE scenarios. Given the insights from RQ3, we also examine how the manifestations of fairness issues differ across demographic groups within SE contexts. For instance, if we find that being treated politely, with respect, and dignity during evaluations significantly affects job satisfaction, we should investigate which demographic groups this applies to the most. Is it more impactful for female developers, suggesting that gender plays a key role in this relationship?}

\section{Methodology}\label{Methods}
The present study seeks to investigate how perceptions of fairness relate to job satisfaction among software practitioners (RQ1). To achieve this, we focus on two key constructs: perceptions of fairness and job satisfaction. The measurement of these constructs has been comprehensively and exhaustively developed and tested using survey methods, making a survey the appropriate tool for our study.
Additionally, to explore the specific real-world scenarios in which these fairness aspects are most likely to affect job satisfaction (RQ3), we ask participants how they have experienced fairness-related contexts identified in the SE field.

In this section, we outline the design of our survey, detailing the methods used to measure these three constructs. Respondents were asked to answer questions on a 5-point Likert scale, ranging from negative to positive, such as from \say{very dissatisfied} to \say{very satisfied} and \say{to no extent} to \say{to a very large extent.} We also explain our strategy for collecting demographic information and recruiting respondents.

\begin{figure}[]
\centering
\includegraphics[width=\textwidth]{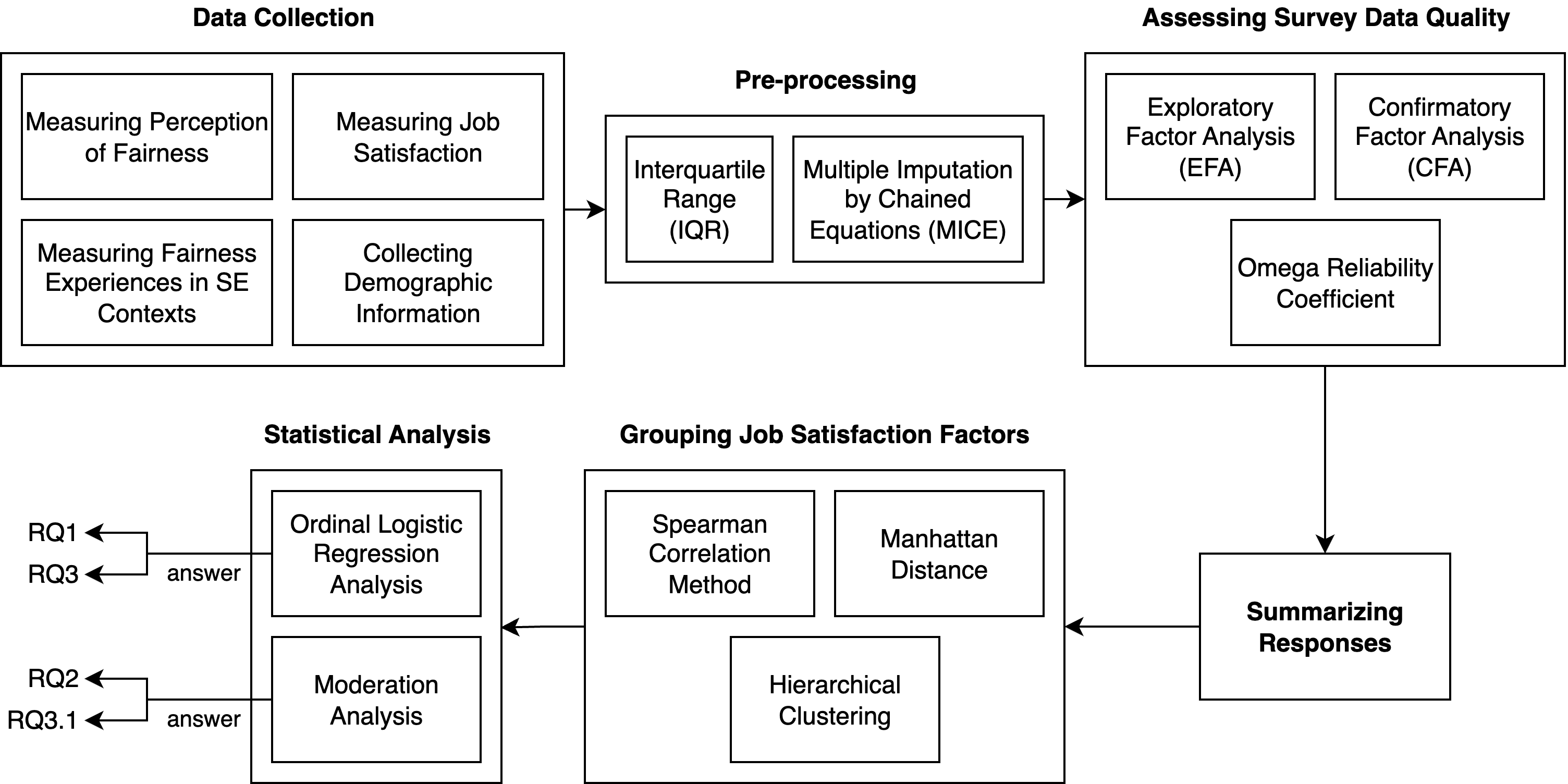}
\captionsetup{font=footnotesize}
\caption{\editss{Overview of Methodology}}
\label{fig:methodology}
\end{figure}

Finally, we detail the steps employed to answer the research questions posed in this study. The scripts for all steps are provided in the supplementary material \cite{figshare} \editss{and Fig. \ref{fig:methodology} illustrates the outline of our methodology}.

\subsection{Data Collection}
\subsubsection{Measuring Perception of Fairness}
\label{sec:measuringfairness}

In order to explore the relationship between fairness and job satisfaction among software practitioners, it is essential to first measure fairness effectively. Fairness, in this context, is understood as a personal judgment or evaluation of organizational activities, particularly regarding whether these activities are perceived as moral and praiseworthy \cite{Goldman2015}. To accurately assess this experience of fairness, we rely on the framework developed by Colquitt in 2001 \cite{Colquitt2001}.

Our survey design draws inspiration from the survey questions developed by Colquitt et al. \cite{Colquitt2001}, which consists of 20 questions. These questions assess various four dimensions of fairness, namely \textit{distributive}, \textit{procedural}, \textit{interpersonal}, and \textit{informational} fairness. Colquitt's measure of fairness was developed through careful research, creating survey items based on key studies in the field of organizational justice, which examines fairness in workplaces. These items were tested in different settings, such as in a university classroom and a manufacturing company, to ensure they worked well across various environments \cite{Colquitt2001}.

By aligning with Colquitt's well-established framework, we ensure that our approach is grounded in a reliable foundation.
The 20 items used in our survey, as developed by Colquitt \cite{Colquitt2001}, are listed in Table \ref{tab:dimensions}.

\begin{table*}[]
\centering
\small
\caption{Questions to Measure Perception of Fairness \cite{Colquitt2001}}
\label{tab:dimensions}
\begin{tabular}{l}
\hline
\multicolumn{1}{c}{Measure Item}                                                                                                           \\ \hline
\multicolumn{1}{c}{\textbf{Distributive Fairness}} \\
\textit{Think about the outcomes you receive from your job (e.g., pay, promotions), to what extent:} \\
 Does your outcome reflect the effort you have put into your work?                                                                        \\
                                 Is your outcome appropriate for the work you have completed?                                                                              \\
                                 Does your outcome reflect what you have contributed to the organization?                                                                  \\
                                 Is your outcome justified, given your performance?                                                                                        \\
                                 \rowcolor[HTML]{EFEFEF}\multicolumn{1}{c}{\textbf{Procedural Fairness}} \\
                                 \rowcolor[HTML]{EFEFEF}\textit{Referring to how decisions are made in your work (e.g., code review, task assignment, performance review), to what extent:}\\
 \rowcolor[HTML]{EFEFEF}Have you been able to express your views and feelings during those procedures?                                                                \\
                                 \rowcolor[HTML]{EFEFEF}Have you had influence over the outcome arrived at by those procedures?                                                                        \\
                                 \rowcolor[HTML]{EFEFEF}Have those procedures been applied consistently?                                                                                                \\
                                 \rowcolor[HTML]{EFEFEF}Have those procedures been free of bias?                                                                                                        \\
                                 \rowcolor[HTML]{EFEFEF}Have those procedures been based on accurate information?                                                                                       \\
                                 \rowcolor[HTML]{EFEFEF}Have you been able to appeal the outcome arrived by those procedures?                                                     \\
                                 \rowcolor[HTML]{EFEFEF}Have those procedures upheld ethical and moral standards?                                                                                       \\
                                 \multicolumn{1}{c}{\textbf{Interpersonal Fairness}}\\
                                 \textit{Regarding interactions with managers or lead in your team, to what extent:}\\
 Has (the authority figure who enacted the procedure) treated you in a polite manner?                                                        \\
                                 Has (the authority figure who enacted the procedure) treated you with dignity? \\
                                 Has (the authority figure who enacted the procedure) treated you with respect?                                                              \\
                                 Has (the authority figure who enacted the procedure) refrained from improper remarks or comments?                                           \\
                                 \rowcolor[HTML]{EFEFEF}\multicolumn{1}{c}{\textbf{Informational Fairness}}\\
                                 \rowcolor[HTML]{EFEFEF}\textit{Regarding communication with managers or leads in your team, to what extent:}\\
 \rowcolor[HTML]{EFEFEF}Has (the authority figure who enacted the procedure) been candid in their communications with you?                     \\
                                 \rowcolor[HTML]{EFEFEF}Has (the authority figure who enacted the procedure) explained the procedures thoroughly?                                                   \\
                                 \rowcolor[HTML]{EFEFEF}Were (the authority figure who enacted the procedure)’s explanations regarding the procedures reasonable?                                   \\
                                 \rowcolor[HTML]{EFEFEF}Has (the authority figure who enacted the procedure) communicated details in a timely manner?                                               \\
                                \rowcolor[HTML]{EFEFEF}Has (the authority figure who enacted the procedure) seemed to tailor their communications to individual’s specific needs?                  \\ \hline
\end{tabular}
\end{table*}


\subsubsection{Measuring Job Satisfaction}
\label{sec:satisfaction}

Given the connection between perceptions of fairness and job satisfaction across various fields, we inquire if the observed link between fairness and job satisfaction apply in the context of SE.
To develop a more tailored approach to measuring job satisfaction in the SE field, we turn to the SE literature. Storey et al. \cite{Storey2021} conducted a comprehensive survey specifically within the field of SE, aiming to identify the factors that most significantly impact job satisfaction among software engineers. Developed through two main iterations and rigorously piloted, the survey incorporated insights from a literature review, onsite observations, and internal survey comments. The final version, which gathered a total of 640 responses, identified 44 factors influencing job satisfaction in SE.

In designing our survey, we draw inspiration from the 44 job satisfaction factors and 1 overall job satisfaction question used by Storey et al. \cite{Storey2021}. However, to ensure that the factors we examine are related to perceptions of fairness, we align our selection with organizational justice literature. 

For example, one of the 44 factors is satisfaction with one's manager. We intuitively believe that a manager's behavior and leadership style play a significant role in how fairness is perceived by employees. For instance, during performance reviews, the way a manager communicates feedback, acknowledges contributions, and makes decisions about promotions or raises can significantly impact whether employees feel they are being treated fairly. Organizational justice literature supports this intuition, highlighting that a manager's leadership style is a crucial factor in shaping employees' perceptions of fairness within an organization \cite{Kiersch2015Is}. Therefore, we believe that satisfaction with one's manager is closely related to perceptions of fairness.

On the other hand, factor such as autonomy, which refers to an individual’s need for independence \cite{Beecham2008}, may not directly relate to perceptions of fairness. Autonomy is more concerned with the freedom to carry out tasks and the ability to make decisions independently \cite{Beecham2008}. 
However, autonomy is primarily about personal control and is less about how fairly an organization’s policies, processes, and decisions are perceived by its employees. As such, autonomy does not align with the fairness constructs that are the focus of our study.

\edits{Consequently, we refined our scope from the original 44 factors to 15 that are more closely aligned with fairness, supported by empirical evidence. 
The complete list of 44 factors is available in the original study by Storey et al. \cite{Storey2021}, and we provide an overview of excluded factors in the supplementary material \cite{figshare}.
The selected factors, presented in Table \ref{tab:satis}, allow us to analyze job satisfaction in a way that is directly linked to perceptions of fairness.}

\begin{table*}[]
\caption{Key Satisfaction Items and Their Rationale in the Context of Organizational Fairness}
\footnotesize
\begin{tabular}{lp{0.825\textwidth}}
\hline
\multicolumn{1}{c}{Satisfaction Items} & \multicolumn{1}{c}{Rationale}                                                                                                                                                                                                           \\ \hline
Manager                                & The authenticity of a manager's leadership style plays a crucial factor in shaping employees' perceptions of fairness within an organization \cite{Kiersch2015Is}.                                                         \\
\rowcolor[HTML]{EFEFEF}Feedback                               & When employees perceive high levels of interactive fairness, the positive impact of developmental feedback on satisfaction and engagement will be amplified \cite{zhang2019influence}.                                  \\
Appreciation                           & Satisfaction with appreciation shown for one's work is closely linked to perceptions of fairness, which in turn may significantly impact an employee's job satisfaction and performance \cite{Shonubi2016Recognition}. \\
\rowcolor[HTML]{EFEFEF}Priorities                             & Employees' perceptions of fairness in the organization can influence how they prioritize and engage with their work tasks \cite{Cropanzano2001}.                                                                       \\
Organization Culture                   & Procedural and distributive fairness can be strong predictors of satisfaction with work schedules  \cite{Zeidler2021The}.                                                                                                 \\
\rowcolor[HTML]{EFEFEF}Team Culture                           & When team members perceive the distribution of equity as fair, it may lead to positive interaction spirals within the team \cite{Breugst2015}.                                                                            \\
Team Collaboration                     & Higher perceptions of procedural and interactional fairness within an organization may be associated with better satisfaction with teamwork \cite{Dayan2008}.                                                                       \\
\rowcolor[HTML]{EFEFEF}Productivity                           & Enhancing organizational fairness can increase staff productivity, recommending that administrators focus on improving components of organizational fairness \cite{Hosseini2016CORRELATION}.                             \\
Ability                                & In an environment where employees perceive high levels of fairness, they are more likely to take initiative and manage themselves effectively \cite{Park2016}.                                                                         \\
\rowcolor[HTML]{EFEFEF}Salary                                 & Larger financial rewards for high performers increase perceptions of distributive and procedural fairness, which in turn foster work motivation \cite{Sebastian2016Motivational}.                                      \\
Benefits                               & Employees' views on whether their compensation is fair are influenced not only by the amount they receive but also by how they perceive the process of determining that compensation \cite{Wu2008}.                    \\
\rowcolor[HTML]{EFEFEF}Rewards                                & Distributive fairness related to pay, benefits, and rewards is significantly linked to job satisfaction \cite{Haar2009How}.                                                                                            \\
Work-Life Balance                      & When decision-making processes are perceived as fair, employees are more likely to be satisfied with their work-life balance and less likely to consider leaving the organization \cite{Cao2013}.                      \\
\rowcolor[HTML]{EFEFEF}Promotion                              & Employees' perceptions of fairness in performance evaluations and promotion decisions are critical to their sense of value and belonging within an organization \cite{Turhan2016The}.                                  \\
Job Security                           & When employees perceive fair treatment in the organization, they tend to feel more secure in their jobs \cite{teimouri2016study}.                                                                                       \\ \hline
\end{tabular}
\label{tab:satis}
\end{table*}

\subsubsection{Measuring Fairness Experiences in SE Contexts}
\label{sec:measuringimpact}

Our study measures perceptions of fairness using survey questions developed by Colquitt \cite{Colquitt2001}, grounded in the organizational justice literature's theoretical framework for understanding fairness in the workplace. However, it is also important to understand how these fairness dimensions manifest in real-world scenarios, particularly within the context of SE. Exploring concrete examples of fairness issues in specific contexts can reveal where and how fairness is challenged, guiding the development of targeted strategies to improve job satisfaction.

A study by Sesari et al. \cite{sesari2024understanding} identifies ten distinct contexts where fairness is a concern within the SE field: \textit{income}, \textit{treatment}, \textit{allocation of work}, \textit{recruitment}, \textit{working hours}, \textit{demand}, \textit{evaluation}, \textit{authorship}, \textit{policy}, and \textit{customer/client relations}. 
These contexts represent primary areas where fairness challenges are most likely to arise, offering valuable insights into where software engineers perceive fairness issues in their daily professional experiences.
To investigate these contexts further, we asked survey respondents to rate the extent to which they have experienced fairness in each of the ten contexts identified by Sesari et al. \cite{sesari2024understanding}. This approach allows us to systematically measure the extent to which fairness perceptions are experienced in various SE scenarios. \edits{Below is the list of fairness experiences in SE contexts faced by practitioners, adapted from the study by Sesari et al. \cite{sesari2024understanding}:
\begin{itemize}
    \item income: fairness in how practitioners are compensated for their contributions;
    \item treatment: fairness in how practitioners are treated and how they treat others in the workplace;
    \item allocation of work: fairness in how tasks and responsibilities are distributed among team members;
    \item recruitment: fairness in the processes used to hire or onboard practitioners;
    \item working hours: fairness in expectations around the amount and distribution of working time;
    \item demand: fairness in the expectations or requests placed on practitioners by those in authority;
    \item evaluation: fairness in how work performance or contributions are assessed;
    \item authorship: fairness in how credit is assigned for contributions to code, design, or other outputs;
    \item policy: fairness in organizational policies or rules that affect practitioners’ work;
    \item customer/client relations: fairness in interactions between practitioners (or their organizations) and external stakeholders such as clients or customers.
\end{itemize}}



\subsubsection{Collecting Demographic Information}
\label{sec:respondentsdemo}
In our survey design, we included a series of questions aimed at collecting demographic information from respondents, such as gender, age, ethnicity, work experience, and role within the organization. The inclusion of these variables allows us to perform subgroup analyses, uncovering potential differences in job satisfaction across diverse demographic groups and enabling the development of fairness-related strategies tailored to each group.

To ensure respondent comfort and privacy, all demographic questions were accompanied by a \say{prefer not to say} option. This approach encourages honest participation while respecting individuals' boundaries \cite{Joinson2008Measuring}. The specific demographic categories included in our survey are as follows:

\textbf{Gender}\textit{: Male, Female}. Given the historical gender biases in the SE field, particularly in areas like hiring and promotion \cite{Wang2019}, we anticipate that gender may play a significant role in shaping job satisfaction. \edits{While we acknowledge that broader or open-text gender categories could capture more diverse identities, such an approach was beyond the scope of this study and would have posed challenges for statistical subgroup analysis given the sample size. Future research could specifically investigate fairness experiences of gender-diverse individuals in SE through targeted sampling and mixed-methods designs.}

\textbf{Age}\textit{: Under 40, 40 or older}. Older employees often report higher satisfaction due to increased experience, stability, and potentially more favorable job conditions \cite{Baltes2020}. This suggests that younger employees might face different challenges which could influence their job satisfaction differently.

\textbf{Ethnicity}\textit{: American Indian/Alaska Native, Asian, Black, Hispanic, Mixed, Native Hawaiian/Pacific Islander, White, Other.} Minority groups might face challenges such as discrimination or lack of representation, which can negatively impact their work experience \cite{Nadri2021,Shameer2023}. This could lead to varying levels of job satisfaction across different ethnicities, with minority groups potentially reporting lower satisfaction.

\textbf{Ethnicity Representation}\textit{: Well-represented, Under-represented, in a Balanced/Diverse Team.} Individuals whose ethnicity is under-represented in their groups may feel marginalized or distinct from their peers \cite{Choi2016}. Those in well-represented or balanced teams may experience higher job satisfaction, while those in under-represented groups might report lower satisfaction due to feelings of isolation or discrimination.

\textbf{Work Experience}\textit{: $\leq1$ year, $1-3$ years, $3-5$ years, $5-7$ years, $>7$ years.} More experienced individuals typically benefit from increased competence, more significant project roles, and stronger workplace relationships \cite{Tarasov2019AssessingJS}. This could imply that those with less experience might face dissatisfaction, potentially due to less job security or fewer opportunities.

\textbf{Role}\textit{: DevOps Engineer, Product Manager, Software Architect, Software Developer/Engineer, Software Tester/QA Engineer, Other.} Roles that involve more autonomy and collaboration, such as Product Management, might contribute to higher job satisfaction \cite{Yilmaz2012ASA}, whereas roles with less autonomy or more routine tasks may see lower satisfaction levels.

\textbf{Role Nature}\textit{: Management, Employee, Other.} Managers might focus on the fairness of their decisions and policies, while employees are more likely to assess fairness in terms of how it affects them personally \cite{gilliland2008justice,barclay2005exploring}. This suggests that job satisfaction might vary between managers and employees.

\textbf{Work Limitation}\textit{: Does a physical, mental, or emotional problem keep you from working? (yes/no) Are you limited in the kind or amount of work you can do because of a physical, mental, or emotional problem? (yes/no).} Individuals with work limitations often perceive lower fairness, particularly in areas such as pay and workplace accommodations \cite{Milner2015}. This might suggest that those with work limitations report lower job satisfaction compared to those without such limitations. The questions used in the survey are based on National Health Interview Survey (NHIS) \cite{Altman2014}.

\subsubsection{Respondent Recruitment}\label{sec:respondent}
To understand how fairness relates to job satisfaction among software practitioners and how this relationship varies across demographic groups, it was important to recruit individuals with experience in the software industry from diverse backgrounds. We employed several strategies to reach these participants.

\edits{We applied purposive sampling \cite{baltes2022sampling}, recruiting respondents who are professional software practitioners via LinkedIn, social media, and professional events. This approach enabled us to reach individuals likely to provide relevant insight into fairness and job satisfaction in software development contexts. Following Baltes and Ralph’s classification \cite{baltes2022sampling}, our sampling strategy corresponds to a form of convenience sampling with elements of expert sampling, as we selectively targeted working professionals in software engineering through professional networks and domain-relevant channels.}

First, we leveraged our professional networks, which include many individuals with substantial experience in SE. We used LinkedIn, a platform designed for professional networking, to connect with these potential respondents. We began by creating a LinkedIn article that explained the purpose and importance of our survey and included a link for easy participation. To broaden our reach beyond our immediate networks, we also shared the LinkedIn article on other social media platforms like Facebook, Instagram, and Twitter (X).

In addition to this online outreach, we directly sent the LinkedIn article to professional contacts who work in software development, ensuring the survey reached those most likely to provide valuable responses. We also promoted the survey in person during a workshop presentation given by the third author. At the workshop, we talked to attendees about the survey and handed out flyers with the survey link, allowing us to engage directly with software practitioners who were interested in our study. Geographically, most of our contacts reside in Europe and Asia.

\subsubsection{Ethical Considerations}
\edits{This study was approved by the relevant ethics committees at the authors’ institutions under approval numbers 97126442 and UCL/CSREC/R/40. All participants were informed of the purpose of the study, their rights, and the handling of their data through a participant information sheet and informed consent form, both of which were presented before the survey began. Participants provided their explicit consent before proceeding with the survey.
Details about data storage, anonymity, and the handling of sensitive information were provided in the Participant Information Sheet, which is included in the supplementary materials \cite{figshare}. The informed consent form is also available.}

\subsection{Pre-processing and Assessing Survey Data Quality}
The survey was made available on January 2, 2024, and remained open for 2 months. During this period, the survey system recorded a total of 111 responses. All respondents provided their consent and responded to the questions. 

We began by cleaning the responses, removing those who were not software practitioners. Respondents who selected \say{other} for their profession but did not specify were labeled as \say{prefer not to say.} 

\edits{We implemented response time analysis as a proxy for detecting inattentive or careless responses, consistent with recommendations in survey methodology literature \cite{read2022racing}. Responses completed in less than one minute were flagged as likely inattentive. While explicit attention-check questions are common in some survey designs, we chose not to include them to maintain engagement and trust among respondents. Prior research indicates that experienced participants may recognize such checks, potentially reducing their effectiveness and introducing bias \cite{pei2020attention, almog2023attention}. }

We established a threshold to detect outliers, focusing on extremely short response times, such as responses completed in less than one minute. Using the Interquartile Range (IQR) method, we identified responses with durations in the first quartile as potential outliers.
The shortest response time was 0.6 minutes, which we deemed too short to yield meaningful data given the expected 10-minute duration. Research supports that very quick survey completions can indicate less attentive responses \cite{Wood2017}. On reviewing the 0.6-minute response, we found missing answers in key areas, reinforcing concerns about data quality for very fast responses.
We also examined responses with durations beyond the third quartile (Q3), ranging from 17 minutes to $\sim$6 days. The extended response times might be due to allowing respondents to return to the survey later, as their progress remains saved even if the tab is left open. Therefore, we did not find it necessary to exclude these, as longer completion times do not necessarily correlate with lower data quality. 

Recognizing that respondents should not be forced to choose an answer that does not accurately reflect their situation, we included a \say{not applicable} option in our survey \cite{DeRouvray2002Designing}. For responses with \say{not applicable} answers, we faced two choices regarding data analysis: either eliminate the entire response or replace the missing values with the closest estimated values. Eliminating responses could reduce the sample size and, in some cases, exclude participants whose input might still be valuable for other parts of the survey. Deleting responses risks introducing bias or underrepresenting certain demographics, which is why we opted for the latter approach of estimation to preserve as much useful data as possible.

To do this, we applied Multiple Imputation by Chained Equations (MICE) using Predictive Mean Matching (PMM). This method fills in missing data by matching them with similar observed values, ensuring completeness and consistency in our dataset. MICE has been demonstrated to be versatile and robust in various research domains \cite{White2011Multiple, Chhabra2017A, Wulff2017Multiple, Azur2011Multiple, Beesley2021Multiple}.

\edits{To ensure the validity and reliability of our survey constructs, we conducted both Exploratory Factor Analysis (EFA) and Confirmatory Factor Analysis (CFA), followed by an assessment of internal consistency using the Omega reliability coefficient \cite{McDonald_1970} \editss{(see Table \ref{tab:fit-summary}).}}

\begin{table}[ht]
\centering
\caption{\editss{Model Fit, Factor Loadings, and Reliability by Construct}}
\begin{tabular}{ll|ccc}
\hline
\textbf{Analysis} & \textbf{Metric} & \textbf{Fairness Perception} & \textbf{Job Satisfaction} & \textbf{Fairness Context in SE} \\
\hline
\multirow{3}{*}{EFA} 
  & RMSEA & 0.06 & 0.08 & 0.08 \\
  & TLI   & 0.94  & 0.87  & 0.90  \\
  & RMSR  & 0.04  & 0.05  & 0.07  \\
  \hline
\multirow{4}{*}{CFA} 
  & CFI   & 0.92  & 0.86  & 0.92  \\
  & TLI   & 0.91  & 0.83  & 0.90  \\
  & RMSEA & 0.08  & 0.10  & 0.09  \\
  & SRMR  & 0.08  & 0.08  & 0.06  \\
  \hline
\multicolumn{2}{l|}{Factor Loadings (range)} & 0.52–0.97 & 0.49–0.87 & 0.41–0.78 \\
\hline
\multicolumn{2}{l|}{Omega Reliability} & 0.94 & 0.93 & 0.89 \\
\hline
\end{tabular}
\label{tab:fit-summary}
\end{table}

\edits{The EFA results aligned well with theoretical expectations. For fairness perceptions (20 items), a four-factor structure emerged, corresponding to Colquitt’s dimensions of distributive, procedural, interpersonal, and informational fairness. The model demonstrated strong fit (RMSEA = 0.059, TLI = 0.937, RMSR = 0.04), with items loading cleanly onto their respective factors and minimal cross-loadings. For job satisfaction (16 items), EFA revealed 
acceptable fit (RMSEA = 0.082, TLI = 0.873, RMSR = 0.05), consistent with conceptual groupings adapted from Storey et al. \cite{Storey2021}. For fairness contexts (10 items), parallel analysis suggested a unidimensional structure, which was supported by good fit indices (RMSEA = 0.081, TLI = 0.903, RMSR = 0.07), indicating that fairness experiences across SE scenarios reflect a single construct.}

\edits{CFA was subsequently performed to confirm these structures. The fairness perceptions model showed good fit (CFI = 0.921, TLI = 0.909, RMSEA = 0.076, SRMR = 0.076), supporting Colquitt’s four-dimensional framework. The job satisfaction model demonstrated acceptable fit (CFI = 0.857, TLI = 0.830, RMSEA = 0.101, SRMR = 0.082), reflecting the exploratory nature of adapting Storey et al.'s factors. The fairness contexts model confirmed the unidimensional structure with strong fit indices (CFI = 0.919, TLI = 0.896, RMSEA = 0.087, SRMR = 0.061).}

\edits{Across all constructs, item-level standardized factor loadings were generally high, ranging from 0.52 to 0.97 for fairness perceptions, 0.49 to 0.87 for job satisfaction, and 0.41 to 0.78 for fairness contexts. These factor loadings indicate that individual items contributed meaningfully to their respective latent constructs.}

\edits{Finally, to assess internal consistency, we applied the Omega reliability coefficient, which is particularly suited for multidimensional constructs and provides a more accurate estimate of reliability than Cronbach’s alpha when factor loadings vary \cite{McDonald_1970}. The Omega values were strong across all constructs, with 0.94 for fairness perceptions, 0.93 for job satisfaction, and 0.89 for fairness contexts. These results indicate a high level of reliability, which confirm that the variance in responses is largely attributable to consistent underlying constructs. Notably, Omega values remained stable before and after multiple imputation, which demonstrate that our handling of missing data did not compromise reliability. After pre-processing, we retained 108 responses for analysis.}

\edits{As mentioned in Section~\ref{sec:respondent}, we employed purposive sampling focused on recruiting software practitioners from diverse backgrounds to align our sampling strategy with our study’s exploratory goals. While our final sample includes 108 respondents, this size is appropriate for our research objective: examining theorized relationships between fairness perceptions and job satisfaction, rather than producing population-level estimates. As Baltes and Ralph emphasize in their critical review of sampling in SE research, “there is no universally correct sample size,” and what matters is how well the sampling strategy supports the intended claims of a study \cite{baltes2022sampling}. They further argue that when a study aims to explore social or organizational constructs, diversity of experience and transparent reporting are often more meaningful than statistical representativeness. Following these recommendations, we prioritized recruiting participants across a range of demographics and clearly report our sampling limitations to avoid overgeneralization.}

\subsection{Summarizing Responses}

To summarize the responses, we employed various data analysis techniques.
First, we calculated descriptive statistics for each survey item to help us understand trends in the responses. We then visualized these responses using box plots and bar charts to provide a clear data distribution. Box plots helped identify outliers and the spread of responses, while bar charts made it easier to compare response frequencies across items.

For the Likert responses to the 20-item fairness scale, analyzing individual items allows us to identify specific aspects of fairness that may need prioritization. However, to understand general trends across the four fairness dimensions, Enoksen \cite{Enoksen2015} recommended aggregating the Likert responses by calculating the mean for each dimension. For example, we calculated the mean of the four items related to distributive fairness to create an aggregated descriptive statistic for this dimension. 

To examine demographic differences in the three perceptions, we used the \textsc{Mann-Whitney U} \cite{Macfarland2016} and \textsc{Kruskal-Wallis} \cite{kruskal1952use} tests, along with \texttt{wilcoxon\_effsize} \cite{tomczak2014need} in R for effect size. We focused on moderate ($0.30 - < 0.5$) and large effect sizes ($>= 0.5$) to highlight meaningful differences \cite{tomczak2014need}.

The \textsc{Mann-Whitney U} test compares responses between two groups \cite{Macfarland2016}, such as those with and without \textit{work limitations}. For example, if the test yielded a p-value less than 0.05 and an effect size of 0.4 for satisfaction with the \textit{manager}, it would indicate a moderate difference between the two groups. We then referred to the bar plot visualizations (see our supplementary material \cite{figshare}) to explore these differences, such as the percentage of respondents with and without \textit{work limitations} reacting negatively to satisfaction with the \textit{manager}.

The \textsc{Kruskal-Wallis} test identifies statistically significant differences among three or more independent groups, such as respondents with varying levels of \textit{work experience} \cite{kruskal1952use}. When significant differences were detected, we conducted post hoc \textsc{Dunn-Bonferroni} tests to pinpoint specific group differences while controlling for Type I errors. For example, if the \textsc{Kruskal-Wallis} test indicated a p-value less than 0.05 for satisfaction with benefits across different work experience levels, and the post hoc test found a significant difference between respondents with \textit{less than 1 year} and \textit{more than 7 years} of experience with an effect size of 0.6, it would suggest a large effect. Similar to the \textsc{Mann-Whitney U} test, we then reviewed visualizations to examine these differences, such as the percentage of respondents with \textit{less than 1 year} of experience versus those with \textit{more than 7 years} reacting to satisfaction with \textit{benefits}.

For the demographic categories of \textit{ethnicity} and \textit{role}, even though they had more than two options, we focused on those with comparable numbers of respondents to avoid false positives. Specifically, for \textit{ethnicity}, we considered only \textit{Asian} and \textit{White} respondents, and for \textit{role}, we focused on \textit{Software Developer/Engineer} and \textit{Software Tester/QA Engineer}. 

\subsection{Grouping Job Satisfaction Factors}

The goal of this step is to identify potential natural groupings or clusters among job satisfaction factors, which can help us treat related factors as cohesive units for more effective analysis. While the job satisfaction factors in our study are designed to be independent to ensure clear differentiation between the factors, it is possible that some factors receive similar reactions from respondents. This similarity suggests that there may be natural groupings or clusters among the factors, which might not be immediately obvious. For example, respondents might perceive satisfaction with \textit{appreciation} and \textit{feedback} as related, reflecting a common underlying sentiment. 

To uncover these natural groupings, we first checked for the independence of factors by calculating the correlations between the satisfaction scores for each pair of factors using the \textsc{Spearman} method. The \textsc{Spearman} correlation method is suitable for ordinal data, like Likert-scale responses, because it assesses how well the relationship between two variables can be described by a monotonic function \cite{Winter2016}. For example, if the satisfaction scores for \textit{feedback} and \textit{appreciation} are highly correlated, it suggests that respondents who rate \textit{feedback} highly also tend to rate \textit{appreciation} highly.

After establishing the relationships between factors through correlation analysis, we computed the dissimilarity matrix using the \textsc{Manhattan} distance, which is appropriate for ordinal data. \editss{Compared to the \textsc{Euclidean} distance, the \textsc{Manhattan} distance provides a more robust measure when dealing with ordinal or non-normally distributed variables, as it does not assume interval-scaled data or linear relationships \cite{vrezankova2019effect}. The \textsc{Manhattan} distance measures the absolute differences between the pairwise correlation coefficients, summing these differences across all pairs \cite{Weiß2019}. This method helps to quantify how dissimilar each pair of job satisfaction factors is based on their correlation values \cite{Weiß2019}. For instance, a low value for the \textsc{Manhattan} distance between satisfaction with \textit{benefits} and satisfaction with \textit{salary} suggests that these factors are closely related in how respondents perceive them.}

\begin{figure}[]
\centering
\includegraphics[width=0.7\textwidth]{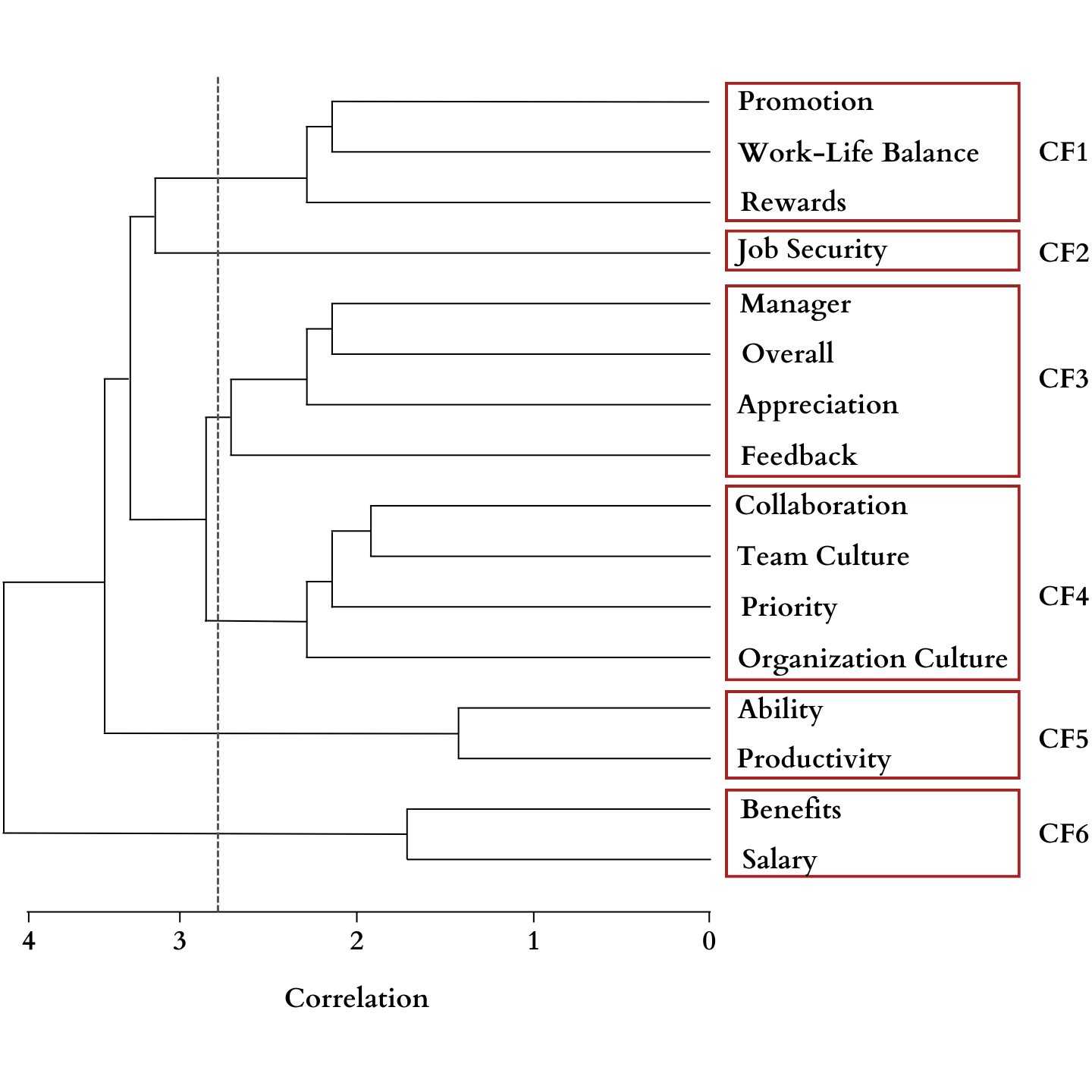}
\captionsetup{font=footnotesize}
\caption{\editss{Dendogram of clustering job satisfaction factors into groups.}}
\label{fig:dendogram}
\end{figure}

We then performed \textsc{hierarchical clustering} using the \textsc{average linkage} method, implemented in \texttt{R} with the \texttt{hclust()} function \cite{Langfelder2012,Guénard2022}. \editss{The \textsc{average linkage} method calculates the average distance between all pairs of factors in two clusters, merging the clusters with the smallest average distance first. This approach ensures that clusters are formed based on the overall similarity of their members, rather than just the closest or farthest points, which helps maintain balanced clusters.}
The result of the \textsc{hierarchical clustering} is visualized in a dendrogram (see Figure \ref{fig:dendogram}), a tree-like diagram that illustrates how clusters are merged at different distance thresholds. For instance, if satisfaction with \textit{benefits} and satisfaction with \textit{salary} are closely related, they will be merged early in the dendrogram.

To identify clusters that were highly correlated, we focused on groups with a correlation of 2.7 or more. 
This threshold allowed us to group the 16 job satisfaction aspects into six distinct clusters (i.e., composite factors (CFs)).

Additionally, we made a slight adjustment by separating the \textit{overall} satisfaction factor from CF3, treating it as a distinct composite factor. This adjustment led to a final set of seven composite factors, as shown in Table \ref{tab:cfs}. For example, satisfaction with \textit{benefits} and \textit{salary} was highly correlated, so we combined these factors into the composite factor \textit{compensation}. When combining factors, we averaged their scores; for instance, if the satisfaction scores for \textit{benefits} and \textit{salary} were $3$ and $4$, respectively, the score for \textit{compensation} would be $(3 + 4) / 2 = 3.5$.

By grouping similar job satisfaction factors into cohesive units, we can better understand how different aspects of job satisfaction are interrelated, leading to more targeted and effective strategies for enhancing job satisfaction among software practitioners.



\begin{table}[]
\centering
\caption{Composite Factors of Practitioners' Job Satisfaction.}
\small
\begin{tabular}{ll}
\hline
\multicolumn{1}{c}{Composite Factor} & \multicolumn{1}{c}{Satisfaction Aspects}                                                               \\ \hline
Overall (ALL)                         & Overall                                                                                                \\
\rowcolor[HTML]{EFEFEF} 
Rewards \& Work-Life Balance (RWB)    & \begin{tabular}[c]{@{}l@{}}Rewards\\ Work-Life Balance\\ Promotion\end{tabular}                        \\
Job Security (SEC)                    & Job Security                                                                                           \\
\rowcolor[HTML]{EFEFEF} 
Manager (MNG)         & \begin{tabular}[c]{@{}l@{}}Feedback\\ Appreciation\\ Manager\end{tabular}                              \\
Work Culture (CLT)                    & \begin{tabular}[c]{@{}l@{}}Organization Culture\\ Priority\\ Team Culture\\ Collaboration\end{tabular} \\
\rowcolor[HTML]{EFEFEF} 
Performance (PRF)                     & \begin{tabular}[c]{@{}l@{}}Productivity\\ Ability\end{tabular}                                         \\
Compensation (COM)                    & \begin{tabular}[c]{@{}l@{}}Salary\\ Benefits\end{tabular}                                              \\ \hline
\end{tabular}
\label{tab:cfs}
\end{table}

\subsection{Statistical Analysis}
\subsubsection{Ordinal Logistic Regression Analysis}
After establishing the job satisfaction clusters, we aimed to understand how fairness perception or fairness contexts are associated with the probability of job satisfaction being in a higher or lower category. To achieve this, we performed an ordinal logistic regression analysis.
We performed an ordinal logistic regression analysis to explore the relationships between survey items, specifically addressing RQ1 and RQ3.
Ordinal logistic regression predicts the probability of an observation falling into one of several ordered categories, making it ideal for analyzing outcomes like job satisfaction levels \cite{Agresti2010}. 

For example, when analyzing job satisfaction levels among software engineers, ordinal logistic regression helps us understand how perceptions of fairness are related to these levels. If job satisfaction is categorized from “very dissatisfied” to “very satisfied,” a positive and significant coefficient for perceptions of fairness would indicate that as the perception of fairness increases, the likelihood of a respondent falling into a higher job satisfaction category also increases. In other words, as perceptions of fairness improve, the odds of a software engineer being more satisfied increase.

\edits{To address RQ1, we applied this analysis to examine the relationship between perceptions of fairness and the seven job satisfaction clusters. For instance, if a practitioner perceives high levels of \textit{distributive fairness}, the ordinal logistic regression might predict a higher likelihood of reporting increased satisfaction in clusters related to \textit{compensation} (COM). This suggests that perceptions of fair distribution of resources and rewards are linked to how satisfied practitioners feel with their compensation.
For RQ3, we aimed to translate the broader understanding of fairness perceptions into actionable strategies by examining how specific fairness perceptions relate to job satisfaction within various SE contexts. To do this, we applied a similar analysis to the one used for RQ1, focusing on the relationship between fairness experiences and job satisfaction in specific scenarios.}

This analysis was conducted using the \texttt{MASS} package in \texttt{R}. Additionally, we calculated odds ratios (OR) to quantify the strength of the associations, providing a clear measure of the effect size for each predictor variable.



\subsubsection{Moderation Analysis} \label{sec:moderation}
After performing the ordinal logistic regression analysis, we explored how the observed relationships differ across demographic groups by employing moderation analysis through RQ2 and RQ3.1. Moderation analysis tests whether a variable, such as a demographic factor, influences the direction and/or strength of the relationship between an independent variable (e.g., perception of fairness) and a dependent variable (e.g., job satisfaction) \cite{baron1986moderator}.
For example, suppose we find that perceptions of procedural fairness are positively associated with job satisfaction. A moderation analysis might examine whether this relationship is stronger for women than for men.

In moderation analysis, the effect size can be positive or negative. A positive effect size indicates that the relationship between fairness perceptions and job satisfaction is stronger for the demographic group being analyzed compared to the reference group, while a negative effect size suggests a weaker relationship. For example, in our analysis of ethnicity representation, the analysis used a \textit{diverse/balanced team} as the reference group. A positive effect size for the \textit{underrepresented} group indicates that the impact of fairness perceptions on job satisfaction is stronger for individuals in \textit{underrepresented} teams compared to those in \textit{balanced teams}.

In our study, we specifically examined the potential moderating effects of various demographic factors, as described in Section \ref{sec:respondentsdemo}. Due to the imbalance in the distribution of demographic categories among our respondents (as shown in Table \ref{tab:demographics}), initially, we were only able to effectively analyze the moderating effects for three demographic groups: \textit{gender}, \textit{ethnicity representation}, and \textit{work limitation}. 
Subsequently, we simplified the categories for the remaining demographic groups to improve the analysis. For instance, \textit{ethnicity} was grouped into \textit{Asian} and \textit{Non-Asian} categories, and \textit{work experience} was divided into \textit{less than 3 years} and \textit{more than 3 years}. This refinement allowed us to effectively analyze the moderating effects for one additional demographic group: \textit{work experience}. 

However, the limited number of respondents in other remaining demographic categories, namely \textit{age}, \textit{ethnicity}, \textit{role} and \textit{role nature} restricted our ability to conduct reliable moderation analyses for these groups. This limitation is due to the potential for biased estimates caused by imbalanced distribution within the sample \cite{Zhang2017,griffin2023tutorial}.

\edits{For RQ2, we sought to identify whether these demographic characteristics alter the impact of fairness perceptions on job satisfaction, and if so, in what ways. For instance, we hypothesized that certain demographic groups, such as those with less representation in the workplace, might experience a stronger relationship between \textit{biased processes} (part of \textit{procedural fairness}) and their satisfaction with \textit{job security} (SEC) compared to more represented groups.}

\edits{For RQ3.1, we aimed to examine how fairness perception influencing job satisfaction differ across demographic groups within the SE context. For instance, if we observe that being \textit{treated politely}, \textit{with respect}, and \textit{dignity} during \textit{evaluations} significantly impacts job satisfaction, it would be important to explore which demographic groups are most affected by this factor. This could suggest that certain groups, such as female developers, find this aspect more impactful, indicating that gender may play a key role in shaping this relationship.}




\section{Findings}\label{sec:findings}
\subsection{Respondents Demographics}
\label{sec:respondentsdemo}
The demographic data is summarized in Table \ref{tab:demographics}. There were 80 male respondents and 25 female respondents. The age distribution shows that 100 respondents were under 40 years old, while 7 were 40 years or older.
Regarding ethnicity, the majority of respondents identified as Asian (79) and White (18). In terms of job roles, most respondents were Software Developers/Engineers (72) and Software Testers/QA Engineers (12).
For work experience, 11 respondents had $\leq$ 1 year of work experience, 40 had $1-3$ years, 26 had $3-5$ years, 14 had $5-7$ years, and 16 had over 7 years of experience. Regarding work limitations, 55 respondents reported no limitations, 43 indicated some form of limitation.

\begin{table}[]
\small
\centering
\caption{Respondents' Demographics}
\begin{tabular}{llc}
\hline
\multicolumn{1}{c}{Type} & \multicolumn{1}{c}{Options}          & N            \\ \hline
Gender                   & \textbf{Male}                        & \textbf{80}  \\
                         & Female                               & 25           \\
                         & Prefer not to say                    & 3            \\ 
\rowcolor[HTML]{EFEFEF}Age                      & \textbf{Under 40 years old}          & \textbf{100} \\\rowcolor[HTML]{EFEFEF}
                         & 40 years old or older                & 7            \\\rowcolor[HTML]{EFEFEF}
                         & Prefer not to say                    & 1            \\
Ethnicity                & \textbf{Asian}                       & \textbf{79}  \\
                         & White                                & 18           \\
                         & Mixed or Multiple Ethnic Groups      & 4            \\
                         & Other Ethnic Group                   & 4            \\
                         & Black                                & 2            \\
                         & Prefer not to say                    & 1            \\
\rowcolor[HTML]{EFEFEF}Ethnicity Representation & \textbf{Well-represented}            & \textbf{44}  \\\rowcolor[HTML]{EFEFEF}
                         & Balanced and Diverse Team            & 40           \\\rowcolor[HTML]{EFEFEF}
                         & Under-represented                    & 20           \\\rowcolor[HTML]{EFEFEF}
                         & Prefer not to say                    & 4            \\
Role                     & \textbf{Software Developer/Engineer} & \textbf{72}  \\
                         & Software Tester/QA Engineer          & 12           \\
                         & DevOps Engineer                      & 5            \\
                         & Software Architect                   & 4            \\
                         & UI/UX Designer                       & 3            \\
                         & Product Manager                      & 2            \\
                         & R\&D Software Engineer               & 2            \\
                         & AI Engineer                          & 1            \\
                         & Application Security Engineer        & 1            \\
                         & Tech Lead                            & 1            \\
                         & Project Manager                      & 1            \\
                         & Prefer not to say                    & 4            \\
\rowcolor[HTML]{EFEFEF}Role Nature              & \textbf{Employee}                    & \textbf{92}  \\\rowcolor[HTML]{EFEFEF}
                         & Management                           & 14           \\\rowcolor[HTML]{EFEFEF}
                         & Outsource                            & 1            \\\rowcolor[HTML]{EFEFEF}
                         & Prefer not to say                    & 1            \\
Work Experience          & \textless{}= 1 year                  & 11           \\
                         & \textbf{1-3 years}                   & \textbf{40}  \\
                         & 3-5 years                            & 26           \\
                         & 5-7 years                            & 14           \\
                         & \textgreater 7 years                 & 16           \\
                         & Prefer not to say                    & 1            \\
\rowcolor[HTML]{EFEFEF}Work Limitation          & \textbf{No}                          & \textbf{55}  \\\rowcolor[HTML]{EFEFEF}
                         & Yes                                  & 43           \\\rowcolor[HTML]{EFEFEF}
                         & Prefer not to say                    & 10           \\ \hline
\end{tabular}
\label{tab:demographics}
\end{table}

\begin{longtable}{lcccclcccc}
\caption{\editss{Fairness perception, job satisfaction, and fairness experience in SE contexts. Significant differences found in various demographic groups: age (~\protect\faClockO~), ethnicity (~\protect\faFlag~), ethnicity representation (~\protect\faGlobe~), work experience (~\protect\faBriefcase~), and work limitation (~\protect\faWheelchair~). All with moderate effect size, and asterisk (*) for large effect size.}}\\
\hline
\multicolumn{1}{c}{\multirow{2}{*}{Survey Items}} & \multirow{2}{*}{Median} & \multirow{2}{*}{IQR} & \multirow{2}{*}{Mean} & \multirow{2}{*}{SD} & \multicolumn{5}{c}{Distribution}      \\ \cline{6-10} 
\multicolumn{1}{c}{}                                &                         &                      &                       &                     & \multicolumn{1}{c}{1} & 2 & 3 & 4 & 5 \\ \hline
\endhead
\rowcolor[HTML]{DFDFDF}\multicolumn{10}{c}{\textit{\textbf{Perception of Fairness}}} \\ 
\rowcolor[HTML]{EFEFEF}\multicolumn{1}{l}{D: Distributive}                    & 3                       & 0.8                 & 3.1                 & 0.7               & \multicolumn{5}{l}{\fbox{\includegraphics[width=0.13\textwidth, height=2mm]{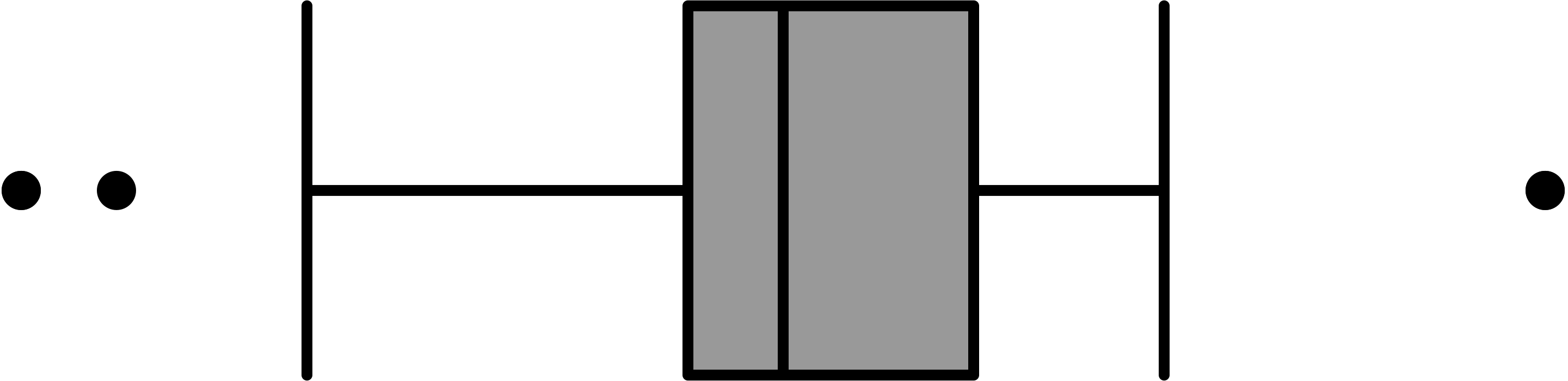}}}                  \\ 
D1: Outcome is justified by the performance       & 3      & 1.3 & 3.1 & 0.9 & \multicolumn{5}{l}{\fbox{\includegraphics[width=0.13\textwidth, height=2mm]{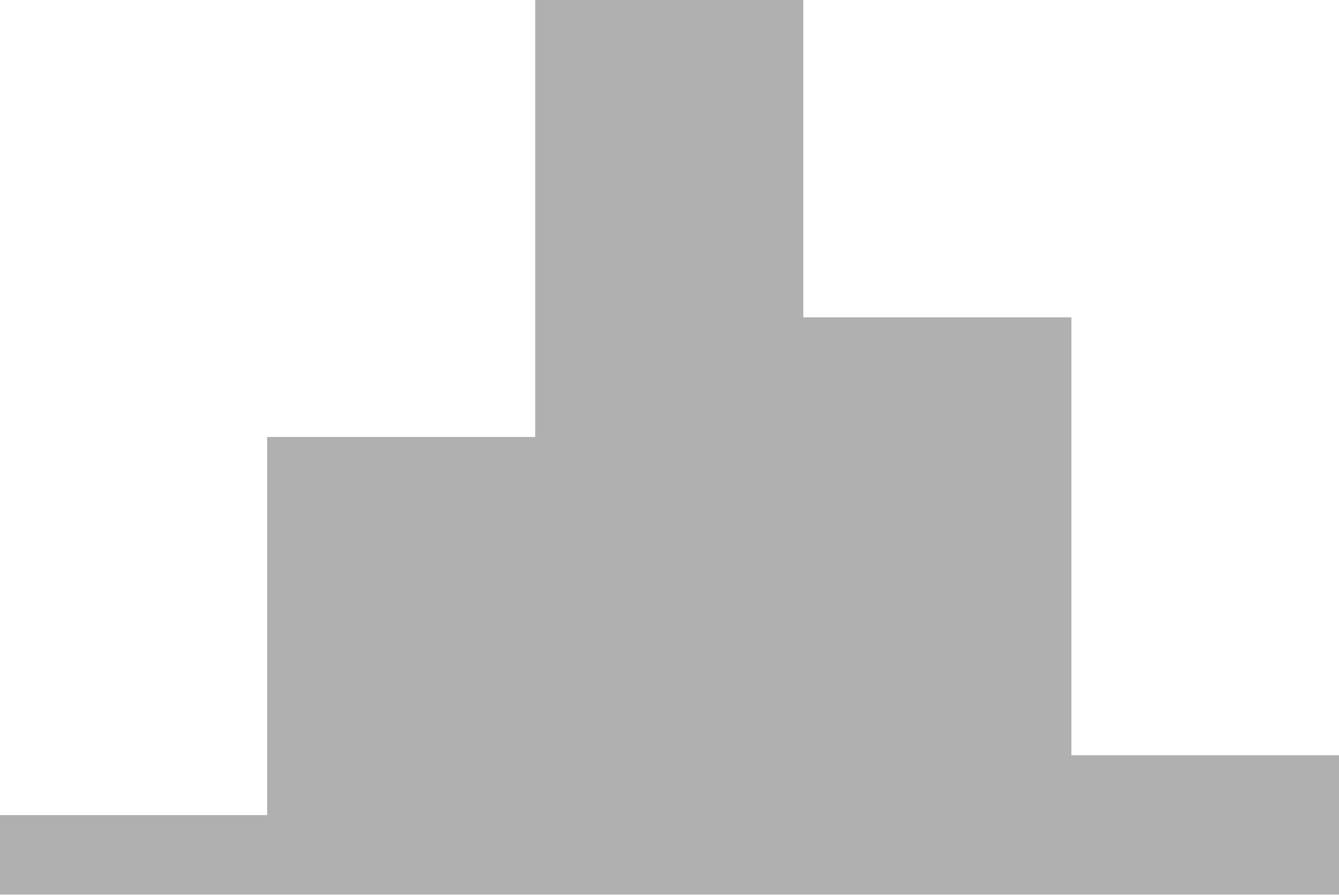}}}                           \\
D2: Outcome is appropriate for the completed work 
& 3      & 1    & 3.1 & 0.9 & \multicolumn{5}{l}{\fbox{\includegraphics[width=0.13\textwidth, height=2mm]{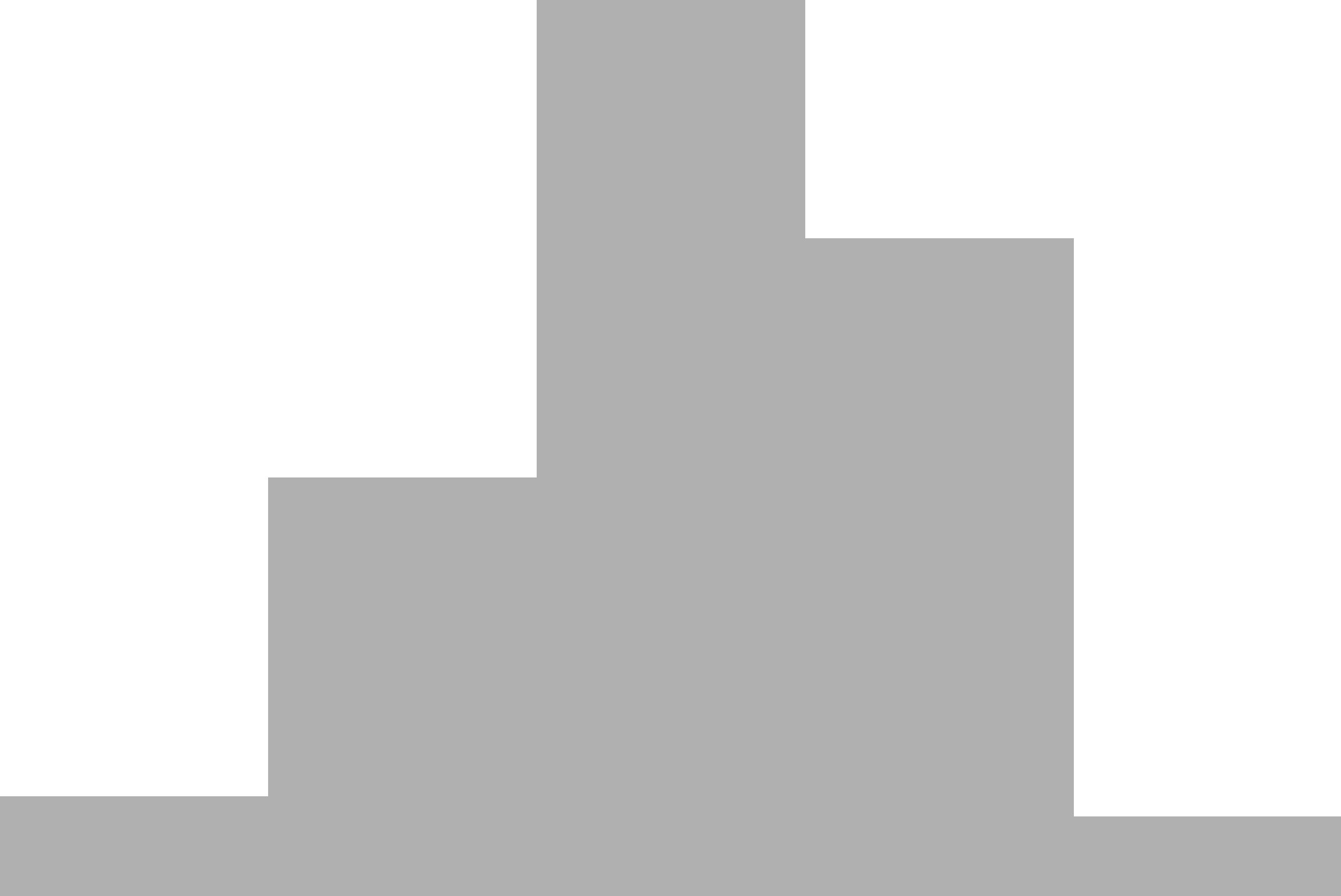}}}                           \\
D3: Outcome reflecting contribution               & 3      & 1    & 3.1 & 0.8 & \multicolumn{5}{l}{\fbox{\includegraphics[width=0.13\textwidth, height=2mm]{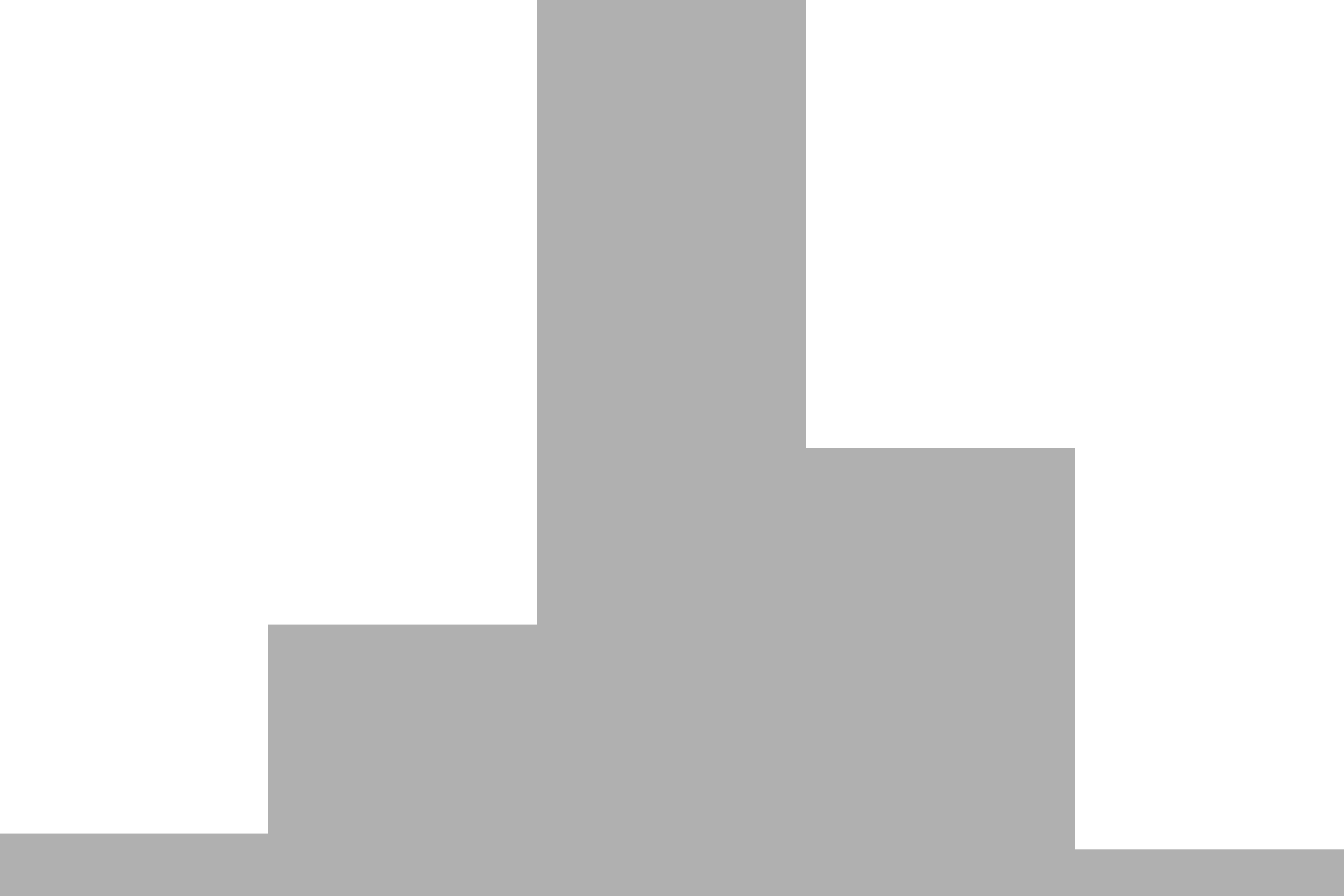}}}                           \\
D4: Outcome reflecting effort                     & 3      & 1    & 3.1 & 0.9 & \multicolumn{5}{l}{\fbox{\includegraphics[width=0.13\textwidth, height=2mm]{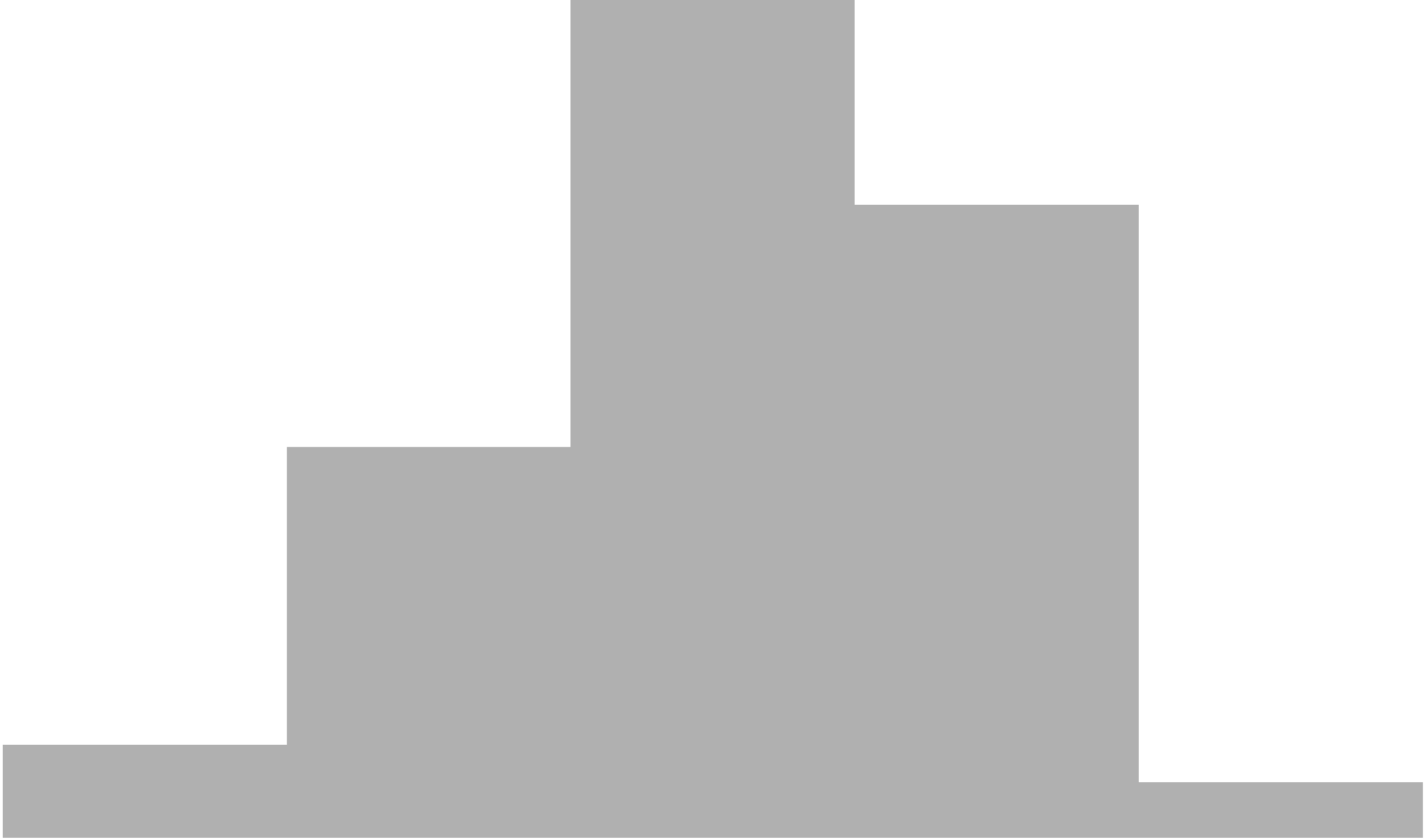}}}                           \\
\rowcolor[HTML]{EFEFEF}\multicolumn{1}{l}{P: Procedural}                & 3.3  & 1    & 3.2 & 0.8 & \multicolumn{5}{l}{\fbox{\includegraphics[width=0.13\textwidth, height=2mm]{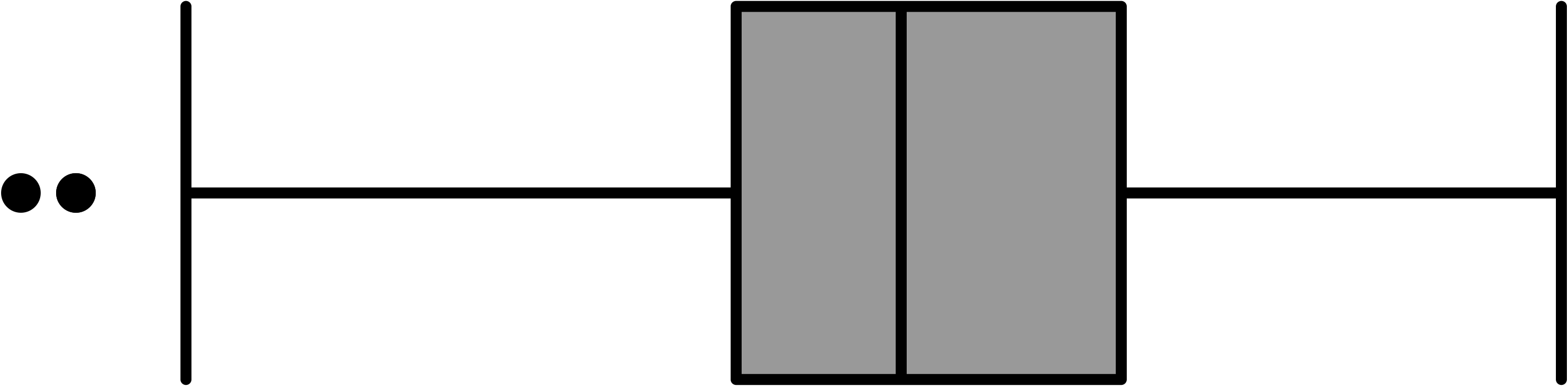}}}                           \\ 
P1: Ethical and moral processes 
& 4      & 1    & 3.5 & 1 & \multicolumn{5}{l}{\fbox{\includegraphics[width=0.13\textwidth, height=2mm]{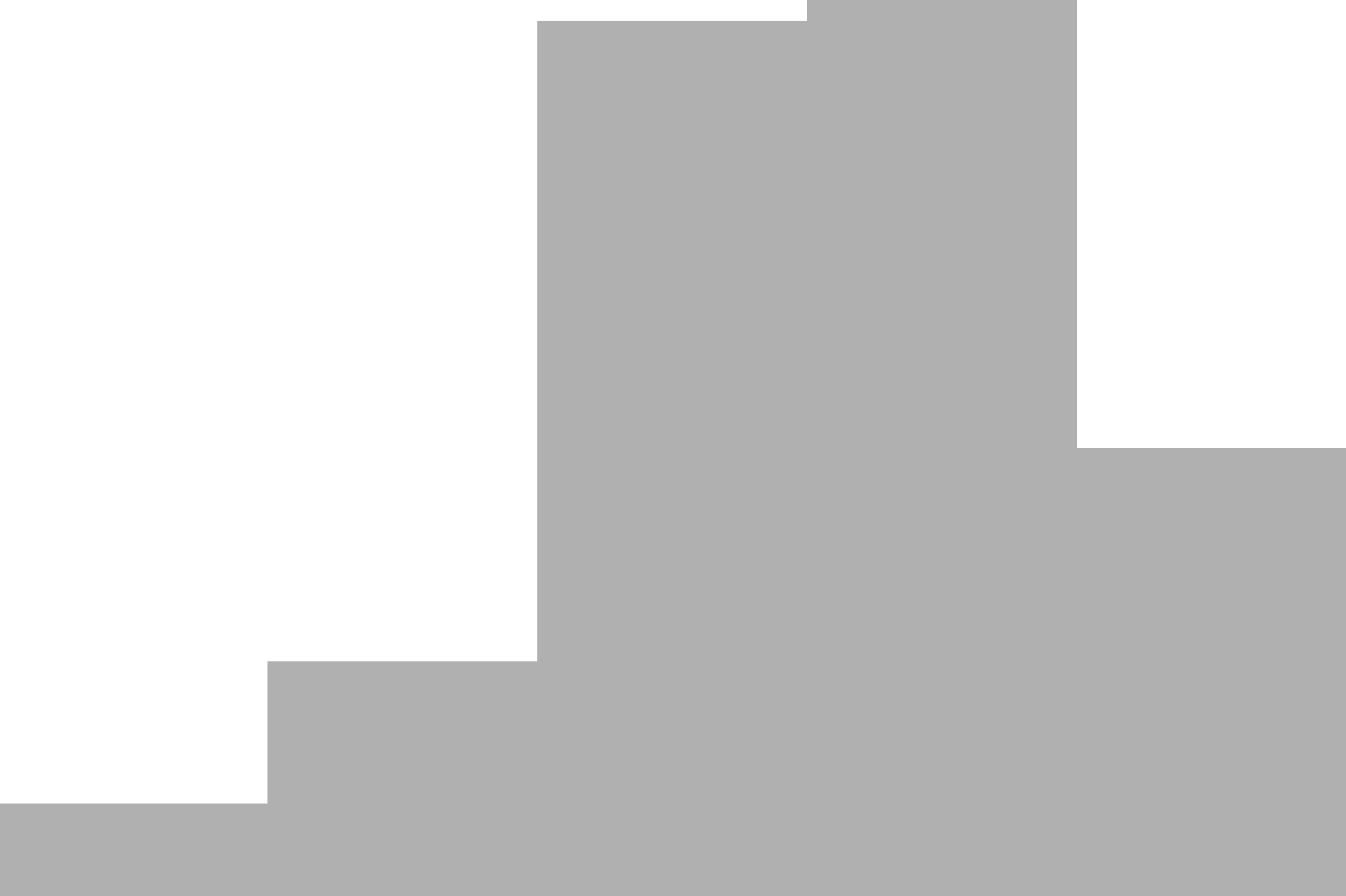}}}                           \\
P2: Voicing opinions 
\protect\faBriefcase                              & 4      & 1    & 3.3 & 1.2 & \multicolumn{5}{l}{\fbox{\includegraphics[width=0.13\textwidth, height=2mm]{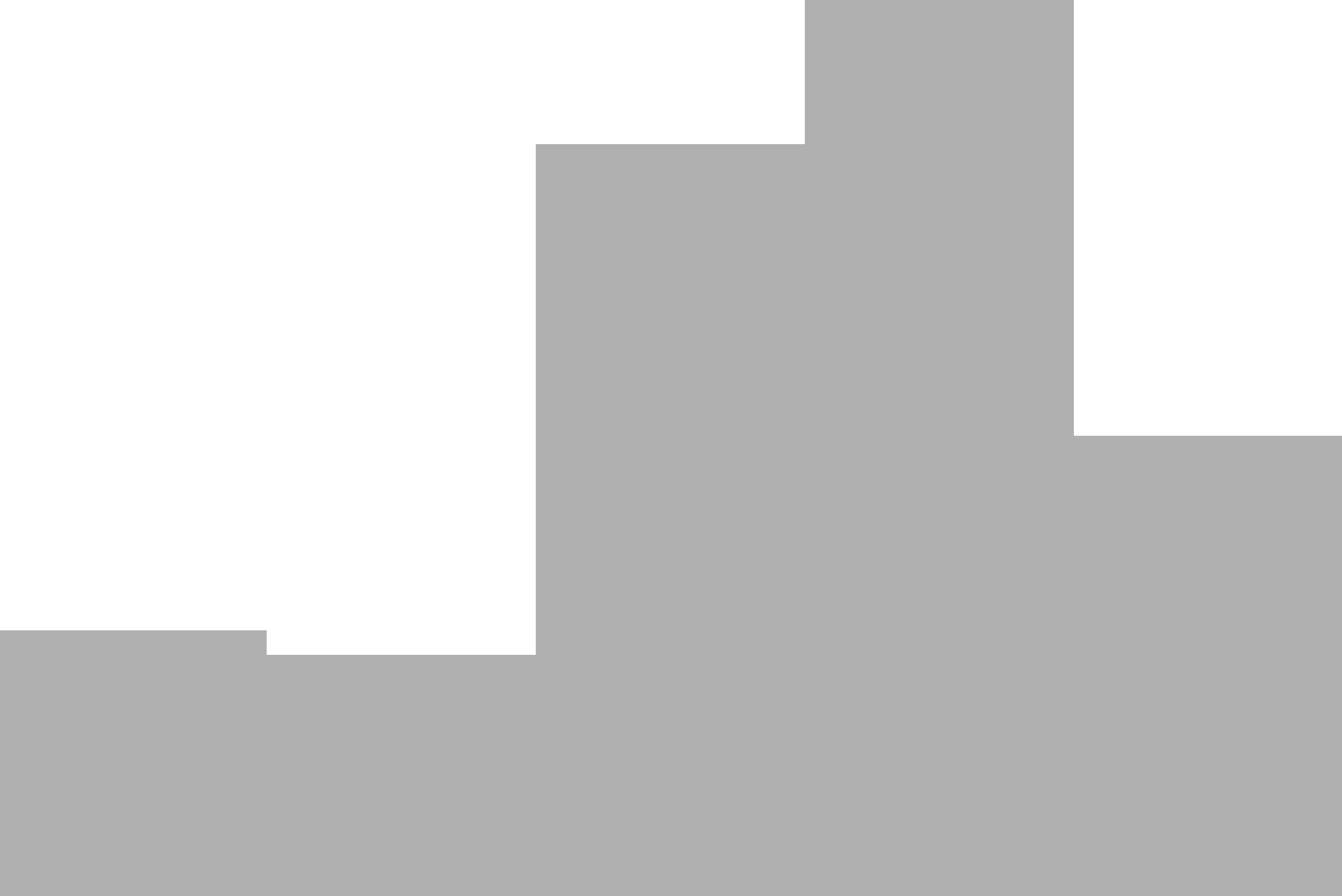}}}                           \\
P3: Accurate information \protect\faGlobe~ \protect\faBriefcase                         & 3      & 1    & 3.3 & 0.9 & \multicolumn{5}{l}{\fbox{\includegraphics[width=0.13\textwidth, height=2mm]{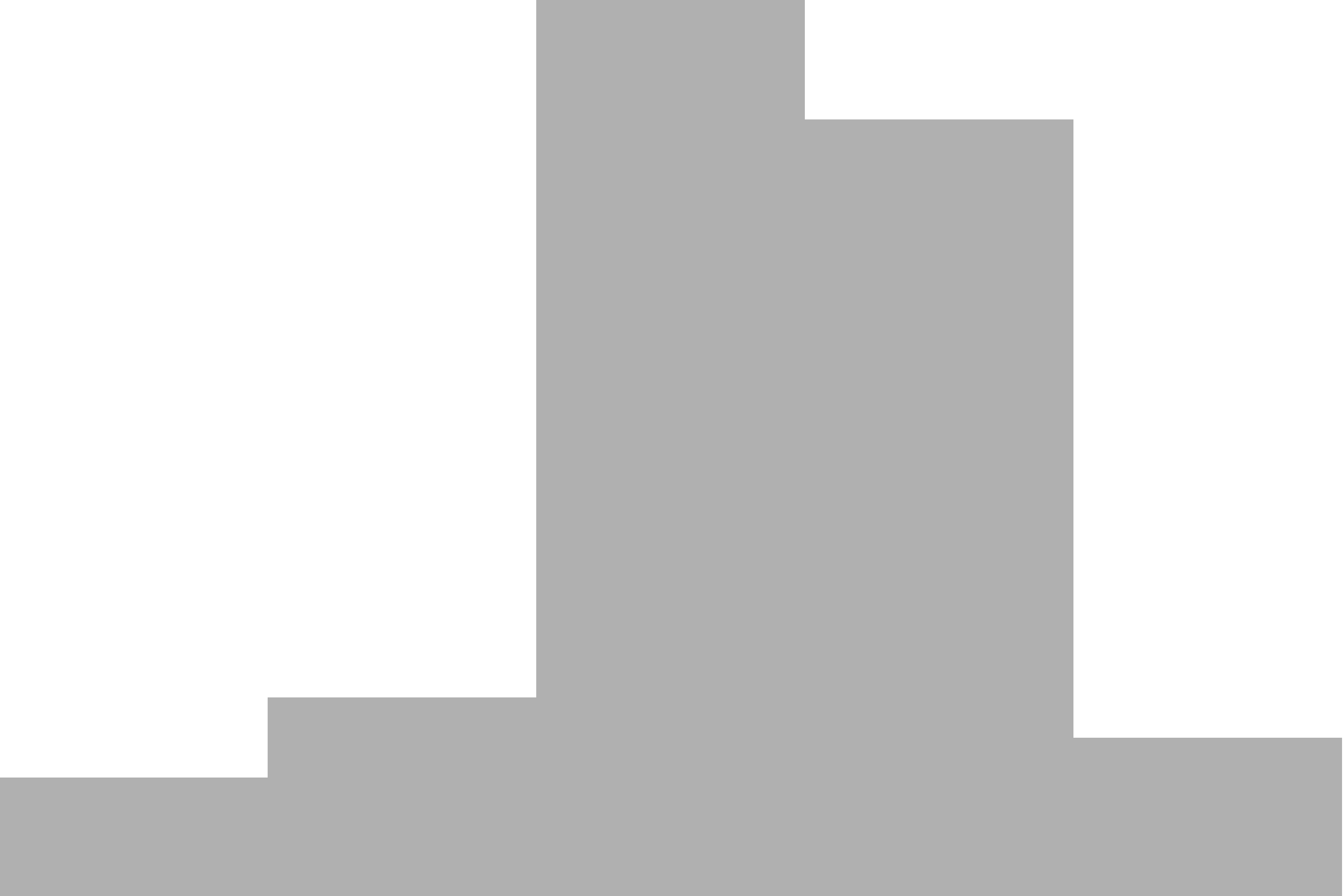}}}                           \\
P4: Free of bias processes \protect\faGlobe~ \protect\faBriefcase                       & 3      & 1    & 3.2 & 1.1 & \multicolumn{5}{l}{\fbox{\includegraphics[width=0.13\textwidth, height=2mm]{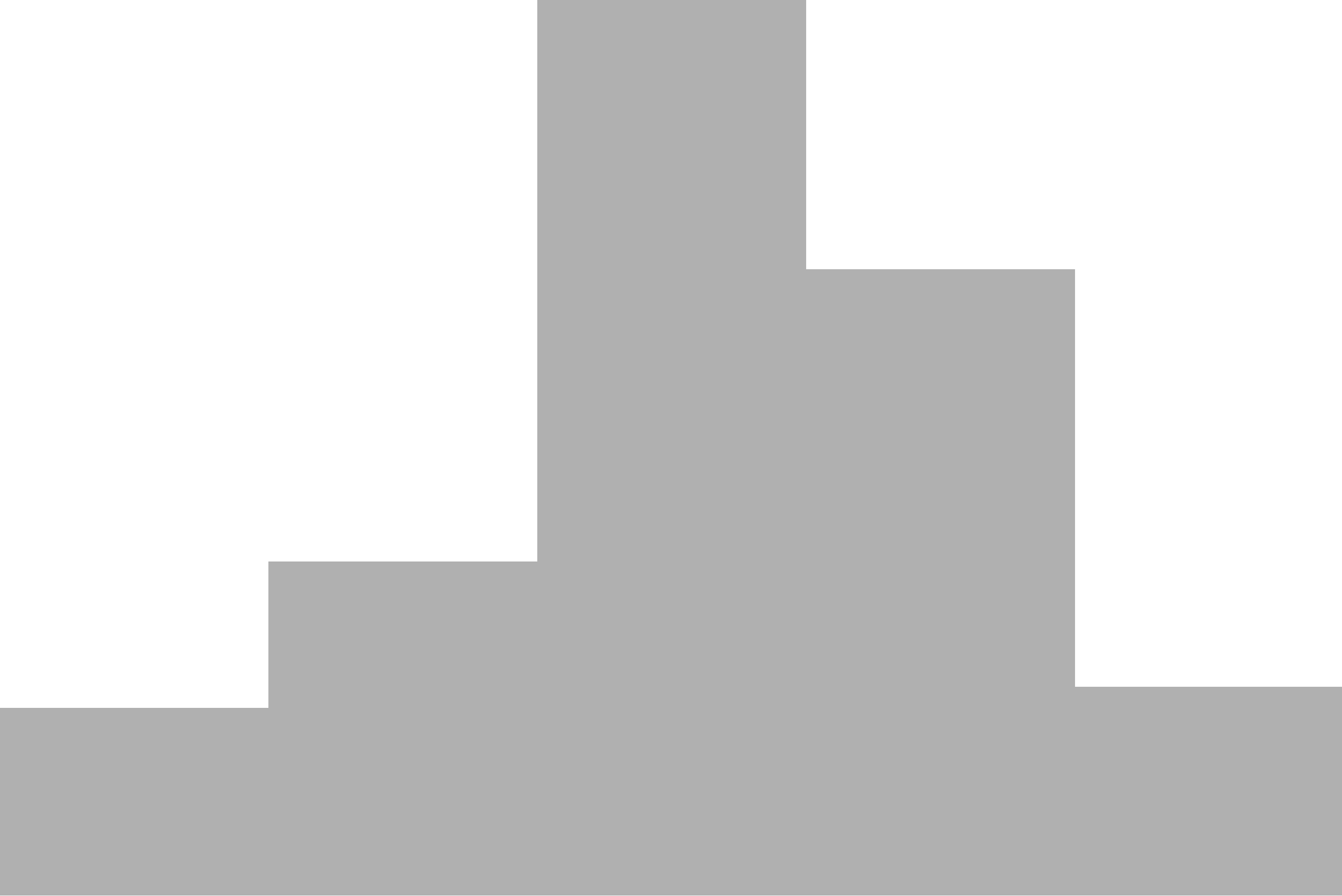}}}                           \\
P5: Have a say in making decisions \protect\faGlobe               & 3      & 2    & 3.1 & 1.1 & \multicolumn{5}{l}{\fbox{\includegraphics[width=0.13\textwidth, height=2mm]{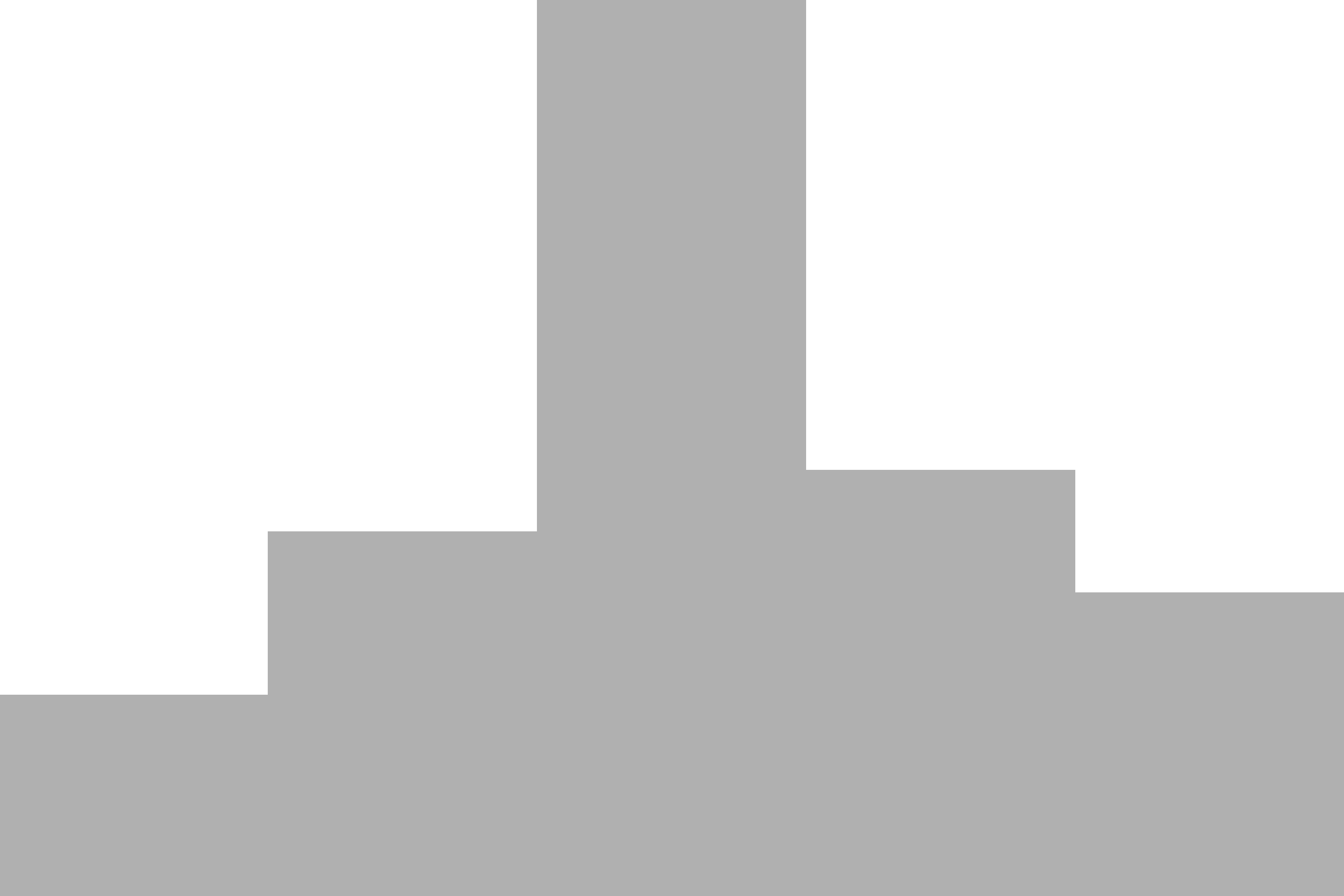}}}                           \\
P6: Consistency in applying processes 
\protect\faBriefcase*            & 3      & 2    & 3.1 & 1.2 & \multicolumn{5}{l}{\fbox{\includegraphics[width=0.13\textwidth, height=2mm]{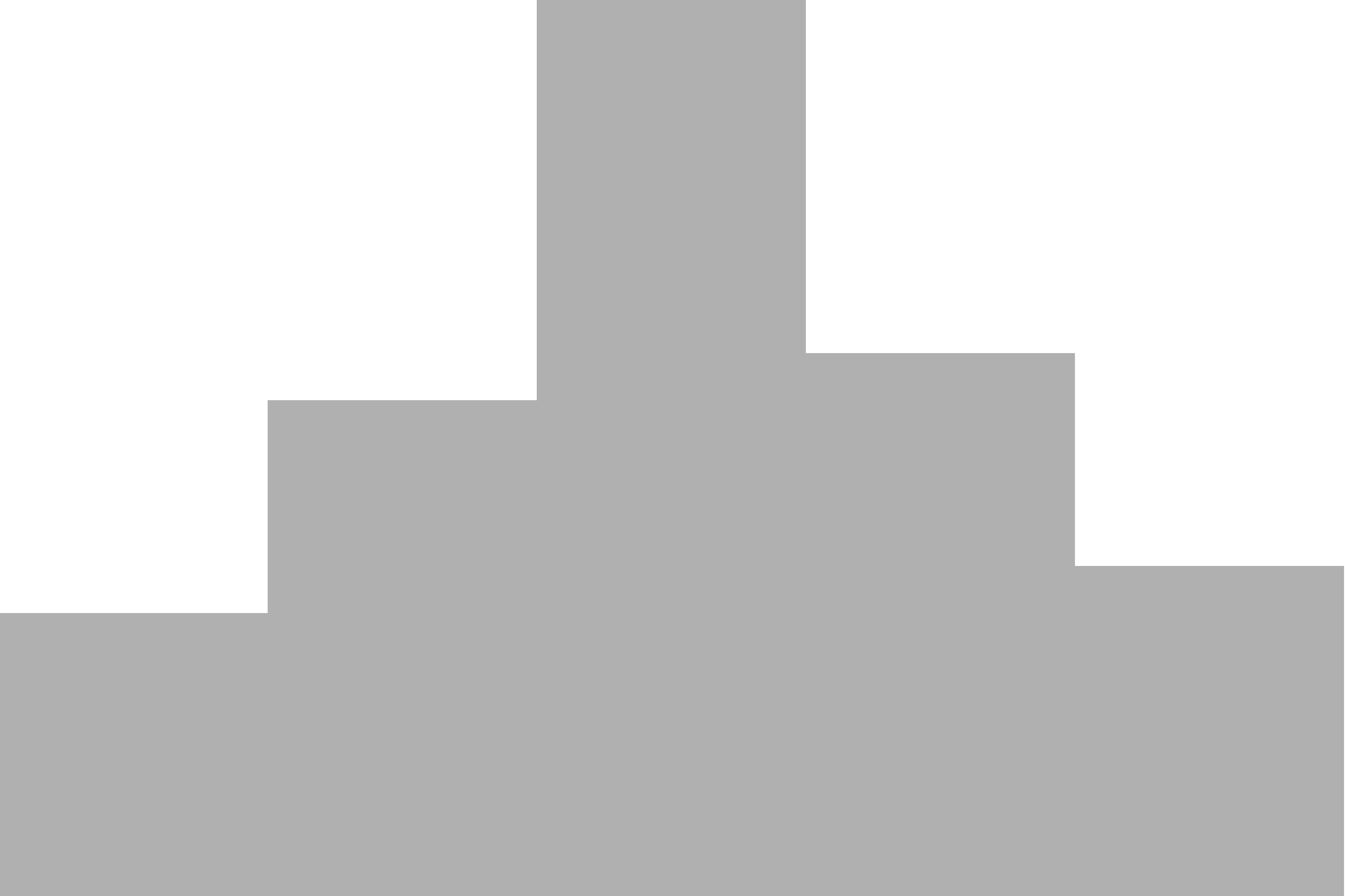}}}                           \\
P7: Appealing decisions \protect\faGlobe~ \protect\faBriefcase                          & 3      & 2    & 3 & 1.2 & \multicolumn{5}{l}{\fbox{\includegraphics[width=0.13\textwidth, height=2mm]{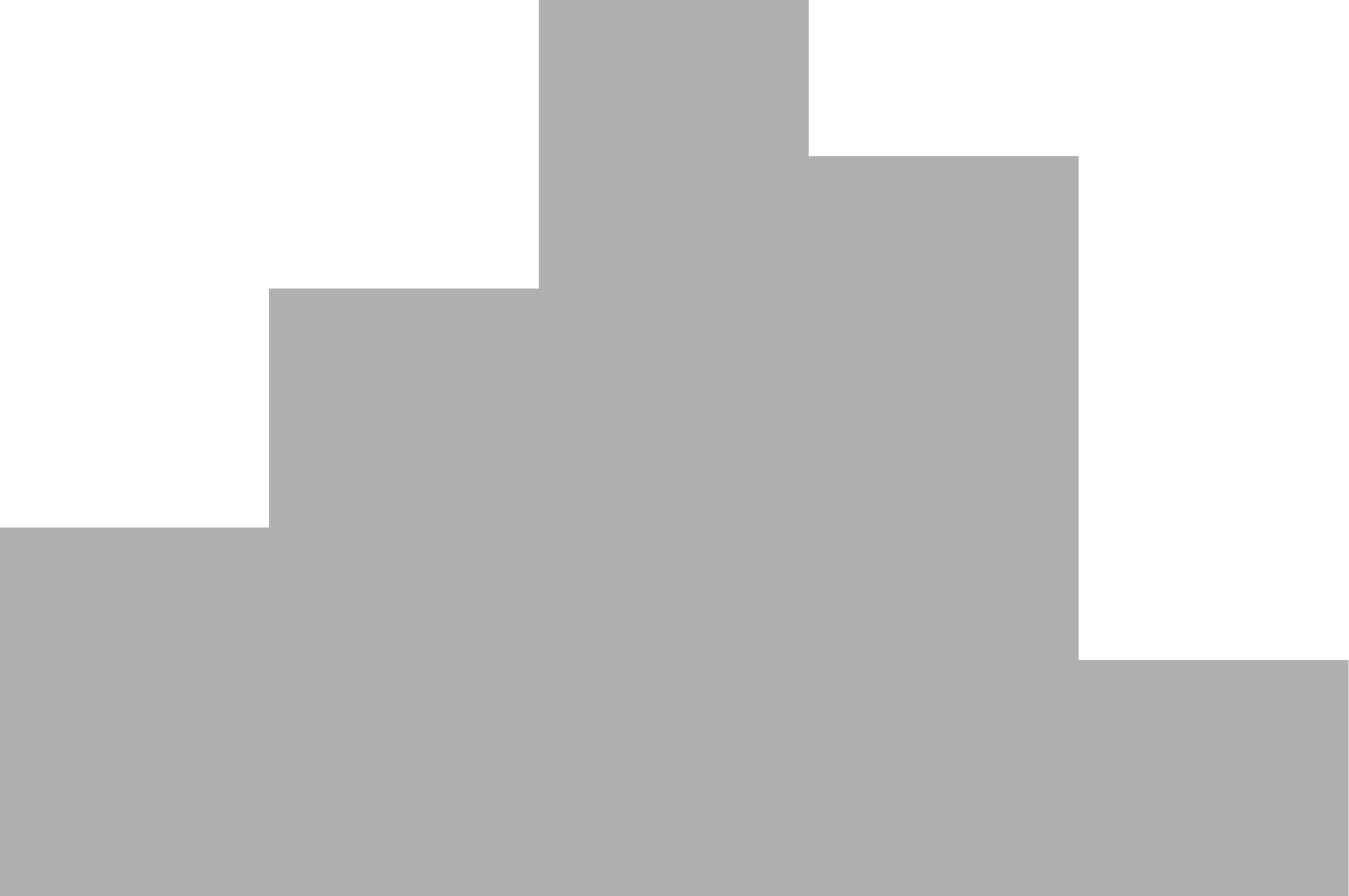}}}                           \\ 
\rowcolor[HTML]{EFEFEF}\multicolumn{1}{l}{Int: Interpersonal}             & 4      & 1.5  & 4.0  & 1 & \multicolumn{5}{l}{\fbox{\includegraphics[width=0.13\textwidth, height=2mm]{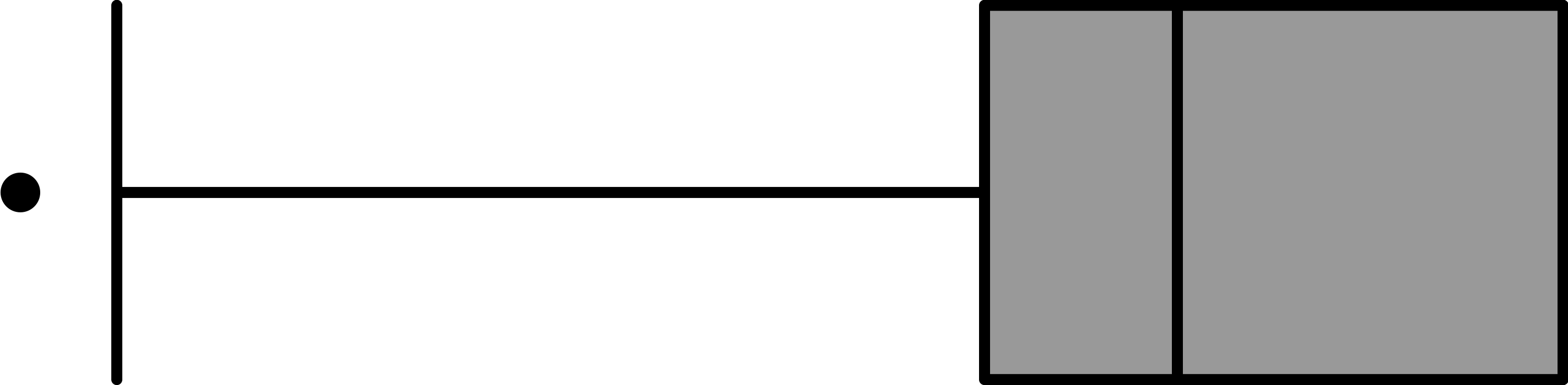}}}                           \\ 
Int1: Treated politely 
& 5      & 1    & 4.3 & 1 & \multicolumn{5}{l}{\fbox{\includegraphics[width=0.13\textwidth, height=2mm]{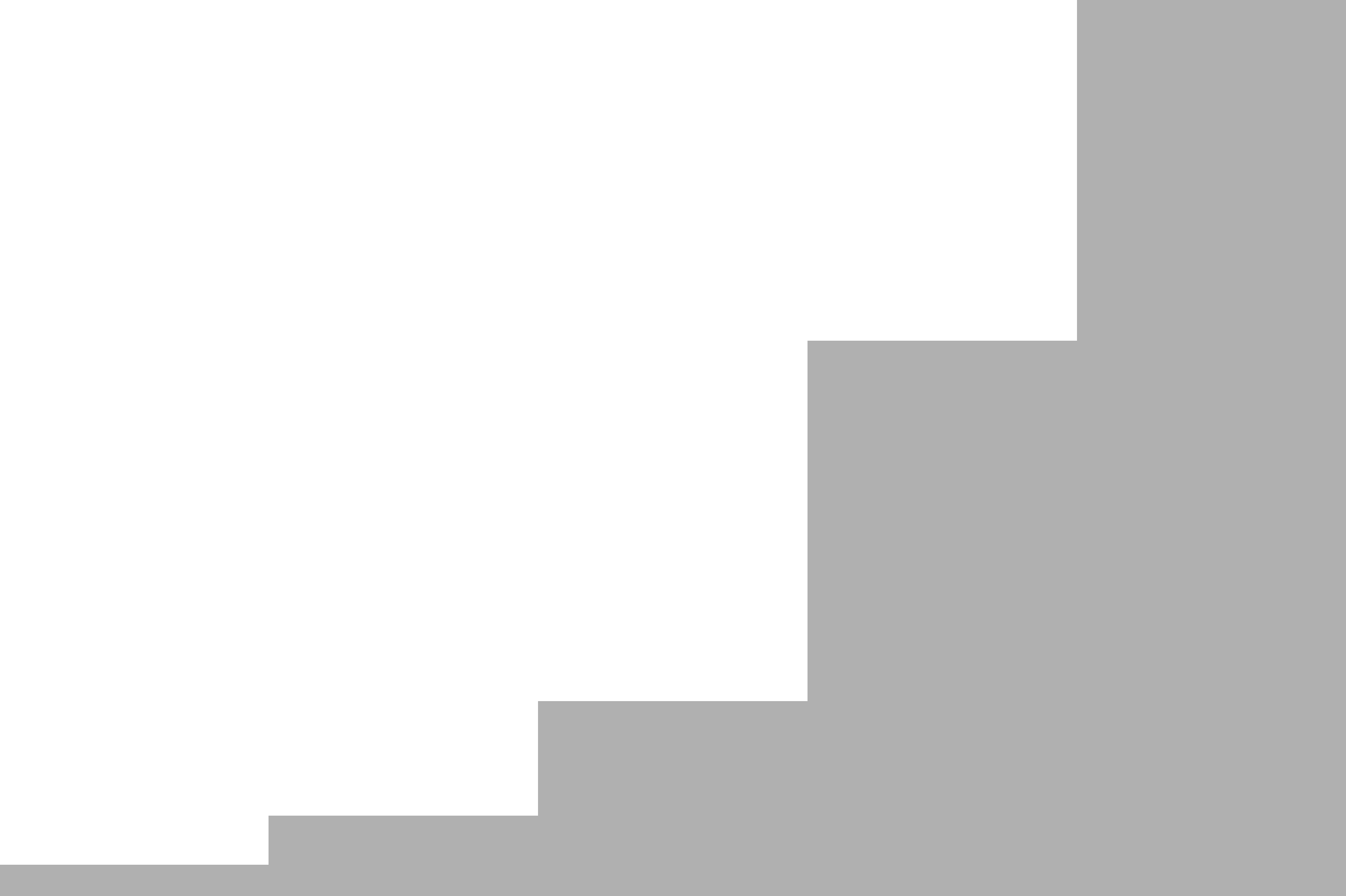}}}                           \\
Int2: Treated with respect 
& 4      & 1    & 4.2 & 1 & \multicolumn{5}{l}{\fbox{\includegraphics[width=0.13\textwidth, height=2mm]{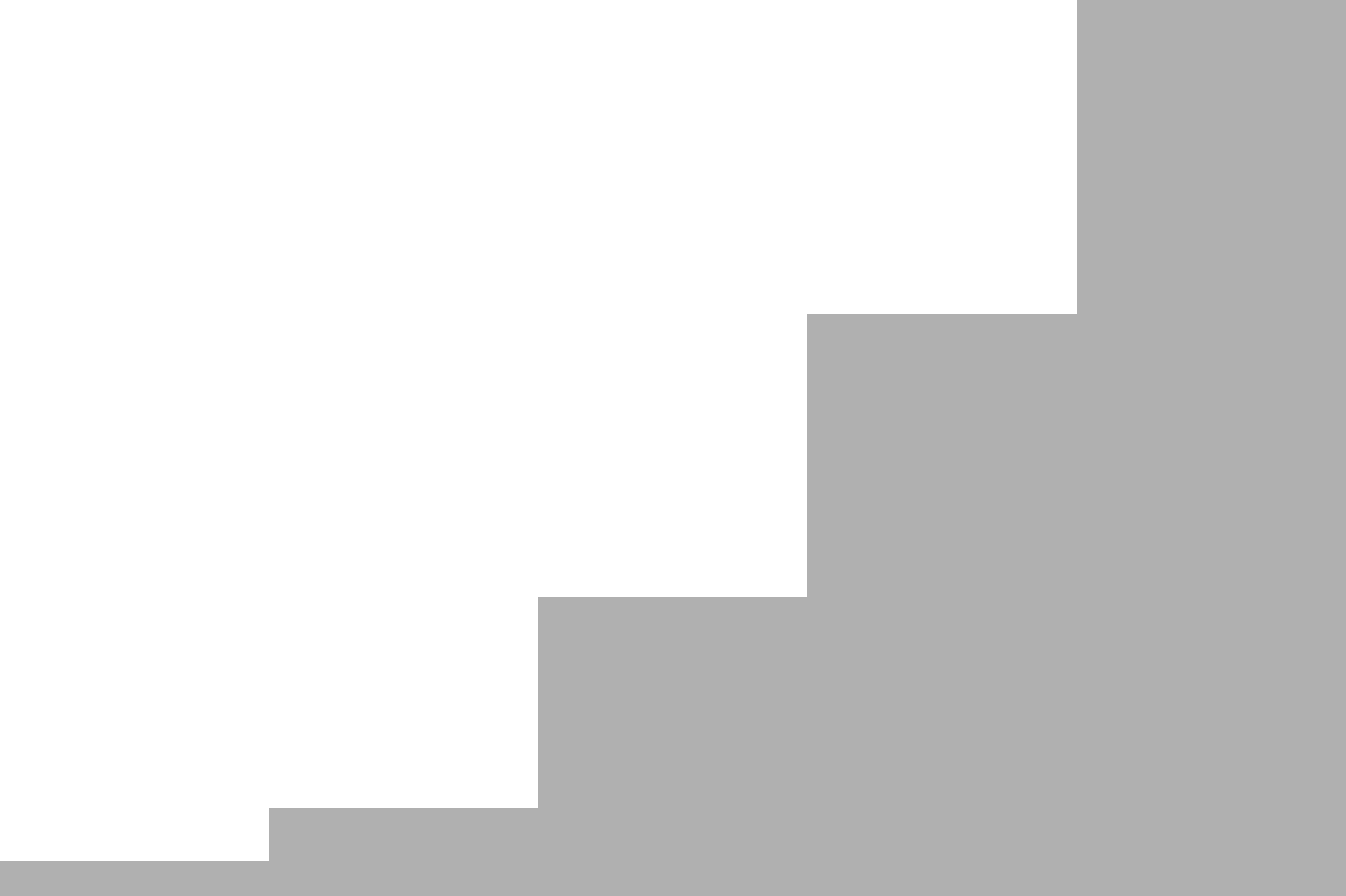}}}                           \\
Int3: Treated with dignity 
& 4      & 2    & 4.1 & 1.1 & \multicolumn{5}{l}{\fbox{\includegraphics[width=0.13\textwidth, height=2mm]{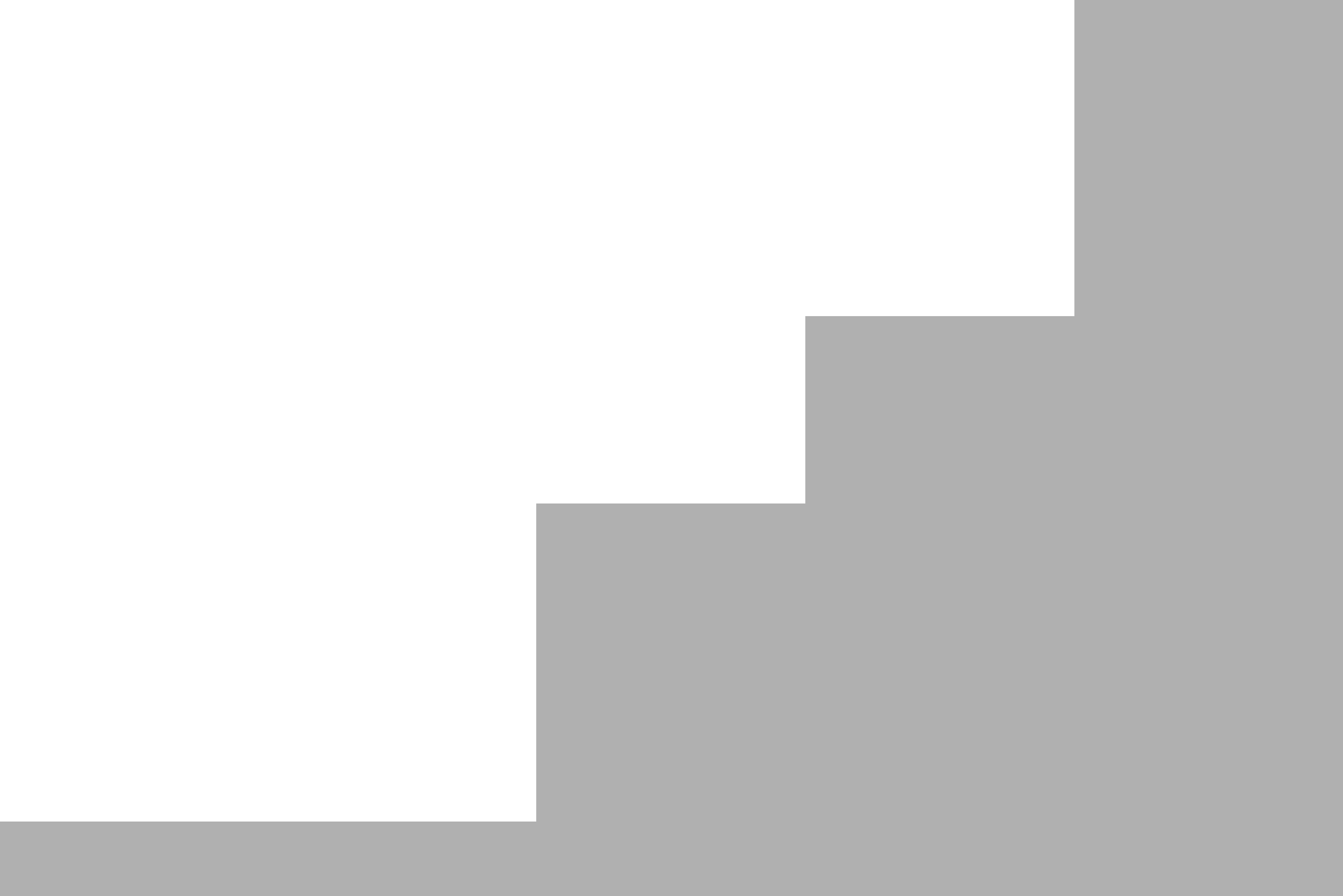}}}                           \\
Int4: No inappropriate comments or behaviour 
& 4      & 2    & 3.7 & 1.3 & \multicolumn{5}{l}{\fbox{\includegraphics[width=0.13\textwidth, height=2mm]{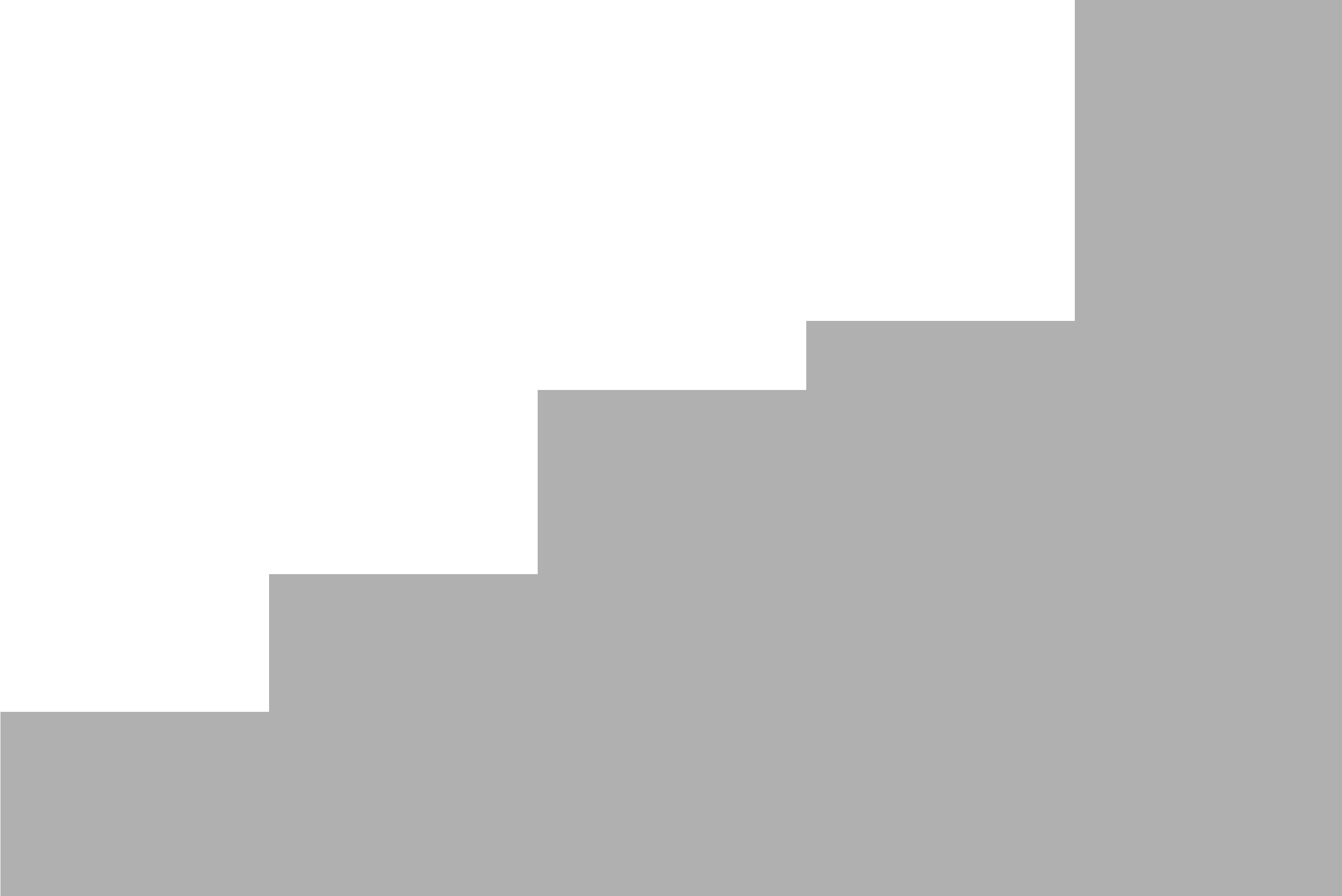}}}                           \\ 
\rowcolor[HTML]{EFEFEF}\multicolumn{1}{l}{Inf: Informational}             & 3.4    & 1    & 3.4 & 0.9 & \multicolumn{5}{l}{\fbox{\includegraphics[width=0.13\textwidth, height=2mm]{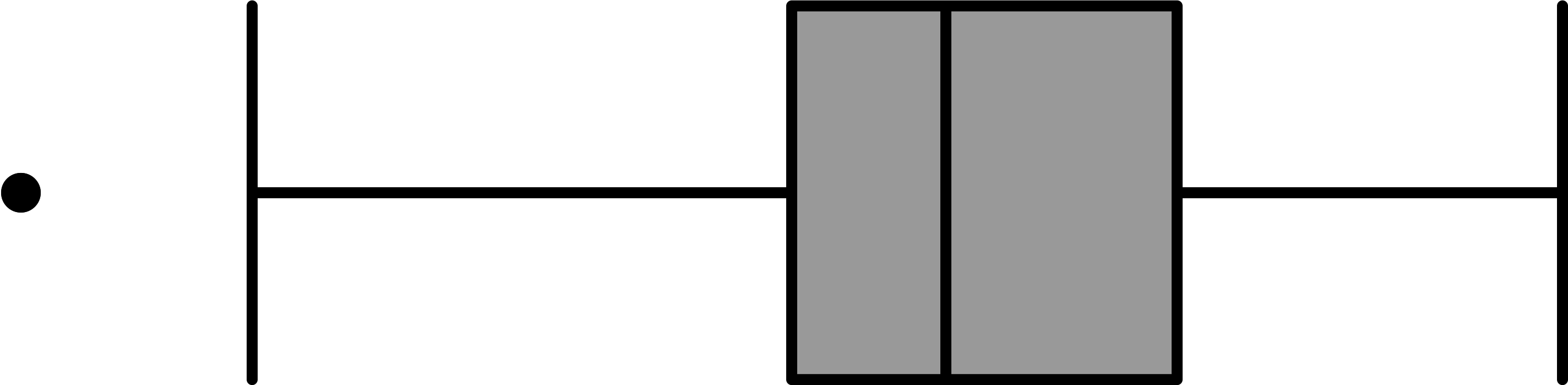}}}                           \\ 
Inf1: Direct communication                          & 4      & 1    & 3.6 & 1.1 & \multicolumn{5}{l}{\fbox{\includegraphics[width=0.13\textwidth, height=2mm]{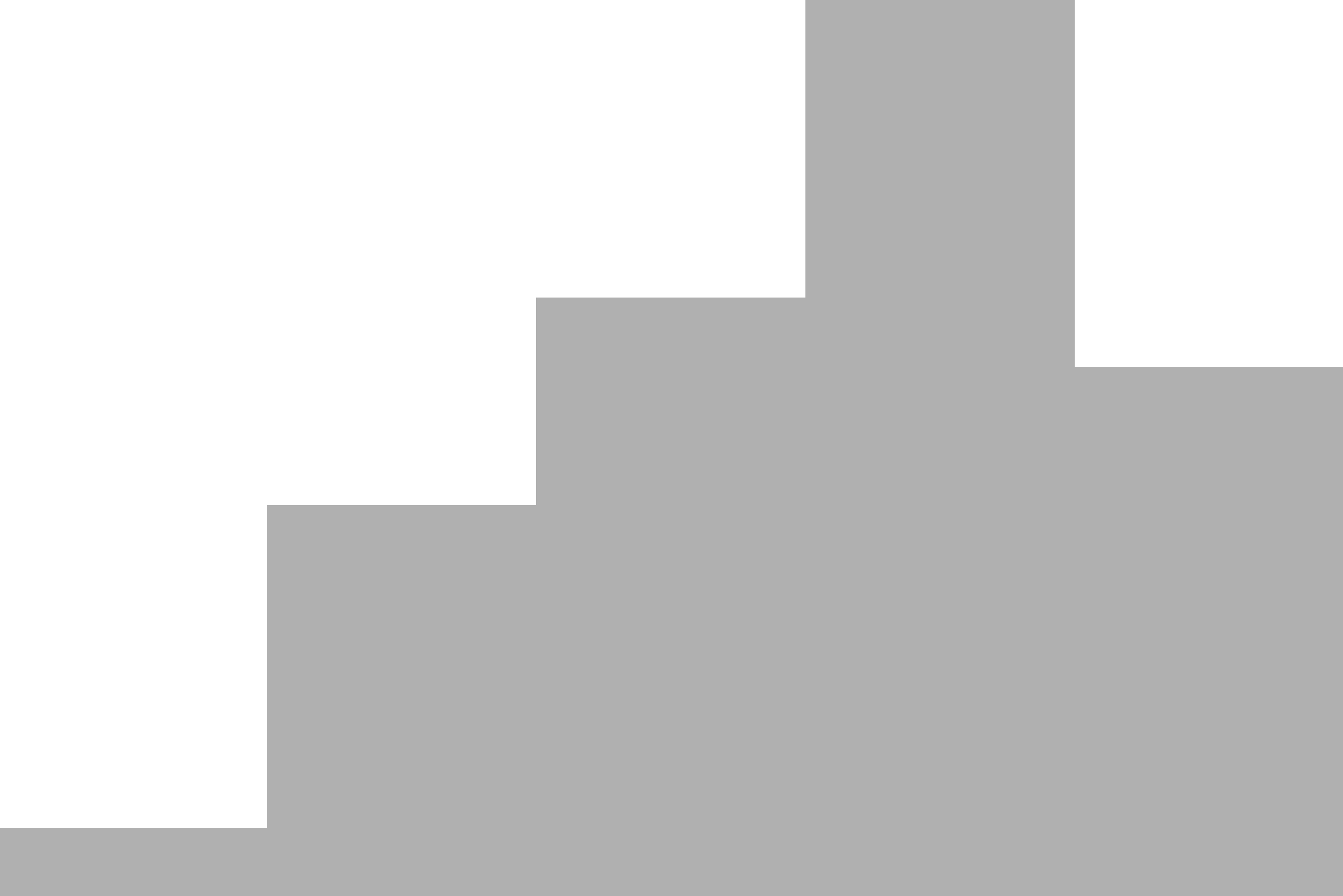}}}                           \\
Inf2: Reasonable explanations                       & 4      & 1    & 3.4 & 1.1 & \multicolumn{5}{l}{\fbox{\includegraphics[width=0.13\textwidth, height=2mm]{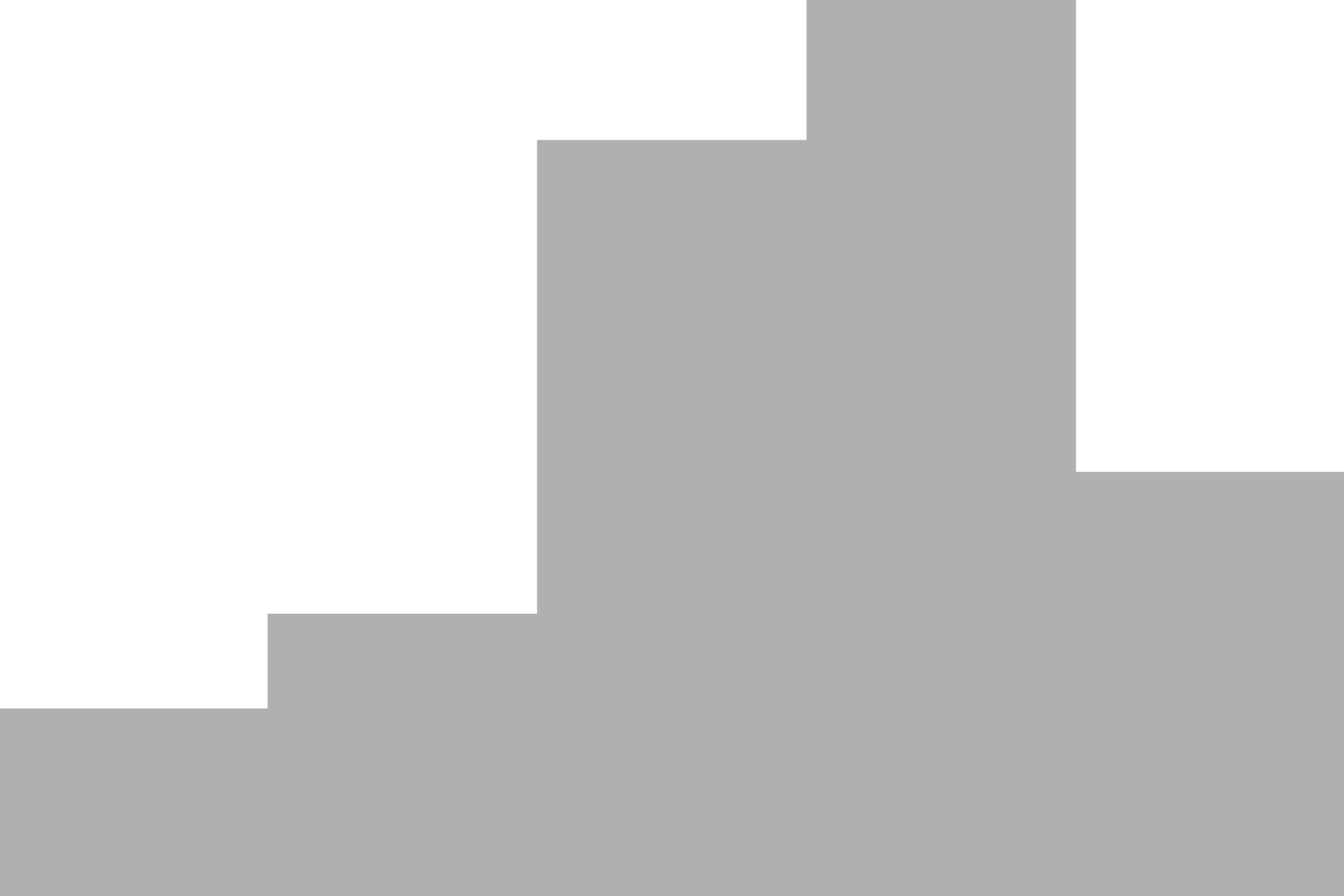}}}                           \\
Inf3: Timely communication                          & 3      & 1    & 3.3 & 1.2 & \multicolumn{5}{l}{\fbox{\includegraphics[width=0.13\textwidth, height=2mm]{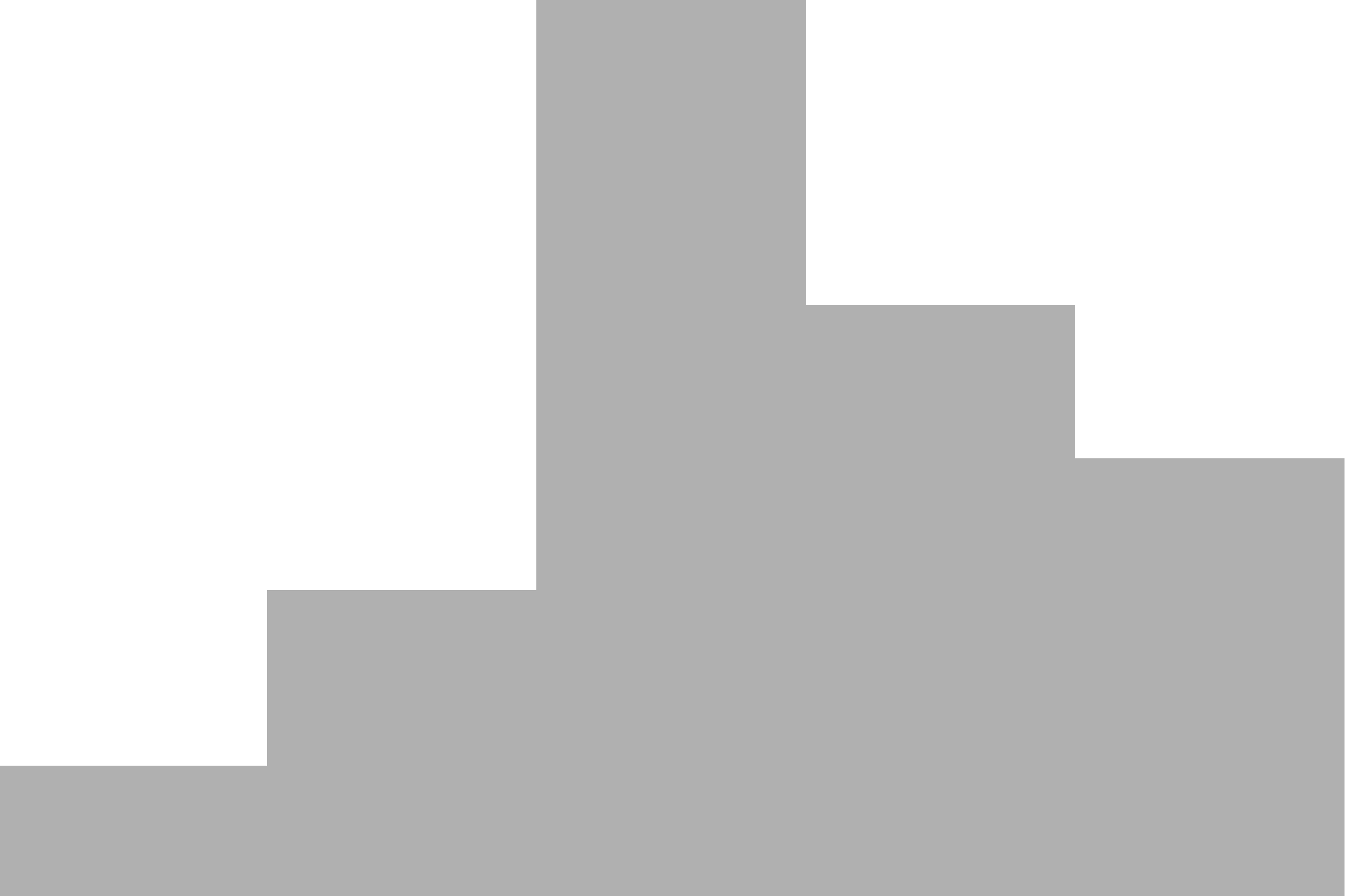}}}                           \\
Inf4: Tailored communication                        & 3      & 1    & 3.3 & 1.1 & \multicolumn{5}{l}{\fbox{\includegraphics[width=0.13\textwidth, height=2mm]{Informational_5.png}}}                           \\
Inf5: Comprehensive explanations                    & 3      & 1    & 3.3 & 1.2 & \multicolumn{5}{l}{\fbox{\includegraphics[width=0.13\textwidth, height=2mm]{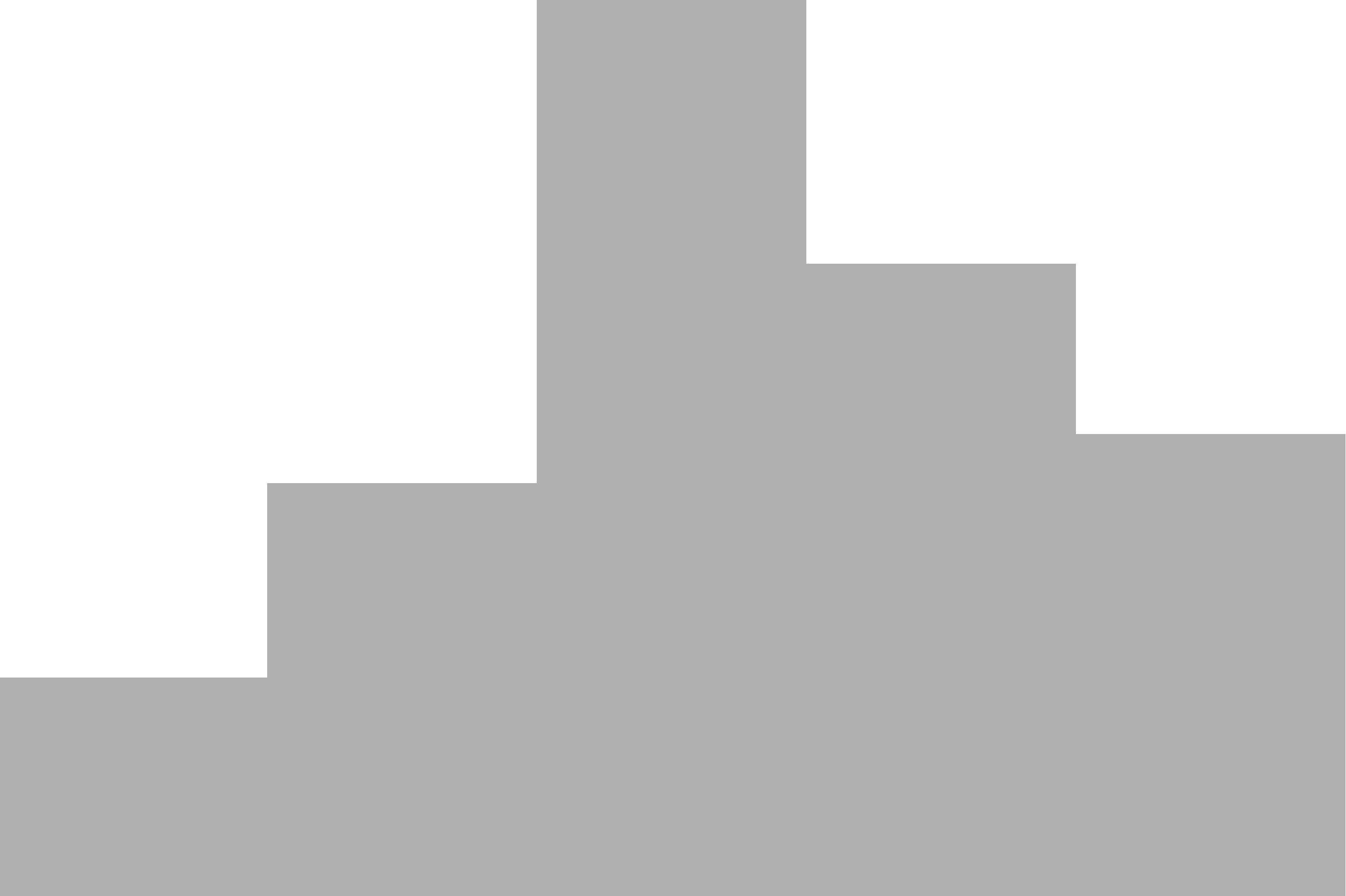}}}                           \\
\rowcolor[HTML]{DFDFDF}\multicolumn{10}{c}{\textit{\textbf{Job Satisfaction}}} \\ 
S1: Overall 
& 4                       & 1                    & 3.5                 & 1               & \multicolumn{5}{l}{\fbox{\includegraphics[width=0.13\textwidth, height=2mm]{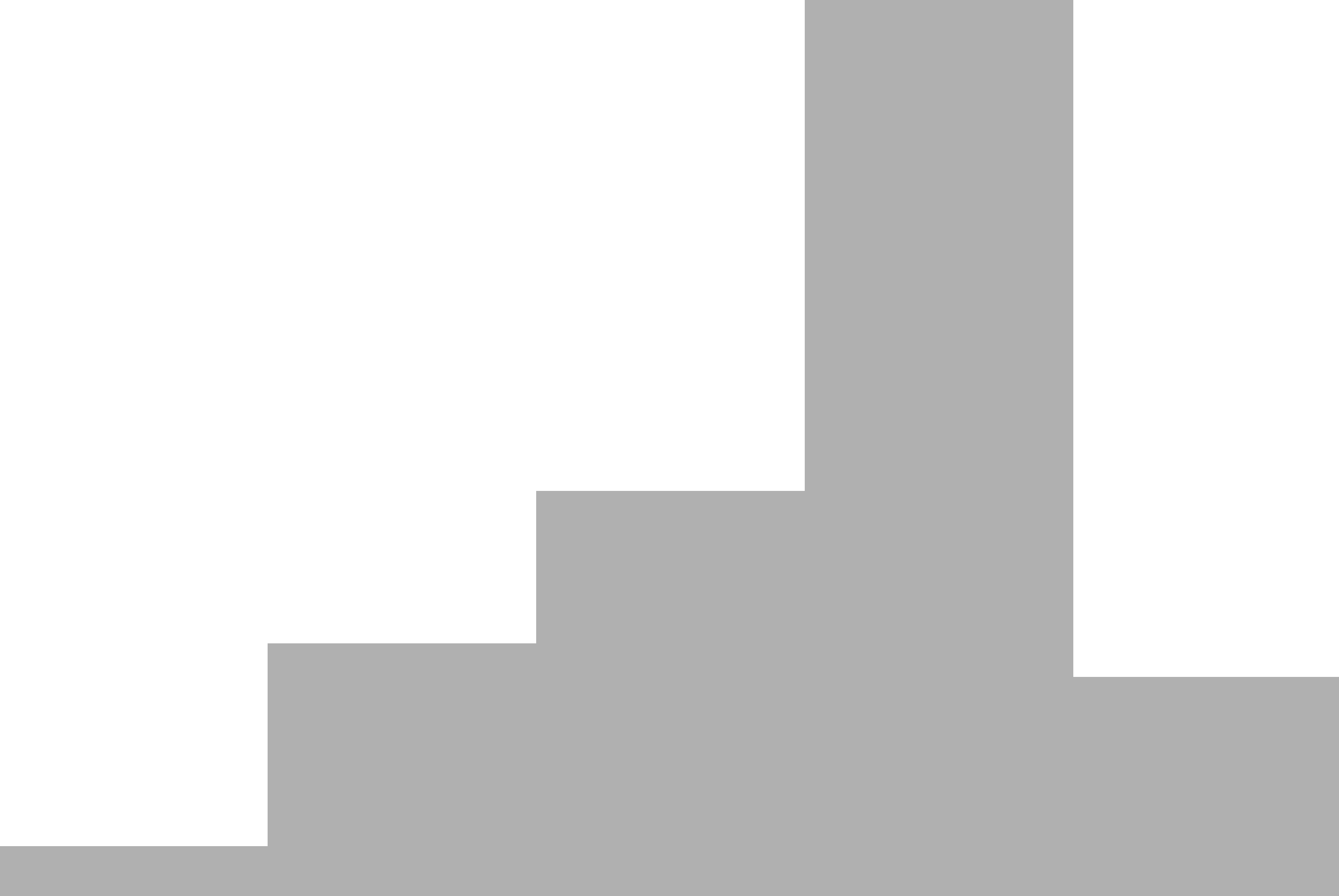}}}                  \\ 
S2: Collaboration 
\protect\faClockO 
& 5                       & 1                    & 4.2                 & 1.2               & \multicolumn{5}{l}{\fbox{\includegraphics[width=0.13\textwidth, height=2mm]{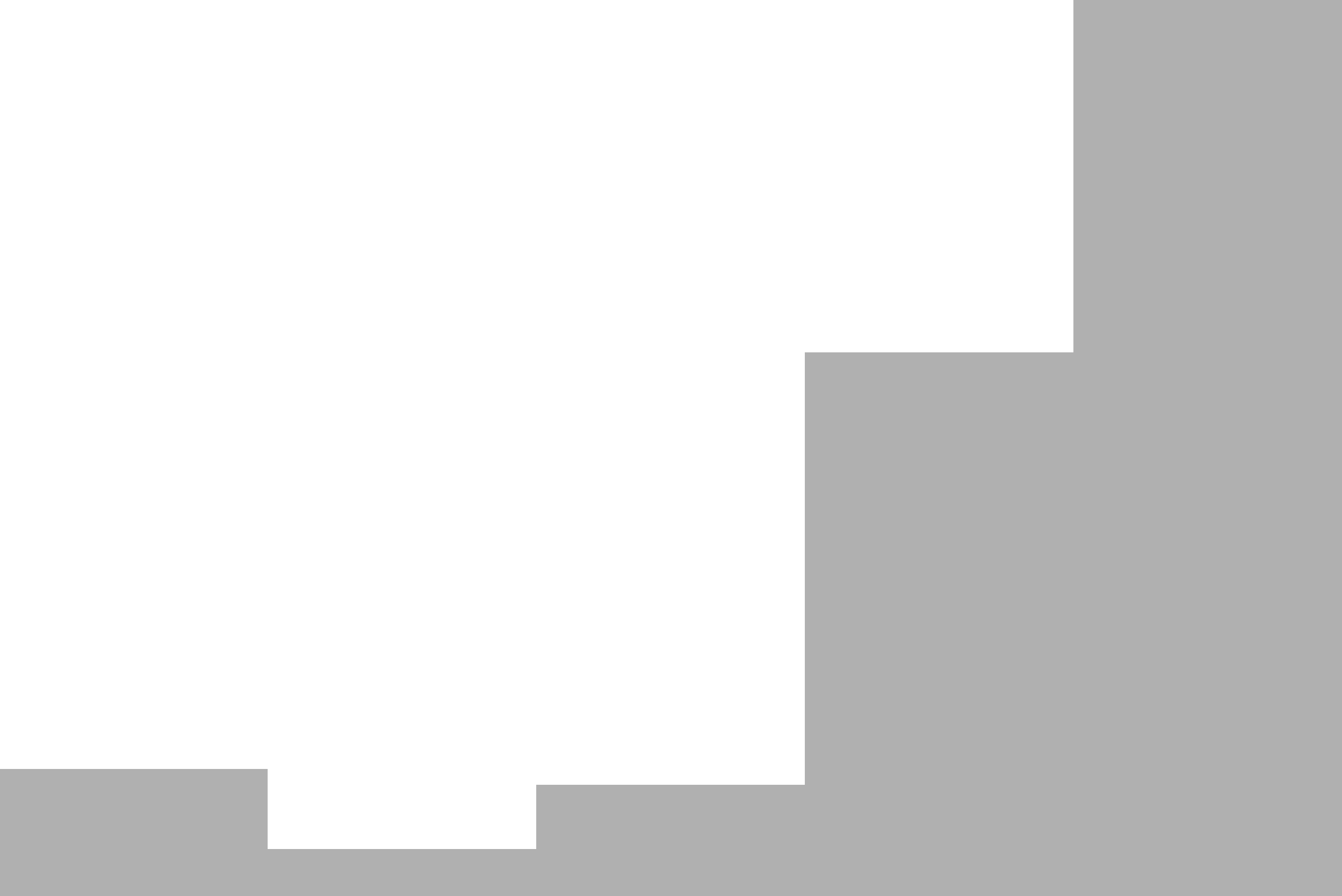}}}                  \\
S3: Ability 
& 4                       & 1                    & 4.1                 & 1               & \multicolumn{5}{l}{\fbox{\includegraphics[width=0.13\textwidth, height=2mm]{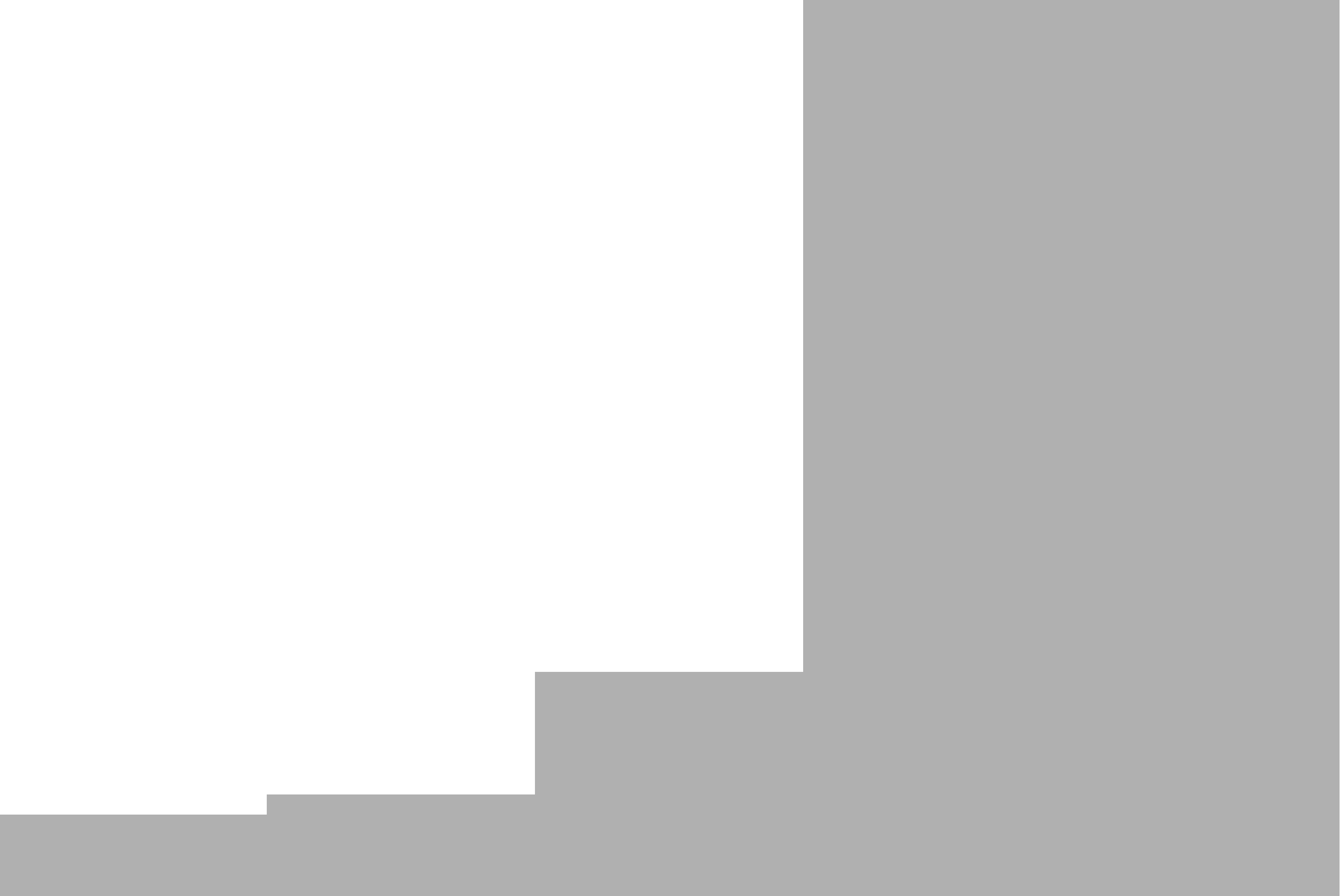}}}                  \\
S4: Productivity 
& 4                       & 1.3                 & 4                 & 1.1               & \multicolumn{5}{l}{\fbox{\includegraphics[width=0.13\textwidth, height=2mm]{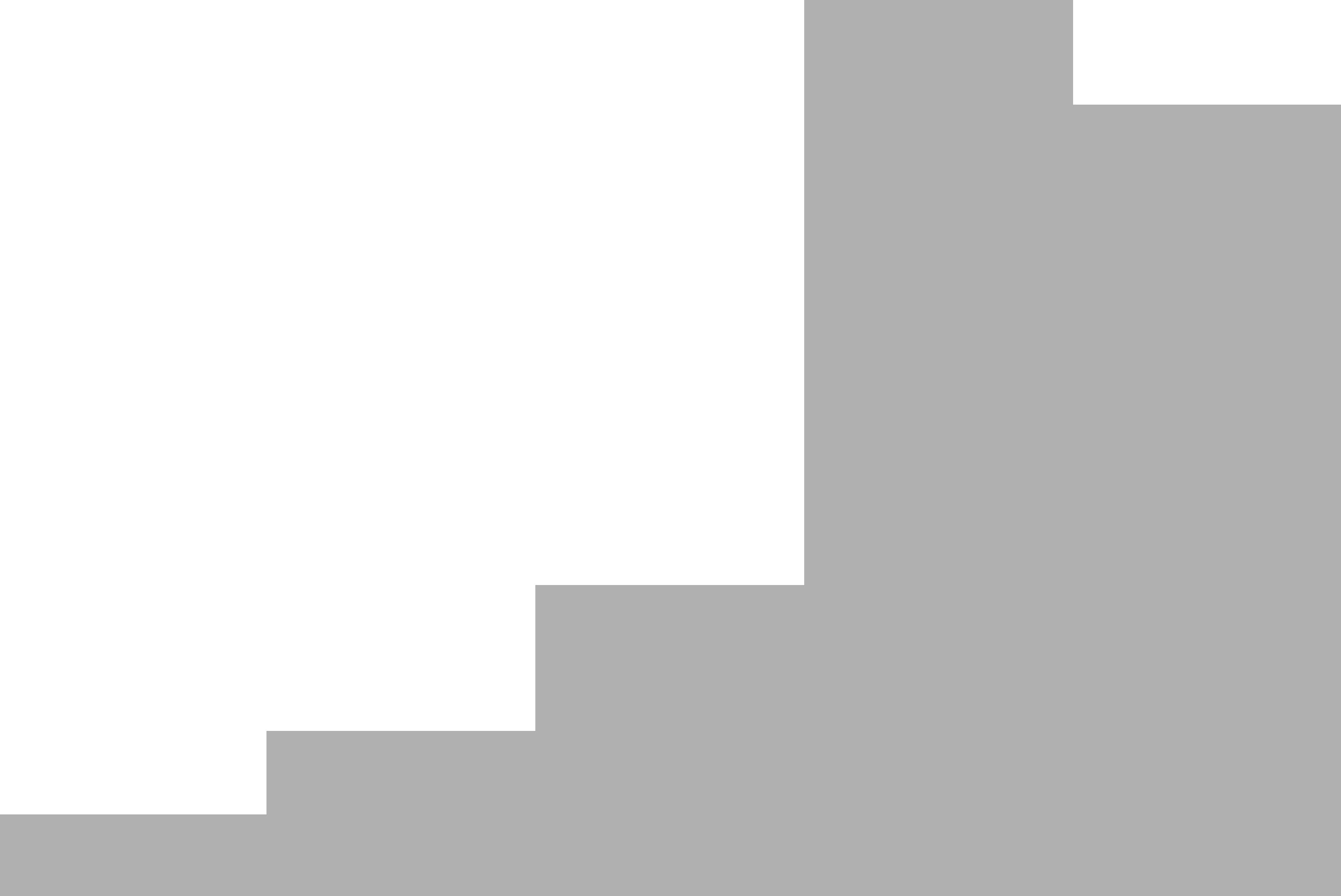}}}                  \\
S5: Team Culture                                      & 4                       & 2                    & 3.9                 & 1.2               & \multicolumn{5}{l}{\fbox{\includegraphics[width=0.13\textwidth, height=2mm]{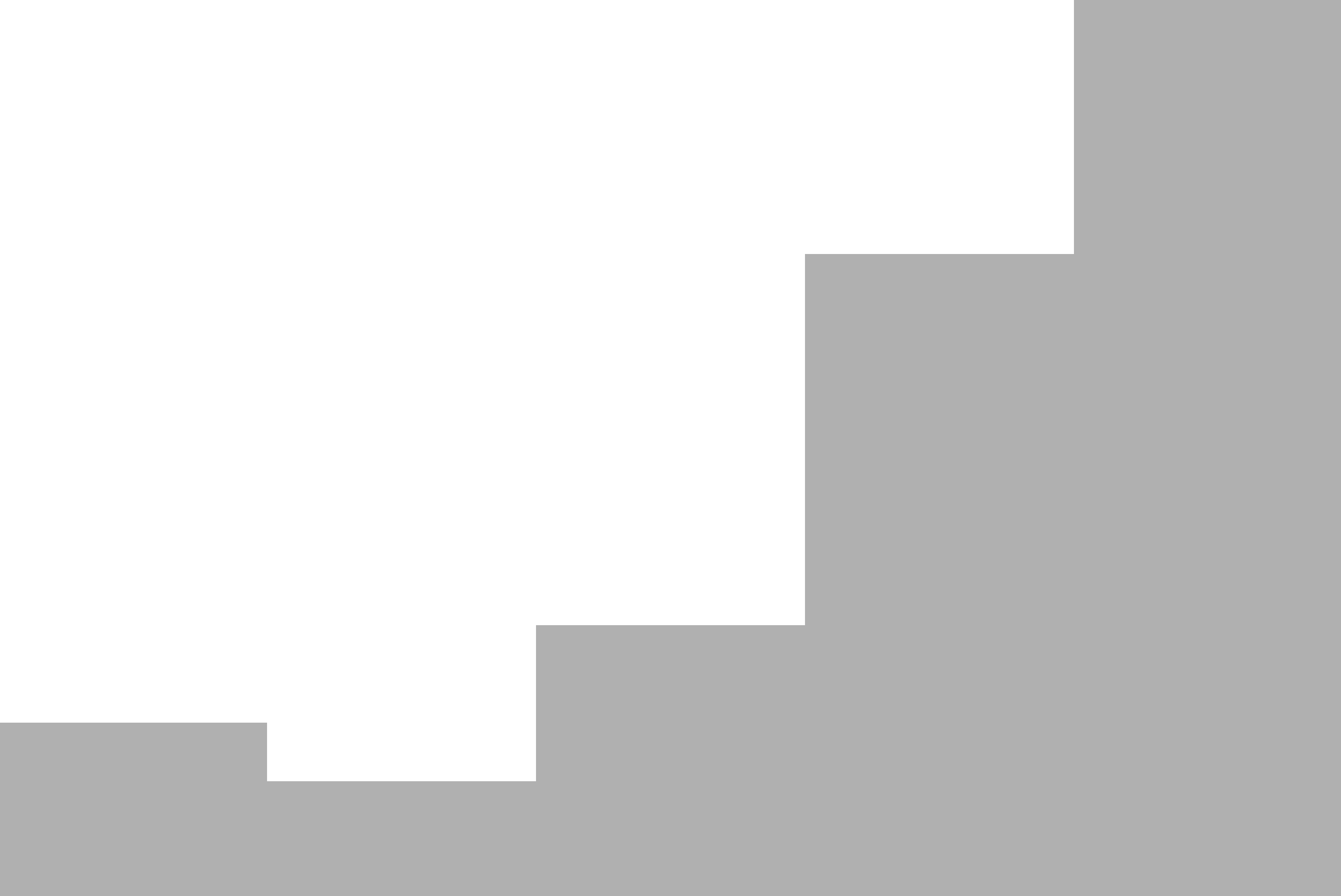}}}                  \\
S6: Manager                                           & 4                       & 2                    & 3.9                 & 1.3               & \multicolumn{5}{l}{\fbox{\includegraphics[width=0.13\textwidth, height=2mm]{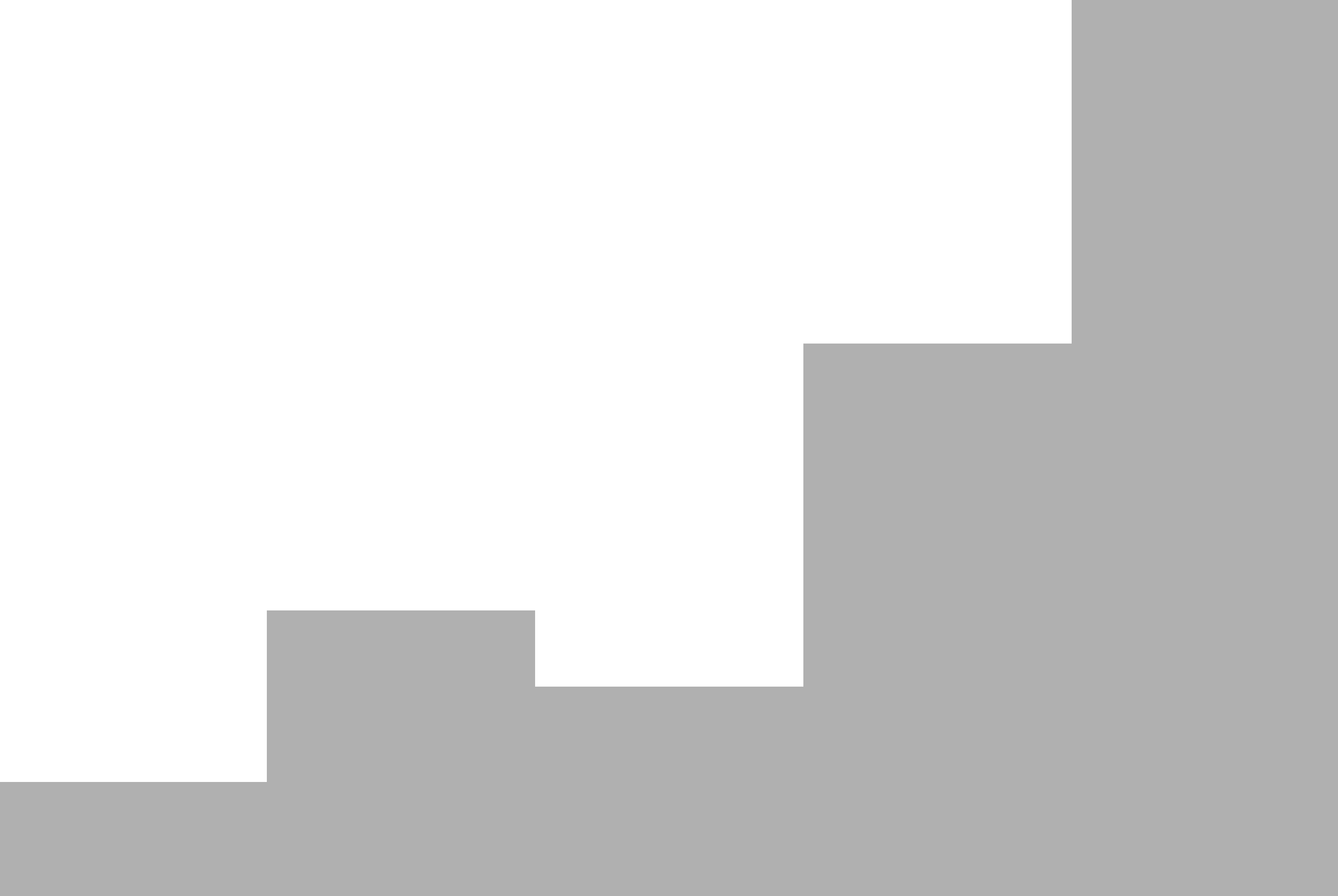}}}                  \\
S7: Feedback 
& 4                       & 2                    & 3.8                 & 1.2               & \multicolumn{5}{l}{\fbox{\includegraphics[width=0.13\textwidth, height=2mm]{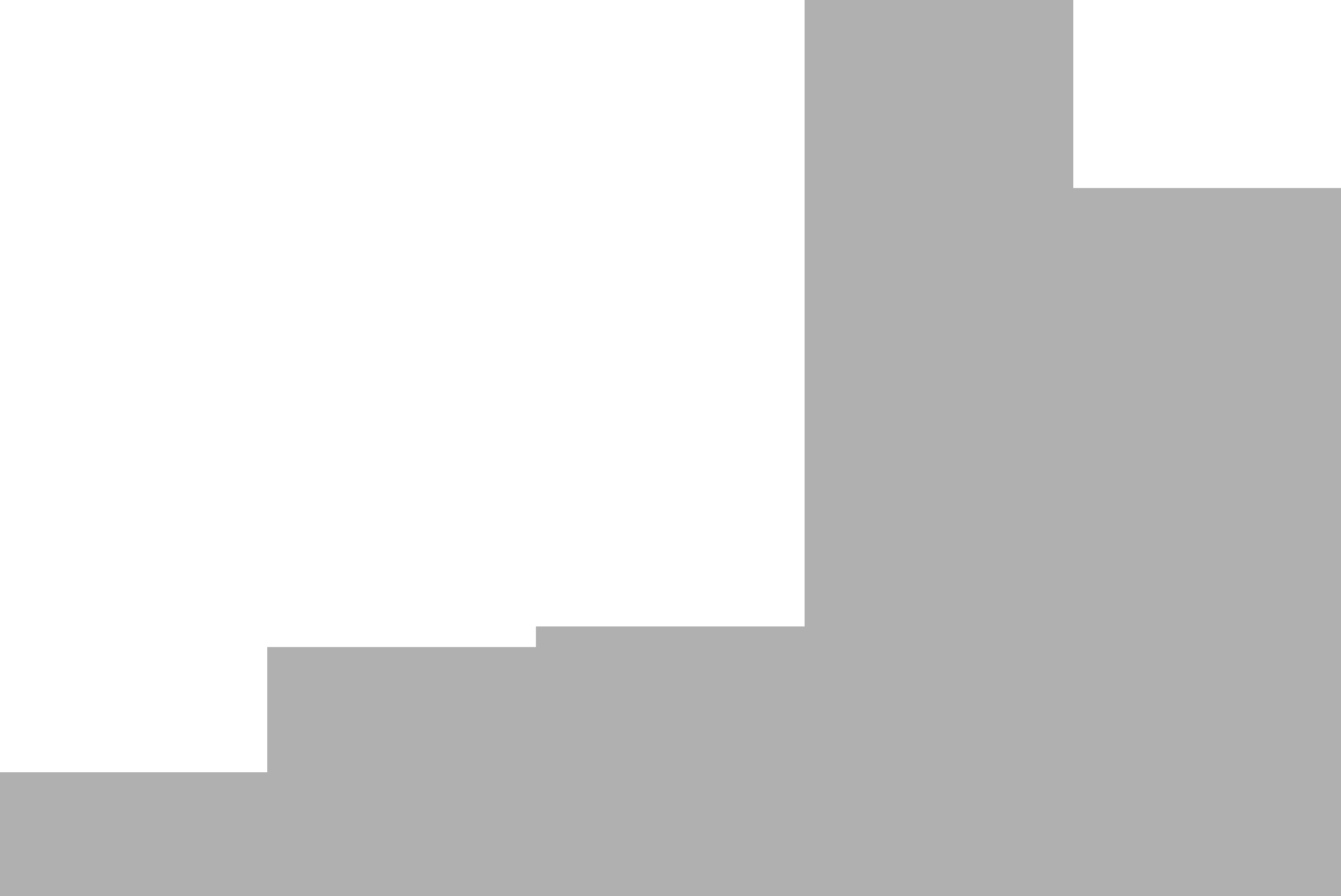}}}                  \\
S8: Benefits \protect\faBriefcase                                          & 4                       & 2                    & 3.7                 & 1.2               & \multicolumn{5}{l}{\fbox{\includegraphics[width=0.13\textwidth, height=2mm]{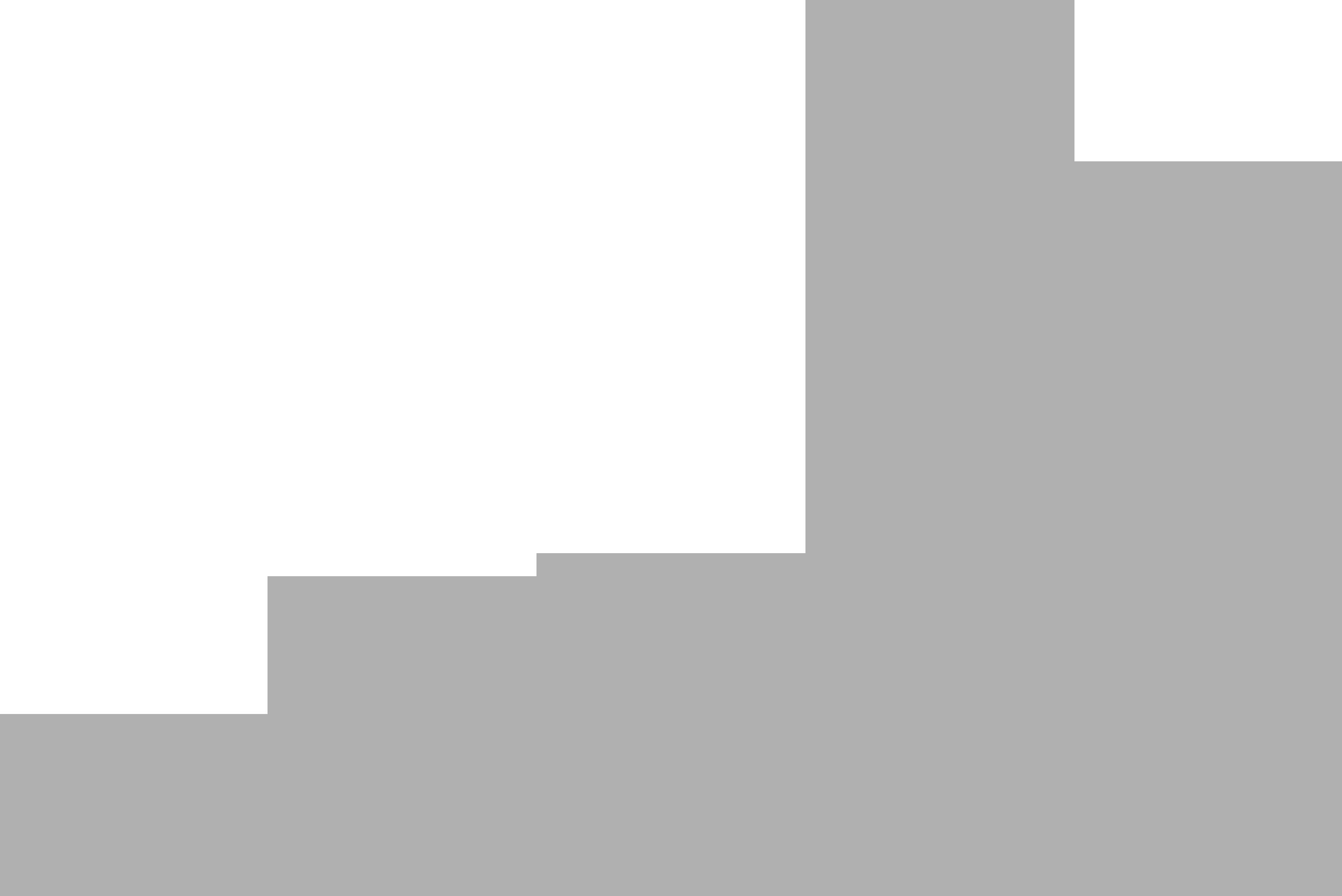}}}                  \\
S9: Appreciation 
& 4                       & 2                    & 3.7                 & 1.2               & \multicolumn{5}{l}{\fbox{\includegraphics[width=0.13\textwidth, height=2mm]{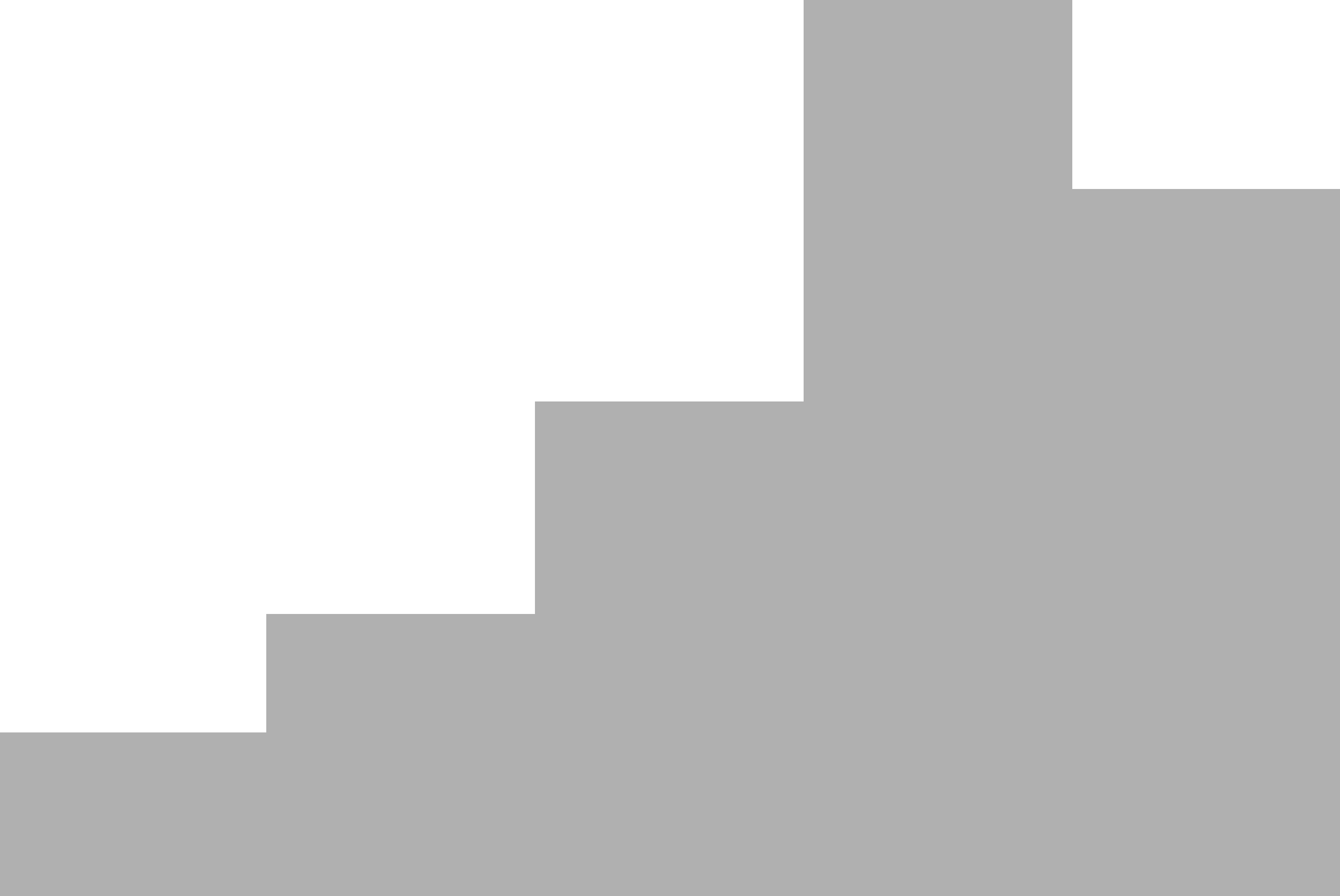}}}                  \\
S10: Priority                                          & 4                       & 1                    & 3.7                 & 1.2               & \multicolumn{5}{l}{\fbox{\includegraphics[width=0.13\textwidth, height=2mm]{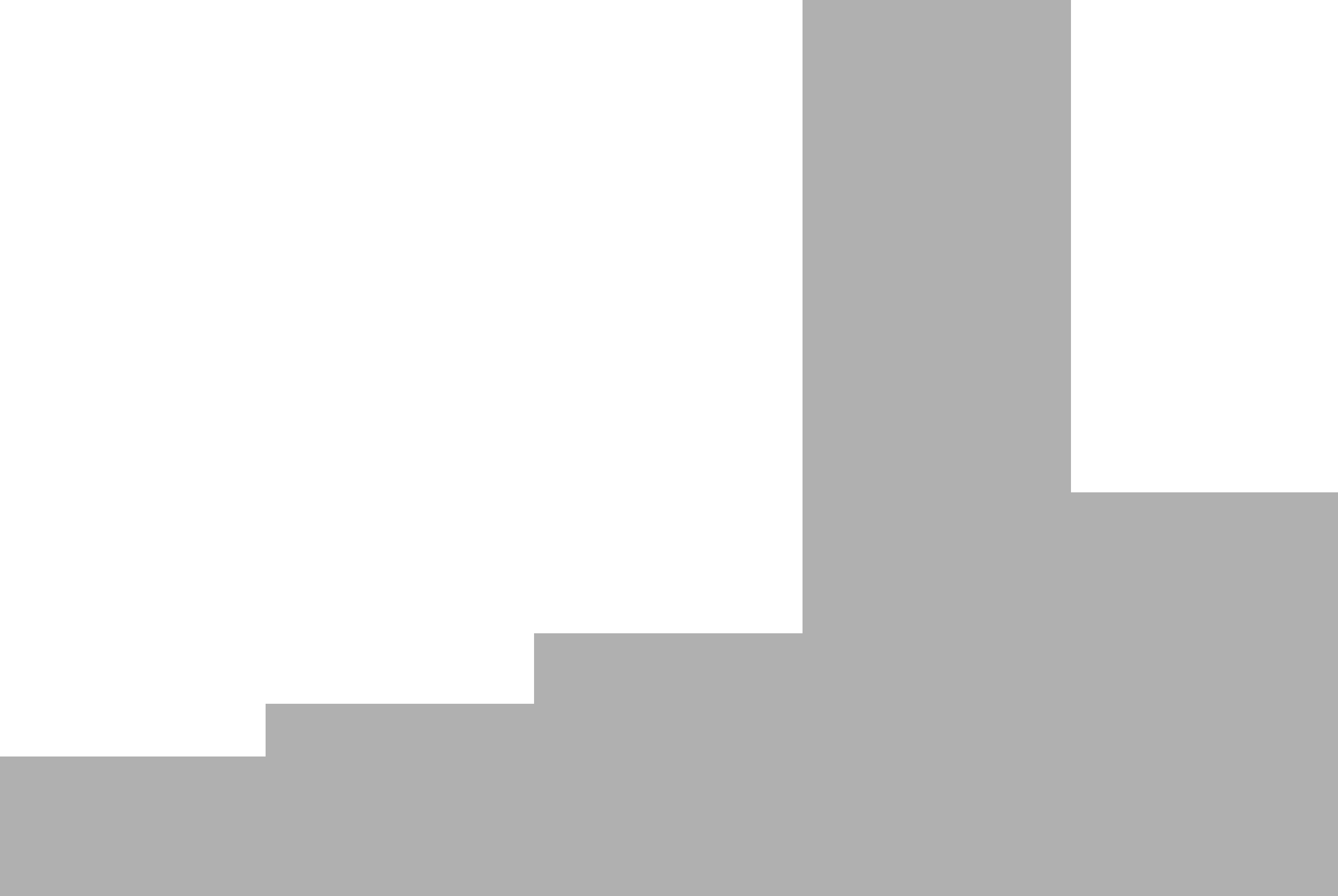}}}                  \\
S11: Work-Life Balance 
& 4                       & 2                    & 3.6                 & 1.4               & \multicolumn{5}{l}{\fbox{\includegraphics[width=0.13\textwidth, height=2mm]{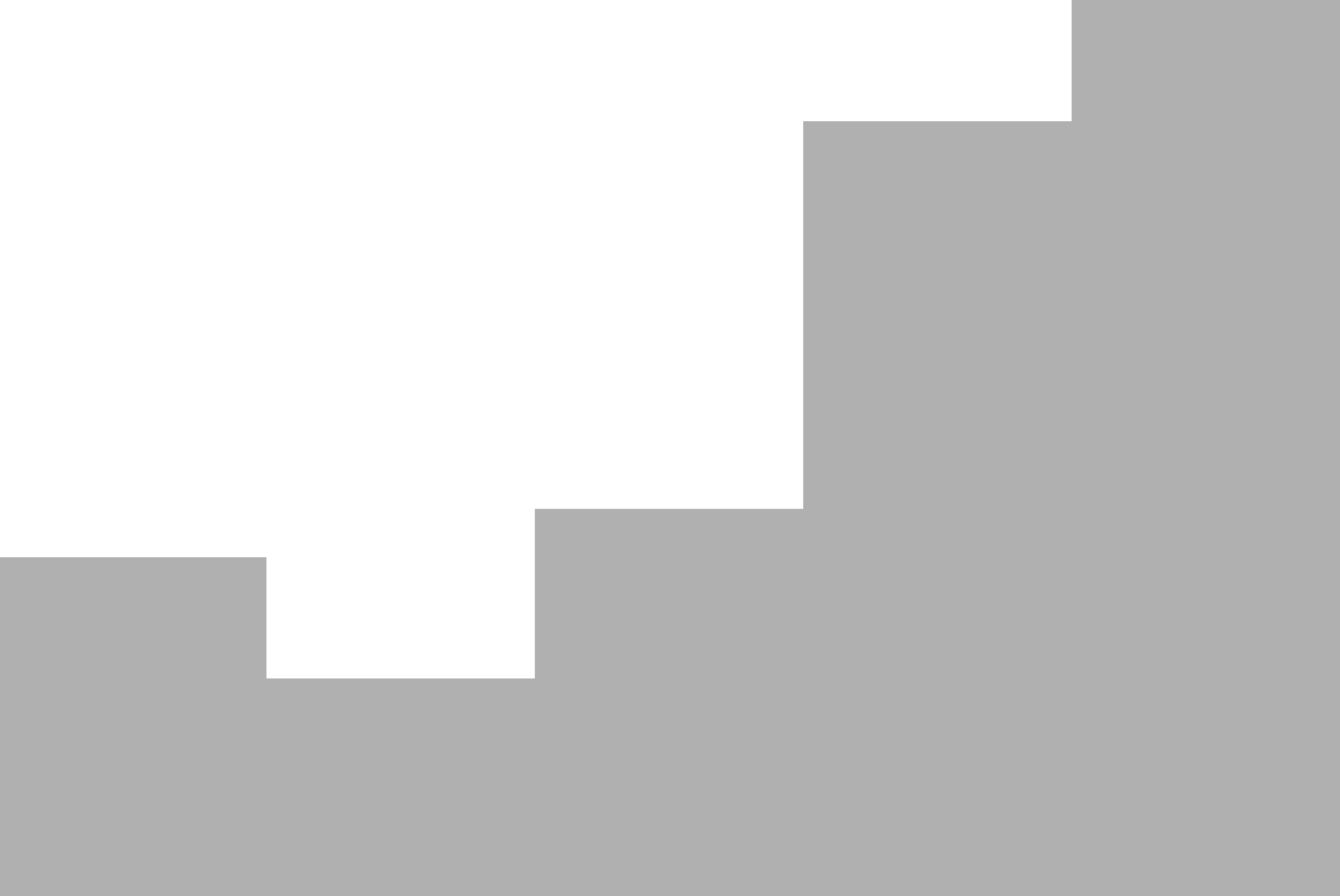}}}                  \\
S12: Organization Culture                              & 4                       & 2                    & 3.6                 & 1.2               & \multicolumn{5}{l}{\fbox{\includegraphics[width=0.13\textwidth, height=2mm]{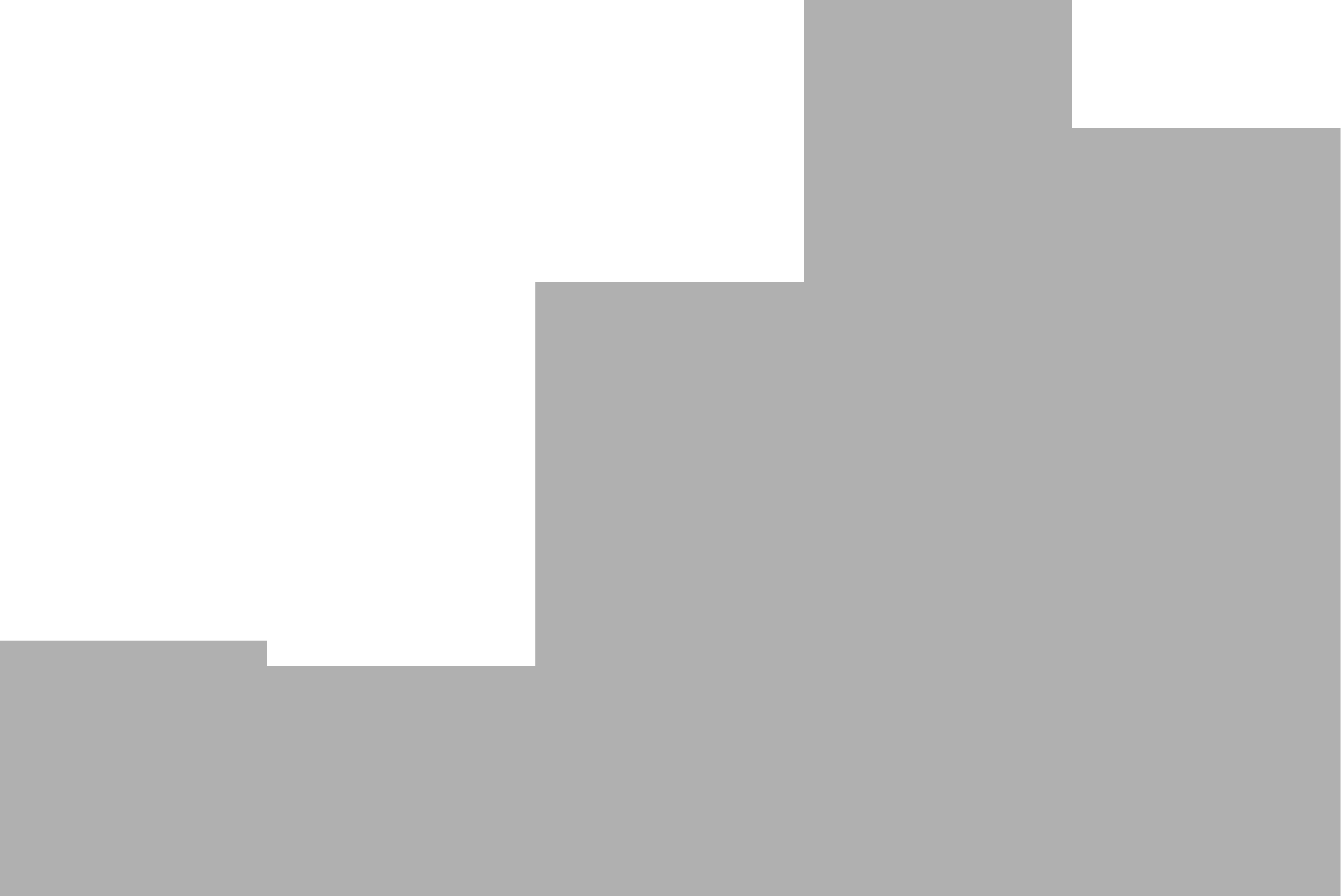}}}                  \\
S13: Rewards                                           & 4                       & 1                    & 3.5                 & 1.2               & \multicolumn{5}{l}{\fbox{\includegraphics[width=0.13\textwidth, height=2mm]{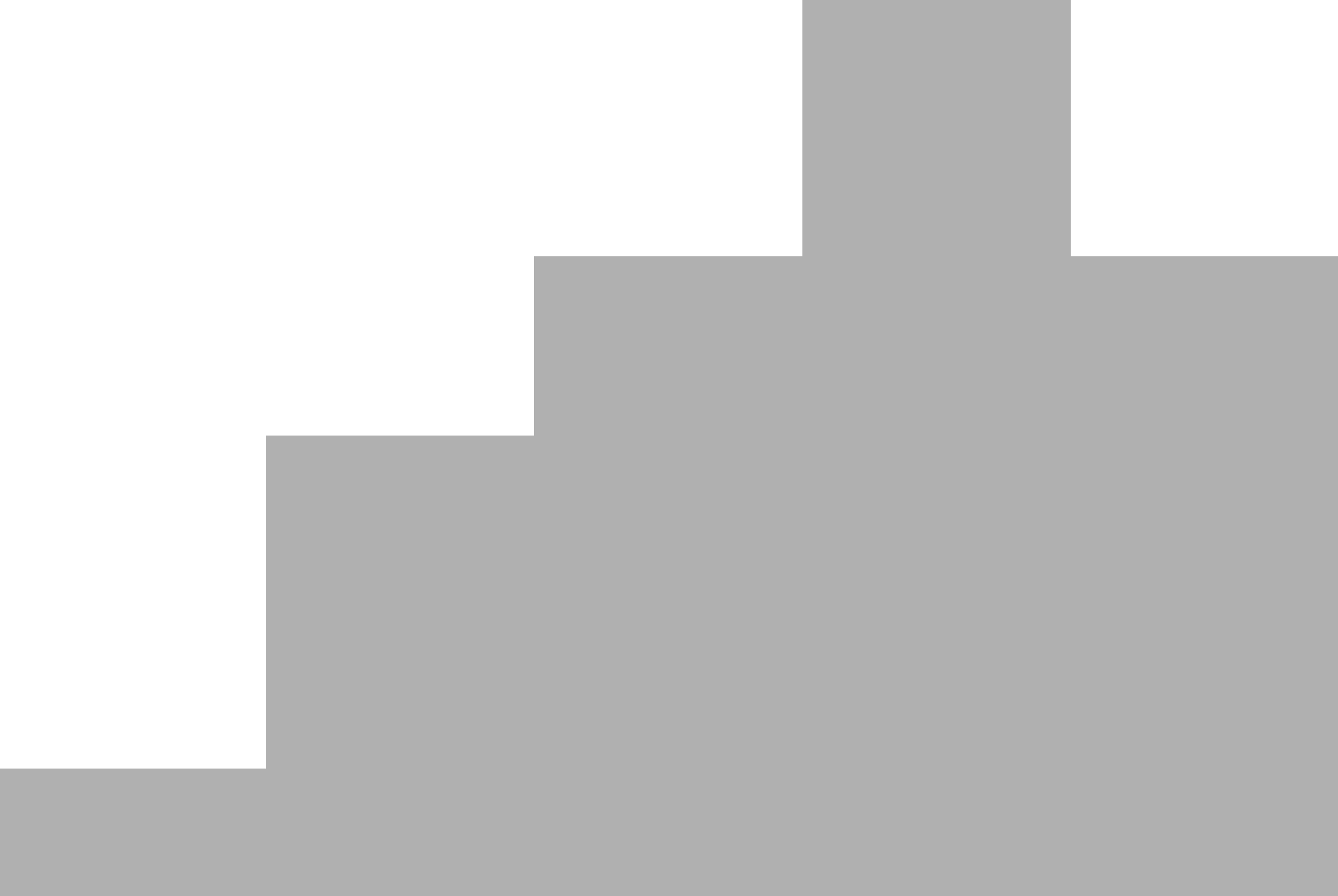}}}                  \\
S14: Salary 
& 4                       & 2.3                 & 3.5                 & 1.3              & \multicolumn{5}{l}{\fbox{\includegraphics[width=0.13\textwidth, height=2mm]{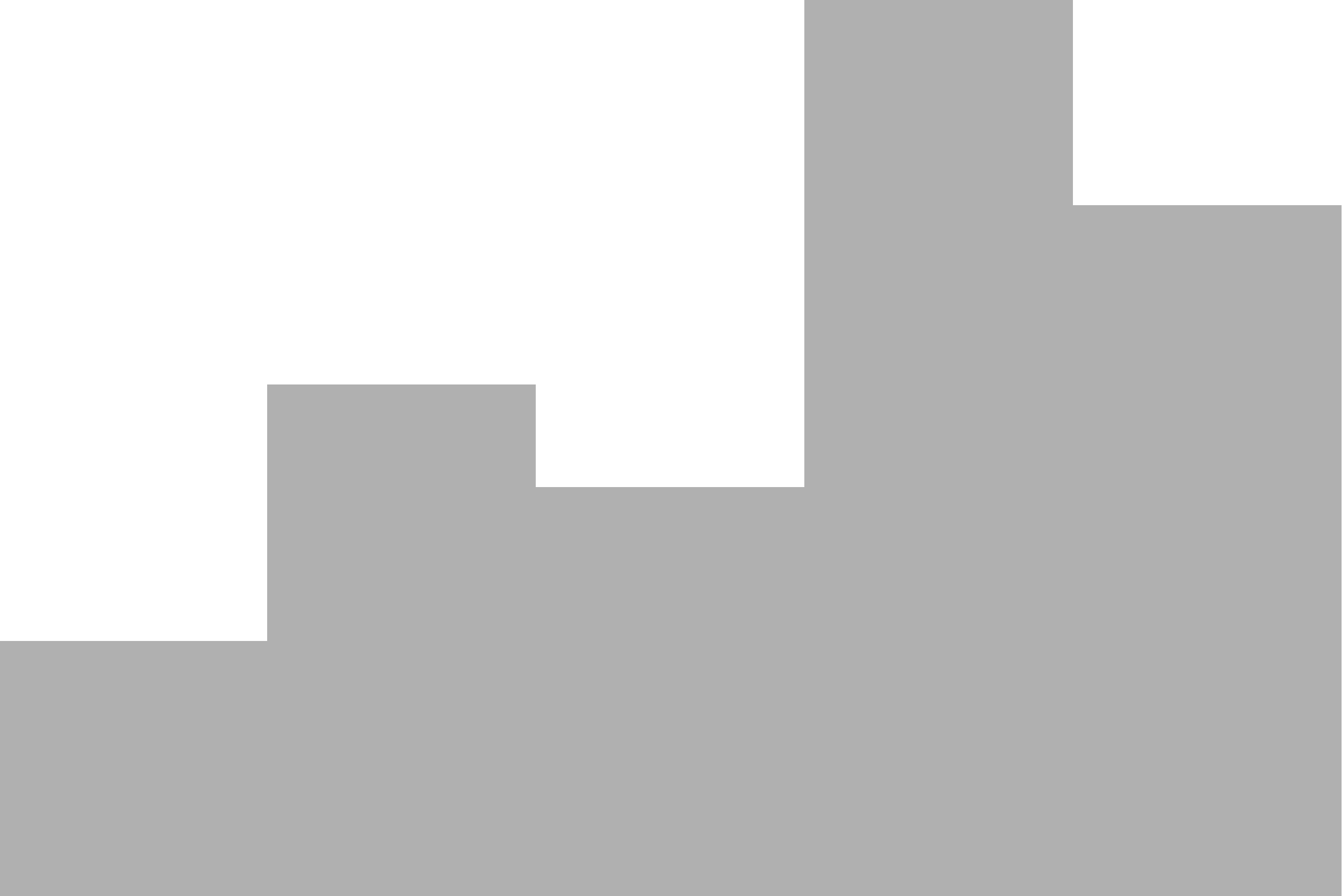}}}                  \\
S15: Job Security 
& 4                       & 1.3                 & 3.4                 & 1.2               & \multicolumn{5}{l}{\fbox{\includegraphics[width=0.13\textwidth, height=2mm]{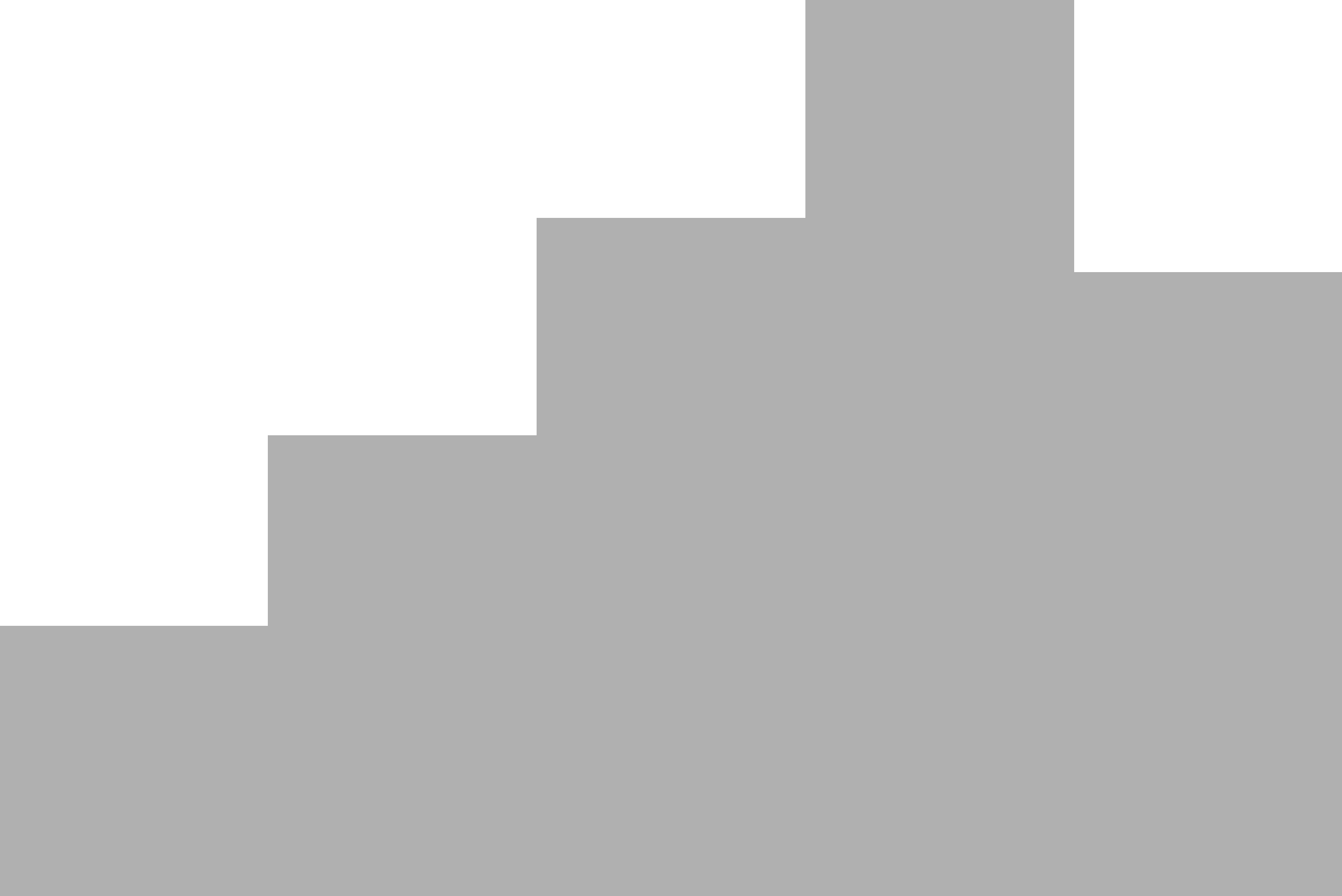}}}                  \\
S16: Promotion 
& 3                       & 2                    & 2.8                 & 1.3               & \multicolumn{5}{l}{\fbox{\includegraphics[width=0.13\textwidth, height=2mm]{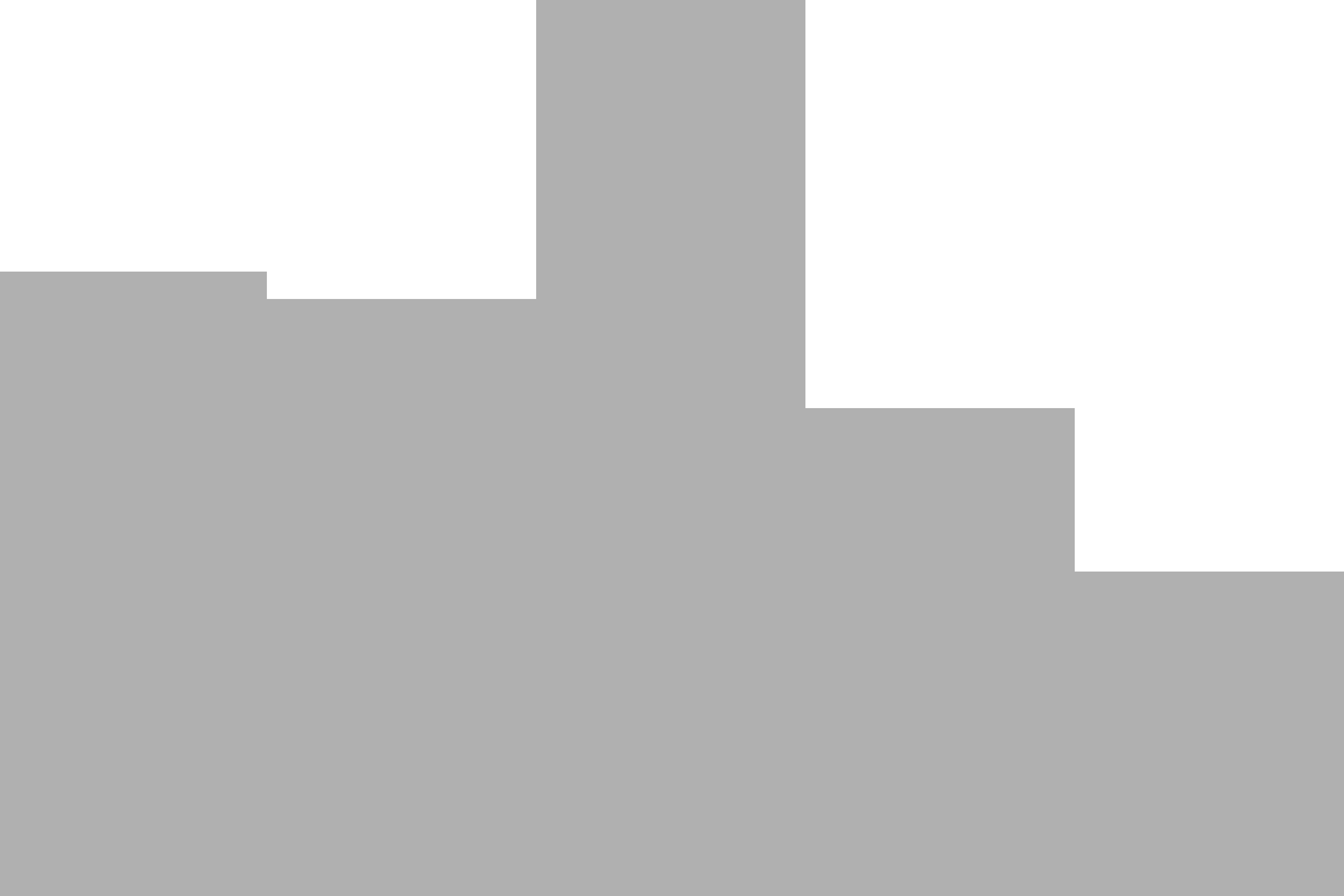}}}                  \\ 
\rowcolor[HTML]{DFDFDF}\multicolumn{10}{c}{\textit{\textbf{Fairness Experience in SE Contexts}}} \\ 
C1: Customer/client Relations \protect\faWheelchair                     & 4                                           & 2                                        & 3.9                                     & 1.0                                  & \multicolumn{5}{l}{\fbox{\includegraphics[width=0.13\textwidth, height=2mm]{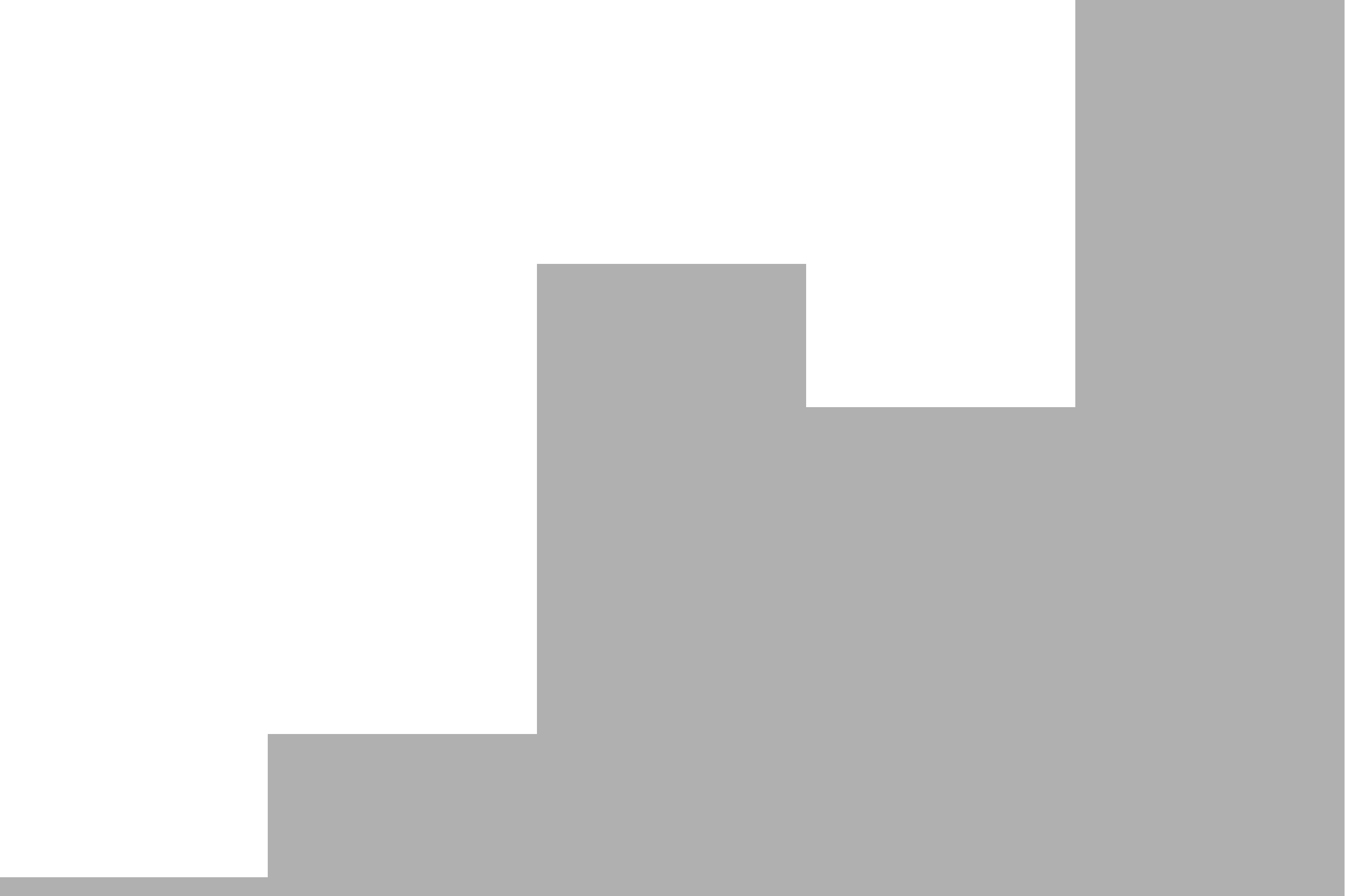}}}                  \\
C2: Recruitment 
& 4                                           & 2                                        & 3.8                                     & 1.2                                   & \multicolumn{5}{l}{\fbox{\includegraphics[width=0.13\textwidth, height=2mm]{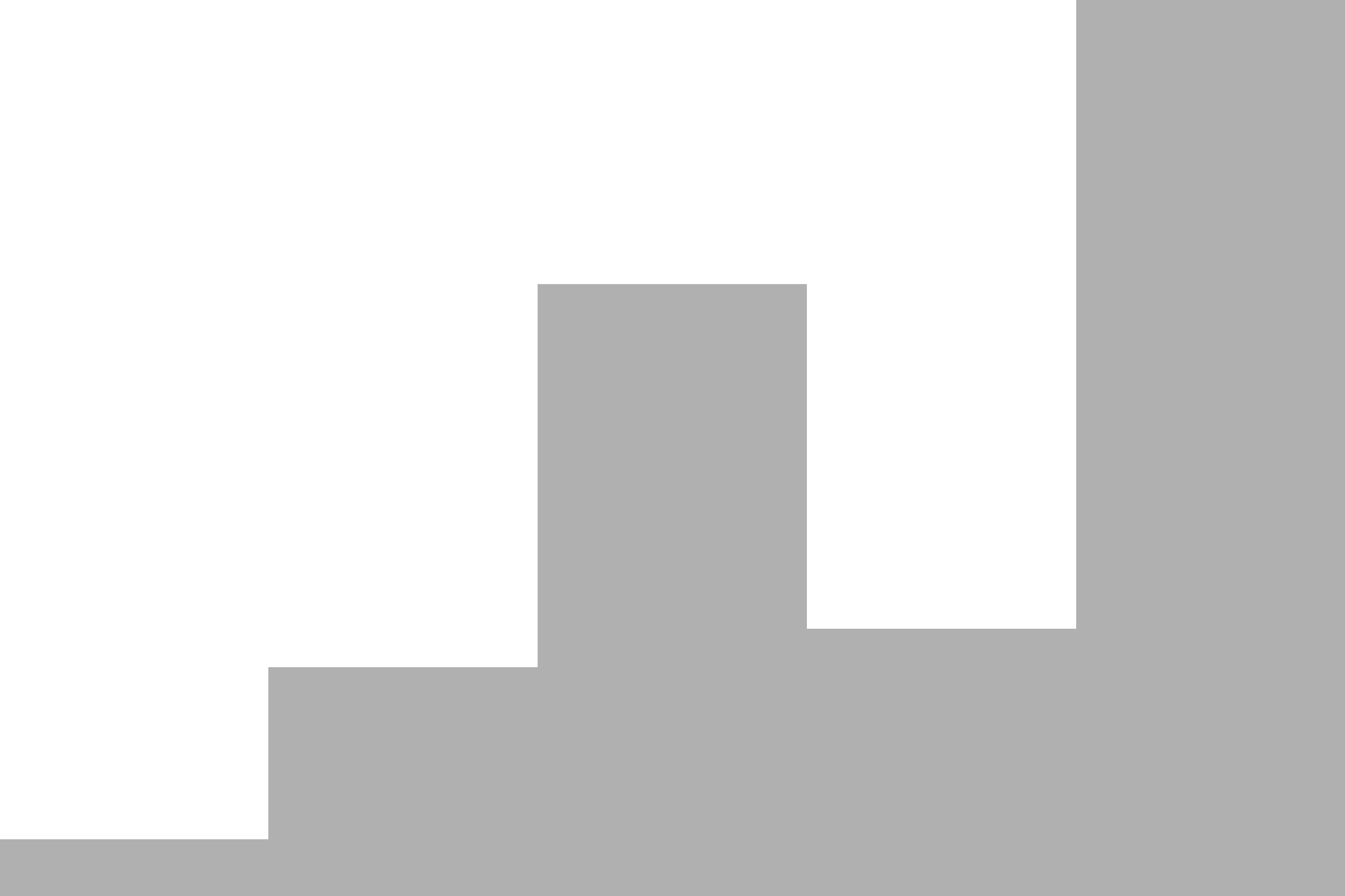}}}                  \\
C3: Treatment 
& 4                                           & 2                                        & 3.8                                     & 1.2                                  & \multicolumn{5}{l}{\fbox{\includegraphics[width=0.13\textwidth, height=2mm]{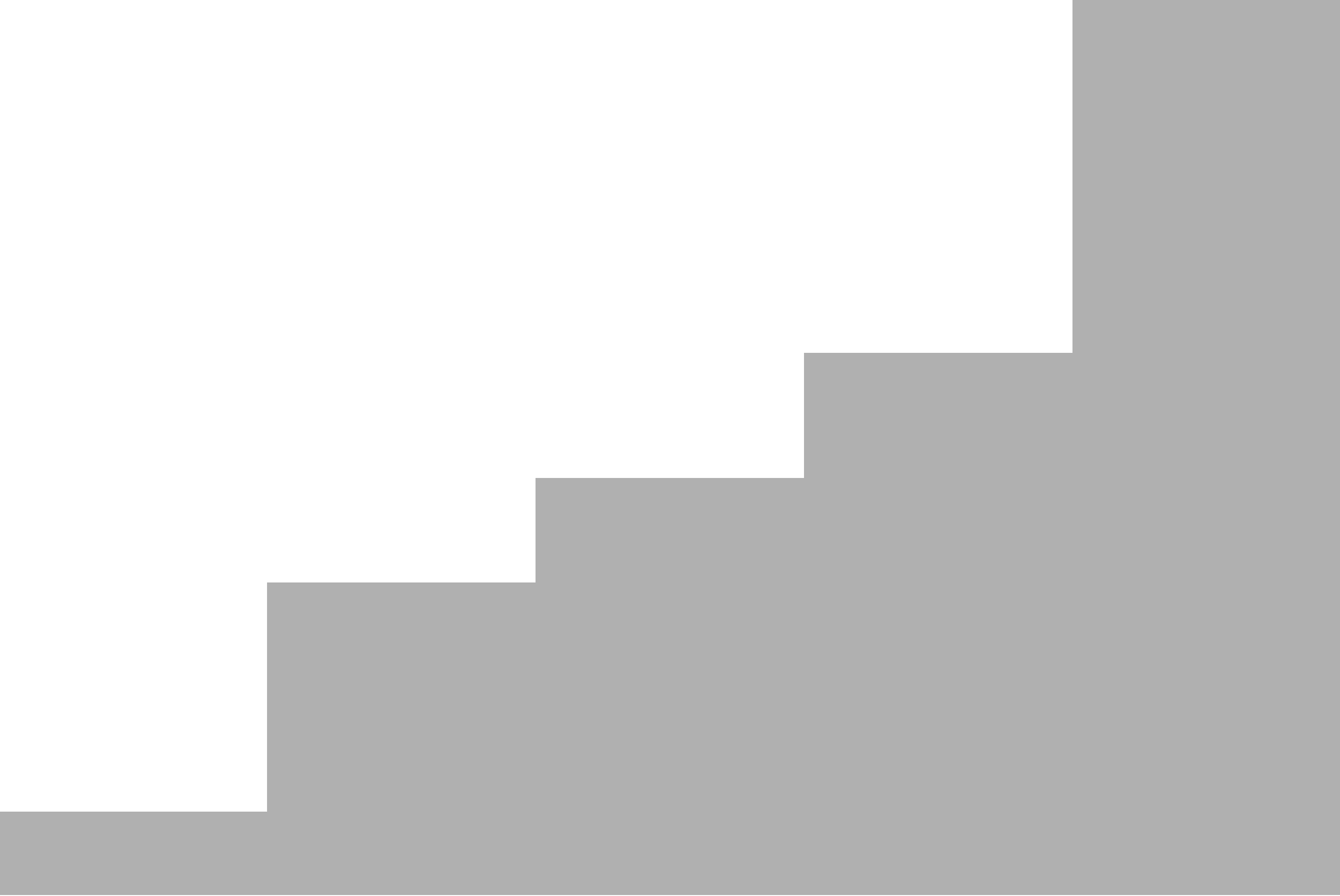}}}                  \\
C4: Policy \protect\faFlag~ \protect\faWheelchair                         & 4                                           & 2                                        & 3.7                                     & 1.3                                   & \multicolumn{5}{l}{\fbox{\includegraphics[width=0.13\textwidth, height=2mm]{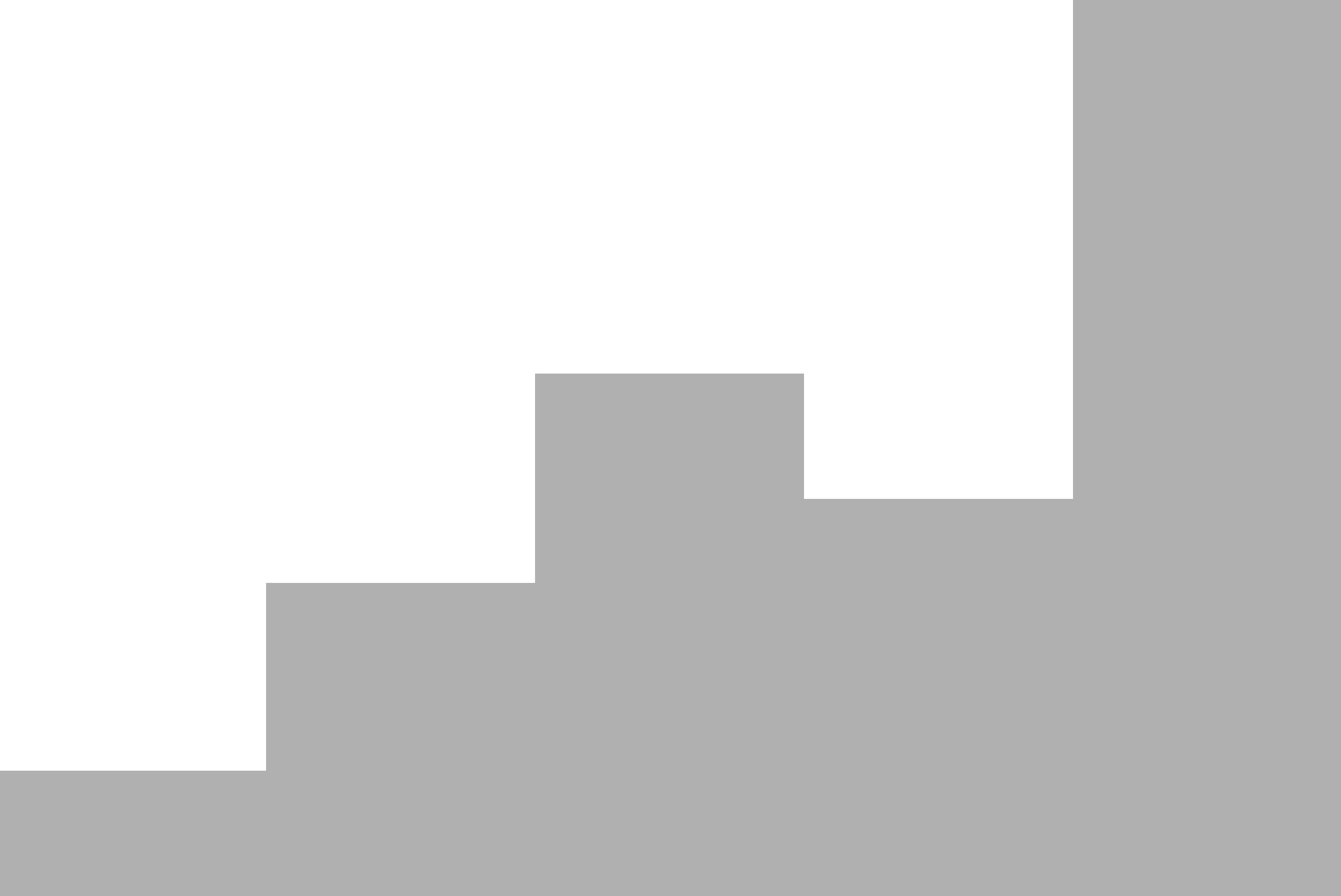}}}                  \\
C5: Evaluation 
\protect\faWheelchair                                          & 3.5                                         & 2                                        & 3.5                                     & 1.3                                   & \multicolumn{5}{l}{\fbox{\includegraphics[width=0.13\textwidth, height=2mm]{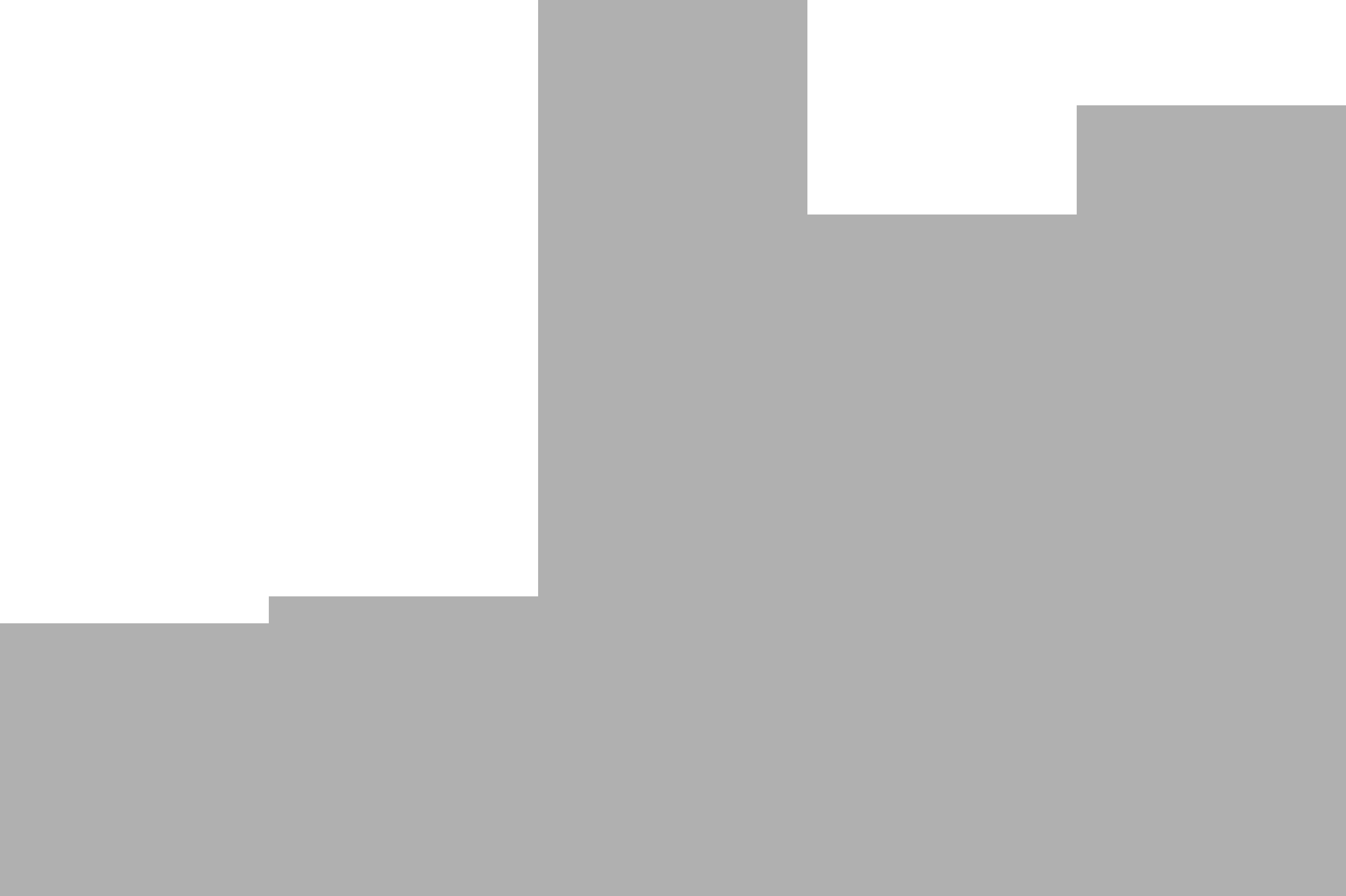}}}                  \\
C6: Working Hours 
& 4                                           & 2                                        & 3.5                                     & 1.3                                   & \multicolumn{5}{l}{\fbox{\includegraphics[width=0.13\textwidth, height=2mm]{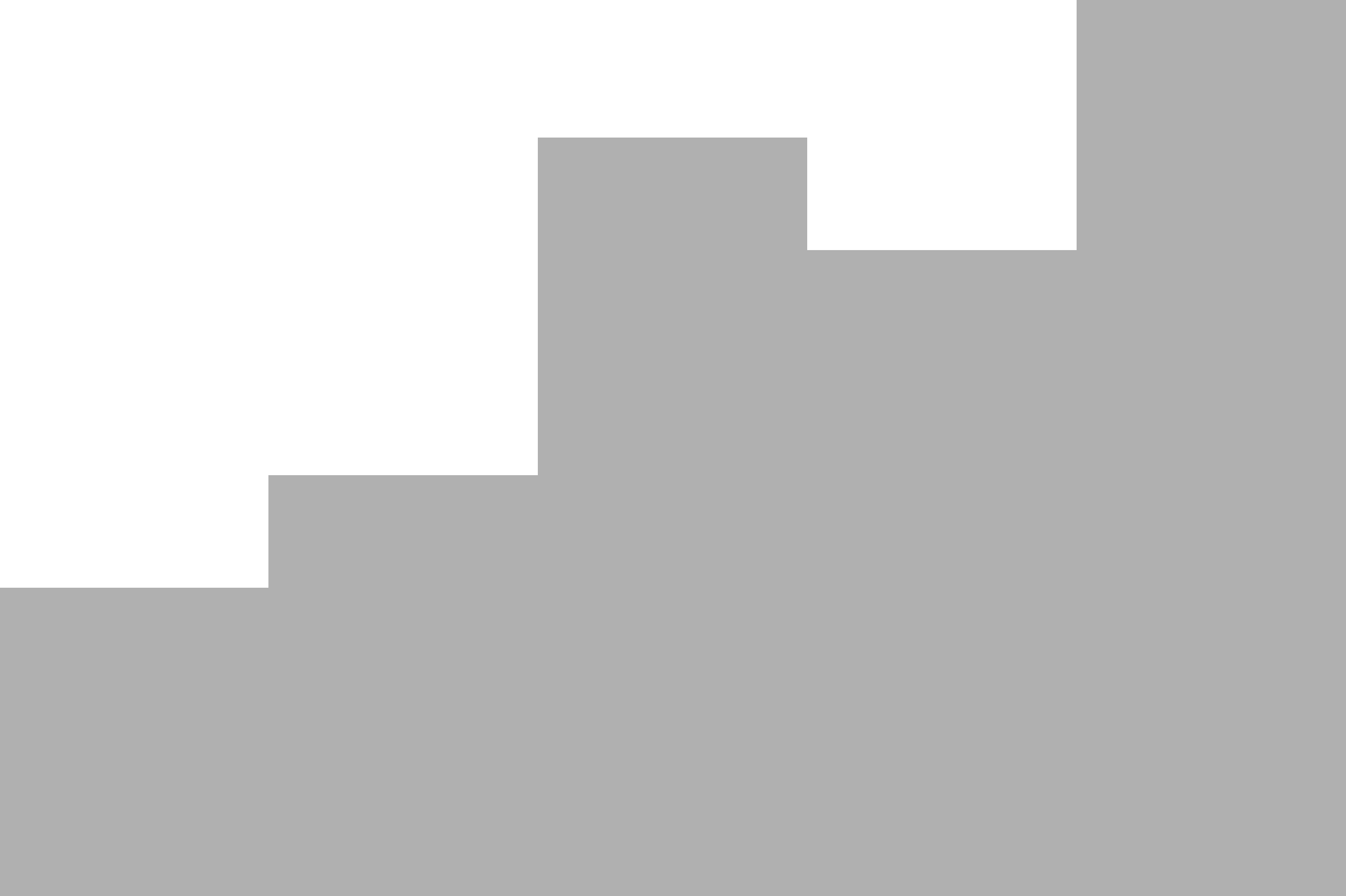}}}                  \\
C7: Allocation of Work 
& 3                                           & 1                                        & 3.4                                     & 1.2                                   & \multicolumn{5}{l}{\fbox{\includegraphics[width=0.13\textwidth, height=2mm]{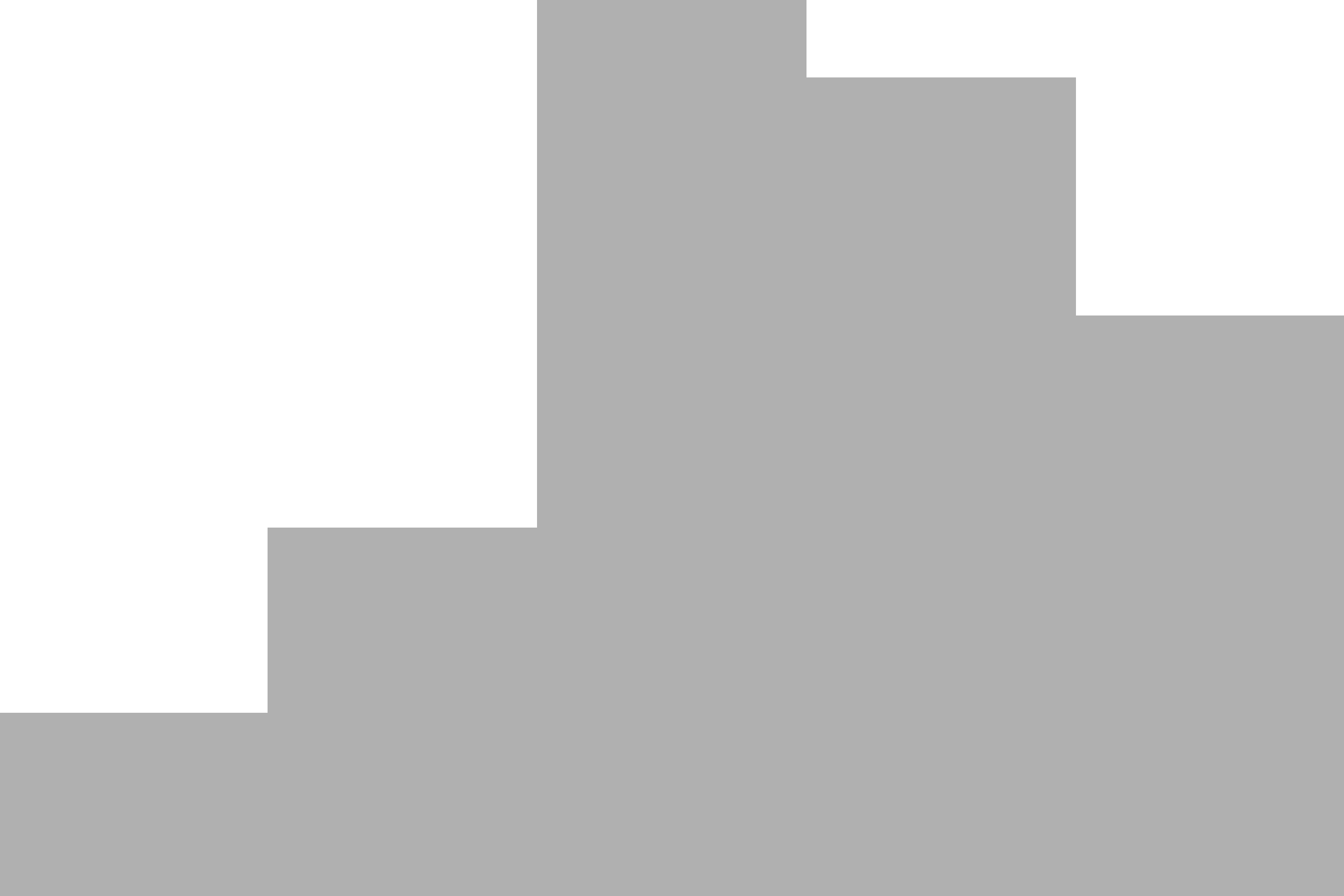}}}                  \\
C8: Income 
& 3                                           & 1                                        & 3.4                                     & 1.2                                   & \multicolumn{5}{l}{\fbox{\includegraphics[width=0.13\textwidth, height=2mm]{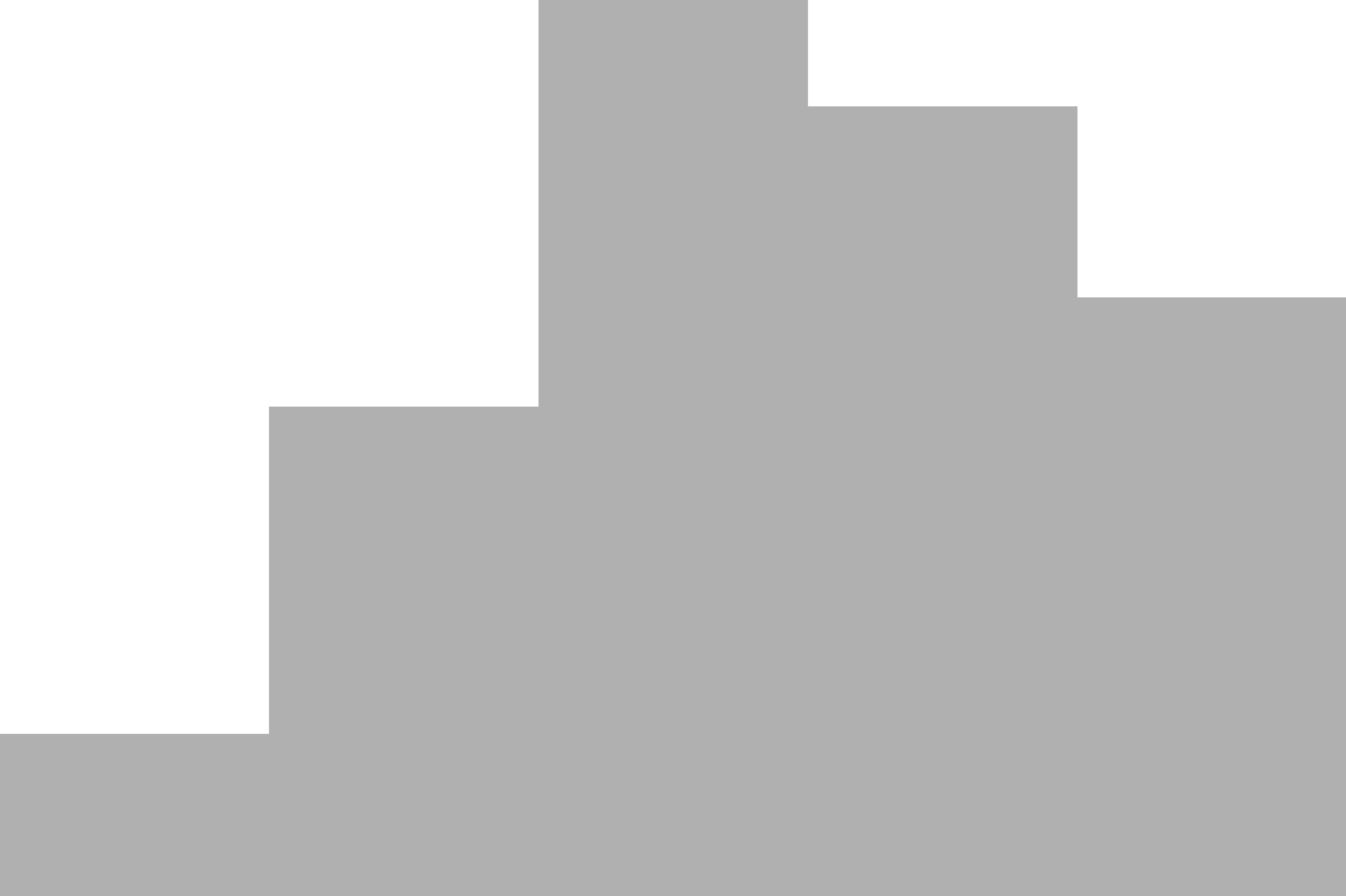}}}                  \\
C9: Authorship 
& 3.5                                         & 1                                        & 3.3                                     & 1.2                                   & \multicolumn{5}{l}{\fbox{\includegraphics[width=0.13\textwidth, height=2mm]{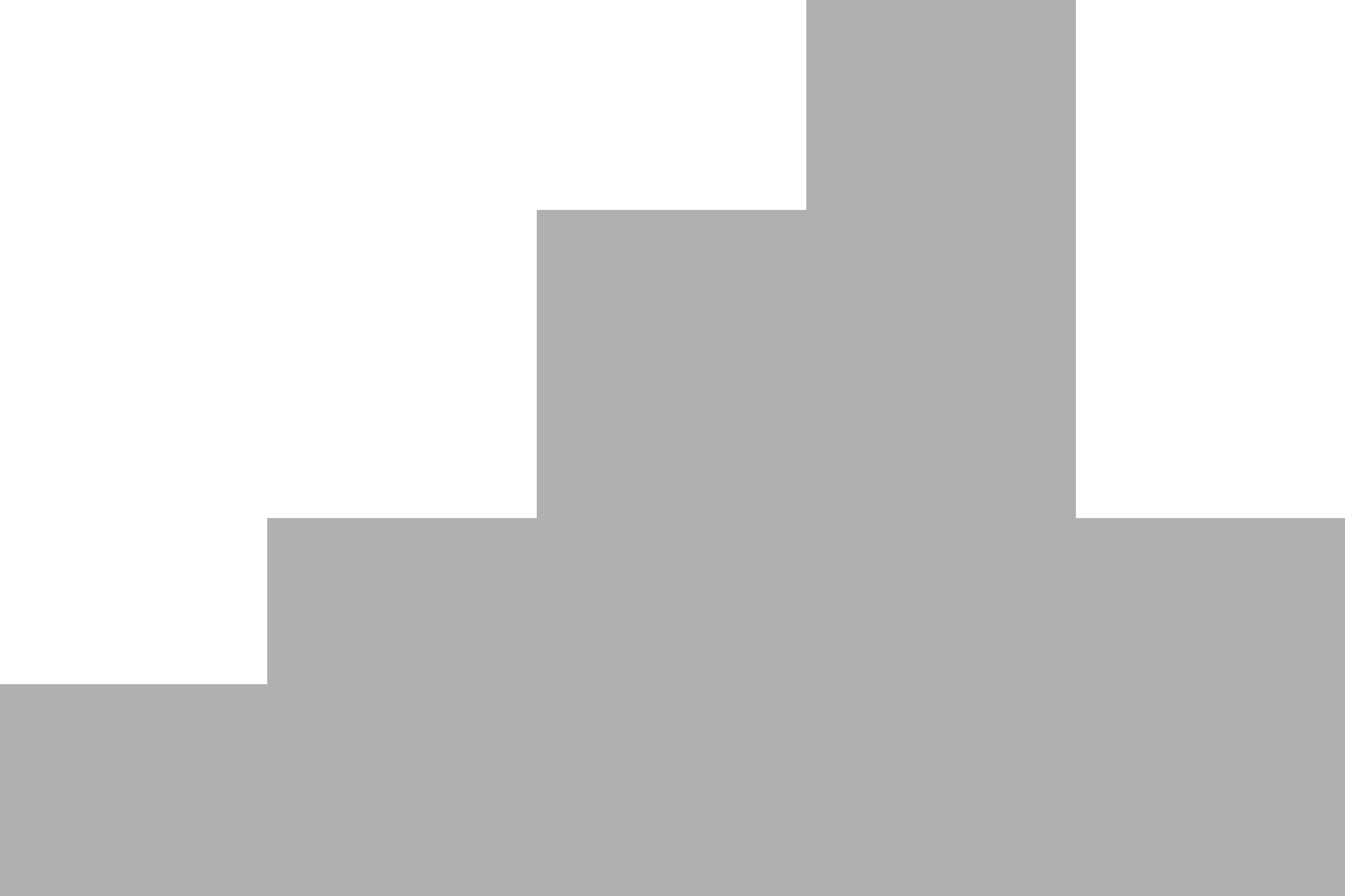}}}                  \\
C10: Demand \protect\faWheelchair                                             & 3                                           & 1                                        & 3.3                                     & 1.2                                   & \multicolumn{5}{l}{\fbox{\includegraphics[width=0.13\textwidth, height=2mm]{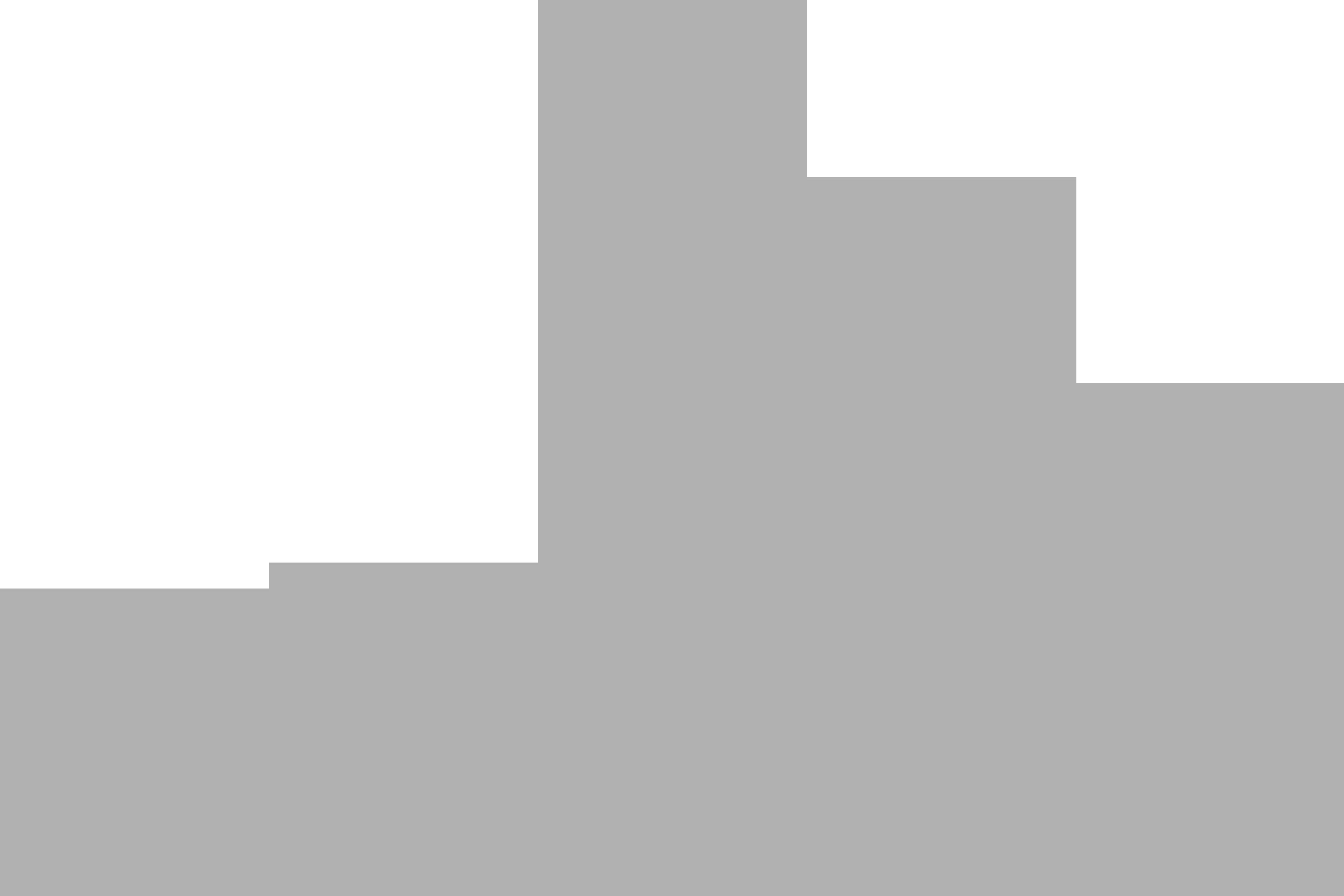}}}                  \\\hline
\label{tab:itemlevel}
\end{longtable}

\subsection{Survey Responses Summary}
Our survey examined software practitioners’ perceptions of fairness, job satisfaction, and their experiences of fairness in various SE contexts. The survey results (Table \ref{tab:itemlevel}) indicate a mixed perception of fairness across different dimensions among software practitioners. While \textit{interpersonal fairness} is generally perceived positively by 90\% respondents, there seems to be moderate fairness with \textit{informational} (79\%), \textit{procedural} (77\%), and \textit{distributive fairness} (77\%).

\edits{It is worth noting that while \textit{interpersonal fairness} received generally positive responses (median and mean = 4), the other dimensions tended to cluster around a median and mean of 3, indicating a more neutral stance. This neutrality could mean several points. It might suggest that respondents view the situation as fair enough and see no major problems, especially if they have not encountered obvious unfairness. But it could also point to uncertainty or disengagement. Some respondents may not have thought much about fairness in their workplace or may lack opportunities to reflect on or discuss it openly.}

Further analysis revealed specific areas within each fairness dimension where perceptions varied significantly. For instance, perceptions of \textit{procedural fairness} were more variable concerning \textit{bias-free processes} (P4), \textit{having a say in decision-making} (P5), \textit{consistency in applying processes} (P6), and the ability to \textit{appeal decisions} (P7). Similarly, mixed perceptions were observed in \textit{informational fairness} items, such as \textit{timely communication}, \textit{tailored communication}, and \textit{comprehensive explanations}.

We also identified notable differences in perceptions of fairness based on \textit{work experience} and \textit{ethnicity representation} within teams. These findings suggest that both work experience and ethnic diversity within teams significantly influence how fairness is perceived in various aspects of the workplace.

In terms of job satisfaction, the majority of respondents expressed overall satisfaction, although there was notable dissatisfaction with specific factors such as \textit{promotion} opportunities. Demographic analysis highlighted moderate differences in job satisfaction influenced by \textit{age} and \textit{work experience}. Younger practitioners and those with less work experience generally reported higher satisfaction levels with aspects like \textit{team collaboration} and \textit{benefits} received.

Regarding fairness experiences in SE contexts, respondents reported moderate experiences with various challenges, with \textit{fairness in demand} being experienced the least. This means that issues related to fairness in demand are perceived as the most unfair among software practitioners.

Significant differences in perceptions of fairness experiences were noted based on \textit{work limitations} and \textit{ethnicity}. Respondents with work limitations and those from underrepresented ethnic groups, particularly Asian respondents, were more likely to experience certain fairness issues, such as \textit{unfair evaluations} and \textit{unfair policy}.

Overall, these findings highlight the varied perceptions of fairness and job satisfaction among software practitioners, emphasizing the importance of considering demographic differences and specific fairness experiences in SE to develop more inclusive and targeted strategies for improving satisfaction.

\subsection{RQ1: Does fairness perception shape job satisfaction among software practitioners? If so, how?}
The results of the analysis are summarized in the first section of Table \ref{tab:generalsummary}. We considered relationships to be impactful if they were statistically significant ($p<0.05$) and had OR exceeding 50\%, to ensure that the identified factors not only demonstrate a reliable statistical relationship but also have a practically meaningful impact on job satisfaction (see \cite{figshare}). \edits{We express OR as percentage increases in the likelihood of higher job satisfaction associated with stronger perceptions of fairness.}
Our analysis demonstrates that perceptions of fairness across various dimensions significantly shape job satisfaction among software practitioners.

All four dimensions of fairness, \textit{distributive}, \textit{procedural}, \textit{interpersonal}, and \textit{informational} fairness, were found to significantly influence overall job satisfaction (ALL) and satisfaction with job security (SEC). This indicates that being fair across all four dimensions leads to higher overall job satisfaction and greater job security among software practitioners. This means that when employees perceive their workplace as fair in terms of distributing outcomes, decision-making processes, interpersonal relations, and communication, they are not only more satisfied with their jobs but also feel more secure in their roles. 

For example, among \textit{interpersonal fairness} items, being \textit{treated with respect} (Int2) had the greatest effect, increasing the likelihood of higher overall job satisfaction by 240\% \edits{(OR = 3.40, 95\% CI[2.19, 5.37])}. This suggests that when employees feel respected, they are more than twice as likely to report higher levels of job satisfaction. Similarly, within \textit{procedural fairness}, \textit{ethical and moral processes} (P1) increased the odds of higher overall job satisfaction by 128\% \edits{(OR = 2.28, 95\% CI[1.54, 3.37])}, indicating that ethical practices are crucial for building trust between employees and management.

Additionally, \textit{accurate information} (P3), another procedural fairness item, was shown to increase satisfaction with job security by 113\% \edits{(OR = 2.13, 95\% CI[1.43, 3.16])}. This finding suggests that when software practitioners receive clear, reliable, and accurate information about organizational processes and decisions, they feel more secure in their roles.



Fairness perceptions also significantly impact other aspects of job satisfaction, such as \textit{rewards and work-life balance} (RWB) and \textit{compensation} (COM). For instance, when outcomes \textit{reflect contribution} (D3) and \textit{effort} (D4), satisfaction with rewards and work-life balance is boosted by 80\% and 53\%, respectively. Similarly, when outcomes are considered \textit{appropriate} (D2) and \textit{reflect effort} (D4), satisfaction with compensation is enhanced by 90\% and 132\%, respectively. This suggests that when organizations ensure that outcomes reflect contribution and effort, practitioners are more likely to be satisfied with their rewards and work-life balance. Additionally, when outcomes are perceived as appropriate and reflect effort, satisfaction with compensation is significantly enhanced.


Both \textit{procedural} and \textit{interpersonal} fairness significantly influence satisfaction with \textit{work culture} (CLT). Specifically, \textit{accurate information} (P3), \textit{ethical and moral processes} (P1) and \textit{voicing opinions} (P2), contributed to positive perceptions of \textit{work culture}, enhancing satisfaction by 100\%, 54\% and 61\%, respectively. Additionally, treatment with \textit{politeness} (Int1), \textit{respect} (Int2), and \textit{dignity} (Int3) boosted satisfaction by 76\%, 67\%, and 55\%, respectively.
This indicates that when organizations ensure accurate information, uphold ethical and moral processes, and encourage voicing opinions, while managers interact with employees with respect, dignity, and politeness, practitioners are more likely to be satisfied with their work culture.

Lastly, both \textit{distributive} and \textit{procedural} fairness significantly affect satisfaction with \textit{performance} (PRF). Factors such as \textit{outcomes justified by performance} (D1), \textit{outcomes appropriate for the completed work} (D2), \textit{effort} (D4), \textit{accurate information} (P3), and \textit{consistency in applying processes} (P6) enhance satisfaction with performance by 71\%, 63\%, 83\%, 86\%, and 72\%, respectively. This means that when organizations ensure that practitioners' performance outcomes are fair and aligned with their contributions and efforts, practitioners are more likely to be satisfied with their performance evaluations. Additionally, when performance evaluations are based on transparent and consistent criteria, it fosters trust in the fairness of the system.



Interestingly, the analysis revealed no significant perception of fairness factors influencing satisfaction with the \textit{manager} (MNG). This is surprising, as one might expect that fair \textit{interpersonal} and \textit{informational} relations would affect satisfaction with a manager. A possible explanation is that this overall finding may mask the experiences of specific demographic subgroups, for whom fairness in interactions with managers may play a more significant role in their satisfaction.

\begin{table}[]
\centering
\caption{\editss{Ordinal Regression Analysis Result. Abbreviations in the first row refer to job satisfaction composite factors and are explained in Table \ref{tab:cfs}.}}
\begin{tabular}{lcllllll}
\hline
\multicolumn{1}{l|}{}     & ALL                                          & \multicolumn{1}{c}{RWB}                       & \multicolumn{1}{c}{SEC}                       & \multicolumn{1}{c}{MNG}                       & \multicolumn{1}{c}{CLT}                      & \multicolumn{1}{c}{PRF}                     & \multicolumn{1}{c}{COM}                         \\ \hline
\multicolumn{8}{c}{\textit{Perception of Fairness}}                                                                                                                                                                                                                                                                                                                       \\ \hline
\multicolumn{1}{l|}{\edits{D1: Outcome is justified by the performance}}   & \cellcolor[HTML]{e5f5e0}+                    &                                               & \multicolumn{1}{c}{\cellcolor[HTML]{e5f5e0}+} &                                               &                                               & \multicolumn{1}{c}{\cellcolor[HTML]{e5f5e0}+} &                                                    \\
\multicolumn{1}{l|}{\edits{D2: Outcome is appropriate for the completed work}}   & \cellcolor[HTML]{e5f5e0}+                    &                                               & \multicolumn{1}{c}{\cellcolor[HTML]{e5f5e0}+} &                       &                                               & \multicolumn{1}{c}{\cellcolor[HTML]{e5f5e0}+} &   \multicolumn{1}{c}{\cellcolor[HTML]{e5f5e0}+}                                                 \\
\multicolumn{1}{l|}{\edits{D3: Outcome reflecting contribution}}   & \multicolumn{1}{l}{}                         & \multicolumn{1}{c}{\cellcolor[HTML]{e5f5e0}+}                                              &                       &                                               &                                               &                                               &                                                    \\
\multicolumn{1}{l|}{\edits{D4: Outcome reflecting effort}}   & \multicolumn{1}{c}{\cellcolor[HTML]{e5f5e0}+}                    &  \multicolumn{1}{c}{\cellcolor[HTML]{e5f5e0}+}                                             &                       &                                               &                                               & \multicolumn{1}{c}{\cellcolor[HTML]{e5f5e0}+} & \multicolumn{1}{c}{\cellcolor[HTML]{a1d99b}+}      \\ \hline
\multicolumn{1}{l|}{\edits{P1: Ethical and moral processes}}   & \cellcolor[HTML]{a1d99b}+                    &                                               & \multicolumn{1}{c}{\cellcolor[HTML]{e5f5e0}+} &                                               & \multicolumn{1}{c}{\cellcolor[HTML]{e5f5e0}+} &                                               &                                                    \\
\multicolumn{1}{l|}{\edits{P2: Voicing opinions}}   & \cellcolor[HTML]{e5f5e0}+                    &                                               &                                               &                       & \multicolumn{1}{c}{\cellcolor[HTML]{e5f5e0}+} &                                               &                                                    \\
\multicolumn{1}{l|}{\edits{P3: Accurate information}}   & \cellcolor[HTML]{e5f5e0}+                    &                                               & \multicolumn{1}{c}{\cellcolor[HTML]{a1d99b}+} &                                               & \multicolumn{1}{c}{\cellcolor[HTML]{a1d99b}+} & \multicolumn{1}{c}{\cellcolor[HTML]{e5f5e0}+} &                                                    \\
\multicolumn{1}{l|}{\edits{P4: Free of bias processes}}   & \cellcolor[HTML]{e5f5e0}+                    &                                               & \multicolumn{1}{c}{\cellcolor[HTML]{e5f5e0}+} &                       &                                               &                                               &                                                    \\
\multicolumn{1}{l|}{\edits{P5: Have a say in making decisions}}   & \cellcolor[HTML]{e5f5e0}+                    &                                               &                                               &                                               &                                               &                                               &                                                    \\
\multicolumn{1}{l|}{\edits{P6: Consistency in applying processes}}   & \cellcolor[HTML]{e5f5e0}+                    &                                               &                       &                                               &                                               & \multicolumn{1}{c}{\cellcolor[HTML]{e5f5e0}+} &                                                    \\
\multicolumn{1}{l|}{\edits{P7: Appealing decisions}}   & \cellcolor[HTML]{e5f5e0}+                    &                                               &                                               &                                               &                                               &                                               &                                                    \\ \hline
\multicolumn{1}{l|}{\edits{Int1: Treated politely}} & \cellcolor[HTML]{a1d99b}+                    &                                               & \multicolumn{1}{c}{\cellcolor[HTML]{e5f5e0}+} &                                               & \multicolumn{1}{c}{\cellcolor[HTML]{e5f5e0}+} &                                               &                                                    \\
\multicolumn{1}{l|}{\edits{Int2: Treated with respect}} & \cellcolor[HTML]{31a354}+                    &                                               & \multicolumn{1}{c}{\cellcolor[HTML]{e5f5e0}+} &                                               & \multicolumn{1}{c}{\cellcolor[HTML]{e5f5e0}+} &                                               &                                                    \\
\multicolumn{1}{l|}{\edits{Int3: Treated with dignity}} & \cellcolor[HTML]{a1d99b}+                    &                                               & \multicolumn{1}{c}{\cellcolor[HTML]{e5f5e0}+} &                                               & \multicolumn{1}{c}{\cellcolor[HTML]{e5f5e0}+} &                                               &                                                    \\
\multicolumn{1}{l|}{\edits{Int4: No inappropriate comments or behaviour}} & \cellcolor[HTML]{e5f5e0}+                    &                                               &                                               &                                               &                                               &                                               &                                                    \\ \hline
\multicolumn{1}{l|}{\edits{Inf1: Direct communication}} &                     &                                               &                                               &                                               &                                               &                                               &                                                    \\
\multicolumn{1}{l|}{\edits{Inf2: Reasonable explanations}} & \cellcolor[HTML]{e5f5e0}+                    &                                               & \multicolumn{1}{c}{\cellcolor[HTML]{e5f5e0}+} &                                               &                                               &                                               &                                                    \\
\multicolumn{1}{l|}{\edits{Inf3: Timely communication}} & \cellcolor[HTML]{e5f5e0}+                    &                                               &                                               &                                               &                                               &                                               &                                                    \\
\multicolumn{1}{l|}{\edits{Inf4: Tailored communication}} & \cellcolor[HTML]{e5f5e0}+                    &                                               &                                               &                                               &                                               &                                               &                                                    \\
\multicolumn{1}{l|}{\edits{Inf5: Comprehensive explanations}} & \cellcolor[HTML]{e5f5e0}+                    &                                               &                                               &                                               &                                               &                                               &                                                    \\ \hline
\multicolumn{8}{c}{\textit{SE Context}}                                                                                                                                                                                                                                                                                                                                        \\ \hline
\multicolumn{1}{l|}{\edits{C1: Customer/client Relations}}  & \multicolumn{1}{l}{}                         &                                               &                                               &                                               &                                               &                                               &                                                    \\
\multicolumn{1}{l|}{\edits{C2: Recruitment}}   & \multicolumn{1}{l}{}                         &                                               &                                               &                                               &                                               &                                               &                                                    \\
\multicolumn{1}{l|}{\edits{C3: Treatment}}   &  &                       &                                               &                                               &                                               &                                               &                                                    \\
\multicolumn{1}{l|}{\edits{C4: Policy}}   & \cellcolor[HTML]{e5f5e0}+                    & \multicolumn{1}{c}{\cellcolor[HTML]{e5f5e0}+} &                                               &                                               &                                               &                                               &                                                    \\
\multicolumn{1}{l|}{\edits{C5: Evaluation}}   & \multicolumn{1}{l}{}                         &                                               & \multicolumn{1}{c}{\cellcolor[HTML]{e5f5e0}+} & \multicolumn{1}{c}{\cellcolor[HTML]{e5f5e0}+} &                                               &                                               &                                                    \\
\multicolumn{1}{l|}{\edits{C6: Working Hours}}   & \multicolumn{1}{l}{}                         & \multicolumn{1}{c}{\cellcolor[HTML]{e5f5e0}+} &                                               &                                               &                       &                                               &                                                    \\
\multicolumn{1}{l|}{\edits{C7: Allocation of Work}}   & \cellcolor[HTML]{e5f5e0}+                    & \multicolumn{1}{c}{\cellcolor[HTML]{e5f5e0}+} &                       &                                               & \multicolumn{1}{c}{\cellcolor[HTML]{e5f5e0}+} &                                               &                                                    \\
\multicolumn{1}{l|}{\edits{C8: Income}}   & \multicolumn{1}{l}{}                         &                                               &                       &                                               &                                               &                                               & \multicolumn{1}{c}{\cellcolor[HTML]{e5f5e0}+}      \\
\multicolumn{1}{l|}{\edits{C9: Authorship}}   & \cellcolor[HTML]{e5f5e0}+                    & \multicolumn{1}{c}{\cellcolor[HTML]{e5f5e0}+} & \multicolumn{1}{c}{\cellcolor[HTML]{e5f5e0}+} & \multicolumn{1}{c}{\cellcolor[HTML]{e5f5e0}+} & \multicolumn{1}{c}{\cellcolor[HTML]{e5f5e0}+}  &                                               &                                                    \\
\multicolumn{1}{l|}{\edits{C10: Demand}}   & \cellcolor[HTML]{e5f5e0}+                    & \multicolumn{1}{c}{\cellcolor[HTML]{e5f5e0}+} &                                               &                                               & \multicolumn{1}{c}{\cellcolor[HTML]{e5f5e0}+} &                                               &                                                    \\
 \hline
                          & \multicolumn{1}{l}{}                         &                                               &                                               &                                               &                                               &                                               &                                                    \\
& \multicolumn{2}{c}{\cellcolor[HTML]{e5f5e0}$50\% \leq \text{OR} < 100\%$ }                   &                                                \multicolumn{3}{c}{\cellcolor[HTML]{a1d99b}$100\% \leq \text{OR} < 200\%$}
                                                                                      &  \multicolumn{2}{c}{\cellcolor[HTML]{31a354}$\text{OR} \geq 200\%$}
\end{tabular}
\label{tab:generalsummary}
\end{table}

\editss{\framedtext{\textit{Takeaway 1}: Fairness across all four dimensions (namely \textit{distributive}, \textit{procedural}, \textit{interpersonal}, and \textit{informational}) significantly enhances software practitioners' overall job satisfaction and satisfaction with job security.}}

\subsection{RQ2: Does the relationship between fairness perceptions and job satisfaction change based on demographic groups? If so, how?}

In this RQ, we aim to determine whether a software practitioner’s demographic background influences how their perceptions of fairness affect job satisfaction using moderation analysis. The demographic groups analyzed include \textit{gender}, \textit{ethnicity representation}, \textit{work experience} and \textit{work limitation}.
Similar to the previous analysis, we considered a demographic group to be influential if the effect was statistically significant ($p<0.05$) and the effect size exceeded 50\% (see \cite{figshare}). 

Our analysis reveals that the relationship between fairness perceptions and job satisfaction does indeed vary across demographic groups.
Table \ref{tab:demosummary} (first section) presents the key findings, with different emojis indicating how the perception of fairness influences job satisfaction among practitioners in the demographic groups represented by those emojis.

\textbf{Gender.} The analysis shows that for \textit{male} practitioners, all significant fairness factors have negative effect sizes, indicating a weaker impact on job satisfaction compared to \textit{female} practitioners (i.e., reference group). Conversely, these factors have a stronger impact on \textit{female} practitioners.

Our analysis also unveils that \textit{distributive}, \textit{procedural} and \textit{informational fairness} have a significant impact on \textit{female} developers' satisfaction with their \textit{manager} (MNG). All \textit{distributive fairness} items (D1-D4) have an average 62\% stronger effect on female practitioners compared to male practitioners. Additionally, other items related to \textit{procedural fairness}, such as \textit{accurate information} (P3), \textit{free of bias processes} (P4), \textit{consistency in applying processes} (P6), and \textit{appealing decisions} (P7), as well as \textit{tailored communication} (Inf4) from \textit{informational fairness}, also contribute to female developers' satisfaction with their manager.
This suggests that to increase female developers' satisfaction with their managers, organizations should focus on ensuring fair outcomes, accurate information, unbiased processes, and tailored communication.

The analysis also revealead that female developers' satisfaction with \textit{job security} (SEC) is influenced by all four dimensions of fairness perceptions. Key factors include receiving outcomes that accurately \textit{reflect their work} (D2) and \textit{effort} (D4), the ability to \textit{voice opinions} (P2), access to \textit{accurate information} (P3), \textit{free of bias processes} (P4), the ability to \textit{appeal decisions} (P7), being \textit{treated with respect} (Int2) and \textit{dignity} (Int3), and receiving \textit{comprehensive explanations} (Inf5).
To improve job security satisfaction for female developers, organizations should ensure that performance outcomes are fair and reflective of their efforts, foster a respectful environment, and maintain clear, unbiased communication.


This difference between \textit{male} and \textit{female} could be due to varying expectations or experiences regarding fairness and recognition between genders. \textit{Female} developers may place greater importance on fairness across all dimensions, possibly due to encountering more significant challenges related to treatment and recognition in their careers.


\textbf{Ethnicity Representation.} The analysis also reveals that developers whose ethnicity is underrepresented in their team, compared to those in diverse team, feel significantly stronger about their satisfaction with their \textit{manager} (MNG), with an increase of over 200\% on average. This is particularly evident in the ability of \textit{voicing opinions} (P2), receiving \textit{accurate information} (P3), experiencing \textit{unbiased processes} (P4), \textit{consistent application of processes} (P6), and the ability to \textit{appeal decisions} (P7).

This suggests that, to increase the satisfaction of underrepresented developers with their managers, organizations should prioritize creating an inclusive environment where these developers feel heard and valued. 
Developers from \textit{underrepresented} groups feel more strongly about these aspects of fairness compared to those in \textit{balanced or diverse teams}, likely because they may face greater challenges in having their voices heard. Minority workers may be more sensitive to supervisory support, given their experiences of being underrepresented. Therefore, they tend to closely associate their perception of fairness with their manager's actions and behaviors.

\textbf{Work Experience.}
The analysis reveals that for practitioners with \textit{more than 3 years} of experience, most significant fairness factors show negative effect sizes. This indicates that the impact of fairness on job satisfaction is weaker for these more experienced practitioners compared to those with \textit{less than 3 years} of experience (the reference group). Conversely, fairness factors have a stronger influence on the job satisfaction of practitioners with \textit{less than 3 years} of experience.

For practitioners with \textit{less than 3 years} of experience, there is a significant correlation between all \textit{interpersonal fairness} items and their satisfaction with \textit{performance} (PRF). This means that these less experienced practitioners feel more satisfied with their performance when they are treated with politeness, dignity, and respect, and when inappropriate behavior or comments are absent. This may be because early-career practitioners are still developing their professional identity and skills. As a result, they may place a greater emphasis on interpersonal interactions to validate their contributions and build confidence in their abilities.

Additionally, there is a correlation between \textit{outcome reflecting effort} (D4), \textit{being treated with respect} (Int2), and \textit{overall job satisfaction} (ALL) for practitioners with \textit{less than 3 years} of experience. This indicates that these practitioners are more satisfied with their overall job when their outcomes, such as promotions or compensation, are aligned with the effort they have put in, and when they are treated with respect. For less experienced practitioners, having their hard work recognized and rewarded is crucial to their sense of achievement and professional growth. When their outcomes reflect their effort, they feel valued, which increases their overall job satisfaction. Additionally, respectful treatment from colleagues and managers reinforces their sense of belonging within the organization.

Finally, the less experienced developers also feel significantly stronger about their satisfaction with their \textit{job security} (SEC). This is particularly evident in receiving \textit{accurate information} (P3), and \textit{consistent application of processes} (P6). This means that less experienced developers place a strong emphasis on receiving accurate information, as well as seeing processes applied consistently, when it comes to their sense of job security. This might be possible because when they are kept well-informed and see that organizational processes are applied consistently, it helps them feel more secure in their roles.

For practitioners with \textit{more than 3 years} of experience, there is a significant correlation between \textit{tailored communication} (Inf4) and satisfaction with \textit{compensation} (COM). This suggests that more experienced practitioners highly value personalized and specific communication regarding their compensation. Tailored communication that acknowledges their unique role and efforts helps them feel that their contributions are properly valued, leading to greater satisfaction with their compensation.

\textbf{Work Limitation.} The table also shows that software practitioners with work limitations feel significantly stronger about their satisfaction with \textit{compensation} (COM), with an increase of over 100\% on average compared to those without limitations. This is particularly evident in the impact of being \textit{treated politely} (Int1), with \textit{dignity} (Int3), avoiding \textit{inappropriate comments or behavior} (Int4), and receiving \textit{comprehensive explanations} (Inf5).

This suggests that practitioners with work limitations are more satisfied with their compensation when they are treated with politeness and dignity, are not subjected to inappropriate comments or behavior, and receive comprehensive explanations from their managers. These factors help create a supportive and respectful work environment that acknowledges their needs. For developers with work limitations, compensation may include not only monetary benefits but also accommodations, flexible working conditions, and support services. These non-monetary aspects can significantly influence their satisfaction.



\editss{\framedtext{\textit{Takeaway 2}: The impact of fairness perceptions on job satisfaction is stronger for female practitioners, underrepresented practitioners, less experienced practitioners, and those with work limitations. This suggests that different demographic groups have distinct needs.}}

\begin{table}[]
\caption{\editss{The influence of developers' demographic backgrounds—male \protect\usym{2642}~, under-represented ethnicity \protect\usym{1F314}~, well-represented ethnicity \protect\usym{1F312}~, more than 3 years work experience \protect\faBriefcase~and work limitations \protect\faWheelchair~—on practitioners' satisfaction. Abbreviations in the first row refer to job satisfaction composite factors and are explained in Table \ref{tab:cfs}.}}
\small
\begin{tabular}{lllllcclcllcl}
\hline
\multicolumn{1}{l|}{}     & \multicolumn{3}{c|}{ALL}                                                                                & \multicolumn{1}{c|}{RWB} & \multicolumn{2}{c|}{SEC}                                                            & \multicolumn{2}{c|}{MNG}                                                                                & \multicolumn{1}{c|}{CLT}                           & \multicolumn{2}{c|}{PRF}                                                                                & \multicolumn{1}{c}{COM}                          \\ \hline
\multicolumn{13}{c}{\textit{Perception of Fairness}}                                                                                                                                                                                                                                                                                                                                                                                                                                                                                                                      \\ \hline
\multicolumn{1}{l|}{\edits{D1: Outcome is justified by the performance}}   & \multicolumn{3}{l|}{}                                                                                   & \multicolumn{1}{l|}{}    & \multicolumn{2}{l|}{}                                                               & \multicolumn{2}{c|}{\cellcolor[HTML]{ffeda0}\usym{2642}}                                                     & \multicolumn{1}{l|}{}                               & \multicolumn{2}{l|}{}                                                                                     & \multicolumn{1}{c}{\cellcolor[HTML]{31a354}\usym{1F314}}   \\
\multicolumn{1}{l|}{\edits{D2: Outcome is appropriate for the completed work}}   & \multicolumn{3}{c|}{\cellcolor[HTML]{ffeda0}\usym{2642}}                                                     & \multicolumn{1}{l|}{}    & \multicolumn{2}{c|}{\cellcolor[HTML]{ffeda0}\usym{2642}}                                 & \multicolumn{2}{c|}{\cellcolor[HTML]{ffeda0}\usym{2642}}                                                     & \multicolumn{1}{l|}{}                               & \multicolumn{2}{l|}{}                                                                                     &                                                     \\
\multicolumn{1}{l|}{\edits{D3: Outcome reflecting contribution}}   & \multicolumn{3}{l|}{}                                                                                   & \multicolumn{1}{l|}{}    & \multicolumn{2}{c|}{\cellcolor[HTML]{31a354}\usym{1F314}}                                  & \multicolumn{2}{c|}{\cellcolor[HTML]{ffeda0}\usym{2642}}                                                     & \multicolumn{1}{l|}{}                               & \multicolumn{2}{c|}{\cellcolor[HTML]{a1d99b}\faWheelchair}                                                      & \multicolumn{1}{c}{\cellcolor[HTML]{31a354}\usym{1F314}}   \\
\multicolumn{1}{l|}{\edits{D4: Outcome reflecting effort}}   & \multicolumn{3}{c|}{\cellcolor[HTML]{ffeda0}\faBriefcase}                                                                                   & \multicolumn{1}{l|}{}    & \cellcolor[HTML]{ffeda0}\usym{2642} & \multicolumn{1}{c|}{\cellcolor[HTML]{a1d99b}\usym{1F314}} & \multicolumn{2}{c|}{\cellcolor[HTML]{ffeda0}\usym{2642}}                                                     & \multicolumn{1}{l|}{}                               & \multicolumn{2}{l|}{}                                                                                     &                                                     \\ \hline
\multicolumn{1}{l|}{\edits{P1: Ethical and moral processes}}   & \multicolumn{3}{l|}{}                                                                                   & \multicolumn{1}{l|}{}    & \multicolumn{2}{l|}{}                                                               & \multicolumn{2}{l|}{}                                                                                   & \multicolumn{1}{l|}{}                               & \multicolumn{2}{l|}{}                                                                                     &                                                     \\
\multicolumn{1}{l|}{\edits{P2: Voicing opinions}}   & \multicolumn{3}{l|}{}                                                                                   & \multicolumn{1}{l|}{}    & \cellcolor[HTML]{ffeda0}\usym{2642} & \multicolumn{1}{c|}{\cellcolor[HTML]{31a354}\usym{1F314}} & \multicolumn{2}{c|}{\cellcolor[HTML]{31a354}\usym{1F314}}                                                      & \multicolumn{1}{l|}{}                               & \multicolumn{2}{l|}{}                                                                                     &                                                     \\
\multicolumn{1}{l|}{\edits{P3: Accurate information}}   & \multicolumn{3}{l|}{}                                                                                   & \multicolumn{1}{l|}{}    & \cellcolor[HTML]{ffeda0}\faBriefcase & \multicolumn{1}{c|}{\cellcolor[HTML]{ffeda0}\usym{2642}}                                 & \multicolumn{1}{c}{\cellcolor[HTML]{ffeda0}\usym{2642}} & \multicolumn{1}{c|}{\cellcolor[HTML]{31a354}\usym{1F314}} & \multicolumn{1}{l|}{}                               & \multicolumn{2}{l|}{}                                                                                     &                                                     \\
\multicolumn{1}{l|}{\edits{P4: Free of bias processes}}   & \multicolumn{3}{c|}{\cellcolor[HTML]{a1d99b}\usym{1F314}}                                                      & \multicolumn{1}{l|}{}    & \cellcolor[HTML]{ffeda0}\usym{2642} & \multicolumn{1}{c|}{\cellcolor[HTML]{31a354}\usym{1F314}} & \multicolumn{1}{c}{\cellcolor[HTML]{ffeda0}\usym{2642}} & \multicolumn{1}{c|}{\cellcolor[HTML]{31a354}\usym{1F314}} & \multicolumn{1}{c|}{\cellcolor[HTML]{ffeda0}\usym{2642}} & \multicolumn{1}{c}{\cellcolor[HTML]{ffeda0}\usym{2642}} & \multicolumn{1}{c|}{\cellcolor[HTML]{31a354}\usym{1F314}}   &                                                     \\
\multicolumn{1}{l|}{\edits{P5: Have a say in making decisions}}   & \multicolumn{3}{l|}{}                                                                                   & \multicolumn{1}{l|}{}    & \multicolumn{2}{c|}{\cellcolor[HTML]{a1d99b}\usym{1F312}}                                   & \multicolumn{2}{l|}{}                                                                                   & \multicolumn{1}{l|}{}                               & \multicolumn{2}{l|}{}                                                                                     &                                                     \\
\multicolumn{1}{l|}{\edits{P6: Consistency in applying processes}}   & \multicolumn{3}{l|}{}                                                                                   & \multicolumn{1}{l|}{}    & \multicolumn{2}{c|}{\cellcolor[HTML]{ffeda0}\faBriefcase}                                                               & \multicolumn{1}{c}{\cellcolor[HTML]{ffeda0}\usym{2642}} & \multicolumn{1}{c|}{\cellcolor[HTML]{31a354}\usym{1F314}} & \multicolumn{1}{l|}{}                               & \multicolumn{2}{c|}{\cellcolor[HTML]{31a354}\usym{1F314}}                                                        &                                                     \\
\multicolumn{1}{l|}{\edits{P7: Appealing decisions}}   & \multicolumn{3}{c|}{\cellcolor[HTML]{ffeda0}\usym{2642}}                                                     & \multicolumn{1}{l|}{}    & \multicolumn{2}{c|}{\cellcolor[HTML]{ffeda0}\usym{2642}}                                 & \multicolumn{1}{c}{\cellcolor[HTML]{ffeda0}\usym{2642}} & \multicolumn{1}{c|}{\cellcolor[HTML]{31a354}\usym{1F314}} & \multicolumn{1}{c|}{\cellcolor[HTML]{ffeda0}\usym{2642}} & \multicolumn{1}{c}{\cellcolor[HTML]{ffeda0}\usym{2642}} & \multicolumn{1}{c|}{\cellcolor[HTML]{a1d99b}\faWheelchair} &                                                     \\ \hline
\multicolumn{1}{l|}{\edits{Int1: Treated politely}} & \multicolumn{3}{l|}{}                                                                                   & \multicolumn{1}{l|}{}    & \multicolumn{2}{l|}{}                                                               & \multicolumn{2}{l|}{}                                                                                   & \multicolumn{1}{l|}{}                               & \multicolumn{2}{c|}{\cellcolor[HTML]{ffeda0}\faBriefcase}                                                                                     & \multicolumn{1}{c}{\cellcolor[HTML]{a1d99b}\faWheelchair} \\
\multicolumn{1}{l|}{\edits{Int2: Treated with respect}} & \multicolumn{3}{c|}{\cellcolor[HTML]{ffeda0}\faBriefcase}                                                                                   & \multicolumn{1}{l|}{}    & \multicolumn{2}{c|}{\cellcolor[HTML]{ffeda0}\usym{2642}}                                 & \multicolumn{2}{l|}{}                                                                                   & \multicolumn{1}{l|}{}                               & \multicolumn{2}{c|}{\cellcolor[HTML]{ffeda0}\faBriefcase}                                                                                     &                                                     \\
\multicolumn{1}{l|}{\edits{Int3: Treated with dignity}} & \multicolumn{3}{l|}{}                                                                                   & \multicolumn{1}{l|}{}    & \multicolumn{2}{c|}{\cellcolor[HTML]{ffeda0}\usym{2642}}                                 & \multicolumn{2}{l|}{}                                                                                   & \multicolumn{1}{l|}{}                               & \multicolumn{2}{c|}{\cellcolor[HTML]{ffeda0}\faBriefcase}                                                                                     & \multicolumn{1}{c}{\cellcolor[HTML]{a1d99b}\faWheelchair} \\
\multicolumn{1}{l|}{\edits{Int4: No inappropriate comments or behaviour}} & \multicolumn{3}{l|}{}                                                                                   & \multicolumn{1}{l|}{}    & \multicolumn{2}{l|}{}                                                               & \multicolumn{2}{l|}{}                                                                                   & \multicolumn{1}{l|}{}                               & \multicolumn{2}{c|}{\cellcolor[HTML]{ffeda0}\faBriefcase}                                                                                     & \multicolumn{1}{c}{\cellcolor[HTML]{a1d99b}\faWheelchair} \\ \hline
\multicolumn{1}{l|}{\edits{Inf1: Direct communication}} & \multicolumn{3}{l|}{}                                                                                   & \multicolumn{1}{l|}{}    & \multicolumn{2}{l|}{}                                                               & \multicolumn{2}{l|}{}                                                                                   & \multicolumn{1}{l|}{}                               & \multicolumn{2}{l|}{}                                                                                     &                                                     \\
\multicolumn{1}{l|}{\edits{Inf2: Reasonable explanations}} & \multicolumn{3}{l|}{}                                                                                   & \multicolumn{1}{l|}{}    & \multicolumn{2}{l|}{}                                                               & \multicolumn{2}{l|}{}                                                                                   & \multicolumn{1}{l|}{}                               & \multicolumn{2}{l|}{}                                                                                     &                                                     \\
\multicolumn{1}{l|}{\edits{Inf3: Timely communication}} & \multicolumn{3}{l|}{}                                                                                   & \multicolumn{1}{l|}{}    & \multicolumn{2}{l|}{}                                                               & \multicolumn{2}{l|}{}                                                                                   & \multicolumn{1}{l|}{}                               & \multicolumn{2}{l|}{}                                                                                     &                                                     \\
\multicolumn{1}{l|}{\edits{Inf4: Tailored communication}} & \multicolumn{3}{l|}{}                                                                                   & \multicolumn{1}{l|}{}    & \multicolumn{2}{l|}{}                                                               & \multicolumn{2}{c|}{\cellcolor[HTML]{ffeda0}\usym{2642}}                                                     & \multicolumn{1}{l|}{}                               & \multicolumn{2}{c|}{}                                                                                     &  \multicolumn{1}{c}{\cellcolor[HTML]{a1d99b}\faBriefcase}                                                   \\
\multicolumn{1}{l|}{\edits{Inf5: Comprehensive explanations}} & \multicolumn{3}{l|}{}                                                                                   & \multicolumn{1}{l|}{}    & \multicolumn{2}{c|}{\cellcolor[HTML]{ffeda0}\usym{2642}}                                 & \multicolumn{2}{l|}{}                                                                                   & \multicolumn{1}{l|}{}                               & \multicolumn{2}{l|}{}                                                                                     & \multicolumn{1}{c}{\cellcolor[HTML]{a1d99b}\faWheelchair} \\ \hline
\multicolumn{12}{c}{\textit{SE Context}}                                                                                                                                                                                                                                                                                                                                                                                                                                                                                                                                       \\ \hline
\multicolumn{1}{l|}{\edits{C1: Customer/client Relations}}  & \multicolumn{3}{l|}{}                                                                                   & \multicolumn{1}{l|}{}    & \multicolumn{2}{l|}{}                                                               & \multicolumn{2}{l|}{}                                                                                   & \multicolumn{1}{l|}{}                               & \multicolumn{2}{l|}{}                                                                                     &                                                     \\
\multicolumn{1}{l|}{\edits{C2: Recruitment}}   & \multicolumn{3}{c|}{\cellcolor[HTML]{ffeda0}\usym{1F314}}                                                      & \multicolumn{1}{l|}{}    & \multicolumn{2}{c|}{\cellcolor[HTML]{ffeda0}\usym{1F314}}                                  & \multicolumn{2}{l|}{}                                                                                   & \multicolumn{1}{l|}{}                               & \multicolumn{2}{l|}{}                                                                                     &                                                     \\
\multicolumn{1}{l|}{\edits{C3: Treatment}}   & \multicolumn{3}{l|}{}                                                                                   & \multicolumn{1}{l|}{}    & \multicolumn{2}{c|}{\cellcolor[HTML]{ffeda0}\usym{1F312}}                                   & \multicolumn{2}{l|}{}                                                                                   & \multicolumn{1}{l|}{}                               & \multicolumn{2}{c|}{\cellcolor[HTML]{ffeda0}\usym{2642}}                                                       & \multicolumn{1}{c}{\cellcolor[HTML]{ffeda0}\usym{1F312}}    \\
\multicolumn{1}{l|}{\edits{C4: Policy}}   & \cellcolor[HTML]{ffeda0}\usym{2642} & \cellcolor[HTML]{a1d99b}\usym{1F314} & \multicolumn{1}{c|}{\cellcolor[HTML]{ffeda0}\faBriefcase} & \multicolumn{1}{l|}{}    & \multicolumn{2}{c|}{\cellcolor[HTML]{ffeda0}\usym{2642}}                                 & \multicolumn{2}{c|}{\cellcolor[HTML]{ffeda0}\usym{2642}}                                                     & \multicolumn{1}{l|}{}                               & \multicolumn{1}{c}{\cellcolor[HTML]{ffeda0}\usym{2642}} & \multicolumn{1}{c|}{\cellcolor[HTML]{31a354}\usym{1F314}}   &                                                     \\
\multicolumn{1}{l|}{\edits{C5: Evaluation}}   & \multicolumn{3}{c|}{\cellcolor[HTML]{ffeda0}\usym{2642}}                                                     & \multicolumn{1}{l|}{}    & \multicolumn{2}{l|}{}                                                               & \multicolumn{2}{c|}{\cellcolor[HTML]{ffeda0}\usym{2642}}                                                     & \multicolumn{1}{l|}{}                               & \multicolumn{2}{l|}{}                                                                                     &                                                     \\
\multicolumn{1}{l|}{\edits{C6: Working Hours}}   & \multicolumn{3}{c|}{\cellcolor[HTML]{ffeda0}\faBriefcase}                                                                                   & \multicolumn{1}{c|}{\cellcolor[HTML]{ffeda0}\faBriefcase}    & \multicolumn{1}{c}{\cellcolor[HTML]{a1d99b}\faWheelchair} & \multicolumn{1}{c|}{\cellcolor[HTML]{ffeda0}\faBriefcase}                                & \multicolumn{2}{l|}{}                                                                                   & \multicolumn{1}{l|}{}                               & \multicolumn{2}{c|}{\cellcolor[HTML]{ffeda0}\faBriefcase}                                                                                     &                                                     \\
\multicolumn{1}{l|}{\edits{C7: Allocation of Work}}   & \multicolumn{3}{c|}{\cellcolor[HTML]{ffeda0}\faBriefcase}                                                                                   & \multicolumn{1}{l|}{}    & \multicolumn{2}{c|}{\cellcolor[HTML]{ffeda0}\usym{2642}}                                 & \multicolumn{2}{l|}{}                                                                                   & \multicolumn{1}{l|}{}                               & \multicolumn{1}{c}{\cellcolor[HTML]{ffeda0}\usym{2642}} & \multicolumn{1}{c|}{\cellcolor[HTML]{31a354}\usym{1F314}}                                                       &                                                     \\
\multicolumn{1}{l|}{\edits{C8: Income}}   & \multicolumn{3}{c|}{\cellcolor[HTML]{ffeda0}\usym{1F312}}                                                       & \multicolumn{1}{l|}{}    & \multicolumn{2}{l|}{}                                                               & \multicolumn{2}{l|}{}                                                                                   & \multicolumn{1}{l|}{}                               & \multicolumn{2}{l|}{}                                                                                     &                                                     \\
\multicolumn{1}{l|}{\edits{C9: Authorship}}   & \multicolumn{3}{c|}{\cellcolor[HTML]{ffeda0}\usym{2642}}                                                     & \multicolumn{1}{l|}{}    & \multicolumn{2}{l|}{}                                                               & \multicolumn{2}{l|}{}                                                                                   & \multicolumn{1}{l|}{}                               & \multicolumn{2}{c|}{\cellcolor[HTML]{ffeda0}\usym{2642}}                                                       &                                                     \\
\multicolumn{1}{l|}{\edits{C10: Demand}}   & \multicolumn{3}{c|}{\cellcolor[HTML]{ffeda0}\faBriefcase}                                                                                   & \multicolumn{1}{l|}{}    & \multicolumn{2}{c|}{\cellcolor[HTML]{ffeda0}\usym{2642}}                                 & \multicolumn{2}{l|}{}                                                                                   & \multicolumn{1}{l|}{}                               & \multicolumn{2}{l|}{}                                                        & \multicolumn{1}{c}{\cellcolor[HTML]{a1d99b}\faWheelchair} \\
 \hline
                          &                                                    &                                                    &                          & \multicolumn{1}{l}{}           & \multicolumn{1}{l}{}                               &                                                    & \multicolumn{1}{l}{}                               &                                                     &                                                    & \multicolumn{1}{l}{}                                 &                                                     \\
\multicolumn{1}{c}{}      & \multicolumn{4}{c}{\cellcolor[HTML]{e5f5e0}$50\% \leq \text{OR} < 100\%$}                                                                &                                & \multicolumn{4}{c}{\cellcolor[HTML]{a1d99b}$100\% \leq \text{OR} < 200\%$}                                                                                         & \multicolumn{1}{c}{}                                & \multicolumn{2}{c}{\cellcolor[HTML]{31a354}$\text{OR} \geq 200\%$} \\
\\
\multicolumn{1}{c}{}      &                                                                 &                                & & & & \multicolumn{4}{c}{\cellcolor[HTML]{ffeda0}$-100\% < \text{OR} \leq -50\%$}           
\end{tabular}
\label{tab:demosummary}
\end{table}

\subsection{RQ3: How do fairness perceptions influence job satisfaction in specific SE contexts?} 
The results of the regression analysis are summarized in the third section of Table \ref{tab:generalsummary}. We found positive and significant correlations between job satisfaction and fairness experiences in specific SE contexts (see \cite{figshare}). 

Overall job satisfaction is significantly correlated with a high level of fairness experience in several specific SE contexts, particularly in the areas of \textit{demand} (C10) \edits{(OR = 1.73, 95\% CI[1.27, 2.37])}, \textit{authorship} (C9) \edits{(OR = 1.59, 95\% CI[1.15, 2.18])}, \textit{allocation of work} (C7) \edits{(OR = 1.56, 95\% CI[1.13, 2.17])}, and \textit{policy} (C4) \edits{(OR = 1.54, 95\% CI[1.14, 2.08])}. Additionally, as observed in RQ1, All four dimensions of fairness, \textit{distributive}, \textit{procedural}, \textit{interpersonal}, and \textit{informational} fairness are linked to higher \textit{overall} satisfaction.
This suggests that activities associated with all four fairness dimensions in these contexts contribute to higher overall job satisfaction. 

Further analysis indicates that, as noted in RQ1, these four dimensions also play a crucial role in enhancing \textit{job security} (SEC). In the context of RQ3, we found a positive and significant correlation in the contexts of \textit{authorship} \edits{(OR = 1.77, 95\% CI[1.29, 2.42])} and \textit{evaluation} \edits{(OR = 1.77, 95\% CI[1.30, 2.41])}. This means that software practitioners feel more secure in their jobs when, during evaluations such as performance reviews, or when their contributions, like coding solutions or innovative ideas, are properly acknowledged in project meetings or official documentation, all four fairness dimensions are fulfilled.

There is also a notable impact of \textit{distributive fairness} in various contexts, such as \textit{working hours} (C6) and \textit{income} (C8), on developers' satisfaction with \textit{rewards and work-life balance} (RWB) and satisfaction with \textit{compensation} (COM). This finding is expected because fairness in how working hours and income are managed directly affects developers' perceptions of being fairly rewarded for their contributions and efforts.

Moreover, \textit{procedural} and \textit{interpersonal fairness} in the contexts of \textit{allocation of work} (C7), \textit{authorship} (C9), and \textit{demand} (C10) significantly influence one's satisfaction with \textit{work culture} (CLT). When work is allocated based on transparent criteria and procedures, developers perceive the process as fair and are more likely to be satisfied with the overall work culture. Additionally, interpersonal fairness in how managers communicate these demands and provide support can significantly affect how employees perceive the fairness of the workload.

Finally, the context of \textit{authorship} (C9) seems particularly well-suited for applying fairness principles. This suggests that practices related to recognizing contributions, giving credit, and ensuring that the right individuals are acknowledged for their work offer opportunities to implement fairness.

\editss{\framedtext{\textit{Takeaway 3}: Fairness experiences across all four dimensions in contexts such as \textit{authorship} significantly enhance \textit{overall} job satisfaction and \textit{job security} among software practitioners.}}

\nocolonsubsection{RQ3.1: How does fairness perceptions influencing job satisfaction differ across demographic groups in the SE contexts?}

Our analysis confirms that fairness perceptions influencing job satisfaction do indeed differ across demographic groups within SE contexts (see \cite{figshare}).
Table \ref{tab:demosummary} (last section) illustrates that fairness perceptions impacting job satisfaction do vary across demographic groups in SE contexts. Different emojis in the table indicate how perceptions of fairness influence job satisfaction among practitioners within the demographic groups represented by those emojis.

\textbf{Gender.}
As noted in RQ2, \textit{distributive}, \textit{procedural} and \textit{informational fairness} were found to have the significant impact on \textit{female} developers' satisfaction with their \textit{manager} (MNG). While in RQ3.1, we found that the contexts of \textit{policy} (C4) \edits{(OR = 0.34, 95\% CI[0.13, 0.77])} and \textit{evaluation} (C5) \edits{(OR = 0.40, 95\% CI[0.17, 0.89])} also significantly affects female developers' satisfaction with their \textit{manager} (MNG).
This suggests that to increase female developers' satisfaction with their managers, organizations should focus on ensuring fair outcomes, accurate information, unbiased processes, and tailored communication during policy implementation and evaluations. 

It was also found in RQ2 that female developers' satisfaction with \textit{job security} (SEC) is shaped by all four dimensions of fairness perceptions. Important elements include receiving outcomes that accurately \textit{reflect their work} (D2) and \textit{effort} (D4), having the ability to \textit{voice opinions} (P2), access to \textit{accurate information} (P3), experiencing \textit{free of bias processes} (P4), having the opportunity to \textit{appeal decisions} (P7), being \textit{treated with respect} (Int2) and \textit{dignity} (Int3), and receiving \textit{comprehensive explanations} (Inf5). These principles are particularly relevant in contexts such as demand (C10), allocation of work (C7), and policy (C4).

This suggests that female developers feel more secure in their jobs when organizations ensure that performance outcomes are fair and reflective of their contributions \cite{Lucht2016Job}, foster a respectful work environment \cite{O'farrell1982Craftworkers}, and maintain clear and unbiased communication \cite{Iseke2019Female}. These can be applied for example, when a female developer is assigned to work (i.e., \textit{allocation of work}) on a critical project with tight deadlines (i.e., \textit{demand}), and that workload distribution is monitored (i.e., \textit{policy}) \cite{Pater2009Gender}.

Finally, looking at the table, it seems that fairness in the context of \textit{policy} (C4) likely impacts female developers' various satisfaction factors more strongly than it does for male developers. This could be because female developers may have historically experienced more inconsistencies or biases in policy implementation in the workplace. As a result, female practitioners may place a higher value on how rules are implemented in the software industry, as they have often faced more challenges.

\textbf{Ethnicity Representation.}
We found that \textit{unbiased processes} (P4) and  \textit{consistent processes} (P6) are particularly important for underrepresented practitioners' satisfaction with \textit{performance} (PERF), especially in the contexts of \textit{policy} (C4) \edits{(OR = 3.61, 95\% CI[1.34, 8.91])} and \textit{allocation of work} (C7) \edits{(OR = 3.08, 95\% CI[1.14, 7.77])}. Our findings suggest that for underrepresented developers, having unbiased and consistent processes is especially important for their satisfaction with performance, particularly in the context of policy decisions and work allocation. When these processes are in place, it appears to support a stronger sense of fairness, which may contribute to greater performance satisfaction.

We also found that \textit{unbiased processes} (P4) is particularly important for underrepresented practitioners' \textit{overall satisfaction} (ALL), especially in the context of \textit{policy} (C4). This suggests that ensuring unbiased processes likely helps underrepresented developers feel that they are being treated fairly, which can mitigate feelings of marginalization or exclusion \cite{Dover2019Mixed}.

\textbf{Work Experience.}
In RQ2, we found that less experienced practitioners feel more satisfied with their performance when treated with politeness, dignity, and respect, and when inappropriate behavior or comments are absent. In the current analysis, satisfaction with \textit{performance} (PRF) seems to be particularly influenced by their \textit{working hours} (C6) \edits{(OR = 0.43, 95\% CI[0.21, 0.89])}.

This suggests that how working hours are managed plays a significant role in shaping the performance satisfaction of less experienced practitioners. It could reflect how an organization or manager considers and respects a practitioner’s time and personal boundaries in relation to their workload. This includes balancing work and personal life, ensuring reasonable working hours, and handling periods of overtime with mutual respect. Less experienced practitioners may be more sensitive to these factors, as they are still establishing themselves professionally \cite{Rastogi2015Ramp-Up,Ando2015}, making respectful management of their time essential to their perception of fairness and performance satisfaction.

We also found that contexts such as \textit{policy} (C4), \textit{working hours} (C6), \textit{allocation of work} (C7), and \textit{demand} (C10) significantly impact the \textit{overall satisfaction} (ALL) of practitioners with less than 3 years of experience. Combining this with the findings from RQ2, where we observed that these practitioners are more satisfied when their outcomes, such as promotions or compensation, align with their effort and when they are treated with respect, we can infer that in each of these contexts, clear recognition of effort and respectful treatment significantly shape the job satisfaction of less experienced practitioners.

This could be because, at this early stage in their careers, less experienced practitioners are seeking affirmation that their contributions are valued, and they rely more heavily on external validation, such as feedback, promotions, or compensation, to gauge their progress \cite{Smith2013The,Li2011The}. In contexts such as policy, working hours, allocation of work, and demand, these practitioners are still learning how the workplace functions and what is expected of them. When their efforts lead to clear, positive outcomes, they are more likely to feel valued and motivated. On the other hand, if their contributions are not recognized or they feel disrespected, it may lead to dissatisfaction, as they might interpret this as a lack of growth.

It is observed from RQ2 that less experienced developers place a strong emphasis on receiving accurate information, as well as seeing processes applied consistently, when it comes to their sense of job security. In this analysis, where we see that context \textit{working hours} (C6) is correlated with \textit{job security} (SEC), we can say that if managers clearly communicate how many hours a project will require and any expectations for overtime or flexibility, and this information remains consistent over time, it helps less experienced practitioners understand what is expected of them. By receiving accurate and consistent updates about their time, these developers are more likely to feel secure that they can meet expectations without sudden changes or surprises.

Finally, it seems that the context of \textit{working hours} (C6) likely impacts less experienced developers’
various satisfaction factors more strongly than it does for
more experienced ones. it affects their satisfaction with \textit{job security} (SEC), with \textit{performance} (PRF), and \textit{overall} (ALL). This likely occurs because less experienced developers are still navigating the expectations of their roles. In contrast, more experienced developers may have developed strategies to manage working hours more effectively.

\textbf{Work Limitation.}
The analysis in RQ2 also shows that software practitioners with work limitations feel significantly stronger about their satisfaction with \textit{compensation} (COM), particularly when it comes to being \textit{treated politely} (Int1), with \textit{dignity} (Int3), avoiding \textit{inappropriate comments or behavior} (Int4), and receiving \textit{comprehensive explanations} (Inf5). In the current analysis, we see that the context of \textit{demand} (C10) affects practitioners with work limitations more strongly than those without when it comes to their satisfaction with \textit{compensation} (COM) \edits{(OR = 2.92, 95\% CI[1.24, 6.83])}.

This suggests that software practitioners with work limitations place a higher value on fair treatment and clear communication, especially in situations where they are required to face high expectations.
These aspects of fairness are likely more important for them because they directly address the challenges these individuals face in the workplace. Practitioners with work limitations may often feel more stress or worry when asked to handle heavy workloads or learn new skills. When managers recognize these limitations and provide clear communication and support, these practitioners are more likely to feel understood and appreciated. Fair treatment in these situations can greatly improve their job satisfaction, particularly regarding compensation, as it meets their need for a supportive and understanding work environment. 

\editss{\framedtext{\textit{Takeaway 4}: Regarding job satisfaction, female practitioners are more impacted by fairness in policy implementation and evaluations. Underrepresented practitioners value fairness in policy implementation and during work allocation. Less experienced practitioners place greater emphasis on fairness when navigating working hours, while practitioners with work limitations prioritize fairness during demanding situations.}}

\section{Discussion}\label{sec:discussion}
\subsection{On Overall Satisfaction and Job Security}
\label{discussion: A}
This study found that all four dimensions of fairness, namely \textit{distributive}, \textit{procedural}, \textit{interpersonal}, and \textit{informational} fairness, significantly influence \textit{overall job satisfaction} and satisfaction with \textit{job security}. \textit{Job security} in this context relates to the perceived stability of one’s employment.


In the SE field, prior research has emphasized the importance of various factors such as autonomy, task variety, and feedback in influencing job satisfaction, particularly in agile environments \cite{Tessem2007}. Autonomy, the degree to which software engineers can control how they work, and feedback on performance are particularly important in shaping their sense of satisfaction and productivity. Similarly, Storey et al. \cite{Storey2021} highlighted appreciation, rewards, impactful work, and work culture as important factors contributing to overall job satisfaction for developers. 

The closest concept to job security in the SE literature is turnover. Turnover in software development refers to the resignation or replacement of software developers, which may lead to severe problems like project postponement and failure \cite{Ma2019A}. In relation to turnover, studies have found that factors such as satisfaction with supervision, workplace morale, and managerial support significantly impact turnover intentions. Research on the impact of developer turnover in open-source software projects highlights the negative impact of high developer turnover, especially among external newcomers, on software quality \cite{Foucault2015Impact}. 

Agile methods have also been linked to higher job satisfaction and lower turnover rates. A study showed that team members using agile methods reported greater job satisfaction and lower turnover intentions compared to those on non-agile teams \cite{melnik2006comparative}. Finally, job satisfaction, onboarding success, and workplace relationships are key factors influencing turnover intentions among software professionals. Training and support for new hires are crucial for retention in these contexts \cite{Sharma2020Exploring}.

This study’s findings add to the existing literature that fairness is a key factor in contributing to software practitioners' job satisfaction and job security. For software practitioners to feel more satisfied with their jobs and more committed to their roles, software organizations should foster fairness across all dimensions. When practitioners perceive fairness in how outcomes are distributed, decisions are made, interpersonal interactions are conducted, and communication is handled, their satisfaction and sense of security in the workplace are significantly enhanced. For researchers, now that we know the importance of fairness in software teams, future studies could explore how fairness remedies can be tailored to address fairness challenges in SE that have already occurred. 

\subsection{Interpersonal Fairness at the Core}
\label{discussion: B}
Among all the dimensions influencing overall job satisfaction, we found that \textit{interpersonal fairness} has the greatest effect. In particular, developers who feel respected are more than twice as likely to report high job satisfaction. Following closely were \textit{treatment with dignity}, \textit{polite treatment}, and avoidance of \textit{inappropriate comments or behavior}. This suggests that out of four fairness dimensions influencing overall job satisfaction, \textit{interpersonal fairness} is at the core.

Fair treatment fosters an environment where individuals can thrive, and this environment is often characterized by high levels of psychological safety \cite{Newman2017Psychological}. According to Google's Project Aristotle, which analyzed data from 180 software teams and 35 statistical models, psychological safety is the most important factor for team success \cite{Osmani_2024}. Software teams that exhibit high psychological safety are not only more innovative and effective but also experience lower turnover rates \cite{Osmani_2024}.

In the fast-paced and collaborative world of SE, psychological safety allows team members to share knowledge, admit mistakes, and propose innovative solutions without fear of negative consequences \cite{Osmani_2024}. This open and supportive atmosphere is closely linked to how individuals are treated within their teams. This highlights the critical role that \textit{interpersonal fairness} plays in fostering psychological safety.
Our findings are also consistent with prior research. Westlund \cite{Westlund2011Leading} found that leadership styles incorporating respect and recognition strongly correlate with increased job satisfaction among software developers.

This suggests that, of the four fairness dimensions, software organizations should prioritize \textit{interpersonal fairness} as a primary lever for improving developer job satisfaction. Managers and team leaders can have a substantial impact on job satisfaction by ensuring respectful, polite, and dignified treatment of their teams. For researchers, future studies could explore how specific \textit{interpersonal fairness} is fostered in diverse teams.

\subsection{Acknowledging Contributions}
\label{discussion: C}
The context of \textit{authorship} seems particularly well-suited for applying fairness principles, as fairness in this context significantly influences \textit{overall satisfaction}, \textit{rewards}, \textit{work-life balance}, \textit{job security}, \textit{manager}, and \textit{team culture satisfaction}. 
In software development, it has been studied that not all contributions are equally recognized. While adding new features is often celebrated, critical but less visible tasks, like bug fixing and code reviews, frequently go unacknowledged \cite{casari2021}. This issue is particularly prevalent in open-source ecosystems, where credit attribution remains a challenge. 

Peripheral developers, those who engage in tasks like bug fixing, are often underappreciated in traditional attribution systems, which tend to favor core contributors \cite{casari2021}. This uneven distribution of credit reinforces power dynamics and can lead to dissatisfaction among those who contribute but remain unrecognized \cite{casari2021}.
Geiger et al. \cite{Geiger2021The} further highlighted that maintainers of open source projects perform significant invisible labor, which is critical for sustaining projects but often goes unrecognized.
Moreover, Kononenko et al. \cite{Kononenko2015Investigating} have shown that code reviews are sometimes underappreciated. Despite their importance in maintaining software quality, the contributions of those who conduct thorough code reviews can go unnoticed. 

To address these imbalances, some open-source ecosystems have begun adopting more equitable models for recognizing contributions. For instance, the \textit{All Contributors} model seeks to fairly acknowledge work beyond just code, including bug reporting, documentation, and community management \cite{Young2021Which}. This ensures that contributors beyond the core developers are properly recognized for their efforts.
At the same time, several studies have explored methods for identifying authorship in source code for both legal purposes and ensuring fair credit for contributions. Bogomolov et al. \cite{Bogomolov2020Authorship} introduced a language-agnostic method for code authorship attribution to detect plagiarism and resolve legal disputes.
A more recent study by Czibula et al.\cite{Czibula2022Enhancing} presented an ensemble of deep autoencoders to improve software authorship attribution to enhance the accuracy in identifying developers behind a code.

Adding to the existing literature, this study underscores the importance of applying the \textit{distributive}, \textit{procedural}, \textit{interpersonal}, and \textit{informational fairness} when recognizing contributions in SE. 
For example, German et al. \cite{German2018} applied Colquitt's fairness theory \cite{Colquitt2001,Colquitt2002} to create a framework for assessing fairness in modern code review processes. This framework provides actionable guidance on implementing fairness principles across the four dimensions in code reviews.

In terms of \textit{distributive fairness}, organizations should determine how to measure contributions and decide whether individuals who contribute more should receive greater recognition or rewards \cite{German2018}. This is to ensure that compensation reflects the value and effort of contributions. \textit{Procedural fairness} involves establishing clear standards for successful contributions and maintaining transparent, accessible review processes \cite{German2018}. This includes, for example, allowing contributors to comment on reviews and providing an appeal mechanism for rejected work \cite{German2018}.

\textit{Interpersonal fairness} is key to maintaining respectful interactions during reviews \cite{German2018}. By defining and enforcing a code of conduct, organizations can ensure contributors are treated respectfully, addressing any inappropriate behavior quickly \cite{German2018}. \textit{Informational fairness}, meanwhile, requires ongoing transparency in the review process. This can involve, for instance, keeping all parties informed, offering constructive feedback, and clearly communicating delays or unexpected outcomes \cite{German2018}. By embedding these principles into code reviews and other recognition processes, software teams can create a more fair system for acknowledging contributions.

On the other hand, researchers can investigate the impact of various  contribution attribution model on team satisfaction, project success, and contributor retention \cite{Lim2002Model,Mcdonald2014Modeling}. By analyzing real-world implementations of fairness frameworks, researchers could identify which approaches yield the most positive outcomes and how to adapt them to different team dynamics.

\subsection{Satisfaction with Managers}
\label{discussion: D}
Interestingly, our analysis did not reveal significant fairness factors influencing overall satisfaction with managers (MNG) for software practitioners as a whole. This is surprising, as one might expect that fair \textit{interpersonal} and \textit{informational} interactions would play a key role in shaping employees’ satisfaction with their managers. Given that managers are typically responsible for overseeing work distribution, providing feedback, and facilitating communication, it is reasonable to assume that fairness in these areas would be crucial to how developers perceive their managers.
However, this broader finding likely masks the experiences of specific demographic subgroups, where fairness plays a more important role in manager satisfaction.

When we looked more closely at the developers' backgrounds, we found that the impact of fairness on satisfaction with managers is stronger for female and underrepresented developers. Specifically, \textit{distributive}, \textit{procedural}, and \textit{informational fairness} significantly impact female developers' satisfaction with their managers.

Previous studies support these findings. Suh \& Hijal-Moghrabi \cite{Suh2021The} demonstrated that fair HR practices and fairness from superiors strongly impacted women’s job satisfaction. Choi \& Rainey \cite{Choi2014Organizational} also found that diversity management and organizational fairness positively influenced job satisfaction, particularly for women, who reported higher satisfaction when fairness was perceived. Choi \cite{Choi2017Workforce} further discovered that racial/ethnic minorities experienced higher job satisfaction when they held majority status in their organization, while Whites reported greater satisfaction in predominantly White settings. This highlights how ethnicity representation can impact one's job satisfaction.

Gender and ethnic biases are prevalent SE as well. Gunawardena et al. \cite{Gunawardena2022DestructiveCI} showed that destructive criticism during code reviews negatively affects female developers' moods and willingness to collaborate. Trinkenreich et al. \cite{Trinkenreich2022} found that women are sometimes assigned to projects based on gender rather than ability, while Paul et al. \cite{Paul2019} revealed that male developers are more likely to write negative comments and offer less assistance to female colleagues.
Our study builds on this body of work, which suggests that managers can play a crucial role in addressing gender bias. By prioritizing fair outcomes, clear communication, and unbiased decision-making processes, managers can significantly improve female developers' satisfaction. 

Similarly, developers from underrepresented ethnic groups, who are often a minority in their teams, face additional challenges in feeling heard and valued. Our findings show that for these developers, satisfaction with managers is strongly influenced by how inclusive and transparent managerial practices are, particularly in decision-making and providing opportunities to voice concerns or appeal decisions. When underrepresented developers feel they lack equal opportunities to participate, they may perceive this as unfair treatment, leading to dissatisfaction with their managers. This sense of disadvantage may stem from a lack of inclusion and transparency \cite{Behfar2006Managing}, which can cause these developers to feel excluded from important team decisions.

This aligns with research by Hawes \cite{Hawes2022The}, which found that minority workers' job satisfaction improves significantly with strong supervisory support. In the predominantly White-dominated field of SE, Nadri et al. \cite{Nadri2021} reported that less than 10\% of contributions to open-source projects come from visibly non-White developers. These systemic barriers further emphasize the need for inclusive and transparent managerial practices to ensure underrepresented developers feel empowered and valued in their roles.

Furthermore, how fairness affects satisfaction with managers across demographic groups highlights the need for further research into how factors like gender and ethnicity representation influence perceptions of management. Researchers can explore how different subgroups experience management differently and what targeted managerial practices can improve their experience. This could include studying the effects of mentorship and sponsorship programs for female and minority developers.

\subsection{Policy Barriers for Women and Minorities}
\label{discussion: E}
Similar to the findings in the previous subsection, fairness in the context of \textit{policy} appears to impact female and underrepresented developers more strongly. This may stem from the fact that certain rules and policies in SE may unintentionally disadvantage female and underrepresented developers, making fairness in policy application crucial for improving their overall job satisfaction.

Policies that lack transparency or rely on subjective criteria can disproportionately disadvantage female and underrepresented developers. These groups often lack access to informal networks that help advance careers, facing a \textit{glass ceiling} despite their qualifications \cite{Daley1996Paths}. Even when women and minorities meet formal criteria like performance ratings, they often struggle to advance because they are excluded from the informal mentorship or sponsorship opportunities that typically benefit others, particularly white males \cite{Daley1996Paths}. Unconscious biases continue to hinder their progress in Science, Technology, Engineering, and Mathematics (STEM) fields, limiting their opportunities for promotion and career growth, even in workplaces that have diversity policies in place \cite{Greider2019Increasing,Villablanca2017Evaluating}.

Inflexible policies around working hours, parental leave, or caregiving responsibilities can also disproportionately impact female developers. Research from sectors like healthcare shows that work-life balance strongly affects organizational commitment and job satisfaction for women \cite{Shabir2020Impact}. Similarly, remote work policies pose challenges if they are not equally accessible or if there is stigma attached to using them. Women, particularly those using flexible work options, often face a \textit{flexibility stigma}, which can lead to perceptions of being less committed \cite{Chung2018Gender, Rogier2004The, Vandello2013When}.
The COVID-19 pandemic has reduced some of the stigma around remote work, offering an opportunity to create more inclusive environments for women and minority employees \cite{Barhate2021Emerging}. 

For organizations aiming to improve satisfaction among female and underrepresented developers, the primary focus should be on ensuring the fairness of policy implementation. Policies around promotion, project allocation, and flexible work arrangements must be applied transparently and without bias. Regularly auditing these policies for fairness, especially in how they impact different demographic groups, is essential for creating an inclusive and supportive work environment. 

\subsection{Managing Work Hours for the Less Experienced}
\label{discussion: F}
Our analysis indicates that the context of \textit{working hours} likely has a stronger impact on less experienced developers' satisfaction across various factors, \textit{job security}, \textit{performance}, and \textit{overall satisfaction}, compared to more experienced developers. Due to their limited familiarity with their job, less experienced developers often face longer working hours as they navigate the learning curve, which can lead to increased dissatisfaction when compared to their more experienced colleagues who manage tasks more efficiently within regular hours.


Several factors may contribute to the challenges faced by less experienced developers. One key challenge is the experience gap. Junior employees often struggle with unfamiliarity with company processes, technical skills, and organizational culture, which delays their performance and puts them at a disadvantage compared to senior employees who have already navigated these hurdles \cite{Rastogi2015Ramp-Up}.
Experienced developers tend to introduce fewer defects, even when working on similar tasks, and are generally more efficient in their performance \cite{Tsunoda2017Evaluating}. Their experience and knowledge of the project help them handle complex tasks more efficiently, often completing them faster and with better results \cite{Fronza2011ProfilingTE}. In contrast, less experienced developers spend more time searching for information and understanding code, leading to slower work processes \cite{Ando2015}. As less experienced developers gain more experience, their performance gradually become more like those of experienced developers \cite{Fronza2011}.

Research shows that long and irregular working hours negatively impact mental capacity and work quality for all developers, regardless of experience. Extended periods of continuous work reduce efficiency \cite{Rodriguez2018}. Meyer et al. suggest developing better tools to help developers plan their workday and boost personal productivity \cite{Meyer2017The}.  Additionally, Rastogi et al. highlight the value of mentorship or buddy systems for new hires, which can help them quickly understand company processes and become productive faster \cite{Rastogi2015Ramp-Up}. A supportive environment where new employees feel comfortable asking questions without fear of judgment is also important for their development \cite{Rastogi2015Ramp-Up}.

Our findings reinforce this by emphasizing that the way working hours are managed has an impact on the satisfaction of less experienced developers. Respecting their time and personal boundaries, especially concerning overtime, is crucial to their sense of fairness. Managers who provide clear communication about project time requirements, including overtime and flexibility help less experienced practitioners feel more secure. Most importantly, respecting developers' personal boundaries by not expecting them to consistently work late or during personal time fosters a fairer work environment.

Moreover, researchers could examine how mentorship or \textit{buddy} systems improve the onboarding process and reduce the learning curve. Studies could assess the effectiveness of these systems in helping new hires manage their time better and become productive more quickly, and how these systems impact satisfaction.

\subsection{Supporting Work-Limited Developers}
\label{discussion: G}
Our analysis shows that developers with work limitations are often disadvantaged, especially when it comes to compensation in high-pressure situations. According to the National Health Interview Survey (NHIS), a work limitation is defined as a condition that limits a person's ability to perform certain tasks or activities required for their job \cite{Altman2014}. For software engineers and developers, this could mean physical, cognitive, or psychological challenges that may affect their performance, such as chronic illnesses, disabilities, or other long-term health conditions.


A study examining work limitations due to chronic health conditions found that such limitations often led to reduced productivity and lower wages \cite{Choe2017Duration}. However, workers whose jobs were better matched to their limitations experienced better outcomes, including higher wages and longer job tenure \cite{Choe2017Duration}. Similarly, Siu et al. found that employees with chronic diseases valued fair treatment, clear communication, and flexible accommodations, all of which positively impacted their job satisfaction and work performance \cite{Siu2013Work}.

Providing accommodations and adjusting work conditions for employees with chronic physical conditions has been shown to improve their self-efficacy and reduce fatigue \cite{Varekamp2011Effect}. Lerner et al. further confirmed that employees with chronic health conditions face challenges in managing time, physical demands, and interpersonal tasks, often resulting in productivity loss \cite{Lerner2003Relationship}. This aligns with the importance of non-monetary compensation, such as flexible work hours and support services, as essential factors for job satisfaction among these employees \cite{Lerner2003Relationship}.

Developers with disabilities face additional challenges in the workplace, including issues with project management practices, workplace dynamics, and the accessibility of development tools. For example, visually impaired developers often struggle with the accessibility of development environments and tools, which negatively impacts their productivity \cite{Johnson2022Program-L:,Huff2020Examining}. Programs like \textit{Catalisa}, which train and hire people with disabilities as software developers, underscore the importance of tailored training to improve inclusion and accessibility within the industry \cite{Cardoso-Pereira2023Supporting}. 

Our findings shed light on how developers with work limitations are particularly affected by fairness problems when faced with high expectations. These developers may experience additional stress or anxiety when managing heavy workloads or learning new tasks. To mitigate these challenges, managers and organizations should prioritize clear communication, offer tailored accommodations, and ensure fair treatment, especially in high-pressure situations. Tailored compensation, both monetary and non-monetary, can greatly enhance their job satisfaction by creating a workplace that meets their needs. By addressing the specific needs of work-limited practitioners through flexible work conditions and support services, organizations can foster an environment where these developers feel more satisfied.

Future research could explore the types of accommodations that are most effective in improving job satisfaction for software developers with work limitations. This could include case studies of specific organizations that implement flexible working conditions \cite{Conradie2019To,Mulay2017A,Scholarios2004Work-life}, and how different types of accommodations \cite{Joyce2010Flexible} (e.g., flexible hours, ergonomic tools, or assistive technology) impact the developers.

\section{Threats to Validity}\label{sec:validity}
\textit{Construct Validity.} Since perceptions of fairness and job satisfaction are subjective, individual interpretations of survey questions may vary based on personal experiences or biases. To minimize this, we used well-established frameworks, including Colquitt’s fairness theory \cite{Colquitt2001} and Storey et al.'s job satisfaction model \cite{Storey2021}. These scales have been rigorously tested and help ensure that the questions measure what they are intended to. We also ensured anonymity in the survey and conducted a pilot test with six respondents to refine any unclear questions and adjust wording based on feedback.

\edits{Like many survey-based studies, our findings rely on self-reported responses, which may be subject to biases such as social desirability or recall bias. While we mitigated some of these risks by ensuring anonymity and using validated survey instruments, we recognize that participants may still over- or under-report certain experiences. Future studies could strengthen construct validity by combining self-reported data with qualitative methods such as interviews or focus groups to triangulate findings and provide deeper insights.}

Another construct-related concern is whether all respondents were legitimate software practitioners. Since the survey was anonymous and not linked to participants’ online profiles, we could not verify professional identity. However, given the absence of monetary incentives, and the use of recruitment channels (e.g., LinkedIn, SE workshops), we consider the risk of identity falsification to be minimal. Respondents had no clear motivation to misrepresent their role, and the targeted nature of our outreach makes it unlikely that the survey reached individuals without relevant SE experience.

In addition, we considered the practical significance of the relationships between fairness and job satisfaction, beyond just statistical significance in the ordinal regression analysis. To address this, we calculated percentage-based odds ratios as effect sizes and only focused on relationships with an odds ratio of 50\% or more. This threshold ensured that the results we considered had substantial practical relevance, filtering out statistically significant results with limited practical implications.

Lastly, perceptions of fairness can vary across cultural contexts. Since SE teams are often diverse and global, fairness constructs rooted in Western organizational justice literature may not fully apply in non-Western settings. By conducting a moderation analysis in RQ2 and RQ3.1, we accounted for potential demographic differences, ensuring that our findings reflect the varying impacts of fairness perceptions on job satisfaction across diverse groups.

\textit{Internal Validity.} While our study highlights a relationship between fairness perceptions and job satisfaction, we acknowledge that it does not conclusively establish causality. However, this limitation is mitigated by both our calculation of the effect size and the extensive body of organizational justice literature, consistent with prior research across various industries and contexts \cite{Mossholder1998, Wesolowski1997, Masterson2000, judge2004organizational, konovsky1991perceived, Moorman1991, colquitt2001justice}. Given this robust theoretical grounding and the significant effect sizes of minimum 50\% effect observed in our study, we argue that the direction of the relationship identified is aligned with well-established patterns in organizational behavior.

\edits{Another potential threat is the imbalance in the distribution of certain demographic categories, which limited our ability to conduct reliable moderation analyses across all groups. To mitigate this, we focused our analyses on demographic variables with sufficiently large subgroup sizes to ensure statistical reliability and reduce the risk of biased estimates, as recommended in prior research \cite{Zhang2017, griffin2023tutorial}. Specifically, we analyzed \textit{gender}, \textit{ethnicity representation}, \textit{work limitation}, and \textit{work experience}, where each category included an adequate number of respondents. Although we were unable to analyze other categories, such as \textit{age}, \textit{ethnicity}, \textit{role}, and \textit{role nature}, this approach ensured that our moderation analyses were grounded in robust data, and to minimize the likelihood of skewed conclusions due to small sample sizes. Further clarification on this methodological decision is available in Section \ref{sec:moderation}.}

\textit{External Validity.} A potential threat to external validity is whether fairness perceptions in SE are simply broader management issues applicable to any field. Our findings show that the four dimensions of fairness affect overall job satisfaction and satisfaction with job security in software teams. Out of the 4 dimensions, \textit{interpersonal fairness} emerged as having the greatest impact on overall job satisfaction. This highlights that, beyond technical aspects, the human and social aspects of software development play a crucial role.

Our findings also show that the four dimensions of fairness not only affect overall job satisfaction and job security in software teams but also have specific implications within the SE context. 
We found that fairness concerns are particularly relevant in SE contexts such as recognizing contributions (i.e., authorship), where developers often feel underappreciated. This demonstrates that fairness issues are not just general management concerns but are integral to improving the working environment in SE.

Our findings also indicate that fairness perceptions impact developers differently depending on their demographic backgrounds. This shows that fairness in SE is not just a collective issue but one that affects different groups in specific ways. This is particularly important for software development teams that are often globally distributed and diverse but also dominated by certain demographics. These dynamics underscore the need for tailored approaches to address fairness disparities across diverse teams.
In essence, our study emphasizes that fairness in SE cannot be reduced to general management practices, but rather it demands context-specific solutions that recognize the unique challenges faced by software practitioners.

\edits{Finally, while our sample is not representative in a statistical sense, we ensured heterogeneity by recruiting participants across gender, age, team composition, roles, experience levels. This diversity improves the credibility of our findings in line with SE research norms for theory-building studies \cite{baltes2022sampling}. We acknowledge that our findings may not generalize to all software practitioners globally, especially those outside our sample’s distribution. As Baltes and Ralph caution \cite{baltes2022sampling}, non-probability samples may limit statistical generalization. We address this by clearly defining our target population and transparently reporting recruitment procedures. Future studies with larger and more representative samples can build on our findings to deepen and extend this line of research.}

\section{Conclusion}\label{sec:conclusion}
This study deepens our understanding of how fairness perceptions significantly influence job satisfaction and job security among software practitioners across all four dimensions: distributive, procedural, interpersonal, and informational fairness. Importantly, we found that the strength of this relationship varies across demographic groups, being notably stronger for female, underrepresented, less experienced practitioners, and those with work limitations.

In specific SE contexts, such as authorship, fairness practices like recognizing contributions and giving proper credit are crucial for enhancing job satisfaction and job security. Fairness perceptions also vary by demographic group: female developers are more impacted by fairness in policy implementation and evaluations, underrepresented practitioners value fairness during high-demand situations, less experienced practitioners emphasize fairness in managing working hours, and practitioners with work limitations prioritize fairness in demanding situations.

Among the four dimensions, interpersonal fairness has the greatest effect on job satisfaction, suggesting that software organizations should prioritize interpersonal fairness as a key lever for improving satisfaction. In particular, authorship stands out as an important area where fairness principles should be applied, given the lack of recognition for all contributions in software development.
Our findings also emphasize that fairness in SE is not just a collective issue but one that affects different groups in specific ways.

Organizations should focus on applying all four fairness dimensions, particularly interpersonal fairness, within SE contexts like authorship, while tailoring practices to meet the needs of diverse demographic groups. Specific strategies have been outlined to address fairness concerns for software practitioners collectively (see Section \ref{discussion: A}–\ref{discussion: C}) and based on their backgrounds (see Section \ref{discussion: D}–\ref{discussion: G}).
\editss{While these findings offer valuable insights, they should be interpreted with caution due to the study’s reliance on self-reported perceptions and its sampling frame as discussed in Section~\ref{sec:validity}.}
Future research could explore remedies for fairness problems that have already occurred, examine how fairness manifests in SE contexts, evaluate the effectiveness of suggested strategies, and bridge the gap between collective and demographic-specific fairness concerns.

\section*{Data Availability}
\label{sec:data}
Our supplementary materials \cite{figshare} contain source code used to retrieve and analyze data, results for all our RQs, and instructions how to replicate the results.

\section*{Acknowledgments}
The survey was approved by the ethics committee at the authors' affiliations under ethics approval numbers 97126442 and UCL/CSREC/R/40.
\balance
\bibliographystyle{ACM-Reference-Format}
\bibliography{sample}


\begin{thebibliography}{145}


\ifx \showCODEN    \undefined \def \showCODEN     #1{\unskip}     \fi
\ifx \showDOI      \undefined \def \showDOI       #1{#1}\fi
\ifx \showISBNx    \undefined \def \showISBNx     #1{\unskip}     \fi
\ifx \showISBNxiii \undefined \def \showISBNxiii  #1{\unskip}     \fi
\ifx \showISSN     \undefined \def \showISSN      #1{\unskip}     \fi
\ifx \showLCCN     \undefined \def \showLCCN      #1{\unskip}     \fi
\ifx \shownote     \undefined \def \shownote      #1{#1}          \fi
\ifx \showarticletitle \undefined \def \showarticletitle #1{#1}   \fi
\ifx \showURL      \undefined \def \showURL       {\relax}        \fi
\providecommand\bibfield[2]{#2}
\providecommand\bibinfo[2]{#2}
\providecommand\natexlab[1]{#1}
\providecommand\showeprint[2][]{arXiv:#2}

\bibitem[fig(2024)]%
        {figshare}
 \bibinfo{year}{2024}\natexlab{}.
\newblock \bibinfo{title}{Supplementary Materials for the Study on Fairness
  Perceptions and Job Satisfaction in Software Engineering}.
\newblock
\newblock
\urldef\tempurl%
\url{https://figshare.com/s/93a86ee9e5fc921f3772}
\showURL{%
\tempurl}


\bibitem[Acuña et~al\mbox{.}(2009)]%
        {Acuña2009}
\bibfield{author}{\bibinfo{person}{S.~T. Acuña}, \bibinfo{person}{M.~N.
  Gómez}, {and} \bibinfo{person}{N. Juristo}.}
  \bibinfo{year}{2009}\natexlab{}.
\newblock \showarticletitle{How do personality, team processes and task
  characteristics relate to job satisfaction and software quality?}
\newblock \bibinfo{journal}{\emph{Inf. and Softw. Technol.}}
  \bibinfo{volume}{51} (\bibinfo{year}{2009}), \bibinfo{pages}{627--639}.
\newblock
Issue 3.
\urldef\tempurl%
\url{https://doi.org/10.1016/j.infsof.2008.08.006}
\showDOI{\tempurl}


\bibitem[Adams(1965)]%
        {ADAMS1965267}
\bibfield{author}{\bibinfo{person}{J.~Stacy Adams}.}
  \bibinfo{year}{1965}\natexlab{}.
\newblock \showarticletitle{Inequity In Social Exchange}.
\newblock \bibinfo{series}{Advances in Experimental Social Psychology},
  Vol.~\bibinfo{volume}{2}. \bibinfo{publisher}{Academic Press},
  \bibinfo{pages}{267--299}.
\newblock
\showISSN{0065-2601}
\urldef\tempurl%
\url{https://doi.org/10.1016/S0065-2601(08)60108-2}
\showDOI{\tempurl}


\bibitem[Agresti(2010)]%
        {Agresti2010}
\bibfield{author}{\bibinfo{person}{A. Agresti}.}
  \bibinfo{year}{2010}\natexlab{}.
\newblock \showarticletitle{Analysis of ordinal categorical data}.
\newblock \bibinfo{journal}{\emph{Wiley Series in Probability and Statistics}}
  (\bibinfo{year}{2010}).
\newblock
\urldef\tempurl%
\url{https://doi.org/10.1002/9780470594001}
\showDOI{\tempurl}


\bibitem[Almog et~al\mbox{.}(2023)]%
        {almog2023attention}
\bibfield{author}{\bibinfo{person}{Shahar Almog}, \bibinfo{person}{Andrea
  V{\'a}squez~Ferreiro}, \bibinfo{person}{Meredith~S Berry}, {and}
  \bibinfo{person}{Jillian~M Rung}.} \bibinfo{year}{2023}\natexlab{}.
\newblock \showarticletitle{Are the attention checks embedded in delay
  discounting tasks a valid marker for data quality?}
\newblock \bibinfo{journal}{\emph{Experimental and clinical
  psychopharmacology}} \bibinfo{volume}{31}, \bibinfo{number}{5}
  (\bibinfo{year}{2023}), \bibinfo{pages}{908}.
\newblock


\bibitem[Altman(2014)]%
        {Altman2014}
\bibfield{author}{\bibinfo{person}{B.~M. Altman}.}
  \bibinfo{year}{2014}\natexlab{}.
\newblock \showarticletitle{Definitions, concepts, and measures of disability}.
\newblock \bibinfo{journal}{\emph{Annals of Epidemiology}}
  \bibinfo{volume}{24} (\bibinfo{year}{2014}), \bibinfo{pages}{2--7}.
\newblock
Issue 1.
\urldef\tempurl%
\url{https://doi.org/10.1016/j.annepidem.2013.05.018}
\showDOI{\tempurl}


\bibitem[Ando et~al\mbox{.}(2015)]%
        {Ando2015}
\bibfield{author}{\bibinfo{person}{Reou Ando}, \bibinfo{person}{Seiji Sato},
  \bibinfo{person}{Chihiro Uchida}, \bibinfo{person}{Hironori Washizaki},
  \bibinfo{person}{Yoshiaki Fukazawa}, \bibinfo{person}{Sakae Inoue},
  \bibinfo{person}{Hiroyuki Ono}, \bibinfo{person}{Yoshiiku Hanai},
  \bibinfo{person}{Masanobu Kanazawa}, \bibinfo{person}{Kazutaka Sone},
  \bibinfo{person}{Katsushi Namba}, {and} \bibinfo{person}{Mikihiko Yamamoto}.}
  \bibinfo{year}{2015}\natexlab{}.
\newblock \showarticletitle{How Does Defect Removal Activity of Developer Vary
  with Development Experience?}
\newblock
\urldef\tempurl%
\url{https://doi.org/10.18293/SEKE2015-221}
\showDOI{\tempurl}


\bibitem[Azur et~al\mbox{.}(2011)]%
        {Azur2011Multiple}
\bibfield{author}{\bibinfo{person}{Melissa~J. Azur}, \bibinfo{person}{E.
  Stuart}, \bibinfo{person}{C. Frangakis}, {and} \bibinfo{person}{P. Leaf}.}
  \bibinfo{year}{2011}\natexlab{}.
\newblock \showarticletitle{Multiple imputation by chained equations: what is
  it and how does it work?}
\newblock \bibinfo{journal}{\emph{Int. J. of Methods in Psychiatric Research}}
  \bibinfo{volume}{20} (\bibinfo{year}{2011}).
\newblock
\urldef\tempurl%
\url{https://doi.org/10.1002/mpr.329}
\showDOI{\tempurl}


\bibitem[Baltes et~al\mbox{.}(2020)]%
        {Baltes2020}
\bibfield{author}{\bibinfo{person}{Sebastian Baltes}, \bibinfo{person}{George
  Park}, {and} \bibinfo{person}{Alexander Serebrenik}.}
  \bibinfo{year}{2020}\natexlab{}.
\newblock \showarticletitle{Is 40 the New 60? How Popular Media Portrays the
  Employability of Older Software Developers}.
\newblock \bibinfo{journal}{\emph{IEEE Softw.}} \bibinfo{volume}{37},
  \bibinfo{number}{6} (\bibinfo{year}{2020}), \bibinfo{pages}{26--31}.
\newblock
\urldef\tempurl%
\url{https://doi.org/10.1109/MS.2020.3014178}
\showDOI{\tempurl}


\bibitem[Baltes and Ralph(2022)]%
        {baltes2022sampling}
\bibfield{author}{\bibinfo{person}{Sebastian Baltes} {and}
  \bibinfo{person}{Paul Ralph}.} \bibinfo{year}{2022}\natexlab{}.
\newblock \showarticletitle{Sampling in software engineering research: A
  critical review and guidelines}.
\newblock \bibinfo{journal}{\emph{Empirical Software Engineering}}
  \bibinfo{volume}{27}, \bibinfo{number}{4} (\bibinfo{year}{2022}),
  \bibinfo{pages}{94}.
\newblock


\bibitem[Barclay et~al\mbox{.}(2005)]%
        {barclay2005exploring}
\bibfield{author}{\bibinfo{person}{Laurie~J Barclay}, \bibinfo{person}{Daniel~P
  Skarlicki}, {and} \bibinfo{person}{S~Douglas Pugh}.}
  \bibinfo{year}{2005}\natexlab{}.
\newblock \showarticletitle{Exploring the role of emotions in injustice
  perceptions and retaliation.}
\newblock  \bibinfo{volume}{90}, \bibinfo{number}{4} (\bibinfo{year}{2005}),
  \bibinfo{pages}{629}.
\newblock


\bibitem[Barhate and Hirudayaraj(2021)]%
        {Barhate2021Emerging}
\bibfield{author}{\bibinfo{person}{B. Barhate} {and} \bibinfo{person}{Malar
  Hirudayaraj}.} \bibinfo{year}{2021}\natexlab{}.
\newblock \showarticletitle{Emerging Career Realities during the Pandemic: What
  Does it Mean for Women’s Career Development?}
\newblock \bibinfo{journal}{\emph{Advances in Developing Human Resources}}
  \bibinfo{volume}{23} (\bibinfo{year}{2021}), \bibinfo{pages}{253 -- 266}.
\newblock
\urldef\tempurl%
\url{https://doi.org/10.1177/15234223211017851}
\showDOI{\tempurl}


\bibitem[Baron and Kenny(1986)]%
        {baron1986moderator}
\bibfield{author}{\bibinfo{person}{Reuben~M Baron} {and}
  \bibinfo{person}{David~A Kenny}.} \bibinfo{year}{1986}\natexlab{}.
\newblock \showarticletitle{The moderator--mediator variable distinction in
  social psychological research: Conceptual, strategic, and statistical
  considerations.}
\newblock \bibinfo{journal}{\emph{J. of personality and social psychology}}
  \bibinfo{volume}{51}, \bibinfo{number}{6} (\bibinfo{year}{1986}),
  \bibinfo{pages}{1173}.
\newblock


\bibitem[Beecham et~al\mbox{.}(2008)]%
        {Beecham2008}
\bibfield{author}{\bibinfo{person}{S. Beecham}, \bibinfo{person}{N. Baddoo},
  \bibinfo{person}{T. Hall}, \bibinfo{person}{H. Robinson}, {and}
  \bibinfo{person}{H. Sharp}.} \bibinfo{year}{2008}\natexlab{}.
\newblock \showarticletitle{Motivation in software engineering: a systematic
  literature review}.
\newblock \bibinfo{journal}{\emph{Inf. and Softw. Technol.}}
  \bibinfo{volume}{50} (\bibinfo{year}{2008}), \bibinfo{pages}{860--878}.
\newblock
Issue 9-10.
\urldef\tempurl%
\url{https://doi.org/10.1016/j.infsof.2007.09.004}
\showDOI{\tempurl}


\bibitem[Beesley et~al\mbox{.}(2021)]%
        {Beesley2021Multiple}
\bibfield{author}{\bibinfo{person}{Lauren~J. Beesley}, \bibinfo{person}{I.
  Bondarenko}, \bibinfo{person}{Michael~R. Elliot}, \bibinfo{person}{A.
  Kurian}, \bibinfo{person}{S. Katz}, {and} \bibinfo{person}{Jeremy M.~G.
  Taylor}.} \bibinfo{year}{2021}\natexlab{}.
\newblock \showarticletitle{Multiple imputation with missing data indicators}.
\newblock \bibinfo{journal}{\emph{Statistical Methods in Medical Research}}
  \bibinfo{volume}{30} (\bibinfo{year}{2021}), \bibinfo{pages}{2685 -- 2700}.
\newblock
\urldef\tempurl%
\url{https://doi.org/10.1177/09622802211047346}
\showDOI{\tempurl}


\bibitem[Behfar et~al\mbox{.}(2006)]%
        {Behfar2006Managing}
\bibfield{author}{\bibinfo{person}{K. Behfar}, \bibinfo{person}{M.~C. Kern},
  {and} \bibinfo{person}{J.~M. Brett}.} \bibinfo{year}{2006}\natexlab{}.
\newblock \showarticletitle{Managing challenges in multicultural teams}.
\newblock \bibinfo{journal}{\emph{Research on Managing Groups and Teams}}
  (\bibinfo{year}{2006}), \bibinfo{pages}{233--262}.
\newblock
\urldef\tempurl%
\url{https://doi.org/10.1016/s1534-0856(06)09010-4}
\showDOI{\tempurl}


\bibitem[Bogomolov et~al\mbox{.}(2020)]%
        {Bogomolov2020Authorship}
\bibfield{author}{\bibinfo{person}{Egor Bogomolov}, \bibinfo{person}{V.
  Kovalenko}, \bibinfo{person}{Alberto Bacchelli}, {and} \bibinfo{person}{T.
  Bryksin}.} \bibinfo{year}{2020}\natexlab{}.
\newblock \showarticletitle{Authorship attribution of source code: a
  language-agnostic approach and applicability in software engineering}.
\newblock \bibinfo{journal}{\emph{Proc. of the 29th ACM Joint Meeting on
  European Softw. Eng. Conf. and Symp. on the Foundations of Softw. Eng.}}
  (\bibinfo{year}{2020}).
\newblock
\urldef\tempurl%
\url{https://doi.org/10.1145/3468264.3468606}
\showDOI{\tempurl}


\bibitem[Breugst et~al\mbox{.}(2015)]%
        {Breugst2015}
\bibfield{author}{\bibinfo{person}{N. Breugst}, \bibinfo{person}{H. Patzelt},
  {and} \bibinfo{person}{P. Rathgeber}.} \bibinfo{year}{2015}\natexlab{}.
\newblock \showarticletitle{How should we divide the pie? equity distribution
  and its impact on entrepreneurial teams}.
\newblock \bibinfo{journal}{\emph{J. of Business Venturing}}
  \bibinfo{volume}{30} (\bibinfo{year}{2015}), \bibinfo{pages}{66--94}.
\newblock
Issue 1.
\urldef\tempurl%
\url{https://doi.org/10.1016/j.jbusvent.2014.07.006}
\showDOI{\tempurl}


\bibitem[Canedo et~al\mbox{.}(2019)]%
        {Canedo2019}
\bibfield{author}{\bibinfo{person}{Edna~Dias Canedo},
  \bibinfo{person}{Heloise~Acco Tives}, \bibinfo{person}{Madianita~Bogo
  Marioti}, \bibinfo{person}{Fabiano Fagundes}, {and} \bibinfo{person}{José
  Antonio~Siqueira de Cerqueira}.} \bibinfo{year}{2019}\natexlab{}.
\newblock \showarticletitle{Barriers Faced by Women in Software Development
  Projects}.
\newblock \bibinfo{journal}{\emph{Inf. 2019, Vol. 10, Page 309}}
  \bibinfo{volume}{10} (\bibinfo{date}{10} \bibinfo{year}{2019}),
  \bibinfo{pages}{309}.
\newblock
Issue 10.
\showISSN{2078-2489}
\urldef\tempurl%
\url{https://doi.org/10.3390/INFO10100309}
\showDOI{\tempurl}


\bibitem[Cao et~al\mbox{.}(2013)]%
        {Cao2013}
\bibfield{author}{\bibinfo{person}{Z. Cao}, \bibinfo{person}{J. Chen}, {and}
  \bibinfo{person}{Y. Song}.} \bibinfo{year}{2013}\natexlab{}.
\newblock \showarticletitle{Does total rewards reduce the core employees’
  turnover intention?}
\newblock \bibinfo{journal}{\emph{Int. J. of Business and Manage.}}
  \bibinfo{volume}{8} (\bibinfo{year}{2013}).
\newblock
Issue 20.
\urldef\tempurl%
\url{https://doi.org/10.5539/ijbm.v8n20p62}
\showDOI{\tempurl}


\bibitem[Cardoso-Pereira et~al\mbox{.}(2023)]%
        {Cardoso-Pereira2023Supporting}
\bibfield{author}{\bibinfo{person}{I. Cardoso-Pereira},
  \bibinfo{person}{Geraldo Gomes}, \bibinfo{person}{D. Ribeiro},
  \bibinfo{person}{A. Souza}, \bibinfo{person}{Danilo Lucena}, {and}
  \bibinfo{person}{Gustavo Pinto}.} \bibinfo{year}{2023}\natexlab{}.
\newblock \showarticletitle{Supporting the Careers of Developers With
  Disabilities: Lessons From Zup Innovation}.
\newblock \bibinfo{journal}{\emph{IEEE Softw.}}  \bibinfo{volume}{40}
  (\bibinfo{year}{2023}), \bibinfo{pages}{58--65}.
\newblock
\urldef\tempurl%
\url{https://doi.org/10.1109/MS.2023.3282544}
\showDOI{\tempurl}


\bibitem[Casari et~al\mbox{.}(2021)]%
        {casari2021}
\bibfield{author}{\bibinfo{person}{A. Casari}, \bibinfo{person}{K. McLaughlin},
  \bibinfo{person}{M.~Z. Trujillo}, \bibinfo{person}{J. Young}, {and}
  \bibinfo{person}{J.~P. Bagrow}.} \bibinfo{year}{2021}\natexlab{}.
\newblock \showarticletitle{Open source ecosystems need equitable credit across
  contributions}.
\newblock \bibinfo{journal}{\emph{Nature Computational Science}}
  \bibinfo{volume}{1} (\bibinfo{year}{2021}), \bibinfo{pages}{2--2}.
\newblock
Issue 1.
\urldef\tempurl%
\url{https://doi.org/10.1038/s43588-020-00011-w}
\showDOI{\tempurl}


\bibitem[Chhabra et~al\mbox{.}(2017)]%
        {Chhabra2017A}
\bibfield{author}{\bibinfo{person}{Geeta Chhabra}, \bibinfo{person}{Vasudha
  Vashisht}, {and} \bibinfo{person}{Jayanthi Ranjan}.}
  \bibinfo{year}{2017}\natexlab{}.
\newblock \showarticletitle{A Comparison of Multiple Imputation Methods for
  Data with Missing Values}.
\newblock \bibinfo{journal}{\emph{Indian journal of science and technology}}
  \bibinfo{volume}{10} (\bibinfo{year}{2017}), \bibinfo{pages}{1--7}.
\newblock
\urldef\tempurl%
\url{https://doi.org/10.17485/IJST/2017/V10I19/110646}
\showDOI{\tempurl}


\bibitem[Choe and Baldwin(2017)]%
        {Choe2017Duration}
\bibfield{author}{\bibinfo{person}{Chung Choe} {and} \bibinfo{person}{M.
  Baldwin}.} \bibinfo{year}{2017}\natexlab{}.
\newblock \showarticletitle{Duration of disability, job mismatch and employment
  outcomes}.
\newblock \bibinfo{journal}{\emph{Applied Economics}}  \bibinfo{volume}{49}
  (\bibinfo{year}{2017}), \bibinfo{pages}{1001 -- 1015}.
\newblock
\urldef\tempurl%
\url{https://doi.org/10.1080/00036846.2016.1210767}
\showDOI{\tempurl}


\bibitem[Choi(2016)]%
        {Choi2016}
\bibfield{author}{\bibinfo{person}{S. Choi}.} \bibinfo{year}{2016}\natexlab{}.
\newblock \showarticletitle{Workforce diversity and job satisfaction of the
  majority and the minority}.
\newblock \bibinfo{journal}{\emph{Rev. of Public Personnel Administration}}
  \bibinfo{volume}{37} (\bibinfo{year}{2016}), \bibinfo{pages}{84--107}.
\newblock
Issue 1.
\urldef\tempurl%
\url{https://doi.org/10.1177/0734371x15623617}
\showDOI{\tempurl}


\bibitem[Choi(2017)]%
        {Choi2017Workforce}
\bibfield{author}{\bibinfo{person}{Sungjoo Choi}.}
  \bibinfo{year}{2017}\natexlab{}.
\newblock \showarticletitle{Workforce Diversity and Job Satisfaction of the
  Majority and the Minority}.
\newblock \bibinfo{journal}{\emph{Review of Public Personnel Administration}}
  \bibinfo{volume}{37} (\bibinfo{year}{2017}), \bibinfo{pages}{107 -- 84}.
\newblock
\urldef\tempurl%
\url{https://doi.org/10.1177/0734371X15623617}
\showDOI{\tempurl}


\bibitem[Choi and Rainey(2014)]%
        {Choi2014Organizational}
\bibfield{author}{\bibinfo{person}{Sungjoo Choi} {and} \bibinfo{person}{H.
  Rainey}.} \bibinfo{year}{2014}\natexlab{}.
\newblock \showarticletitle{Organizational Fairness and Diversity Management in
  Public Organizations}.
\newblock \bibinfo{journal}{\emph{Review of Public Personnel Administration}}
  \bibinfo{volume}{34} (\bibinfo{year}{2014}), \bibinfo{pages}{307 -- 331}.
\newblock
\urldef\tempurl%
\url{https://doi.org/10.1177/0734371X13486489}
\showDOI{\tempurl}


\bibitem[Chung(2018)]%
        {Chung2018Gender}
\bibfield{author}{\bibinfo{person}{Heejung Chung}.}
  \bibinfo{year}{2018}\natexlab{}.
\newblock \showarticletitle{Gender, Flexibility Stigma and the Perceived
  Negative Consequences of Flexible Working in the UK}.
\newblock \bibinfo{journal}{\emph{Social Indicators Research}}
  (\bibinfo{year}{2018}), \bibinfo{pages}{1--25}.
\newblock
\urldef\tempurl%
\url{https://doi.org/10.1007/S11205-018-2036-7}
\showDOI{\tempurl}


\bibitem[Colquitt(2001)]%
        {Colquitt2001}
\bibfield{author}{\bibinfo{person}{Jason~A. Colquitt}.}
  \bibinfo{year}{2001}\natexlab{}.
\newblock \showarticletitle{On the dimensionality of organizational justice: a
  construct validation of a measure.}
\newblock \bibinfo{journal}{\emph{The J. of Applied Psychology}}
  \bibinfo{volume}{86 3} (\bibinfo{year}{2001}), \bibinfo{pages}{386--400}.
\newblock


\bibitem[Colquitt and Chertkoff(2002)]%
        {Colquitt2002}
\bibfield{author}{\bibinfo{person}{Jason~A. Colquitt} {and}
  \bibinfo{person}{Jerome~M. Chertkoff}.} \bibinfo{year}{2002}\natexlab{}.
\newblock \showarticletitle{Explaining Injustice: The Interactive Effect of
  Explanation and Outcome on Fairness Perceptions and Task Motivation}.
\newblock \bibinfo{journal}{\emph{J. of Manage.}} \bibinfo{volume}{28},
  \bibinfo{number}{5} (\bibinfo{year}{2002}), \bibinfo{pages}{591--610}.
\newblock
\urldef\tempurl%
\url{https://doi.org/10.1177/014920630202800502}
\showDOI{\tempurl}


\bibitem[Colquitt et~al\mbox{.}(2001)]%
        {colquitt2001justice}
\bibfield{author}{\bibinfo{person}{Jason~A Colquitt}, \bibinfo{person}{Donald~E
  Conlon}, \bibinfo{person}{Michael~J Wesson}, \bibinfo{person}{Christopher~OLH
  Porter}, {and} \bibinfo{person}{K~Yee Ng}.} \bibinfo{year}{2001}\natexlab{}.
\newblock \showarticletitle{Justice at the millennium: a meta-analytic review
  of 25 years of organizational justice research.}
\newblock \bibinfo{journal}{\emph{J. of applied psychology}}
  \bibinfo{volume}{86}, \bibinfo{number}{3} (\bibinfo{year}{2001}),
  \bibinfo{pages}{425}.
\newblock


\bibitem[Colquitt and Rodell(2015)]%
        {Colquitt2015}
\bibfield{author}{\bibinfo{person}{Jason~A. Colquitt} {and}
  \bibinfo{person}{Jessica~B. Rodell}.} \bibinfo{year}{2015}\natexlab{}.
\newblock \showarticletitle{{Measuring Justice and Fairness}}.
\newblock In \bibinfo{booktitle}{\emph{{The Oxford Handbook of Justice in the
  Workplace}}}. \bibinfo{publisher}{Oxford University Press}.
\newblock
\showISBNx{9780199981410}


\bibitem[Conradie and Klerk(2019)]%
        {Conradie2019To}
\bibfield{author}{\bibinfo{person}{W.~J. Conradie} {and}
  \bibinfo{person}{J.~J.~d. Klerk}.} \bibinfo{year}{2019}\natexlab{}.
\newblock \showarticletitle{To flex or not to flex? Flexible work arrangements
  amongst software developers in an emerging economy}.
\newblock \bibinfo{journal}{\emph{SA J. of Human Resource Management}}
  (\bibinfo{year}{2019}).
\newblock
\urldef\tempurl%
\url{https://doi.org/10.4102/sajhrm.v17i0.1175}
\showDOI{\tempurl}


\bibitem[Cropanzano et~al\mbox{.}({[n.\,d.]})]%
        {Cropanzano2001}
\bibfield{author}{\bibinfo{person}{R. Cropanzano}, \bibinfo{person}{D.~E.
  Rupp}, \bibinfo{person}{C.~J. Mohler}, {and} \bibinfo{person}{M. Schminke}.}
  \bibinfo{year}{[n.\,d.]}\natexlab{}.
\newblock \showarticletitle{Three roads to organizational justice}.
\newblock \bibinfo{journal}{\emph{Research in Personnel and Human Resources
  Manage.}} (\bibinfo{year}{[n.\,d.]}), \bibinfo{pages}{1--113}.
\newblock
\urldef\tempurl%
\url{https://doi.org/10.1016/s0742-7301(01)20001-2}
\showDOI{\tempurl}


\bibitem[Cugueró-Escofet and Rosanas(2013)]%
        {Cuguero2013}
\bibfield{author}{\bibinfo{person}{Natàlia Cugueró-Escofet} {and}
  \bibinfo{person}{Josep~M. Rosanas}.} \bibinfo{year}{2013}\natexlab{}.
\newblock \showarticletitle{The just design and use of management control
  systems as requirements for goal congruence}.
\newblock \bibinfo{journal}{\emph{Manage. Accounting Research}}
  \bibinfo{volume}{24}, \bibinfo{number}{1} (\bibinfo{year}{2013}),
  \bibinfo{pages}{23--40}.
\newblock
\showISSN{1044-5005}


\bibitem[Czibula et~al\mbox{.}(2022)]%
        {Czibula2022Enhancing}
\bibfield{author}{\bibinfo{person}{G. Czibula}, \bibinfo{person}{M. Lupea},
  {and} \bibinfo{person}{Anamaria Briciu}.} \bibinfo{year}{2022}\natexlab{}.
\newblock \showarticletitle{Enhancing the Performance of Software Authorship
  Attribution Using an Ensemble of Deep Autoencoders}.
\newblock \bibinfo{journal}{\emph{Mathematics}} (\bibinfo{year}{2022}).
\newblock
\urldef\tempurl%
\url{https://doi.org/10.3390/math10152572}
\showDOI{\tempurl}


\bibitem[Daley(1996)]%
        {Daley1996Paths}
\bibfield{author}{\bibinfo{person}{Dennis~M. Daley}.}
  \bibinfo{year}{1996}\natexlab{}.
\newblock \showarticletitle{Paths of Glory and the Glass Ceiling: Differing
  Patterns of Career Advancement among Women and Minority Federal Employees}.
\newblock \bibinfo{journal}{\emph{Public Administration Quarterly}}
  \bibinfo{volume}{20} (\bibinfo{year}{1996}), \bibinfo{pages}{143}.
\newblock


\bibitem[Dayan and Benedetto(2008)]%
        {Dayan2008}
\bibfield{author}{\bibinfo{person}{M. Dayan} {and} \bibinfo{person}{A.~D.
  Benedetto}.} \bibinfo{year}{2008}\natexlab{}.
\newblock \showarticletitle{Procedural and interactional justice perceptions
  and teamwork quality}.
\newblock \bibinfo{journal}{\emph{J. of Business \& Industrial Marketing}}
  \bibinfo{volume}{23} (\bibinfo{year}{2008}), \bibinfo{pages}{566--576}.
\newblock
Issue 8.
\urldef\tempurl%
\url{https://doi.org/10.1108/08858620810913371}
\showDOI{\tempurl}


\bibitem[DeRouvray and Couper(2002)]%
        {DeRouvray2002Designing}
\bibfield{author}{\bibinfo{person}{Cristel DeRouvray} {and} \bibinfo{person}{M.
  Couper}.} \bibinfo{year}{2002}\natexlab{}.
\newblock \showarticletitle{Designing a Strategy for Reducing “No Opinion”
  Responses in Web-Based Surveys}.
\newblock \bibinfo{journal}{\emph{Social Science Computer Rev.}}
  \bibinfo{volume}{20} (\bibinfo{year}{2002}), \bibinfo{pages}{3 -- 9}.
\newblock
\urldef\tempurl%
\url{https://doi.org/10.1177/089443930202000101}
\showDOI{\tempurl}


\bibitem[Deutsch(1975)]%
        {Deutsch1975}
\bibfield{author}{\bibinfo{person}{Morton Deutsch}.}
  \bibinfo{year}{1975}\natexlab{}.
\newblock \showarticletitle{Equity, Equality, and Need: What Determines Which
  Value Will Be Used as the Basis of Distributive Justice?}
\newblock \bibinfo{journal}{\emph{J. of Social Issues}} \bibinfo{volume}{31},
  \bibinfo{number}{3} (\bibinfo{year}{1975}), \bibinfo{pages}{137--149}.
\newblock
\urldef\tempurl%
\url{https://doi.org/10.1111/j.1540-4560.1975.tb01000.x}
\showDOI{\tempurl}


\bibitem[Dover et~al\mbox{.}(2019)]%
        {Dover2019Mixed}
\bibfield{author}{\bibinfo{person}{Tessa~L. Dover}, \bibinfo{person}{Cheryl~R.
  Kaiser}, {and} \bibinfo{person}{B. Major}.} \bibinfo{year}{2019}\natexlab{}.
\newblock \showarticletitle{Mixed Signals: The Unintended Effects of Diversity
  Initiatives}.
\newblock \bibinfo{journal}{\emph{Social Issues and Policy Review}}
  (\bibinfo{year}{2019}).
\newblock
\urldef\tempurl%
\url{https://doi.org/10.1111/sipr.12059}
\showDOI{\tempurl}


\bibitem[Enoksen(2015)]%
        {Enoksen2015}
\bibfield{author}{\bibinfo{person}{E. Enoksen}.}
  \bibinfo{year}{2015}\natexlab{}.
\newblock \showarticletitle{Examining the dimensionality of colquitt's
  organizational justice scale in a public health sector context}.
\newblock \bibinfo{journal}{\emph{Psychological Reports}}
  \bibinfo{volume}{116} (\bibinfo{year}{2015}), \bibinfo{pages}{723--737}.
\newblock
Issue 3.
\urldef\tempurl%
\url{https://doi.org/10.2466/01.pr0.116k26w0}
\showDOI{\tempurl}


\bibitem[Fagerholm et~al\mbox{.}(2014)]%
        {Fagerholm2014}
\bibfield{author}{\bibinfo{person}{Fabian Fagerholm}, \bibinfo{person}{Marko
  Ikonen}, \bibinfo{person}{Petri Kettunen}, \bibinfo{person}{J\"{u}rgen
  M\"{u}nch}, \bibinfo{person}{Virpi Roto}, {and} \bibinfo{person}{Pekka
  Abrahamsson}.} \bibinfo{year}{2014}\natexlab{}.
\newblock \showarticletitle{How do software developers experience team
  performance in lean and agile environments?}. In
  \bibinfo{booktitle}{\emph{Proc. of the 18th Int. Conf. on Evaluation and
  Assessment in Softw. Eng.}} (London, England, United Kingdom)
  \emph{(\bibinfo{series}{EASE '14})}. \bibinfo{publisher}{Assoc. for Comput.
  Machinery}, \bibinfo{address}{New York, NY, USA}, Article
  \bibinfo{articleno}{7}, \bibinfo{numpages}{10}~pages.
\newblock
\showISBNx{9781450324762}
\urldef\tempurl%
\url{https://doi.org/10.1145/2601248.2601285}
\showDOI{\tempurl}


\bibitem[Foucault et~al\mbox{.}(2015)]%
        {Foucault2015Impact}
\bibfield{author}{\bibinfo{person}{Matthieu Foucault}, \bibinfo{person}{Marc
  Palyart}, \bibinfo{person}{Xavier Blanc}, \bibinfo{person}{G. Murphy}, {and}
  \bibinfo{person}{Jean-Rémy Falleri}.} \bibinfo{year}{2015}\natexlab{}.
\newblock \showarticletitle{Impact of developer turnover on quality in
  open-source software}.
\newblock \bibinfo{journal}{\emph{Proc. of the 2015 10th Joint Meeting on
  Foundations of Softw. Eng.}} (\bibinfo{year}{2015}).
\newblock
\urldef\tempurl%
\url{https://doi.org/10.1145/2786805.2786870}
\showDOI{\tempurl}


\bibitem[Franca et~al\mbox{.}(2013)]%
        {Franca2013}
\bibfield{author}{\bibinfo{person}{A.~C. Franca}, \bibinfo{person}{A.~L. de
  Araujo}, {and} \bibinfo{person}{F.~B. da Silva}.}
  \bibinfo{year}{2013}\natexlab{}.
\newblock \showarticletitle{Motivation of software engineers: A qualitative
  case study of a research and development organisation}. In
  \bibinfo{booktitle}{\emph{2013 6th Int. Workshop on Cooperative and Human
  Aspects of Softw. Eng. (CHASE)}}. \bibinfo{publisher}{IEEE Computer Soc.},
  \bibinfo{address}{Los Alamitos, CA, USA}, \bibinfo{pages}{9--16}.
\newblock
\urldef\tempurl%
\url{https://doi.org/10.1109/CHASE.2013.6614726}
\showDOI{\tempurl}


\bibitem[Fran{\c{c}}a et~al\mbox{.}(2018)]%
        {francca2018motivation}
\bibfield{author}{\bibinfo{person}{C{\'e}sar Fran{\c{c}}a},
  \bibinfo{person}{Fabio~QB Da~Silva}, {and} \bibinfo{person}{Helen Sharp}.}
  \bibinfo{year}{2018}\natexlab{}.
\newblock \showarticletitle{Motivation and satisfaction of software engineers}.
\newblock \bibinfo{journal}{\emph{IEEE Trans. on Softw. Eng.}}
  \bibinfo{volume}{46}, \bibinfo{number}{2} (\bibinfo{year}{2018}),
  \bibinfo{pages}{118--140}.
\newblock


\bibitem[França et~al\mbox{.}(2012)]%
        {França2012}
\bibfield{author}{\bibinfo{person}{A.~César~C. França},
  \bibinfo{person}{David E.~S. Carneiro}, {and} \bibinfo{person}{Fabio Q. B.~da
  Silva}.} \bibinfo{year}{2012}\natexlab{}.
\newblock \showarticletitle{Towards an Explanatory Theory of Motivation in
  Software Eng.: A Qualitative Case Study of a Small Software Company}. In
  \bibinfo{booktitle}{\emph{2012 26th Brazilian Symp. on Softw. Eng.}}
  \bibinfo{pages}{61--70}.
\newblock
\urldef\tempurl%
\url{https://doi.org/10.1109/SBES.2012.28}
\showDOI{\tempurl}


\bibitem[França et~al\mbox{.}(2014)]%
        {França2014}
\bibfield{author}{\bibinfo{person}{A.~C.~C. França}, \bibinfo{person}{F.~Q.
  B.~d. Silva}, \bibinfo{person}{A.~d. L.~C. Felix}, {and}
  \bibinfo{person}{D.~E.~S. Carneiro}.} \bibinfo{year}{2014}\natexlab{}.
\newblock \showarticletitle{Motivation in software engineering industrial
  practice: a cross-case analysis of two software organisations}.
\newblock \bibinfo{journal}{\emph{Inf. and Softw. Technol.}}
  \bibinfo{volume}{56} (\bibinfo{year}{2014}), \bibinfo{pages}{79--101}.
\newblock
Issue 1.
\urldef\tempurl%
\url{https://doi.org/10.1016/j.infsof.2013.06.006}
\showDOI{\tempurl}


\bibitem[Fronza and Vlasenko(2011a)]%
        {Fronza2011ProfilingTE}
\bibfield{author}{\bibinfo{person}{Ilenia Fronza} {and} \bibinfo{person}{Jelena
  Vlasenko}.} \bibinfo{year}{2011}\natexlab{a}.
\newblock \showarticletitle{Profiling the Effort of Novices in Software
  Development Teams - An Analysis using Data Collected Non Invasively}. In
  \bibinfo{booktitle}{\emph{Int. Conf. on Enterprise Inf. Syst.}}
\newblock


\bibitem[Fronza and Vlasenko(2011b)]%
        {Fronza2011}
\bibfield{author}{\bibinfo{person}{Ilenia Fronza} {and} \bibinfo{person}{Jelena
  Vlasenko}.} \bibinfo{year}{2011}\natexlab{b}.
\newblock \showarticletitle{Profiling the Effort of Novices in Software
  Development Teams - An Analysis using Data Collected Non Invasively.}
\newblock \bibinfo{journal}{\emph{ICEIS 2011 - Proc. of the 13th Int. Conf. on
  Enterprise Inf. Syst.}}  \bibinfo{volume}{3}, \bibinfo{pages}{392--395}.
\newblock


\bibitem[Geiger et~al\mbox{.}(2021)]%
        {Geiger2021The}
\bibfield{author}{\bibinfo{person}{R. Geiger}, \bibinfo{person}{Dorothy
  Howard}, {and} \bibinfo{person}{L. Irani}.} \bibinfo{year}{2021}\natexlab{}.
\newblock \showarticletitle{The Labor of Maintaining and Scaling Free and
  Open-Source Software Projects}.
\newblock \bibinfo{journal}{\emph{Proc. of the ACM on Human-Computer
  Interaction}}  \bibinfo{volume}{5} (\bibinfo{year}{2021}), \bibinfo{pages}{1
  -- 28}.
\newblock
\urldef\tempurl%
\url{https://doi.org/10.1145/3449249}
\showDOI{\tempurl}


\bibitem[German et~al\mbox{.}(2018)]%
        {German2018}
\bibfield{author}{\bibinfo{person}{Daniel~M. German}, \bibinfo{person}{Gregorio
  Robles}, \bibinfo{person}{Germán Poo-Caamaño}, \bibinfo{person}{Xin Yang},
  \bibinfo{person}{Hajimu Iida}, {and} \bibinfo{person}{Katsuro Inoue}.}
  \bibinfo{year}{2018}\natexlab{}.
\newblock \showarticletitle{"Was My Contribution Fairly Rev.ed?" A Framework to
  Study the Perception of Fairness in Modern Code Rev.s}. In
  \bibinfo{booktitle}{\emph{2018 IEEE/ACM 40th Int. Conf. on Softw. Eng.
  (ICSE)}}. \bibinfo{pages}{523--534}.
\newblock
\urldef\tempurl%
\url{https://doi.org/10.1145/3180155.3180217}
\showDOI{\tempurl}


\bibitem[Gilliland et~al\mbox{.}(2008)]%
        {gilliland2008justice}
\bibfield{author}{\bibinfo{person}{Stephen~W Gilliland},
  \bibinfo{person}{Dirk~D Steiner}, {and} \bibinfo{person}{Daniel~P
  Skarlicki}.} \bibinfo{year}{2008}\natexlab{}.
\newblock \bibinfo{booktitle}{\emph{Justice, morality, and social
  responsibility}}.
\newblock \bibinfo{publisher}{IAP}.
\newblock


\bibitem[Goldman and Cropanzano(2015)]%
        {Goldman2015}
\bibfield{author}{\bibinfo{person}{Barry Goldman} {and}
  \bibinfo{person}{Russell Cropanzano}.} \bibinfo{year}{2015}\natexlab{}.
\newblock \showarticletitle{“Justice” and “fairness” are not the same
  thing}.
\newblock \bibinfo{journal}{\emph{J. of Organizational Behavior}}
  \bibinfo{volume}{36}, \bibinfo{number}{2} (\bibinfo{year}{2015}),
  \bibinfo{pages}{313--318}.
\newblock
\urldef\tempurl%
\url{https://doi.org/10.1002/job.1956}
\showDOI{\tempurl}


\bibitem[Greenberg(1993)]%
        {Greenberg1993}
\bibfield{author}{\bibinfo{person}{Jerald Greenberg}.}
  \bibinfo{year}{1993}\natexlab{}.
\newblock \showarticletitle{{Stealing in the Name of Justice: Inf.al and
  Interpersonal Moderators of Theft Reactions to Underpayment Inequity}}.
\newblock \bibinfo{journal}{\emph{Organizational Behavior and Human Decision
  Processes}} \bibinfo{volume}{54}, \bibinfo{number}{1}
  (\bibinfo{date}{February} \bibinfo{year}{1993}), \bibinfo{pages}{81--103}.
\newblock


\bibitem[Greider et~al\mbox{.}(2019)]%
        {Greider2019Increasing}
\bibfield{author}{\bibinfo{person}{C. Greider}, \bibinfo{person}{J. Sheltzer},
  \bibinfo{person}{N.~C. Cantalupo}, \bibinfo{person}{W. Copeland},
  \bibinfo{person}{N. Dasgupta}, \bibinfo{person}{Nancy Hopkins},
  \bibinfo{person}{Jaclyn~M. Jansen}, \bibinfo{person}{L. Joshua-Tor},
  \bibinfo{person}{Gary Mcdowell}, \bibinfo{person}{J. Metcalf},
  \bibinfo{person}{BethAnn McLaughlin}, \bibinfo{person}{Ann Olivarius},
  \bibinfo{person}{Erin~K O'Shea}, \bibinfo{person}{Jennifer Raymond},
  \bibinfo{person}{D. Ruebain}, \bibinfo{person}{J. Steitz},
  \bibinfo{person}{B. Stillman}, \bibinfo{person}{S. Tilghman},
  \bibinfo{person}{Virginia Valian}, \bibinfo{person}{L. Villa-komaroff}, {and}
  \bibinfo{person}{J. Wong}.} \bibinfo{year}{2019}\natexlab{}.
\newblock \showarticletitle{Increasing gender diversity in the STEM research
  workforce}.
\newblock \bibinfo{journal}{\emph{Science}}  \bibinfo{volume}{366}
  (\bibinfo{year}{2019}), \bibinfo{pages}{692 -- 695}.
\newblock
\urldef\tempurl%
\url{https://doi.org/10.1126/science.aaz0649}
\showDOI{\tempurl}


\bibitem[Griffin et~al\mbox{.}(2023)]%
        {griffin2023tutorial}
\bibfield{author}{\bibinfo{person}{Beth~Ann Griffin}, \bibinfo{person}{Megan~S
  Schuler}, \bibinfo{person}{Matt Cefalu}, \bibinfo{person}{Lynsay Ayer},
  \bibinfo{person}{Mark Godley}, \bibinfo{person}{Noah Greifer},
  \bibinfo{person}{Donna~L Coffman}, {and} \bibinfo{person}{Daniel~F
  McCaffrey}.} \bibinfo{year}{2023}\natexlab{}.
\newblock \showarticletitle{A tutorial for propensity score weighting for
  moderation analysis with categorical variables: an application examining
  smoking disparities among sexual minority adults}.
\newblock \bibinfo{journal}{\emph{Medical care}} \bibinfo{volume}{61},
  \bibinfo{number}{12} (\bibinfo{year}{2023}), \bibinfo{pages}{836--845}.
\newblock


\bibitem[Gunawardena et~al\mbox{.}(2022)]%
        {Gunawardena2022DestructiveCI}
\bibfield{author}{\bibinfo{person}{Sanuri~Dananja Gunawardena},
  \bibinfo{person}{Peter Devine}, \bibinfo{person}{Isabelle Beaumont},
  \bibinfo{person}{Lola Garden}, \bibinfo{person}{Emerson~Rex Murphy-Hill},
  {and} \bibinfo{person}{Kelly Blincoe}.} \bibinfo{year}{2022}\natexlab{}.
\newblock \showarticletitle{Destructive Criticism in Software Code Review
  Impacts Inclusion}. In \bibinfo{booktitle}{\emph{Computer Supported
  Cooperative Work}}.
\newblock


\bibitem[Guénard and Legendre(2022)]%
        {Guénard2022}
\bibfield{author}{\bibinfo{person}{G. Guénard} {and} \bibinfo{person}{P.
  Legendre}.} \bibinfo{year}{2022}\natexlab{}.
\newblock \showarticletitle{Hierarchical clustering with contiguity constraint
  in r}.
\newblock \bibinfo{journal}{\emph{J. of Statistical Softw.}}
  \bibinfo{volume}{103} (\bibinfo{year}{2022}).
\newblock
Issue 7.
\urldef\tempurl%
\url{https://doi.org/10.18637/jss.v103.i07}
\showDOI{\tempurl}


\bibitem[Haar and Spell(2009)]%
        {Haar2009How}
\bibfield{author}{\bibinfo{person}{J. Haar} {and} \bibinfo{person}{C. Spell}.}
  \bibinfo{year}{2009}\natexlab{}.
\newblock \showarticletitle{How does distributive justice affect work
  attitudes? The moderating effects of autonomy}.
\newblock \bibinfo{journal}{\emph{The Int. J. of Human Resource Manage.}}
  \bibinfo{volume}{20} (\bibinfo{year}{2009}), \bibinfo{pages}{1827 -- 1842}.
\newblock
\urldef\tempurl%
\url{https://doi.org/10.1080/09585190903087248}
\showDOI{\tempurl}


\bibitem[Hawes and Wang(2022)]%
        {Hawes2022The}
\bibfield{author}{\bibinfo{person}{Frances~M Hawes} {and}
  \bibinfo{person}{Shuangshuang Wang}.} \bibinfo{year}{2022}\natexlab{}.
\newblock \showarticletitle{The Impact of Supervisor Support on the Job
  Satisfaction of Immigrant and Minority Long-Term Care Workers}.
\newblock \bibinfo{journal}{\emph{J. of Applied Gerontology}}
  \bibinfo{volume}{41} (\bibinfo{year}{2022}), \bibinfo{pages}{2157 -- 2166}.
\newblock
\urldef\tempurl%
\url{https://doi.org/10.1177/07334648221104088}
\showDOI{\tempurl}


\bibitem[Hosseini et~al\mbox{.}(2016)]%
        {Hosseini2016CORRELATION}
\bibfield{author}{\bibinfo{person}{M. Hosseini}, \bibinfo{person}{L. Moradi},
  \bibinfo{person}{S. Khanjani}, {and} \bibinfo{person}{E. Bakhshi}.}
  \bibinfo{year}{2016}\natexlab{}.
\newblock \showarticletitle{CORRELATION BETWEEN ORGANIZATIONAL JUSTICE AND
  PRODUCTIVITY OF WELFARE ORGANIZATION’S STAFFS}.
\newblock \bibinfo{journal}{\emph{J. of Health Promotion Manage.}}
  \bibinfo{volume}{5} (\bibinfo{year}{2016}), \bibinfo{pages}{70--77}.
\newblock


\bibitem[Huff et~al\mbox{.}(2020)]%
        {Huff2020Examining}
\bibfield{author}{\bibinfo{person}{Earl~W. Huff}, \bibinfo{person}{Kwajo
  Boateng}, \bibinfo{person}{Makayla Moster}, \bibinfo{person}{Paige
  Rodeghero}, {and} \bibinfo{person}{Julian Brinkley}.}
  \bibinfo{year}{2020}\natexlab{}.
\newblock \showarticletitle{Examining The Work Experience of Programmers with
  Visual Impairments}.
\newblock \bibinfo{journal}{\emph{2020 IEEE Int. Conf. on Softw. Maintenance
  and Evolution (ICSME)}} (\bibinfo{year}{2020}), \bibinfo{pages}{707--711}.
\newblock
\urldef\tempurl%
\url{https://doi.org/10.1109/ICSME46990.2020.00077}
\showDOI{\tempurl}


\bibitem[Iseke and Pull(2019)]%
        {Iseke2019Female}
\bibfield{author}{\bibinfo{person}{Anja Iseke} {and} \bibinfo{person}{Kerstin
  Pull}.} \bibinfo{year}{2019}\natexlab{}.
\newblock \showarticletitle{Female Executives and Perceived Employer
  Attractiveness: On the Potentially Adverse Signal of Having a Female CHRO
  Rather Than a Female CFO}.
\newblock \bibinfo{journal}{\emph{J. of Business Ethics}}
  \bibinfo{volume}{156} (\bibinfo{year}{2019}), \bibinfo{pages}{1113--1133}.
\newblock
\urldef\tempurl%
\url{https://doi.org/10.1007/S10551-017-3640-1}
\showDOI{\tempurl}


\bibitem[Johnson et~al\mbox{.}(2022)]%
        {Johnson2022Program-L:}
\bibfield{author}{\bibinfo{person}{Jazette Johnson}, \bibinfo{person}{Andrew
  Begel}, \bibinfo{person}{R. Ladner}, {and} \bibinfo{person}{Denae Ford}.}
  \bibinfo{year}{2022}\natexlab{}.
\newblock \showarticletitle{Program-L: Online Help Seeking Behaviors by Blind
  and Low Vision Programmers}.
\newblock \bibinfo{journal}{\emph{2022 IEEE Symp. on Visual Languages and
  Human-Centric Comput. (VL/HCC)}} (\bibinfo{year}{2022}).
\newblock
\urldef\tempurl%
\url{https://doi.org/10.1109/vl/hcc53370.2022.9833106}
\showDOI{\tempurl}


\bibitem[Joinson et~al\mbox{.}(2008)]%
        {Joinson2008Measuring}
\bibfield{author}{\bibinfo{person}{A. Joinson}, \bibinfo{person}{Carina~B.
  Paine}, \bibinfo{person}{T. Buchanan}, {and} \bibinfo{person}{Ulf-Dietrich
  Reips}.} \bibinfo{year}{2008}\natexlab{}.
\newblock \showarticletitle{Measuring self-disclosure online: Blurring and
  non-response to sensitive items in web-based surveys}.
\newblock \bibinfo{journal}{\emph{Comput. Hum. Behav.}}  \bibinfo{volume}{24}
  (\bibinfo{year}{2008}), \bibinfo{pages}{2158--2171}.
\newblock
\urldef\tempurl%
\url{https://doi.org/10.1016/j.chb.2007.10.005}
\showDOI{\tempurl}


\bibitem[Joyce et~al\mbox{.}(2010)]%
        {Joyce2010Flexible}
\bibfield{author}{\bibinfo{person}{K. Joyce}, \bibinfo{person}{Roman Pabayo},
  \bibinfo{person}{J.~A. Critchley}, {and} \bibinfo{person}{Clare Bambra}.}
  \bibinfo{year}{2010}\natexlab{}.
\newblock \showarticletitle{Flexible working conditions and their effects on
  employee health and wellbeing.}
\newblock \bibinfo{journal}{\emph{The Cochrane database of systematic reviews}}
   \bibinfo{volume}{2} (\bibinfo{year}{2010}), \bibinfo{pages}{CD008009}.
\newblock
\urldef\tempurl%
\url{https://doi.org/10.1002/14651858.CD008009.pub2}
\showDOI{\tempurl}


\bibitem[Judge and Colquitt(2004)]%
        {judge2004organizational}
\bibfield{author}{\bibinfo{person}{Timothy~A Judge} {and}
  \bibinfo{person}{Jason~A Colquitt}.} \bibinfo{year}{2004}\natexlab{}.
\newblock \showarticletitle{Organizational justice and stress: the mediating
  role of work-family conflict.}
\newblock \bibinfo{journal}{\emph{J. of applied psychology}}
  \bibinfo{volume}{89}, \bibinfo{number}{3} (\bibinfo{year}{2004}),
  \bibinfo{pages}{395}.
\newblock


\bibitem[Kiersch and Byrne(2015)]%
        {Kiersch2015Is}
\bibfield{author}{\bibinfo{person}{Christa~E. Kiersch} {and}
  \bibinfo{person}{Zinta~S. Byrne}.} \bibinfo{year}{2015}\natexlab{}.
\newblock \showarticletitle{Is Being Authentic Being Fair? Multilevel
  Examination of Authentic Leadership, Justice, and Employee Outcomes}.
\newblock \bibinfo{journal}{\emph{J. of Leadership \& Organizational Studies}}
  \bibinfo{volume}{22} (\bibinfo{year}{2015}), \bibinfo{pages}{292 -- 303}.
\newblock
\urldef\tempurl%
\url{https://doi.org/10.1177/1548051815570035}
\showDOI{\tempurl}


\bibitem[Kononenko et~al\mbox{.}(2015)]%
        {Kononenko2015Investigating}
\bibfield{author}{\bibinfo{person}{Oleksii Kononenko}, \bibinfo{person}{Olga
  Baysal}, \bibinfo{person}{Latifa Guerrouj}, \bibinfo{person}{Yaxin Cao},
  {and} \bibinfo{person}{Michael~W. Godfrey}.} \bibinfo{year}{2015}\natexlab{}.
\newblock \showarticletitle{Investigating code review quality: Do people and
  participation matter?}
\newblock \bibinfo{journal}{\emph{2015 IEEE Int. Conf. on Softw. Maintenance
  and Evolution (ICSME)}} (\bibinfo{year}{2015}), \bibinfo{pages}{111--120}.
\newblock
\urldef\tempurl%
\url{https://doi.org/10.1109/ICSM.2015.7332457}
\showDOI{\tempurl}


\bibitem[Konovsky and Cropanzano(1991)]%
        {konovsky1991perceived}
\bibfield{author}{\bibinfo{person}{Mary~A Konovsky} {and}
  \bibinfo{person}{Russell Cropanzano}.} \bibinfo{year}{1991}\natexlab{}.
\newblock \showarticletitle{Perceived fairness of employee drug testing as a
  predictor of employee attitudes and job performance.}
\newblock \bibinfo{journal}{\emph{J. of applied psychology}}
  \bibinfo{volume}{76}, \bibinfo{number}{5} (\bibinfo{year}{1991}),
  \bibinfo{pages}{698}.
\newblock


\bibitem[Kruskal and Wallis(1952)]%
        {kruskal1952use}
\bibfield{author}{\bibinfo{person}{William~H Kruskal} {and}
  \bibinfo{person}{W~Allen Wallis}.} \bibinfo{year}{1952}\natexlab{}.
\newblock \showarticletitle{Use of ranks in one-criterion variance analysis}.
\newblock \bibinfo{journal}{\emph{J. of the American statistical Assoc.}}
  \bibinfo{volume}{47}, \bibinfo{number}{260} (\bibinfo{year}{1952}),
  \bibinfo{pages}{583--621}.
\newblock


\bibitem[Lambiase et~al\mbox{.}(2024)]%
        {lambiase2024investigating}
\bibfield{author}{\bibinfo{person}{Stefano Lambiase}, \bibinfo{person}{Gemma
  Catolino}, \bibinfo{person}{Fabio Palomba}, \bibinfo{person}{Filomena
  Ferrucci}, {and} \bibinfo{person}{Daniel Russo}.}
  \bibinfo{year}{2024}\natexlab{}.
\newblock \showarticletitle{Investigating the role of cultural values in
  adopting large language models for software engineering}.
\newblock \bibinfo{journal}{\emph{ACM Transactions on Software Engineering and
  Methodology}} (\bibinfo{year}{2024}).
\newblock


\bibitem[Langfelder and Horvath(2012)]%
        {Langfelder2012}
\bibfield{author}{\bibinfo{person}{P. Langfelder} {and} \bibinfo{person}{S.
  Horvath}.} \bibinfo{year}{2012}\natexlab{}.
\newblock \showarticletitle{Fastrfunctions for robust correlations and
  hierarchical clustering}.
\newblock \bibinfo{journal}{\emph{J. of Statistical Softw.}}
  \bibinfo{volume}{46} (\bibinfo{year}{2012}).
\newblock
Issue 11.
\urldef\tempurl%
\url{https://doi.org/10.18637/jss.v046.i11}
\showDOI{\tempurl}


\bibitem[Lenberg and Feldt(2018)]%
        {Lenberg2018Psychological}
\bibfield{author}{\bibinfo{person}{Per Lenberg} {and} \bibinfo{person}{R.
  Feldt}.} \bibinfo{year}{2018}\natexlab{}.
\newblock \showarticletitle{Psychological Safety and Norm Clarity in Software
  Eng. Teams}.
\newblock \bibinfo{journal}{\emph{2018 IEEE/ACM 11th Int. Workshop on
  Cooperative and Human Aspects of Softw. Eng. (CHASE)}}
  (\bibinfo{year}{2018}), \bibinfo{pages}{79--86}.
\newblock
\urldef\tempurl%
\url{https://doi.org/10.1145/3195836.3195847}
\showDOI{\tempurl}


\bibitem[Lerner et~al\mbox{.}(2003)]%
        {Lerner2003Relationship}
\bibfield{author}{\bibinfo{person}{D. Lerner}, \bibinfo{person}{B. Amick},
  \bibinfo{person}{Jennifer~C. Lee}, \bibinfo{person}{Ted Rooney},
  \bibinfo{person}{W. Rogers}, \bibinfo{person}{Hong Chang}, {and}
  \bibinfo{person}{E. Berndt}.} \bibinfo{year}{2003}\natexlab{}.
\newblock \showarticletitle{Relationship of Employee-Reported Work Limitations
  to Work Productivity}.
\newblock \bibinfo{journal}{\emph{Medical Care}}  \bibinfo{volume}{41}
  (\bibinfo{year}{2003}), \bibinfo{pages}{649--659}.
\newblock
\urldef\tempurl%
\url{https://doi.org/10.1097/01.MLR.0000062551.76504.A9}
\showDOI{\tempurl}


\bibitem[Leventhal(1976)]%
        {LEVENTHAL197691}
\bibfield{author}{\bibinfo{person}{Gerald~S. Leventhal}.}
  \bibinfo{year}{1976}\natexlab{}.
\newblock \showarticletitle{The Distribution of Rewards and Resources in Groups
  and Organizations}.
\newblock \bibinfo{series}{Advances in Experimental Social Psychology},
  Vol.~\bibinfo{volume}{9}. \bibinfo{publisher}{Academic Press},
  \bibinfo{pages}{91--131}.
\newblock
\showISSN{0065-2601}


\bibitem[Leventhal(1980)]%
        {Leventhal1980}
\bibfield{author}{\bibinfo{person}{Gerald~S. Leventhal}.}
  \bibinfo{year}{1980}\natexlab{}.
\newblock \bibinfo{booktitle}{\emph{{What Should Be Done with Equity Theory?}}}
\newblock \bibinfo{publisher}{Springer US}, \bibinfo{address}{Boston, MA},
  \bibinfo{pages}{27--55}.
\newblock
\showISBNx{978-1-4613-3087-5}
\urldef\tempurl%
\url{https://doi.org/10.1007/978-1-4613-3087-5_2}
\showDOI{\tempurl}


\bibitem[Li et~al\mbox{.}(2011)]%
        {Li2011The}
\bibfield{author}{\bibinfo{person}{Ning Li}, \bibinfo{person}{T.~B. Harris},
  \bibinfo{person}{W. Boswell}, {and} \bibinfo{person}{Zhitao Xie}.}
  \bibinfo{year}{2011}\natexlab{}.
\newblock \showarticletitle{The role of organizational insiders' developmental
  feedback and proactive personality on newcomers' performance: an
  interactionist perspective.}
\newblock \bibinfo{journal}{\emph{The J. of applied psychology}}
  \bibinfo{volume}{96 6} (\bibinfo{year}{2011}), \bibinfo{pages}{1317--27}.
\newblock
\urldef\tempurl%
\url{https://doi.org/10.1037/a0024029}
\showDOI{\tempurl}


\bibitem[Lim and Ling(2002)]%
        {Lim2002Model}
\bibfield{author}{\bibinfo{person}{Eng~Hwee Lim} {and} \bibinfo{person}{F.
  Ling}.} \bibinfo{year}{2002}\natexlab{}.
\newblock \showarticletitle{Model for predicting clients' contribution to
  project success}.
\newblock \bibinfo{journal}{\emph{Engineering, Construction and Architectural
  Management}}  \bibinfo{volume}{9} (\bibinfo{year}{2002}),
  \bibinfo{pages}{388--395}.
\newblock
\urldef\tempurl%
\url{https://doi.org/10.1108/EB021233}
\showDOI{\tempurl}


\bibitem[Lucht(2016)]%
        {Lucht2016Job}
\bibfield{author}{\bibinfo{person}{Tracy Lucht}.}
  \bibinfo{year}{2016}\natexlab{}.
\newblock \showarticletitle{Job Satisfaction and Gender}.
\newblock \bibinfo{journal}{\emph{Journalism Practice}}  \bibinfo{volume}{10}
  (\bibinfo{year}{2016}), \bibinfo{pages}{405 -- 423}.
\newblock
\urldef\tempurl%
\url{https://doi.org/10.1080/17512786.2015.1025416}
\showDOI{\tempurl}


\bibitem[Ma et~al\mbox{.}(2019)]%
        {Ma2019A}
\bibfield{author}{\bibinfo{person}{Zifei Ma}, \bibinfo{person}{Ruiyin Li},
  \bibinfo{person}{Tong Li}, \bibinfo{person}{Rui Zhu}, \bibinfo{person}{Rong
  Jiang}, \bibinfo{person}{Juan Yang}, \bibinfo{person}{Mingjing Tang}, {and}
  \bibinfo{person}{Ming Zheng}.} \bibinfo{year}{2019}\natexlab{}.
\newblock \showarticletitle{A data-driven risk measurement model of software
  developer turnover}.
\newblock \bibinfo{journal}{\emph{Soft Computing}}  \bibinfo{volume}{24}
  (\bibinfo{year}{2019}), \bibinfo{pages}{825--842}.
\newblock
\urldef\tempurl%
\url{https://doi.org/10.1007/s00500-019-04540-z}
\showDOI{\tempurl}


\bibitem[MacFarland and Yates(2016)]%
        {Macfarland2016}
\bibfield{author}{\bibinfo{person}{T.~W. MacFarland} {and}
  \bibinfo{person}{J.~M. Yates}.} \bibinfo{year}{2016}\natexlab{}.
\newblock \showarticletitle{Mann–whitney u test}.
\newblock \bibinfo{journal}{\emph{Introduction to Nonparametric Statistics for
  the Biological Sciences Using R}} (\bibinfo{year}{2016}),
  \bibinfo{pages}{103--132}.
\newblock
\urldef\tempurl%
\url{https://doi.org/10.1007/978-3-319-30634-6_4}
\showDOI{\tempurl}


\bibitem[Masterson et~al\mbox{.}(2000)]%
        {Masterson2000}
\bibfield{author}{\bibinfo{person}{Suzanne~S. Masterson}, \bibinfo{person}{Kyle
  Lewis}, \bibinfo{person}{Barry~M. Goldman}, {and} \bibinfo{person}{M.~Susan
  Taylor}.} \bibinfo{year}{2000}\natexlab{}.
\newblock \showarticletitle{Integrating Justice and Social Exchange: The
  Differing Effects of Fair Procedures and Treatment on Work Relationships}.
\newblock \bibinfo{journal}{\emph{The Academy of Manage. J.}}
  \bibinfo{volume}{43}, \bibinfo{number}{4} (\bibinfo{year}{2000}),
  \bibinfo{pages}{738--748}.
\newblock
\showISSN{00014273}


\bibitem[Mcdonald et~al\mbox{.}(2014)]%
        {Mcdonald2014Modeling}
\bibfield{author}{\bibinfo{person}{Nora Mcdonald}, \bibinfo{person}{Kelly
  Blincoe}, \bibinfo{person}{Eva Petakovic}, {and} \bibinfo{person}{S.
  Goggins}.} \bibinfo{year}{2014}\natexlab{}.
\newblock \showarticletitle{Modeling Distributed Collaboration on GitHub}.
\newblock \bibinfo{journal}{\emph{Adv. Complex Syst.}}  \bibinfo{volume}{17}
  (\bibinfo{year}{2014}).
\newblock
\urldef\tempurl%
\url{https://doi.org/10.1142/S0219525914500246}
\showDOI{\tempurl}


\bibitem[McDonald(1970)]%
        {McDonald_1970}
\bibfield{author}{\bibinfo{person}{Roderick~P. McDonald}.}
  \bibinfo{year}{1970}\natexlab{}.
\newblock \showarticletitle{THE THEORETICAL FOUNDATIONS OF PRINCIPAL FACTOR
  ANALYSIS, CANONICAL FACTOR ANALYSIS, AND ALPHA FACTOR ANALYSIS}.
\newblock \bibinfo{journal}{\emph{British J. of Mathematical and Statistical
  Psychology}} \bibinfo{volume}{23}, \bibinfo{number}{1} (\bibinfo{date}{May}
  \bibinfo{year}{1970}), \bibinfo{pages}{1–21}.
\newblock
\urldef\tempurl%
\url{https://doi.org/10.1111/j.2044-8317.1970.tb00432.x}
\showDOI{\tempurl}


\bibitem[Melnik and Maurer(2006)]%
        {melnik2006comparative}
\bibfield{author}{\bibinfo{person}{Grigori Melnik} {and} \bibinfo{person}{Frank
  Maurer}.} \bibinfo{year}{2006}\natexlab{}.
\newblock \showarticletitle{Comparative analysis of job satisfaction in agile
  and non-agile software development teams}. In \bibinfo{booktitle}{\emph{Int.
  conference on extreme programming and agile processes in software
  engineering}}. Springer, \bibinfo{pages}{32--42}.
\newblock


\bibitem[Meyer et~al\mbox{.}(2017)]%
        {Meyer2017The}
\bibfield{author}{\bibinfo{person}{André~N. Meyer}, \bibinfo{person}{Laura~E.
  Barton}, \bibinfo{person}{G. Murphy}, \bibinfo{person}{Thomas Zimmermann},
  {and} \bibinfo{person}{Thomas Fritz}.} \bibinfo{year}{2017}\natexlab{}.
\newblock \showarticletitle{The Work Life of Developers: Activities, Switches
  and Perceived Productivity}.
\newblock \bibinfo{journal}{\emph{IEEE Trans. on Softw. Eng.}}
  \bibinfo{volume}{43} (\bibinfo{year}{2017}), \bibinfo{pages}{1178--1193}.
\newblock
\urldef\tempurl%
\url{https://doi.org/10.1109/TSE.2017.2656886}
\showDOI{\tempurl}


\bibitem[Milner et~al\mbox{.}(2015)]%
        {Milner2015}
\bibfield{author}{\bibinfo{person}{A. Milner}, \bibinfo{person}{Z. Aitken},
  \bibinfo{person}{L. Krnjacki}, \bibinfo{person}{R. Bentley},
  \bibinfo{person}{T. Blakely}, \bibinfo{person}{A.~D. LaMontagne}, {and}
  \bibinfo{person}{A. Kavanagh}.} \bibinfo{year}{2015}\natexlab{}.
\newblock \showarticletitle{Perceived fairness of pay among people with and
  without disabilities: a propensity score matched analysis of working
  australians}.
\newblock \bibinfo{journal}{\emph{Scandinavian J. of Work, Environment \&
  Health}}  \bibinfo{volume}{41} (\bibinfo{year}{2015}),
  \bibinfo{pages}{451--459}.
\newblock
Issue 5.
\urldef\tempurl%
\url{https://doi.org/10.5271/sjweh.3515}
\showDOI{\tempurl}


\bibitem[Moorman(1991)]%
        {Moorman1991}
\bibfield{author}{\bibinfo{person}{Robert Moorman}.}
  \bibinfo{year}{1991}\natexlab{}.
\newblock \showarticletitle{Relationship Between Organizational Justice and
  Organizational Citizenship Behaviors: Do Fairness Perceptions Influence
  Employee Citizenship?}
\newblock \bibinfo{journal}{\emph{J. of Applied Psychology}}
  \bibinfo{volume}{76} (\bibinfo{date}{12} \bibinfo{year}{1991}),
  \bibinfo{pages}{845--855}.
\newblock
\urldef\tempurl%
\url{https://doi.org/10.1037//0021-9010.76.6.845}
\showDOI{\tempurl}


\bibitem[Mossholder et~al\mbox{.}(1998)]%
        {Mossholder1998}
\bibfield{author}{\bibinfo{person}{Kevin~W. Mossholder},
  \bibinfo{person}{Nathan Bennett}, {and} \bibinfo{person}{Christopher~L.
  Martin}.} \bibinfo{year}{1998}\natexlab{}.
\newblock \showarticletitle{A Multilevel Analysis of Procedural Justice
  Context}.
\newblock \bibinfo{journal}{\emph{J. of Organizational Behavior}}
  \bibinfo{volume}{19}, \bibinfo{number}{2} (\bibinfo{year}{1998}),
  \bibinfo{pages}{131--141}.
\newblock
\showISSN{08943796, 10991379}


\bibitem[Mulay(2017)]%
        {Mulay2017A}
\bibfield{author}{\bibinfo{person}{Harshada Mulay}.}
  \bibinfo{year}{2017}\natexlab{}.
\newblock \showarticletitle{A Study of Impact of Flexible Working Hours Improve
  Work Life Balance in Software Industry}.
\newblock \bibinfo{journal}{\emph{Abhinav-Int. Monthly Refereed J. Of Research
  In Management \& Technology}}  \bibinfo{volume}{6} (\bibinfo{year}{2017}),
  \bibinfo{pages}{27--33}.
\newblock


\bibitem[Nadri et~al\mbox{.}(2021)]%
        {Nadri2021}
\bibfield{author}{\bibinfo{person}{Reza Nadri}, \bibinfo{person}{Gema
  Rodr{\'{\i}}guez{-}P{\'{e}}rez}, {and} \bibinfo{person}{Meiyappan Nagappan}.}
  \bibinfo{year}{2021}\natexlab{}.
\newblock \showarticletitle{On the Relationship Between the Developer's
  Perceptible Race and Ethnicity and the Evaluation of Contributions in {OSS}}.
\newblock \bibinfo{journal}{\emph{CoRR}}  \bibinfo{volume}{abs/2104.06143}
  (\bibinfo{year}{2021}).
\newblock
\showeprint[arXiv]{2104.06143}


\bibitem[Newman et~al\mbox{.}(2017)]%
        {Newman2017Psychological}
\bibfield{author}{\bibinfo{person}{Alexander Newman}, \bibinfo{person}{Ross
  Donohue}, {and} \bibinfo{person}{N. Eva}.} \bibinfo{year}{2017}\natexlab{}.
\newblock \showarticletitle{Psychological safety: A systematic review of the
  literature}.
\newblock \bibinfo{journal}{\emph{Human Resource Manage. Rev.}}
  \bibinfo{volume}{27} (\bibinfo{year}{2017}), \bibinfo{pages}{521--535}.
\newblock
\urldef\tempurl%
\url{https://doi.org/10.1016/J.HRMR.2017.01.001}
\showDOI{\tempurl}


\bibitem[O'farrell and Harlan(1982)]%
        {O'farrell1982Craftworkers}
\bibfield{author}{\bibinfo{person}{B. O'farrell} {and} \bibinfo{person}{S.
  Harlan}.} \bibinfo{year}{1982}\natexlab{}.
\newblock \showarticletitle{Craftworkers and Clerks: The Effect of Male
  Co-Worker Hostility on Women's Satisfaction with Non-Traditional Jobs}.
\newblock \bibinfo{journal}{\emph{Social Problems}}  \bibinfo{volume}{29}
  (\bibinfo{year}{1982}), \bibinfo{pages}{252--265}.
\newblock
\urldef\tempurl%
\url{https://doi.org/10.2307/800158}
\showDOI{\tempurl}


\bibitem[Osmani(2024)]%
        {Osmani_2024}
\bibfield{author}{\bibinfo{person}{Addy Osmani}.}
  \bibinfo{year}{2024}\natexlab{}.
\newblock \bibinfo{booktitle}{\emph{Leading Effective Eng. Teams: Lessons for
  individual contributors \& Managers from 10 years at google}}.
\newblock \bibinfo{publisher}{O’Reilly Media, Inc}.
\newblock


\bibitem[Park et~al\mbox{.}(2016)]%
        {Park2016}
\bibfield{author}{\bibinfo{person}{Y. Park}, \bibinfo{person}{J.~H. Song},
  {and} \bibinfo{person}{D.~H. Lim}.} \bibinfo{year}{2016}\natexlab{}.
\newblock \showarticletitle{Organizational justice and work engagement: the
  mediating effect of self-leadership}.
\newblock \bibinfo{journal}{\emph{Leadership \& Organization Development J.}}
  \bibinfo{volume}{37} (\bibinfo{year}{2016}), \bibinfo{pages}{711--729}.
\newblock
Issue 6.
\urldef\tempurl%
\url{https://doi.org/10.1108/lodj-09-2014-0192}
\showDOI{\tempurl}


\bibitem[Pater et~al\mbox{.}(2009)]%
        {Pater2009Gender}
\bibfield{author}{\bibinfo{person}{I.~D. Pater}, \bibinfo{person}{A.~V.
  Vianen}, {and} \bibinfo{person}{M. Bechtoldt}.}
  \bibinfo{year}{2009}\natexlab{}.
\newblock \showarticletitle{Gender Differences in Job Challenge: A Matter of
  Task Allocation}.
\newblock \bibinfo{journal}{\emph{Gender, Work and Organization}}
  \bibinfo{volume}{17} (\bibinfo{year}{2009}), \bibinfo{pages}{433--453}.
\newblock
\urldef\tempurl%
\url{https://doi.org/10.1111/J.1468-0432.2009.00477.X}
\showDOI{\tempurl}


\bibitem[Paul et~al\mbox{.}(2019)]%
        {Paul2019}
\bibfield{author}{\bibinfo{person}{Rajshakhar Paul}, \bibinfo{person}{Amiangshu
  Bosu}, {and} \bibinfo{person}{Kazi~Zakia Sultana}.}
  \bibinfo{year}{2019}\natexlab{}.
\newblock \showarticletitle{Expressions of Sentiments during Code Rev.s: Male
  vs. Female}. In \bibinfo{booktitle}{\emph{2019 IEEE 26th Int. Conf. on Softw.
  Analysis, Evolution and Reengineering (SANER)}}. \bibinfo{pages}{26--37}.
\newblock
\urldef\tempurl%
\url{https://doi.org/10.1109/SANER.2019.8667987}
\showDOI{\tempurl}


\bibitem[Pei et~al\mbox{.}(2020)]%
        {pei2020attention}
\bibfield{author}{\bibinfo{person}{Weiping Pei}, \bibinfo{person}{Arthur
  Mayer}, \bibinfo{person}{Kaylynn Tu}, {and} \bibinfo{person}{Chuan Yue}.}
  \bibinfo{year}{2020}\natexlab{}.
\newblock \showarticletitle{Attention please: Your attention check questions in
  survey studies can be automatically answered}. In
  \bibinfo{booktitle}{\emph{Proceedings of The Web Conference 2020}}.
  \bibinfo{pages}{1182--1193}.
\newblock


\bibitem[Rastogi et~al\mbox{.}(2015)]%
        {Rastogi2015Ramp-Up}
\bibfield{author}{\bibinfo{person}{Ayushi Rastogi}, \bibinfo{person}{Suresh
  Thummalapenta}, \bibinfo{person}{Thomas Zimmermann},
  \bibinfo{person}{Nachiappan Nagappan}, {and} \bibinfo{person}{J. Czerwonka}.}
  \bibinfo{year}{2015}\natexlab{}.
\newblock \showarticletitle{Ramp-Up Journey of New Hires: Tug of War of Aids
  and Impediments}.
\newblock \bibinfo{journal}{\emph{2015 ACM/IEEE Int. Symp. on Empirical Softw.
  Eng. and Measurement (ESEM)}} (\bibinfo{year}{2015}), \bibinfo{pages}{1--10}.
\newblock
\urldef\tempurl%
\url{https://doi.org/10.1109/ESEM.2015.7321212}
\showDOI{\tempurl}


\bibitem[Read et~al\mbox{.}(2022)]%
        {read2022racing}
\bibfield{author}{\bibinfo{person}{Blair Read}, \bibinfo{person}{Lukas
  Wolters}, {and} \bibinfo{person}{Adam~J Berinsky}.}
  \bibinfo{year}{2022}\natexlab{}.
\newblock \showarticletitle{Racing the clock: Using response time as a proxy
  for attentiveness on self-administered surveys}.
\newblock \bibinfo{journal}{\emph{Political Analysis}} \bibinfo{volume}{30},
  \bibinfo{number}{4} (\bibinfo{year}{2022}), \bibinfo{pages}{550--569}.
\newblock


\bibitem[{\v{R}}ezankov{\'a} and Nov{\'a}k(2019)]%
        {vrezankova2019effect}
\bibfield{author}{\bibinfo{person}{Hana {\v{R}}ezankov{\'a}} {and}
  \bibinfo{person}{Richard Nov{\'a}k}.} \bibinfo{year}{2019}\natexlab{}.
\newblock \showarticletitle{Effect of ordinal variable transformations on
  hierarchical clustering results: A case study on the Big Data phenomenon}. In
  \bibinfo{booktitle}{\emph{22nd International Scientific Conference on
  Applications of Mathematics and Statistics in Economics (AMSE 2019)}}.
  Atlantis Press, \bibinfo{pages}{81--90}.
\newblock


\bibitem[Rodriguez et~al\mbox{.}(2018)]%
        {Rodriguez2018}
\bibfield{author}{\bibinfo{person}{Ariel Rodriguez}, \bibinfo{person}{Fumiya
  Tanaka}, {and} \bibinfo{person}{Yasutaka Kamei}.}
  \bibinfo{year}{2018}\natexlab{}.
\newblock \showarticletitle{Empirical study on the relationship between
  developer's working habits and efficiency}. In
  \bibinfo{booktitle}{\emph{Proc. of the 15th Int. Conf. on Mining Softw.
  Repositories}} (Gothenburg, Sweden) \emph{(\bibinfo{series}{MSR '18})}.
  \bibinfo{publisher}{Assoc. for Comput. Machinery}, \bibinfo{address}{New
  York, NY, USA}, \bibinfo{pages}{74–77}.
\newblock
\showISBNx{9781450357166}
\urldef\tempurl%
\url{https://doi.org/10.1145/3196398.3196458}
\showDOI{\tempurl}


\bibitem[Rogier and Padgett(2004)]%
        {Rogier2004The}
\bibfield{author}{\bibinfo{person}{Sara~A. Rogier} {and} \bibinfo{person}{M.
  Padgett}.} \bibinfo{year}{2004}\natexlab{}.
\newblock \showarticletitle{The impact of utilizing a flexible work schedule on
  the perceived career advancement potential of women}.
\newblock \bibinfo{journal}{\emph{Human Resource Development Quarterly}}
  \bibinfo{volume}{15} (\bibinfo{year}{2004}), \bibinfo{pages}{89--106}.
\newblock
\urldef\tempurl%
\url{https://doi.org/10.1002/HRDQ.1089}
\showDOI{\tempurl}


\bibitem[Scholarios and Marks(2004)]%
        {Scholarios2004Work-life}
\bibfield{author}{\bibinfo{person}{D. Scholarios} {and}
  \bibinfo{person}{Abigail Marks}.} \bibinfo{year}{2004}\natexlab{}.
\newblock \showarticletitle{Work-life balance and the software worker}.
\newblock \bibinfo{journal}{\emph{Human Resource Manage. J.}}
  \bibinfo{volume}{14} (\bibinfo{year}{2004}), \bibinfo{pages}{54--74}.
\newblock
\urldef\tempurl%
\url{https://doi.org/10.1111/J.1748-8583.2004.TB00119.X}
\showDOI{\tempurl}


\bibitem[Sebastian(2016)]%
        {Sebastian2016Motivational}
\bibfield{author}{\bibinfo{person}{Urieşi Sebastian}.}
  \bibinfo{year}{2016}\natexlab{}.
\newblock \showarticletitle{Motivational Effects of Pay Dispersion in Pay for
  Performance Programs Implemented in Romanian Companies}.
\newblock \bibinfo{journal}{\emph{Manage. and Marketing}}  \bibinfo{volume}{11}
  (\bibinfo{year}{2016}), \bibinfo{pages}{431--448}.
\newblock
\urldef\tempurl%
\url{https://doi.org/10.1515/MMCKS-2016-0007}
\showDOI{\tempurl}


\bibitem[Sesari et~al\mbox{.}(2024)]%
        {sesari2024understanding}
\bibfield{author}{\bibinfo{person}{Emeralda Sesari}, \bibinfo{person}{Federica
  Sarro}, {and} \bibinfo{person}{Ayushi Rastogi}.}
  \bibinfo{year}{2024}\natexlab{}.
\newblock \showarticletitle{Understanding Fairness in Software Engineering:
  Insights from Stack Exchange Sites}. In \bibinfo{booktitle}{\emph{Proceedings
  of the 18th ACM/IEEE International Symposium on Empirical Software
  Engineering and Measurement}} (Barcelona, Spain) \emph{(\bibinfo{series}{ESEM
  '24})}. \bibinfo{publisher}{Association for Computing Machinery},
  \bibinfo{address}{New York, NY, USA}, \bibinfo{pages}{269–280}.
\newblock
\showISBNx{9798400710476}
\urldef\tempurl%
\url{https://doi.org/10.1145/3674805.3686687}
\showDOI{\tempurl}


\bibitem[Shabir and Gani(2020)]%
        {Shabir2020Impact}
\bibfield{author}{\bibinfo{person}{Sana Shabir} {and} \bibinfo{person}{A.
  Gani}.} \bibinfo{year}{2020}\natexlab{}.
\newblock \showarticletitle{Impact of work–life balance on organizational
  commitment of women health-care workers}.
\newblock \bibinfo{journal}{\emph{Int. J. of Organizational Analysis}}
  \bibinfo{volume}{28} (\bibinfo{year}{2020}), \bibinfo{pages}{917--939}.
\newblock
\urldef\tempurl%
\url{https://doi.org/10.1108/ijoa-07-2019-1820}
\showDOI{\tempurl}


\bibitem[Shameer et~al\mbox{.}(2023)]%
        {Shameer2023}
\bibfield{author}{\bibinfo{person}{S. Shameer}, \bibinfo{person}{G.
  Rodríguez-Pérez}, {and} \bibinfo{person}{M. Nagappan}.}
  \bibinfo{year}{2023}\natexlab{}.
\newblock \showarticletitle{Relationship between diversity of collaborative
  group members’ race and ethnicity and the frequency of their collaborative
  contributions in github}.
\newblock \bibinfo{journal}{\emph{Empirical Softw. Eng.}}  \bibinfo{volume}{28}
  (\bibinfo{year}{2023}).
\newblock
Issue 4.
\urldef\tempurl%
\url{https://doi.org/10.1007/s10664-023-10313-y}
\showDOI{\tempurl}


\bibitem[Sharma and Stol(2020)]%
        {Sharma2020Exploring}
\bibfield{author}{\bibinfo{person}{Gaurav~G. Sharma} {and}
  \bibinfo{person}{Klaas-Jan Stol}.} \bibinfo{year}{2020}\natexlab{}.
\newblock \showarticletitle{Exploring onboarding success, organizational fit,
  and turnover intention of software professionals}.
\newblock \bibinfo{journal}{\emph{J. Syst. Softw.}}  \bibinfo{volume}{159}
  (\bibinfo{year}{2020}).
\newblock
\urldef\tempurl%
\url{https://doi.org/10.1016/j.jss.2019.110442}
\showDOI{\tempurl}


\bibitem[Sharp et~al\mbox{.}(2009)]%
        {sharp2009models}
\bibfield{author}{\bibinfo{person}{Helen Sharp}, \bibinfo{person}{Nathan
  Baddoo}, \bibinfo{person}{Sarah Beecham}, \bibinfo{person}{Tracy Hall}, {and}
  \bibinfo{person}{Hugh Robinson}.} \bibinfo{year}{2009}\natexlab{}.
\newblock \showarticletitle{Models of motivation in software engineering}.
\newblock \bibinfo{journal}{\emph{Inf. and software technology}}
  \bibinfo{volume}{51}, \bibinfo{number}{1} (\bibinfo{year}{2009}),
  \bibinfo{pages}{219--233}.
\newblock


\bibitem[Shonubi et~al\mbox{.}(2016)]%
        {Shonubi2016Recognition}
\bibfield{author}{\bibinfo{person}{Olurotimi~Adebayo Shonubi},
  \bibinfo{person}{Norida Abdullah}, \bibinfo{person}{Rahman Hashim}, {and}
  \bibinfo{person}{N.~A. Hamid}.} \bibinfo{year}{2016}\natexlab{}.
\newblock \showarticletitle{Recognition and Appreciation and the Moderating
  Role of Self-esteem on Job Satisfaction and Performance among IT Employees in
  Melaka}.
\newblock   \bibinfo{volume}{4} (\bibinfo{year}{2016}).
\newblock
\urldef\tempurl%
\url{https://doi.org/10.17265/2328-7136/2016.05.001}
\showDOI{\tempurl}


\bibitem[Siu et~al\mbox{.}(2013)]%
        {Siu2013Work}
\bibfield{author}{\bibinfo{person}{A.~M. Siu}, \bibinfo{person}{A. Hung},
  \bibinfo{person}{Ada Y.~L. Lam}, {and} \bibinfo{person}{A. Cheng}.}
  \bibinfo{year}{2013}\natexlab{}.
\newblock \showarticletitle{Work limitations, workplace concerns, and job
  satisfaction of persons with chronic disease.}
\newblock \bibinfo{journal}{\emph{Work}}  \bibinfo{volume}{45 1}
  (\bibinfo{year}{2013}), \bibinfo{pages}{107--15}.
\newblock
\urldef\tempurl%
\url{https://doi.org/10.3233/WOR-121550}
\showDOI{\tempurl}


\bibitem[Smith et~al\mbox{.}(2013)]%
        {Smith2013The}
\bibfield{author}{\bibinfo{person}{Laura G.~E. Smith},
  \bibinfo{person}{Catherine~E. Amiot}, \bibinfo{person}{Joanne~R Smith},
  \bibinfo{person}{V. Callan}, {and} \bibinfo{person}{D. Terry}.}
  \bibinfo{year}{2013}\natexlab{}.
\newblock \showarticletitle{The Social Validation and Coping Model of
  Organizational Identity Development}.
\newblock \bibinfo{journal}{\emph{J. of Management}}  \bibinfo{volume}{39}
  (\bibinfo{year}{2013}), \bibinfo{pages}{1952 -- 1978}.
\newblock
\urldef\tempurl%
\url{https://doi.org/10.1177/0149206313488212}
\showDOI{\tempurl}


\bibitem[Storey et~al\mbox{.}(2021)]%
        {Storey2021}
\bibfield{author}{\bibinfo{person}{Margaret-Anne Storey},
  \bibinfo{person}{Thomas Zimmermann}, \bibinfo{person}{Christian Bird},
  \bibinfo{person}{Jacek Czerwonka}, \bibinfo{person}{Brendan Murphy}, {and}
  \bibinfo{person}{Eirini Kalliamvakou}.} \bibinfo{year}{2021}\natexlab{}.
\newblock \showarticletitle{Towards a Theory of Software Developer Job
  Satisfaction and Perceived Productivity}.
\newblock \bibinfo{journal}{\emph{IEEE Trans. on Softw. Eng.}}
  \bibinfo{volume}{47}, \bibinfo{number}{10} (\bibinfo{year}{2021}),
  \bibinfo{pages}{2125--2142}.
\newblock
\urldef\tempurl%
\url{https://doi.org/10.1109/TSE.2019.2944354}
\showDOI{\tempurl}


\bibitem[Suh and Hijal-Moghrabi(2021)]%
        {Suh2021The}
\bibfield{author}{\bibinfo{person}{Jiwon Suh} {and} \bibinfo{person}{Imane
  Hijal-Moghrabi}.} \bibinfo{year}{2021}\natexlab{}.
\newblock \showarticletitle{The Effects of Fairness on Female Managers’
  Perception of Career Prospects and Job Satisfaction: A Study across Sectors}.
\newblock \bibinfo{journal}{\emph{Int. J. of Public Administration}}
  \bibinfo{volume}{45} (\bibinfo{year}{2021}), \bibinfo{pages}{644 -- 657}.
\newblock
\urldef\tempurl%
\url{https://doi.org/10.1080/01900692.2021.1876728}
\showDOI{\tempurl}


\bibitem[Tarasov(2019)]%
        {Tarasov2019AssessingJS}
\bibfield{author}{\bibinfo{person}{Aleksandr~V. Tarasov}.}
  \bibinfo{year}{2019}\natexlab{}.
\newblock \showarticletitle{Assessing Job Satisfaction of Software Engineers
  Using GQM Approach}. In \bibinfo{booktitle}{\emph{Int. Conf. on Softw.
  Technol.: Methods and Tools}}.
\newblock


\bibitem[Teimouri et~al\mbox{.}(2016)]%
        {teimouri2016study}
\bibfield{author}{\bibinfo{person}{Hadi Teimouri},
  \bibinfo{person}{Maryam~Goodarzvand Chegini}, \bibinfo{person}{Kouroush
  Jenab}, \bibinfo{person}{Sam Khoury}, {and} \bibinfo{person}{Kim LaFevor}.}
  \bibinfo{year}{2016}\natexlab{}.
\newblock \showarticletitle{Study of the relationship between employee
  engagement and organisational effectiveness}.
\newblock \bibinfo{journal}{\emph{Int. J. of Business Excellence}}
  \bibinfo{volume}{10}, \bibinfo{number}{1} (\bibinfo{year}{2016}),
  \bibinfo{pages}{37--54}.
\newblock


\bibitem[Tessem and Maurer(2007)]%
        {Tessem2007}
\bibfield{author}{\bibinfo{person}{B. Tessem} {and} \bibinfo{person}{F.
  Maurer}.} \bibinfo{year}{2007}\natexlab{}.
\newblock \showarticletitle{Job satisfaction and motivation in a large agile
  team}.
\newblock \bibinfo{journal}{\emph{Lecture Notes in Computer Science}}
  (\bibinfo{year}{2007}), \bibinfo{pages}{54--61}.
\newblock
\urldef\tempurl%
\url{https://doi.org/10.1007/978-3-540-73101-6_8}
\showDOI{\tempurl}


\bibitem[Thibaut and Walker(1975)]%
        {thibaut_walker_1975}
\bibfield{author}{\bibinfo{person}{John Thibaut} {and} \bibinfo{person}{Laurens
  Walker}.} \bibinfo{year}{1975}\natexlab{}.
\newblock \bibinfo{booktitle}{\emph{Procedural justice: A psychological
  analysis}}.
\newblock \bibinfo{publisher}{Erlbaum}.
\newblock


\bibitem[Tomczak and Tomczak(2014)]%
        {tomczak2014need}
\bibfield{author}{\bibinfo{person}{Maciej Tomczak} {and} \bibinfo{person}{Ewa
  Tomczak}.} \bibinfo{year}{2014}\natexlab{}.
\newblock \showarticletitle{The need to report effect size estimates revisited.
  An overview of some recommended measures of effect size}.
\newblock  (\bibinfo{year}{2014}).
\newblock


\bibitem[Trinkenreich et~al\mbox{.}(2022a)]%
        {Trinkenreich2022AnEI}
\bibfield{author}{\bibinfo{person}{Bianca Trinkenreich},
  \bibinfo{person}{Ricardo Britto}, \bibinfo{person}{Marco~Aur{\'e}lio Gerosa},
  {and} \bibinfo{person}{Igor Steinmacher}.} \bibinfo{year}{2022}\natexlab{a}.
\newblock \showarticletitle{An Empirical Investigation on the Challenges Faced
  by Women in the Software Industry: A Case Study}.
\newblock \bibinfo{journal}{\emph{ArXiv}}  \bibinfo{volume}{abs/2203.10555}
  (\bibinfo{year}{2022}).
\newblock


\bibitem[Trinkenreich et~al\mbox{.}(2022b)]%
        {Trinkenreich2022}
\bibfield{author}{\bibinfo{person}{Bianca Trinkenreich}, \bibinfo{person}{Igor
  Wiese}, \bibinfo{person}{Anita Sarma}, \bibinfo{person}{Marco Gerosa}, {and}
  \bibinfo{person}{Igor Steinmacher}.} \bibinfo{year}{2022}\natexlab{b}.
\newblock \showarticletitle{Women’s Participation in Open Source Software: A
  Survey of the Literature}.
\newblock \bibinfo{journal}{\emph{ACM Trans. Softw. Eng. Methodol.}}
  (\bibinfo{date}{jan} \bibinfo{year}{2022}).
\newblock
\showISSN{1049-331X}


\bibitem[Tsunoda et~al\mbox{.}(2017)]%
        {Tsunoda2017Evaluating}
\bibfield{author}{\bibinfo{person}{Taketo Tsunoda}, \bibinfo{person}{H.
  Washizaki}, \bibinfo{person}{Y. Fukazawa}, \bibinfo{person}{S. Inoue},
  \bibinfo{person}{Y. Hanai}, {and} \bibinfo{person}{Masanobu Kanazawa}.}
  \bibinfo{year}{2017}\natexlab{}.
\newblock \showarticletitle{Evaluating the work of experienced and
  inexperienced developers considering work difficulty in sotware development}.
\newblock \bibinfo{journal}{\emph{2017 18th IEEE/ACIS Int. Conf. on Softw.
  Eng., Artificial Intelligence, Networking and Parallel/Distributed Comput.
  (SNPD)}} (\bibinfo{year}{2017}), \bibinfo{pages}{161--166}.
\newblock
\urldef\tempurl%
\url{https://doi.org/10.1109/SNPD.2017.8022717}
\showDOI{\tempurl}


\bibitem[Turhan et~al\mbox{.}(2016)]%
        {Turhan2016The}
\bibfield{author}{\bibinfo{person}{Mithat Turhan}, \bibinfo{person}{O.
  Köprülü}, {and} \bibinfo{person}{I. Helvacı}.}
  \bibinfo{year}{2016}\natexlab{}.
\newblock \showarticletitle{The Relationship between Academic Staff's
  Perception of Organizational Justice and Demographic Factors: A Case Study in
  Foundation Universities in Turkey}.
\newblock \bibinfo{journal}{\emph{Int. Rev. of Manage. and Business Research}}
  \bibinfo{volume}{5} (\bibinfo{year}{2016}), \bibinfo{pages}{1406}.
\newblock


\bibitem[Vandello et~al\mbox{.}(2013)]%
        {Vandello2013When}
\bibfield{author}{\bibinfo{person}{J. Vandello}, \bibinfo{person}{V.
  Hettinger}, \bibinfo{person}{J. Bosson}, {and} \bibinfo{person}{J. Siddiqi}.}
  \bibinfo{year}{2013}\natexlab{}.
\newblock \showarticletitle{When Equal Isn't Really Equal: The Masculine
  Dilemma of Seeking Work Flexibility}.
\newblock \bibinfo{journal}{\emph{J. of Social Issues}}  \bibinfo{volume}{69}
  (\bibinfo{year}{2013}), \bibinfo{pages}{303--321}.
\newblock
\urldef\tempurl%
\url{https://doi.org/10.1111/JOSI.12016}
\showDOI{\tempurl}


\bibitem[Varekamp et~al\mbox{.}(2011)]%
        {Varekamp2011Effect}
\bibfield{author}{\bibinfo{person}{I. Varekamp}, \bibinfo{person}{J. Verbeek},
  \bibinfo{person}{A.~D. de Boer}, {and} \bibinfo{person}{F. van Dijk}.}
  \bibinfo{year}{2011}\natexlab{}.
\newblock \showarticletitle{Effect of job maintenance training program for
  employees with chronic disease - a randomized controlled trial on
  self-efficacy, job satisfaction, and fatigue.}
\newblock \bibinfo{journal}{\emph{Scandinavian journal of work, environment \&
  health}}  \bibinfo{volume}{37 4} (\bibinfo{year}{2011}),
  \bibinfo{pages}{288--97}.
\newblock
\urldef\tempurl%
\url{https://doi.org/10.5271/sjweh.3149}
\showDOI{\tempurl}


\bibitem[Verwijs and Russo(2024)]%
        {verwijs2024agile}
\bibfield{author}{\bibinfo{person}{Christiaan Verwijs} {and}
  \bibinfo{person}{Daniel Russo}.} \bibinfo{year}{2024}\natexlab{}.
\newblock \showarticletitle{Do Agile scaling approaches make a difference? an
  empirical comparison of team effectiveness across popular scaling
  approaches}.
\newblock \bibinfo{journal}{\emph{Empirical Software Engineering}}
  \bibinfo{volume}{29}, \bibinfo{number}{4} (\bibinfo{year}{2024}),
  \bibinfo{pages}{75}.
\newblock


\bibitem[Villablanca et~al\mbox{.}(2017)]%
        {Villablanca2017Evaluating}
\bibfield{author}{\bibinfo{person}{A. Villablanca}, \bibinfo{person}{Yueju Li},
  \bibinfo{person}{L. Beckett}, {and} \bibinfo{person}{L. Howell}.}
  \bibinfo{year}{2017}\natexlab{}.
\newblock \showarticletitle{Evaluating a Medical School's Climate for Women's
  Success: Outcomes for Faculty Recruitment, Retention, and Promotion.}
\newblock \bibinfo{journal}{\emph{J. of women's health}}  \bibinfo{volume}{26
  5} (\bibinfo{year}{2017}), \bibinfo{pages}{530--539}.
\newblock
\urldef\tempurl%
\url{https://doi.org/10.1089/jwh.2016.6018}
\showDOI{\tempurl}


\bibitem[Wang and Redmiles(2019)]%
        {Wang2019}
\bibfield{author}{\bibinfo{person}{Yi Wang} {and} \bibinfo{person}{David
  Redmiles}.} \bibinfo{year}{2019}\natexlab{}.
\newblock \showarticletitle{Implicit Gender Biases in Professional Software
  Development: An Empirical Study}. In \bibinfo{booktitle}{\emph{2019 IEEE/ACM
  41st Int. Conf. on Softw. Eng.: Softw. Eng. in Soc. (ICSE-SEIS)}}.
  \bibinfo{pages}{1--10}.
\newblock
\urldef\tempurl%
\url{https://doi.org/10.1109/ICSE-SEIS.2019.00009}
\showDOI{\tempurl}


\bibitem[Weiß(2019)]%
        {Weiß2019}
\bibfield{author}{\bibinfo{person}{C. Weiß}.} \bibinfo{year}{2019}\natexlab{}.
\newblock \showarticletitle{Distance-based analysis of ordinal data and ordinal
  time series}.
\newblock \bibinfo{journal}{\emph{J. of the American Statistical Assoc.}}
  \bibinfo{volume}{115} (\bibinfo{year}{2019}), \bibinfo{pages}{1189--1200}.
\newblock
Issue 531.
\urldef\tempurl%
\url{https://doi.org/10.1080/01621459.2019.1604370}
\showDOI{\tempurl}


\bibitem[Wesolowski and Mossholder(1997)]%
        {Wesolowski1997}
\bibfield{author}{\bibinfo{person}{Mark~A Wesolowski} {and}
  \bibinfo{person}{Kevin~W Mossholder}.} \bibinfo{year}{1997}\natexlab{}.
\newblock \showarticletitle{Relational demography in supervisor--subordinate
  dyads: Impact on subordinate job satisfaction, burnout, and perceived
  procedural justice}.
\newblock \bibinfo{journal}{\emph{J. of Organizational Behavior: The Int. J. of
  Industrial, Occupational and Organizational Psychology and Behavior}}
  \bibinfo{volume}{18}, \bibinfo{number}{4} (\bibinfo{year}{1997}),
  \bibinfo{pages}{351--362}.
\newblock


\bibitem[Westlund(2011)]%
        {Westlund2011Leading}
\bibfield{author}{\bibinfo{person}{S. Westlund}.}
  \bibinfo{year}{2011}\natexlab{}.
\newblock \showarticletitle{Leading Techies: Assessing Project Leadership
  Styles Most Significantly Related to Software Developer Job Satisfaction}.
\newblock \bibinfo{journal}{\emph{Int. J. Hum. Cap. Inf. Technol. Prof.}}
  \bibinfo{volume}{2} (\bibinfo{year}{2011}), \bibinfo{pages}{1--15}.
\newblock
\urldef\tempurl%
\url{https://doi.org/10.4018/jhcitp.2011040101}
\showDOI{\tempurl}


\bibitem[White et~al\mbox{.}(2011)]%
        {White2011Multiple}
\bibfield{author}{\bibinfo{person}{I. White}, \bibinfo{person}{P. Royston},
  {and} \bibinfo{person}{A. Wood}.} \bibinfo{year}{2011}\natexlab{}.
\newblock \showarticletitle{Multiple imputation using chained equations: Issues
  and guidance for practice}.
\newblock \bibinfo{journal}{\emph{Statistics in Medicine}}
  \bibinfo{volume}{30} (\bibinfo{year}{2011}).
\newblock
\urldef\tempurl%
\url{https://doi.org/10.1002/sim.4067}
\showDOI{\tempurl}


\bibitem[Winter et~al\mbox{.}(2016)]%
        {Winter2016}
\bibfield{author}{\bibinfo{person}{J.~d. Winter}, \bibinfo{person}{S.~D.
  Gosling}, {and} \bibinfo{person}{J. Potter}.}
  \bibinfo{year}{2016}\natexlab{}.
\newblock \showarticletitle{Comparing the pearson and spearman correlation
  coefficients across distributions and sample sizes: a tutorial using
  simulations and empirical data.}
\newblock \bibinfo{journal}{\emph{Psychological Methods}}  \bibinfo{volume}{21}
  (\bibinfo{year}{2016}), \bibinfo{pages}{273--290}.
\newblock
Issue 3.
\urldef\tempurl%
\url{https://doi.org/10.1037/met0000079}
\showDOI{\tempurl}


\bibitem[Wood et~al\mbox{.}(2017)]%
        {Wood2017}
\bibfield{author}{\bibinfo{person}{D. Wood}, \bibinfo{person}{P.~D. Harms},
  \bibinfo{person}{G.~H. Lowman}, {and} \bibinfo{person}{J.~A. DeSimone}.}
  \bibinfo{year}{2017}\natexlab{}.
\newblock \showarticletitle{Response speed and response consistency as mutually
  validating indicators of data quality in online samples}.
\newblock \bibinfo{journal}{\emph{Social Psychological and Personality
  Science}}  \bibinfo{volume}{8} (\bibinfo{year}{2017}),
  \bibinfo{pages}{454--464}.
\newblock
Issue 4.
\urldef\tempurl%
\url{https://doi.org/10.1177/1948550617703168}
\showDOI{\tempurl}


\bibitem[Wright and Cropanzano(2000)]%
        {wright2000psychological}
\bibfield{author}{\bibinfo{person}{Thomas~A Wright} {and}
  \bibinfo{person}{Russell Cropanzano}.} \bibinfo{year}{2000}\natexlab{}.
\newblock \showarticletitle{Psychological well-being and job satisfaction as
  predictors of job performance.}
\newblock \bibinfo{journal}{\emph{J. of occupational health psychology}}
  \bibinfo{volume}{5}, \bibinfo{number}{1} (\bibinfo{year}{2000}),
  \bibinfo{pages}{84}.
\newblock


\bibitem[Wu and Wang(2008)]%
        {Wu2008}
\bibfield{author}{\bibinfo{person}{X. Wu} {and} \bibinfo{person}{C. Wang}.}
  \bibinfo{year}{2008}\natexlab{}.
\newblock \showarticletitle{The impact of organizational justice on employees'
  pay satisfaction, work attitudes and performance in chinese hotels}.
\newblock \bibinfo{journal}{\emph{J. of Human Resources in Hospitality \&
  Tourism}}  \bibinfo{volume}{7} (\bibinfo{year}{2008}),
  \bibinfo{pages}{181--195}.
\newblock
Issue 2.
\urldef\tempurl%
\url{https://doi.org/10.1080/15332840802156923}
\showDOI{\tempurl}


\bibitem[Wulff and Jeppesen(2017)]%
        {Wulff2017Multiple}
\bibfield{author}{\bibinfo{person}{Jesper~N. Wulff} {and} \bibinfo{person}{L.
  Jeppesen}.} \bibinfo{year}{2017}\natexlab{}.
\newblock \showarticletitle{Multiple imputation by chained equations in praxis:
  Guidelines and review}.
\newblock \bibinfo{journal}{\emph{The Electronic J. of Business Research
  Methods}}  \bibinfo{volume}{15} (\bibinfo{year}{2017}),
  \bibinfo{pages}{41--56}.
\newblock


\bibitem[Yilmaz et~al\mbox{.}(2012)]%
        {Yilmaz2012ASA}
\bibfield{author}{\bibinfo{person}{Murat Yilmaz}, \bibinfo{person}{Rory~V.
  O'Connor}, {and} \bibinfo{person}{Paul~M. Clarke}.}
  \bibinfo{year}{2012}\natexlab{}.
\newblock \showarticletitle{A Systematic Approach to the Comparison of Roles in
  the Software Development Processes}. In \bibinfo{booktitle}{\emph{Int. Conf.
  on Softw. Process Improvement and Capability Determination}}.
\newblock


\bibitem[Young et~al\mbox{.}(2021)]%
        {Young2021Which}
\bibfield{author}{\bibinfo{person}{Jean-Gabriel Young}, \bibinfo{person}{Amanda
  Casari}, \bibinfo{person}{Katie McLaughlin}, \bibinfo{person}{Milo~Z.
  Trujillo}, \bibinfo{person}{Laurent H'ebert-Dufresne}, {and}
  \bibinfo{person}{James~P. Bagrow}.} \bibinfo{year}{2021}\natexlab{}.
\newblock \showarticletitle{Which contributions count? Analysis of attribution
  in open source}.
\newblock \bibinfo{journal}{\emph{2021 IEEE/ACM 18th Int. Conf. on Mining
  Softw. Repositories (MSR)}} (\bibinfo{year}{2021}),
  \bibinfo{pages}{242--253}.
\newblock
\urldef\tempurl%
\url{https://doi.org/10.1109/MSR52588.2021.00036}
\showDOI{\tempurl}


\bibitem[Zeidler et~al\mbox{.}(2021)]%
        {Zeidler2021The}
\bibfield{author}{\bibinfo{person}{P. Zeidler}, \bibinfo{person}{R. Sims},
  {and} \bibinfo{person}{Y. Yurova}.} \bibinfo{year}{2021}\natexlab{}.
\newblock \showarticletitle{The importance of organisational justice on
  schedule satisfaction: a study of Latin American call centre employees}.
\newblock \bibinfo{journal}{\emph{Int. J. of Human Resources Development and
  Manage.}}  \bibinfo{volume}{21} (\bibinfo{year}{2021}), \bibinfo{pages}{54}.
\newblock
\urldef\tempurl%
\url{https://doi.org/10.1504/IJHRDM.2021.10037654}
\showDOI{\tempurl}


\bibitem[Zhang and Zhu(2019)]%
        {zhang2019influence}
\bibfield{author}{\bibinfo{person}{Kun Zhang} {and} \bibinfo{person}{Xin Zhu}.}
  \bibinfo{year}{2019}\natexlab{}.
\newblock \showarticletitle{The influence and mechanism of supervisor
  developmental feedback on employee’s work engagement}. In
  \bibinfo{booktitle}{\emph{2018 Int. Symp. on Social Science and Manage.
  Innovation (SSMI 2018)}}. Atlantis Press, \bibinfo{pages}{256--260}.
\newblock


\bibitem[Zhang and Wang(2017)]%
        {Zhang2017}
\bibfield{author}{\bibinfo{person}{Q. Zhang} {and} \bibinfo{person}{L. Wang}.}
  \bibinfo{year}{2017}\natexlab{}.
\newblock \showarticletitle{Moderation analysis with missing data in the
  predictors.}
\newblock \bibinfo{journal}{\emph{Psychological Methods}}  \bibinfo{volume}{22}
  (\bibinfo{year}{2017}), \bibinfo{pages}{649--666}.
\newblock
Issue 4.
\urldef\tempurl%
\url{https://doi.org/10.1037/met0000104}
\showDOI{\tempurl}


\end{thebibliography}

\end{document}